\documentclass[10pt,aps,pra,showpacs,showkeys,twocolumn]{revtex4-2}
\usepackage{amsmath, amssymb}
\usepackage{amsthm}
\usepackage{mathbbol}
\usepackage{bm}
\usepackage{graphicx}
\usepackage{xcolor}
\usepackage{units}
\usepackage{tikz}
\usepackage{multirow}
\usepackage{subfig}
\usepackage{physics}
\usetikzlibrary{automata,positioning,shapes,snakes,decorations}
\usepackage{ulem}

\DeclareMathOperator*{\argmin}{argmin}

\renewcommand{\vec}[1]{\mathbf{#1}}

\begin{document}

\title{Exact rate analysis for quantum repeaters with imperfect memories and \\entanglement swapping as soon as possible}
\author{Lars Kamin}
\email{lkamin@uwaterloo.ca}

\author{Evgeny Shchukin}
\email{evgeny.shchukin@gmail.com}

\author{Frank Schmidt}
\email{fschmi@students.uni-mainz.de}

\author{Peter van Loock}
\email{loock@uni-mainz.de}
\affiliation{Johannes-Gutenberg University of Mainz, Institute of
Physics, Staudingerweg 7, 55128 Mainz, Germany}

\begin{abstract}
We present an exact rate analysis for a secret key that can be shared among two parties employing a linear quantum repeater chain. One of our main motivations is to address the question whether simply placing quantum memories along a quantum communication channel can be beneficial in a realistic setting. The underlying model assumes deterministic entanglement swapping of single-spin quantum memories and it excludes probabilistic entanglement distillation, and thus two-way classical communication, on higher nesting levels. Within this framework, we identify the essential properties of any optimal repeater scheme: entanglement distribution in parallel, entanglement swapping as soon and parallel quantum storage as little as possible. While these features are obvious or trivial for the simplest repeater with one middle station, for more stations they cannot always be combined. We propose an
optimal scheme including channel loss and memory dephasing, proving its optimality for the case of two stations and conjecturing it for the general case. In an even more realistic setting, we consider
additional tools and parameters such as memory cut-offs, multiplexing, initial state and swapping
gate fidelities, and finite link coupling efficiencies in order to identify potential
regimes in memory-assisted quantum key distribution beyond one middle station that exceed the rates of the smallest quantum repeaters as well as those obtainable in all-optical schemes unassisted by stationary memory qubits and two-way classical communication. Our analytical treatment enables us to determine simultaneous trade-offs between various parameters,
their scaling, and their
influence on the performance ordering among different types of protocols, comparing two-photon interference after dual-rail qubit 
transmission with one-photon interference of single-rail qubits or, similarly, optical interference of coherent states.
We find that for experimental parameter values that are highly demanding but not impossible (up to 10s coherence time, about 80\% link coupling, and state or gate infidelities in the regime of 1-2\%), one secret bit can be shared per second over a total distance of 800km with repeater stations placed at every 100km -- a significant improvement over ideal point-to-point or realistic twin-field quantum key distribution at GHz clock rates.

\end{abstract}

\pacs{03.67.Mn, 03.65.Ud, 42.50.Dv}

\keywords{quantum repeaters, quantum memory}

\maketitle

\section{Introduction}\label{sec:Introduction}

Recent progress on quantum computers with tens of qubits led to
experimental demonstrations of quantum devices that are able to solve
specifically adapted problems which are not soluble in an efficient manner with 
the help of classical computers alone. These devices are primarily based upon
solid-state (superconducting) systems \cite{Arute2019,Qiskit}, however, there are also photonics approaches \cite{Pan2020}. 
While these schemes still have to be enhanced in terms of size, i.e. the number
of qubits (scalability), their error robustness and corresponding logical encoding
(fault tolerance), as well as their range of applicability (eventually reaching universality),
this progress represents a threat to common classical communication systems.
Eventually, this may compromise our current key distribution protocols. 

Although there are recent developments in classical cryptography to address
the threat imposed by such quantum devices (``post-quantum cryptography''),  
quantum mechanics also gives a possible solution to
this by means of quantum key distribution (QKD) \cite{NLRMP,PirRMP}. 
Many QKD protocols have been proposed such as the most prominent, so-called BB84 scheme \cite{BB84}.
Indeed among the various quantum technologies that promise to enable their users to fulfil tasks impossible without quantum resources, quantum communication is special. Unlike quantum computers there are already commercially available quantum communication systems intended for costumers who wish to communicate in the classical, real world in a basically unconditionally secure fashion -- independent of mathematically unproven assumptions exploiting the concept of QKD. QKD systems are naturally realized for photonic systems using non-classical optical quantum states such as single-photon, weak \cite{Hwang2002, Lo2004} or even bright coherent states \cite{PirRMP}.

\subsection{Previous works and state of the art}
 
Current point-to-point QKD systems, directly connecting the sender (Alice) and the receiver (Bob) via an optical-fiber channel, are limited in distance due to the exponentially growing transmission loss along the channel. Typical maximal distances are 100-200km. 
A very recent QKD variant, so-called twin-field (TF) QKD\cite{Lucamarini},
allows to push these limits farther (basically doubling the effective distance)
by placing an (untrusted) middle station between Alice and Bob. Remarkably, TF QKD achieves this loss scaling advantage in an all-optical fashion with no need for quantum storage at the middle station and at an, in principle, unlimited clock rate with no need for two-way classical communication. It further inherits the improved security features of measurement-device-independent (MDI) QKD schemes \cite{LoCurty,PirBraun}. However, the original TF QKD concept is not known to be further scalable beyond the effective distance doubling. 

In classical communication, the distance problem is straightforwardly overcome by introducing repeater stations along the fiber channel (about every 50-100km) in order to reamplify (and typically reshape) the optical pulses.
On a fundamental level, the famous No-Cloning-theorem \cite{NoCloning,Dieks1982}, prohibits such solutions for quantum communication. 

As a possible remedy, the concept of quantum repeaters has been developed \cite{BriegelDur,Dur1999,Hartmann2007}.
With the help of sufficiently short-range entanglement distributions,
quantum memories, entanglement distillation and swapping, in principle, 
scalable long-distance, fiber-based quantum communication becomes possible,
including long-range QKD. The original quantum repeater proposals 
assumed small-scale non-universal quantum computers at each repeater node 
in order to perform the necessary gates for the entangled-pair manipulations,
and hence clearly appeared to be technologically less demanding than
a fully-fledged fault-tolerant and universal quantum computer.
Related to this, for QKD applications including those over large distances, 
there are very powerful, classical post-processing techniques which
allow to relax the minimal requirements on the experimental states and gates. 
Nonetheless, as a whole, these original quantum repeater systems would
still have high experimental requirements. 

This led to some quantum repeater proposals specifically adapted
to certain matter memory systems and light-matter interfaces.
Probably the most prominent such proposal is the ``DLCZ'' quantum repeater \cite{DLCZ, Sangouard},
based upon atomic-ensemble nodes that no longer rely upon 
the execution of difficult two-qubit entangling gates,
but instead only require linear-optical state manipulations
and photon detectors. Other schemes rely upon single emitters in solid-state repeater nodes, especially colour centers in diamond \cite{ChildressNV, Humphreys2017}. Alternative proposals employ optical coherent states and their cavity-QED interactions with single-spin-based quantum memory nodes \cite{HybridPRL}. 

These proposals made a possible realization of a large-scale quantum repeater more likely, but as a complete implementation, they would still be fundamentally limited in their achievable (secret) key rates per second.
The reason for this is the need for two-way classical communication
on all, including the highest ``nesting'' levels in order to conduct entanglement distillation and confirm successful entanglement swappings when these are probabilistic.

Today this type of quantum repeater schemes are referred to as 1st-generation quantum repeaters. A memory-assisted QKD scheme was proposed in Ref. \cite{tf_repeater}, extending the TF concept to memory-based quantum repeaters.
In principle, this scheme achieves an effective distance doubling compared with standard quantum repeaters or, equivalently, it exhibits the standard loss scaling with about half as many
memory stations as in a standard quantum repeater (while the other half are all-optical
stations with beam splitter and photon detectors).
Apart from a certain level of memory assistance,
this repeater scheme also relies upon two-way classical communication
(between the nearest stations)
and hence can operate only at a limited clock rate determined
by the classical signalling time per segment.
Moreover, for its large-scale operation the scheme would 
require an additional element for quantum error correction. 

Alternative schemes circumventing the fundamental limitations
are the so-called 2nd- and 3rd-generation quantum repeaters
that exploit quantum error correction codes to suppress
the effect of gate and memory errors or channel loss, respectively \cite{JiangRvw}. A 3rd-generation quantum repeater no longer requires
quantum memories and two-way classical communication and so it can be,
in principle, realized in an all-optical fashion at a clock rate
only limited by the local error correction operations.

It is important to stress that all these quantum repeaters 
are designed to allow for a genuine long-distance quantum state transfer.
In the QKD context, this means that the intermediate stations along
the repeater channel may be untrusted. If instead sufficiently many trusted stations can be 
placed along the communication channel between Alice and Bob,
and the quantum signals can be converted into classical information
at each station (as a whole, effectively corresponding to classically connected, independent,
sufficiently short-range QKD links), large-scale QKD is already possible and being demonstrated \cite{ChinaDaily}.

Conceptually, this also applies to long-range links enabled by satellites \cite{Yin2017, Vallone2015}.
It is only the genuine quantum repeater that incorporates
two main features at the same time:
{\it long-distance scalability and long-distance privacy}.

From a practical point of view, it is expected that global quantum communication
systems will be a combination of both elements: genuine fiber-based quantum repeaters
over intermediate distances (thousands of km) and satellite-based quantum links
bridging even longer distances (tens of thousands of km; the earth's
circumference is about 40000km).
While such truly global quantum communication may eventually lead
to some form of a ``quantum internet'' \cite{WehnerHanson}, 
only the coherent long-distance quantum state transfer 
as enabled by a genuine quantum repeater allows to 
consider applications that go beyond long-range QKD.
In fact, the original quantum repeater proposals 
were not specifically intended for or adapted to long-range QKD.
They can be used for any application that relies upon
the distribution of entangled states over large distances
including large-scale quantum networks.
Obvious applications are distributed quantum tasks
such as distributed quantum computing, coherently
connecting quantum computers which are spatially 
far apart. 
These ultimate long-distance quantum communication applications
will then impose much higher demands on the fault tolerance 
of the experimental quantum states and gates.
In particular, QKD-specifc classical post-processing 
will no longer be applicable.
In this work, we shall consider small to intermediate-scale quantum repeaters
that allow to do QKD or coherently connect quantum nodes
at a corresponding size and at a reasonably practical clock rate.

\subsection{This work}

In this work, we will focus on small-scale or medium-size quantum repeater systems
beyond a single middle station and without probabilistic entanglement distillation
on higher ``nesting levels". This class of quantum repeaters is of great interest
for at least two reasons. 

(i) There are now first experiments
of memory-enhanced quantum communication basically demonstrating memory-assisted MDI QKD \cite{Lukin, Rempe}. Therefore the natural next step for the experimentalists will be to
connect such elementary modules to obtain larger repeater systems
with {\it two or more intermediate stations}, thus
bridging larger distances and, unlike memory-assisted MDI QKD, ultimately relying upon classical communication between the repeater stations \cite{White}.

These next near-term experiments will aim at a distance extension still independent
of additional and more complicated schemes such as entanglement distillation
on ``higher nesting levels''.
Restricting the entanglement manipulations to the level of the elementary
repeater segments will also help to avoid the use of long-distance two-way classical signalling
like in a fully scalable 1st-generation quantum repeater, and hence
allow for still limited but reasonable repeater clock rates.
In this regime, comparing (secret key) rates per second of the quantum repeaters with those 
of an (ideal) point-to-point link or TF QKD scheme is in some way most fair and meaningful.

While the current experimental repeater demonstrations with a single repeater station \cite{Lukin,Rempe} would still suffer from too low clock rates and link coupling efficiencies before giving a practical repeater advantage, an urgent theoretical question
is whether, under practical realistic circumstances, it really helps to place memory stations along a quantum communication channel and execute memory-assisted QKD without extra active quantum error correction. In principle, placing a middle station between Alice and Bob allows to gain a repeater advantage per channel use \cite{NL,WehPar,White}.

Omitting the non-scalable all-optical TF approach,
is there a practical benefit also in terms of secret bits per second
when using a two-segment quantum repeater?
Moreover, and this is the focus of the present work,
is there even a further advantage when adding more stations beyond a single
middle station under realistic assumptions and with no extra
quantum error correction?
We will see that for up to eight repeater segments, covering distances up to around 800km, the quantum repeaters treated in this work, assuming experimental parameter values that are demanding but not impossible to achieve in practice, can exceed the performance limits
of the other schemes. For larger distances, the attainable absolute rates of point-to-point quantum communication become extremely small. However, for quantum repeaters, additional elements
of quantum error correction will be needed, as otherwise the final rates would vanish and no gain can be expected over point-to-point communication.

(ii) The second point refers to the theoretical treatment. Typically, the repeater rates can be calculated either numerically including many protocol variations and (experimental) degrees of freedom \cite{Coopmans} or approximately in certain regimes \cite{Sangouard}  (there are also semi-analytical approaches, see Refs. \cite{Kuzmin2021,Kuzmin2019}).

If errors are neglected an exact and even optimized raw rate calculation is possible
even for non-unit (but constant) entanglement swapping probabilities
using the formalism of Markov chains and decision processes \cite{PvL,Shchukin2021} (see also Refs. \cite{Vinay2019,Khatri2019}).
This approach works well for repeaters up to about ten segments;
for too many repeater segments the resulting linear equation systems become intractable.
Nonetheless, for the smallest repeaters with only a single middle station,
it was shown how to calculate secret key rates even including 
various experimental parameters, though partially also employing approximations for the raw rates \cite{NL,WehPar}.
In this work we will go beyond the case of a single middle station
and present exact calculations of {\it secret key rates obtainable with 
realistic small and intermediate-scale quantum repeaters}.
The theoretical difficulty here is, even already when only channel loss and memory dephasing is considered, that for repeaters beyond a single middle station there are various distribution and swapping strategies and so it becomes non-trivial to determine the optimal ones. 
The usual treatment in this case is based upon the so-called doubling strategy where for a repeater with a power-of-two number of segments only certain pairs of segments will be connected in order to double the distances at each repeater level. As a consequence, sometimes entanglement connections will be postponed even though neighboring pairs may be ready already, thus unnecessarily accumulating more memory dephasing errors. With regards to memory dephasing, the best strategy appears to be {\it to swap as soon as possible} and here we will show how this type of repeater strategy can be exactly and analytically treated. This element is the crucial step that enables us to propose optimal quantum repeater schemes.

On the hardware side, memory-based quantum repeaters require
sufficiently long-lived quantum memories and efficient, typically
light-matter-based interfaces converting flying into stationary qubits.
In the context of our theoretical treatment, the stationary qubits
are assumed to be represented by single spins in a suitable
solid-state quantum node such as colour (NV or SiV) centers in diamond,
usually separately treated as short-lived electronic
and long-lived nuclear spins \cite{LukinSiV, HansonNV}.
As for efficient quantum emitters and short-lived quantum memories
semiconductor quantum dots may be considered too \cite{White}.  
Alternatively, various types of atom or ion qubits could be taken into account \cite{White}.

While all these different hardware platforms
have their own assets and disadvantages (e.g.
the required temperatures which range from room or modestly low temperatures
for atoms/ions/NV to cryogenic temperatures for NV/SiV/quantum dots),
and every one eventually requires a specifically adapted physical model,
to a certain extent the quantum repeater performance based
on these elements and assuming only a single repeater station
can be assessed (or at least qualitatively bounded from above) using a fairly simple
physical model that includes {\it three experimental parameters}:
the link coupling efficiency, the memory coherence time,
and the experimental clock rate \cite{White}.

In order to incorporate an appropriate experimental 
memory coherence time into the model,
qubit dephasing errors can be considered where
the stationary qubit is never lost but subject to
random phase flips with a probability exponentially
growing with the storage time. 
Already this rather simple model is theoretically
non-trivial, because it leads to two distinct impacts
on the final secret key rates.
On the one hand, a finite link coupling efficiency
(including all constant inefficiencies per segment from the
sources, detectors, and interfaces) and a segment-length-dependent
transmission efficiency affect the raw rate
of the qubit transmission (which, if expressed as rate per second,
also directly depends on the repeater clock rate).
Thereby, in logarithmic rate-versus-distance plots (like those frequently shown 
later in this article), a finite link coupling leads to an offset
towards smaller rates at zero distance, while a finite channel transmission
results in a certain (negative) slope.
On the other hand, a finite memory coherence time influences the final Alice-Bob state fidelity
or QKD error rate
(which also indirectly depends on the repeater clock rate,
i.e. the time duration per entanglement distribution attempt
per segment, determining the possible number of distribution attempts
within a given memory coherence time).
This becomes manifest as an increase of the (negative) slope for growing distances,
moving from an initially repeater-like slope towards one corresponding to
a point-to-point transmission.

There are interesting concepts to suppress this latter effect
by introducing more sophisticated memory models such as memory buffers
or cut-offs. Especially a memory cut-off \cite{CollinsPrl}
has turned out to be useful without the need for additional experimental resources.
It means that a maximal storage time is
imposed at every memory node and any loaded stationary qubits waiting
for a longer duration will be reinitialized.
As a result, state fidelities can be kept high at the expense of 
a decreasing raw rate due to the frequently occurring reinitializations
(which implies that a memory cut-off must neither be set too low nor too high). 
Theoretically, including memory cut-offs into the rate analysis
significantly increases the complexity (becoming manifest
in e.g. quickly growing Markov-chain matrices) \cite{PvL}.

For small quantum repeaters, especially those with only one middle station,
a secret key rate analysis remains possible \cite{WehPar, White}.
For larger quantum repeaters, the effective rates may be calculated 
via recursively obtained expressions \cite{Jiang}, via different kinds of approximations and assumptions \cite{Elkouss2021} or with the help of numerical simulations \cite{Coopmans}.
Nonetheless, in our treatment, we shall explicitly include
a memory cut-off in some protocols allowing us to extrapolate 
its positive impact on other schemes. 

We choose to incorporate random dephasing as the 
dominating source of memory errors.
While memory dephasing is generally an error to be taken into account,
it is particularly important for those stationary qubits
encoded into single solid-state spins, e.g. for colour centers
or quantum dots \cite{White}.
We omit (time-dependent) memory decay (loss) which additionally becomes relevant
for atomic memories, either as collective spin modes
of atomic ensembles or in the form of an individual atom in 
a cavity (generally, atoms and trapped ions may be subject to both dephasing and decay) \cite{DLCZ,HybridPRL,HybridNJP,Rempe}. 
It turns out that the effect of memory dephasing can be accurately 
included into the statistical repeater model, since the
total, accumulated dephasing in the final Alice-Bob density operator
follows a simple sum rule \cite{tf_repeater}.
Thus, the statistical averaging can be applied to the final state,
for which we derive a recursive formula that also includes
depolarizing errors from the initially distributed states and
from the imperfect Bell measurement gates in every
entanglement swapping operation. The main complication will be
to determine the correct dephasing variables for the different
swapping strategies and identify the optimal schemes. As a result, we extend the simple model of Ref.~\cite{White}

not only with regards to the repeater's size, but also to include additional experimental parameters:
{\it besides the above three parameters we then have 
one or two extra parameters for the initially distributed states}
(taking into account initial dephasing or depolarization errors
depending on the protocol) {\it and one extra depolarization
parameter for the local gates and Bell measurements.}

Our analytical treatment enables us to identify 
the scaling of the various parameters, their specific impact
onto the repeater performance (for QKD, affecting either the raw rate
or the error-dependent secret key fraction), and the
resulting trade-offs. Most apparent is the trade-off 
for quantum repeaters with $n$ segments and $n-1$ intermediate memory stations 
leading to an improved loss scaling with an $n$-times bigger
effective attenuation distance compared with a point-to-point link ($n=1$), 
but a final state fidelity parameter 
decreasing as the power of $2n-1$ (assuming equal gate and initial state 
error rates).
We will then be able to consider repeater protocol variations
with an improved scaling of the basic loss and fidelity parameters.
Based upon the above-mentioned TF concept with coherent states
or basically replacing two-photon by one-photon interferences at the
beam splitter stations, these repeaters exhibit 
a $2n$-times bigger effective attenuation distance 
while keeping the $2n-1$ power scaling of the final state fidelity parameter
for $n-1$ memory stations. However, they are subject to some
extra intrinsic (dephasing) errors even when only channel loss is considered, which will turn out to be an essential complication that prevents to fully exploit the improved 
scaling of the basic parameters in comparison with the standard repeater protocols that do not suffer from intrinsic dephasing.

Comparing different repeater protocols and incorporating
the optimized memory dephasing from our statistical model into them, 
we find that for experimental parameter values that are highly demanding but not impossible 
(up to 10s coherence time, 80\% link coupling, and state or gate infidelities in the regime of 1-2\%), one secret bit can be shared per second over a total distance of 800km.
This represents a significant improvement over ideal point-to-point or realistic TF QKD at GHz clock rates.
In particular, the repeaterless, point-to-point bound \cite{PLOB}, for e.g. 800km is $3 \times 10^{-16}$ bits per channel use
or $0.3 \mu$bits per second (at GHz clock rate).
We will see that, in order to clearly beat this with those reasonable experimental parameters from above, the number of repeater stations 
must neither be too high nor too low, and so placing a station at every 100km will work well.

As mentioned before, our schemes are generally independent of the typically
used doubling strategies in quantum repeaters
(which are most suitable to incorporate entanglement distillation 
in a systematic way and which are included as a special case in our sets of swapping strategies). Instead we will consider general memory-assisted entanglement distribution with possible QKD applications. Compatible with our analysis are also schemes that aim at an enhanced 
initial state distribution efficiency or fidelity as, for example, in multiplexing-assisted or the above-mentioned 2nd-generation quantum repeaters.
In any case, the subsequent steps after the initial distributions in each repeater segment are simple entanglement swapping steps combined with quantum storage in single spins. For the entanglement swapping we assume unit success probability. This assumption is experimentally justified for systems where Bell measurements or, more generally, gates can be performed in a deterministic fashion, for instance, with atoms or ions or solid-state-based spin qubits \cite{White}.
For a linear quantum repeater chain, this system is still remarkably complex.

The assumption of deterministic entanglement swapping will allow us to calculate
the exact (secret key) rates in a quantum repeater up to eight segments.
We will distinguish schemes with sequential and parallel entanglement distributions and also consider different swapping strategies. Based on {\it two characteristic random variables},
the total repeater waiting time and the accumulated dephasing time of the final state, and their probability generating functions, we will be able to determine exact, optimized secret key rates. 
In principle, this gives us access to the {\it full
statistics of this class of quantum repeaters.}
Optimality here refers to the minimal dephasing among all parallel-distribution (and hence maximal raw-rate)
schemes. For three segments and two intermediate stations, we show that the resulting secret key 
rates are optimal among all schemes. For more segments and stations we conjecture this to hold too, however, there is the loophole that sequential-distribution schemes (generally exhibiting smaller raw rates) may accumulate less dephasing and as a result, in combination, lead to a higher secret key rate. We conclude that our treatment gives evidence for any optimal scheme to distribute entangled pairs in parallel, to swap as soon as possible, and to simultaneously store qubits as little as possible. However, here the first and the third property are not compatible, which leads to another trade-off between
high efficiencies (raw rates) and small state fidelities (high error rates) as commonly encountered for entanglement distribution and quantum repeaters.
The (partially or fully) sequential schemes have the advantage that parallel storage of qubits can be avoided to a certain (or even a full) extent. However, since the sequential schemes are overall slower, their total dephasing may still exceed that of the fastest repeater schemes with parallel storage. 
For up to eight repeater segments, our optimal scheme, exhibiting the smallest total dephasing among all fast repeater schemes, also exhibits a smaller total dephasing than the fully sequential scheme.

The outline of this paper is as follows. 
In Sec.~\ref{sec:QRwonemiddlestation} we will first review the known results
and existing approaches to analyze secret key rates for the smallest possible
quantum repeater based upon a single middle station, including calculations
of the repeater raw rate and physical error models to describe the evolution
of the relevant density operators. The methods for the statistical analysis
-- probability generating functions, and the figure of merit
to quantitatively assess the repeater performance -- a QKD secret key rate,
will be introduced in Sec.~\ref{sec:Methods}.

In Sec.~\ref{sec:Physical Modelling} we will then start introducing our
new, generalized treatment for quantum repeaters beyond a single middle station.
For this, we present two subsections on the two characteristic random variables
-- the waiting time and the dephasing time, which contain the entire statistical
information of the class of quantum repeaters considered in our work. 
In order to be able to take into account optimal strategies
for the initial entanglement distribution and the subsequent entanglement swapping
in more complex quantum repeaters with two or more intermediate repeater stations,
we discuss in detail in various subsections sequential and parallel distribution
as well as optimal swapping schemes. Still in Sec.~\ref{sec:Physical Modelling},
we show how these optimizations can be applied to the statistics of 
various quantum repeaters, explicitly calculating the probability generating functions
of the two basic random variables for two-, three-, four- and eight-segment
quantum repeaters. In particular, for the four- and eight-segment cases we will show 
how and to what extent our optimized and exact treatment of the memory dephasing
will improve the relevant quantities of the final state density operators
as compared with the usually employed, canonical schemes such as ``doubling''.
The interesting case of a three-segment repeater and its optimization will be discussed in more detail in an appendix.

Finally, in Sec.~\ref{sec:Secret Key Rate} we will analyze the secret key rates of all proposed schemes and compare them for various repeater sizes with the ``PLOB" bound \cite{PLOB}. 
For this, we will explicitly consider the extended set of experimental parameters
and insert experimentally meaningful values (representing current and future experimental
capabilities) for them.
A particular focus will be on the initial state and gate parameters
and their impact on the repeater performance.
We shall compare the performances of different schemes, discuss the possibility of including multiplexing, and examine what influence a memory cut-off and what (scaling) advantages the different types of encoding for the flying qubits can have.
For the latter, we discuss in more detail schemes based on the TF concept and, for the comparison between different schemes and encodings, the final secret key rates per second.
Sec.~\ref{sec:Conclusion} concludes the paper with a final summary of the results and their implications. Various additional technical details can be found in the appendices.

\section{Quantum repeaters with one middle station}\label{sec:QRwonemiddlestation}

\begin{figure}[ht]
	\includegraphics{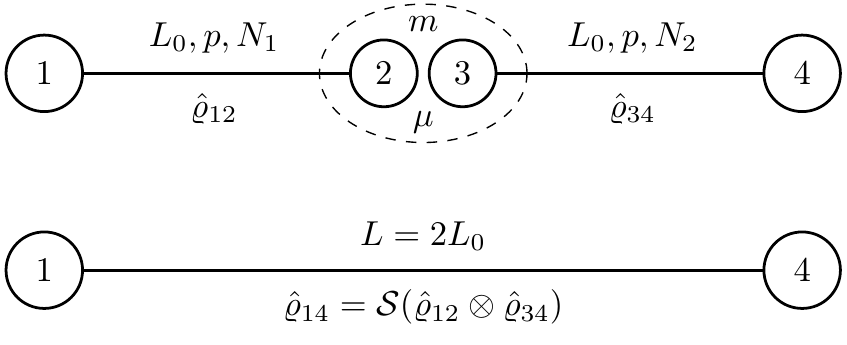}
	\caption{A two-segment quantum repeater. Each segment has length $L_0$
	and is characterized by a distribution success probability $p$,
	a (geometrically distributed) random number of distribution attempts $N$
	(with expectation value $\bar N = 1/p$), and a ``final'' two-qubit state $\hat\rho$ (subscripts denote segments or qubits at the nodes). ``Final'' here means that the, in general, imperfectly distributed states may be further subject to memory dephasing for a maximal number of $m$ time steps (distribution attempts). After an imperfect swapping operation $\mathcal S$ (error parameter $\mu$), the repeater end nodes share an entangled state over distance $2 L_0$.}\label{fig:2seg}
\end{figure}

A small quantum repeater composed of two segments and one middle station,
as schematically shown in Fig.~\ref{fig:2seg}, 
is pretty well understood and it is known how to obtain the secret key rates 
in a QKD scheme assisted by a single memory station,
even including experimental imperfections \cite{NL,WehPar,White,tf_repeater},
including memory cutoffs \cite{WehPar,White,PvL, CollinsPrl},
and for general, probabilistic entanglement swappping \cite{PvL}.
First experimental demonstrations of memory-enhanced quantum communication
are also based on this simplest repeater setting \cite{Lukin}.
In such a small quantum repeater, there is only a single Bell measurement
on the spin memories at the central station, and so the entanglement swapping
``strategy'' is clear. Later we will briefly discuss the two-segment case
as a special case of our more general rate analysis treatment, easily 
deriving the statistical properties of the two basic random repeater variables,
the total waiting and dephasing times, and obtaining the optimal scheme \cite{White, tf_repeater}. 

The smallest, two-segment quantum repeater also serves as a basic building block 
for general, larger quantum repeaters.
In the scheme of Fig.~\ref{fig:2seg},
each segment distributes an entangled pair of (mostly) stationary qubits 
by connecting its end nodes through flying qubits.
The goal is to share entanglement between the two qubits at the end nodes 
of the whole repeater.
The specific entanglement distribution scheme in each segment depends
on the repeater protocol and it may involve memory nodes sending 
or receiving photons \cite{White}.

In the notation of Fig.~\ref{fig:2seg}, from
an entangled state $\hat{\varrho}_{12}$ of qubits 1 and 2 and an entangled state $\hat{\varrho}_{34}$ of qubits 3 and
4, we create an entangled state $\hat{\varrho}_{14}$ of qubits 1 and 4. Here the states $\hat{\varrho}_{12}$ and
$\hat{\varrho}_{34}$ subject to the Bell measurement for the entanglement swapping operation are those quantum states present in the segments at the moment when the swapping is
performed. If, for example, segment 1 generates an entangled state earlier than segment 2, then $\hat{\varrho}_{12}$ enters the swapping step in the form of the initially, distributed state
(which is not necessarily a pure maximally entangled state) after it was subject to memory dephasing while waiting for segment 2.
Thus, our physical model includes state imperfections that originate from
the initial distribution as well as from the storage time,
as we shall discuss in detail below. In addition, we will include
an error parameter for the swapping gate itself.

\subsection{Raw rate}\label{sec:rawrate}

The entanglement distribution in an elementary segment is typically not a deterministic process and
several attempts are necessary to successfully share an
entangled pair of qubits among two neighboring stations.
If the probability of successful generation in each attempt is $p$, then the number of time steps
until success is a geometrically distributed random variable $N$ with success parameter $p$. We denote the failure probability as $q = 1 - p$. 
The parameter $p$ is primarily given by the probability that a photonic qubit is successfully transmitted via a fiber channel of length $L_0$ connecting two stations, $\exp(-L_0/\unit[22]{km})$. It also includes local state preparation/detection, fiber coupling, frequency conversion, and memory ``write-in'' efficiencies. 
The random variables for different segments (in
Fig.~\ref{fig:2seg} denoted as $N_1$ and $N_2$ for the first and the second segment, respectively) are independent and
identically distributed geometric random variables. Only when both segments have generated an entangled state, we perform a
swapping operation on the adjacent ends (nodes 2 and 3) of the segments and, when successful, we will be left with an entangled
state of qubits 1 and 4. 

In general, the swapping operation is also non-deterministic, but here we consider only
the case of deterministic swapping. Under this simple assumption we can still cover
a large class of physically relevant and realistic repeater schemes and obtain exact and optimized rates for them. Moreover, especially for larger repeaters (still with no entanglement distillations), this assumption allows 
to circumvent the need for classical communication times longer than the elementary time $\tau$
(as defined below) in order to confirm
successful entanglement swapping operations on ``higher'' repeater levels
beyond the initial distributions in each segment.
Physically, this assumption requires that in our schemes the Bell measurements
for entanglement swapping (including the memory ``read-out'' operations) can be performed deterministically. Nonetheless, the swapping operations can still be imperfect,
introducing errors in the states, as will be described below.

Due to the non-deterministic nature of the initial entanglement generation, 
the whole process of
entanglement distribution is also non-deterministic and fully described by the number of attempts up to and including
the successful distribution (so, this number is always larger than zero). The real, wall-clock time needed for
entanglement generation or distribution can be obtained from the number of attempts by multiplying it with an
elementary time unit, typically $\tau = L_0/c_f$, where again $L_0$ is the length of the segment and $c_f = c/n_\mathrm{r}$ is the speed of light
in the optical fiber ($c$ is the speed of light in vacuum and $n_\mathrm{r}$ is the index of refraction of the fiber, and depending on the specific distribution protocol there may be an extra factor 2). 
The elementary time unit is actually composed of the classical (and quantum) signalling time
per segment $\tau$ and the local processing time. However, for typical $L_0$ values
as considered here, the former largely dominates over the latter, and so we may neglect 
the local times, as they would hardly change the final secret key rates \cite{White}.

If one of the two segments generates entanglement earlier than the other, then the created state must be kept in memory.
The exact technique employed to implement this quantum memory is irrelevant for our analysis. The simplest model assumes
that the state can be kept in memory for arbitrarily long. A useful assumption in the realistic setting with imperfect quantum memories is to set a certain limit of $m$ time
units on the memory storage time, thus restarting the creation process whenever this threshold is reached.

\subsection{Errors}

When the quantum repeater is employed for long-range QKD, errors will become manifest
in terms of a reduced secret key fraction, as introduced in the subsequent section.
In order to compute this secret key fraction, we need to know the finally distributed state (density operator) of the
complete repeater system, and for this we require a more detailed physical model.
We shall establish a relation between the finally distributed state as a
function of the initial states in each segment and various errors that appear in the process of entanglement
distribution. The physical model is rather common and has been used before in
several works, both analytical and numerical.
Especially, a two-segment quantum repeater can be treated analytically based on simple Pauli errors
representing memory dephasing and gate (Bell measurement) errors.

We address the effect of imperfect quantum storage at a memory node
via a dephasing model where the stored quantum state is waiting for an adjacent
segment to successfully generate or distribute entanglement. 
This kind of memory error can be modelled by a one-qubit dephasing
channel,
\begin{equation}\label{eq:Gl}
    \Gamma_\lambda(\hat{\varrho})=(1 - \lambda) \hat{\varrho} +\lambda Z \hat{\varrho} Z,
\end{equation}
where $Z$ is a qubit Pauli phase flip operator.
We assume that $0 \leqslant \lambda < 1/2$, and any such number can be represented as $\lambda = (1 - e^{-\alpha})/2$
for some $\alpha > 0$. We denote the map in Eq.~\eqref{eq:Gl} also as $\Gamma_\alpha$. To avoid confusion, throughout this work we use the following definition:
\begin{equation}\label{eq:Ga}
	\Gamma_\alpha(\hat{\varrho}) = \frac{1 + e^{-\alpha}}{2} \hat{\varrho} +
	\frac{1 - e^{-\alpha}}{2} Z \hat{\varrho} Z.
\end{equation}
The definition for a dephasing two-qubit channel is obtained from Eqs.~\eqref{eq:Gl}-\eqref{eq:Ga} by the replacement $Z
\to Z \otimes I$ if the dephasing acts on the first qubit and by $Z \to I \otimes Z$ if the dephasing acts on the second
qubit.

Errors may also occur when a Bell state measurement is performed. This kind of errors is modelled by a two-qubit depolarizing channel,
\begin{equation}
    \tilde{\Gamma}_\mu(\hat{\varrho}) = \mu \hat{\varrho} + (1-\mu) \frac{\hat{\mathbb{1}}}{4}.
\end{equation}
We do not consider dark counts of the detectors,
since the optical propagation distances $L_0$ after which
a detection attempt takes place remain sufficiently small in
any quantum relay or repeater.
Thanks  to  recent  technological  developments  typical  dark  count  rates  can  be  reduced 
far below 1 dark count per second. In Ref.~\cite{Schuck2013} they were shown to be in the range of
\(\unit{mHz} \). Dark counts of such a low frequency have no significant impact on the secret key rate in our schemes.

Let us now apply this to the case of a two-segment quantum repeater.
The Bell measurement of qubits 2 and 3 produces from a pair of states $\hat{\varrho}_{12}$ and $\hat{\varrho}_{34}$ a
state $\hat{\varrho}_{14}$, see Fig.~\ref{fig:2seg}. The initial state $\hat{\varrho}_{1234} = \hat{\varrho}_{12}
\otimes \hat{\varrho}_{34}$ of all four qubits 1, 2, 3 and 4 is the product of the states of qubits 1, 2 and qubits 3,
4. After the measurement the state $\hat{\varrho}_{14}$ of qubits 1 and 4 becomes
\begin{equation}\label{eq:Srho}
	\hat{\varrho}_{14} \equiv \mathcal{S}(\hat{\varrho}_{1234}) = \frac{\Tr_{23}(\hat{P}_{23}
	\tilde{\Gamma}_{\mu, 23}(\hat{\varrho}_{1234}) \hat{P}_{23})}{\Tr(\hat{P}_{23}
	\tilde{\Gamma}_{\mu, 23}(\hat{\varrho}_{1234}) \hat{P}_{23})},
\end{equation}
where $\mu$ describes the imperfection of the measurement and $\hat{P}_{23} = |\Psi^+\rangle_{23}\langle\Psi^+|$ is one of the four measurement operators
in the two-qubit Bell state basis of the central subsystem (qubits 2 and 3),
$\{ |\Phi^\pm\rangle_{23}\langle\Phi^\pm|,\,|\Psi^\pm\rangle_{23}\langle\Psi^\pm| \}$,
where $|\Phi^\pm\rangle = (|00\rangle \pm |11\rangle)/\sqrt{2},\,
|\Psi^\pm\rangle = (|10\rangle \pm |01\rangle)/\sqrt{2}$, for qubits
defined via the two $Z$ eigenstates $|0\rangle, \,|1\rangle$
(for any one of the other three Bell measurement outcomes, the analysis below
is similarly applicable). 
In this case, Eq.~\eqref{eq:Srho}
reduces to
\begin{equation}\label{eq:rhod}
	\hat{\varrho}_{14} \equiv \mathcal{S}(\hat{\varrho}_{1234}) =
	\frac{_{23}\langle\Psi^+|\tilde{\Gamma}_{\mu, 23}(\hat{\varrho}_{1234})|\Psi^+\rangle_{23}}
	{\Tr(_{23}\langle\Psi^+|\tilde{\Gamma}_{\mu, 23}(\hat{\varrho}_{1234})|\Psi^+\rangle_{23})}.
\end{equation}
A simple way to compute the right-hand side of this relation for an arbitrary density operator
$\hat{\varrho}_{1234}$ is given in App.~\ref{app:Trace Identities}.

In general, states of the form
\begin{equation}\label{eq:rho0}
	\hat{\varrho}_0 = \tilde{\Gamma}_{\mu_0}\bigl(F_0\dyad{\Psi^+} + (1-F_0)\dyad{\Psi^-}\bigr)
\end{equation}
play an important role in the full theory presented below. It is easy to verify that
\begin{equation}
	(I \otimes Z) \hat{\varrho}_0 (I \otimes Z) = (Z \otimes I) \hat{\varrho}_0 (Z \otimes I),
\end{equation}
so it does not matter whether $\Gamma_\alpha$ acts on the first or second qubit of $\hat{\varrho}_0$ and either
application we simply denote as $\Gamma_\alpha(\hat{\varrho}_0)$. An easily checkable relation is
\begin{equation}\label{eq:Ga2}
	\Gamma_\alpha(\hat{\varrho}_0) = \tilde{\Gamma}_{\mu_0}\bigl(F\dyad{\Psi^+} + (1-F)\dyad{\Psi^-}\bigr),
\end{equation}
where the new parameter $F$ is expressed in terms of the original one, $F_0$, as
\begin{equation}\label{eq:Fprime}
	F = \frac{1}{2}(2F_0-1) e^{-\alpha} + \frac{1}{2}.
\end{equation}

The initial fidelity parameter $F_0$ (describing an initial dephasing 
of the distributed states) combined with the $\mu_0$-dependent initial
depolarization are both included in the initial $\hat \rho_0$ in Eq.~\eqref{eq:rho0},
because later this will allow for an elegant recursive state relation for
larger repeaters. It will also allow to switch between different initial
physical errors depending on the specific repeater realization.
In general, the maps in Eq.~\eqref{eq:Ga} satisfy the relation $\Gamma_\alpha \circ \Gamma_\beta = \Gamma_{\alpha + \beta}$. In
particular, we have $\Gamma_\alpha \circ \ldots \circ \Gamma_\alpha = \Gamma_{k\alpha}$, where $\Gamma_\alpha$ is used
$k$ times on the left-hand side. So, applying $\Gamma_\alpha$ to the state $\hat{\varrho}_0$ given by
Eq.~\eqref{eq:rho0} several times, we have to multiply $\alpha$ in Eq.~\eqref{eq:Fprime} by this number of times.

In a two-segment quantum repeater, if we start with the distributed states $\hat{\varrho}_{12}$ and $\hat{\varrho}_{34}$ of the special form
(similar to Eq.~\eqref{eq:rho0})
\begin{equation}\label{eq:Drho}
\begin{split}
	\hat{\varrho}_{12} &= \tilde{\Gamma}_{\mu_1}\bigl(F_1|\Psi^+\rangle_{12}\langle\Psi^+| + (1 - F_1)|\Psi^-\rangle_{12}\langle\Psi^-|\bigr), \\
	\hat{\varrho}_{34} &= \tilde{\Gamma}_{\mu_2}\bigl(F_2|\Psi^+\rangle_{34}\langle\Psi^+| + (1 - F_2)|\Psi^-\rangle_{34}\langle\Psi^-|\bigr), \\
\end{split}
\end{equation}
then the ``swapped'', finally distributed state $\hat{\varrho}_{14}$, given by Eq.~\eqref{eq:rhod}, is also of the same form,
\begin{equation}\label{eq:rho14}
	\hat{\varrho}_{14} = \tilde{\Gamma}_{\mu_d}\bigl(F_d|\Psi^+\rangle_{14}\langle\Psi^+| + (1 - F_d)|\Psi^-\rangle_{14}\langle\Psi^-|\bigr),
\end{equation}
where $\mu_d = \mu \mu_1 \mu_2$ and $F_d$ reads as
\begin{equation}\label{eq:Fd}
	F_d = \frac{1}{2}(2F_1 - 1)(2F_2 - 1) + \frac{1}{2}.
\end{equation}
We see that the form of the state is preserved by the total distribution procedure
of a two-segment repeater. The same conclusion will be applicable to larger
repeaters as well --- if all segments start in a state of the form given by Eq.~\eqref{eq:rho0}, then the finally distributed
state will also be of the same form.

For the two-segment repeater,
let us now assume that both segments generate the same state as in Eq.~\eqref{eq:rho0}, 
but not necessarily simultaneously, and so generally only after
some waiting time we perform the entanglement swapping and distribute entanglement over the two segments. If the first segment
generates entanglement after $N_1$ time units, and the second segment after $N_2$ time units, and we perform the entanglement swapping
after $N$ time units, with $N \geqslant N_1, N_2$, then the states $\hat{\varrho}_{12}$ and $\hat{\varrho}_{34}$ prior to swapping will be of
the form in Eq.~\eqref{eq:Drho} with $\mu_1 = \mu_2 = \mu_0$ and
\begin{equation}
\begin{split}
	F_1 &= \frac{1}{2}(2F_0 - 1)e^{-(N-N_1)\alpha} + \frac{1}{2}, \\
	F_2 &= \frac{1}{2}(2F_0 - 1)e^{-(N-N_2)\alpha} + \frac{1}{2}. \\
\end{split}
\end{equation}
The final, distributed state is then given by Eq.~\eqref{eq:rho14} where, according to Eq.~\eqref{eq:Fd}, the parameters are $\mu_d
= \mu \mu^2_0$ and
\begin{equation}
	F_d = \frac{1}{2}(2F_0-1)^2 e^{-(2N-N_1-N_2)\alpha} + \frac{1}{2}.
\end{equation}
This distributed state is subject to less dephasing when we swap as early as possible, thus $N = \max(N_1, N_2)$, so the
integer term in front of $\alpha$ is equal to $2\max(N_1, N_2) - N_1 - N_2 = |N_1 - N_2|$. Extra factors depending on the number of spins subject to dephasing in one segment (in particular, a factor of 2 for one spin pair) can be absorbed into $\alpha$. The precise physical meaning of $\alpha$ will be discussed later when we calculate the memory-assisted secret key rates in a quantum repeater. Furthermore, here we omitted explicit factors depending on the number of memory qubits that are subject to dephasing in a single repeater segment (in our model this will be one or two spins).

\section{Methods and figure of merit}\label{sec:Methods}

Before we move to the more general case of more than two segments and more than
just one middle station, we need some general methods and tools from statistics.
This will enable us to derive an analytic, statistical model for larger quantum repeaters
beyond one middle station (the physical model remains basically the same as for the
small, elementary two-segment quantum repeater), where we calculate average values
or moments of two random variables: the total repeater waiting time $K_n$ and the total
(i.e., the totally accumulated) memory dephasing time $D_n$.
As a quantitative figure of merit, it is useful to consider the secret key rate
of QKD, as it combines in a single quantity the two typically competing effects in
a quantum repeater system: the speed at which quantum states can be distributed
over the entire communication distance
and the quality of the totally distributed quantum states.
These two effects are naturally related to the above-mentioned two
random variables.
For our purposes here, throughout we shall rely on asymptotic expressions for the secret key rate
omitting effects of finite key lengths.
Of course, alternatively, one could also treat the total state distribution efficiencies
and qualities (fidelities) separately and individually, and then also consider
quantum repeater applications beyond long-range QKD.

\subsection{Probability generating function}

The method of probability generating functions (PGFs) plays an important role in our treatment of statistical properties
of quantum repeaters. For any random variable $X$ taking integer non-negative values its PGF $G_{X}(t)$ is defined via
\begin{equation}
	G_X(t) = \mathbf{E}[t^X] = \sum^{+\infty}_{k = 0} \mathbf{P}(X = k)t^k.
\end{equation}
The series on the right-hand side converges at least for all complex values of $t$ such that $|t| \leqslant 1$. The PGF
contains all statistical information about $X$, which can be easily extracted if an explicit expression for $G_X(t)$ is
known. For example, the average value of $X$, $\mathbf{E}[X] \equiv \overline{X}$, and its variance $\mathbf{V}[X]
\equiv \sigma^2_X = \mathbf{E}[(X - \overline{X})^2]$, are expressed as follows:
\begin{equation}\label{eq:PGF}
\begin{split}
	\mathbf{E}(X) &= G'_X(1), \\
	\mathbf{V}(X) &= G^{\prime\prime}_X(1) + G'_X(1) - G^{\prime 2}_X(1).
\end{split}
\end{equation}
For any $\alpha \geqslant 0$ the random variable $e^{-\alpha X}$ has a finite average value, which can be computed as
\begin{equation}\label{eq:PGF_2}
	\mathbf{E}[e^{-\alpha X}] = G_X(e^{-\alpha}).
\end{equation}
Note that for this random variable, besides the mean or average value, any statistical moment can be easily obtained and the $k$th-moment simply becomes $\mathbf{E}[e^{-\alpha k X}] = G_X(e^{-k \alpha})$. 

Two kinds of random variables appear in our model of quantum repeaters where one is related to the raw rate
and the other to the secret key fraction of QKD as introduced below. It is not always possible to get a compact expression for the PGF of these
random variables explicitly, but when it is, we use the equations above to obtain statistical properties of the
corresponding random variables.

\subsection{Secret key rate}\label{sec:skr}

The main figure of merit in our study is the quantum repeater secret key rate, which can be defined as the product of two quantities,
\begin{equation}
	S = Rr,
\end{equation}
where $R$ is the raw rate and $r$ is the secret key fraction. The raw rate is simply the inverse average
waiting time,
\begin{equation}
	R = \frac{1}{T},
\end{equation}
where $T = \mathbf{E}[K]$ is the average number of steps $K$ needed to successfully distribute one entangled qubit pair over the entire communication distance between Alice and Bob (giving an average time duration in seconds when multiplied with an appropriate time unit $\tau$). The secret key fraction of the BB84 QKD protocol \cite{BB84,PirRMP}, assuming one-way post-processing, is given by
\begin{equation}\label{eq:skf}
	r = 1 - h(\overline{e_x}) - h(\overline{e_z}),
\end{equation}
where $e_x$ and $e_z$ are the quantum bit error rates (QBERs),
\begin{equation}\label{eq:exez}
\begin{split}
	e_z &= \langle 00|\hat{\varrho}_n|00\rangle + \langle 11|\hat{\varrho}_n|11\rangle, \\
	e_x &= \langle +-|\hat{\varrho}_n|+-\rangle + \langle -+|\hat{\varrho}_n|-+\rangle,
\end{split}
\end{equation}
and $h(p)$ is the binary entropy function,
\begin{equation}
	h(p) = -p\log_2(p) - (1-p)\log_2(1-p).
\end{equation}
The QBERs $e_x$ and $e_z$ in Eq.~\eqref{eq:exez} are obtainable from the final, distributed state $\hat{\varrho}_n$ of an
$n$-segment quantum repeater, which in our case will depend on the dephasing random variable, and so we have to insert average values in Eq.~\eqref{eq:skf}
as indicated by the bars. We thus need a complete model of quantum repeaters to compute the statistical properties of the relevant
random variables associated with the number of steps to distribute entanglement or the density operator of the distributed state.
Given such a model,
the aim of our work is to compute and analyze secret key rates of quantum repeaters with an increasing size, up to eight segments, considering and optimizing different distribution and swapping schemes.
Besides the most common BB84 QKD protocol,
alternatively, we may also consider the six-state protocol 
\cite{sixstate} 
which would slightly improve the secret key rate. 
Assuming again one-way post-processing, the secret key fraction $r$ 
of the six-state protocol is given by $1-H(\boldsymbol{\lambda})$ 
\cite[App. A]{RevModPhys.81.1301} 
where $H(\cdot)$ is the Shannon entropy and the vector $\boldsymbol{\lambda}$ must contain the corresponding weights of the four Bell states in the final density operator $\hat{\varrho}_n$. Throughout this work all secret key rates
are calculated from their asymptotic expressions and hence effects
of finite key lengths are not included here. This simplifies the analytical treatment of a quantum repeater chain, which, as we will see, quickly becomes rather complex for a growing number of stations, involving many distinct choices and strategies for the entanglement manipulations. Moreover, our rate analysis shall also be useful to assess and compare the performances of different quantum repeaters in applications beyond QKD.      

\section{Quantum repeaters beyond one middle station}\label{sec:Physical Modelling}

\begin{figure*}
	\includegraphics{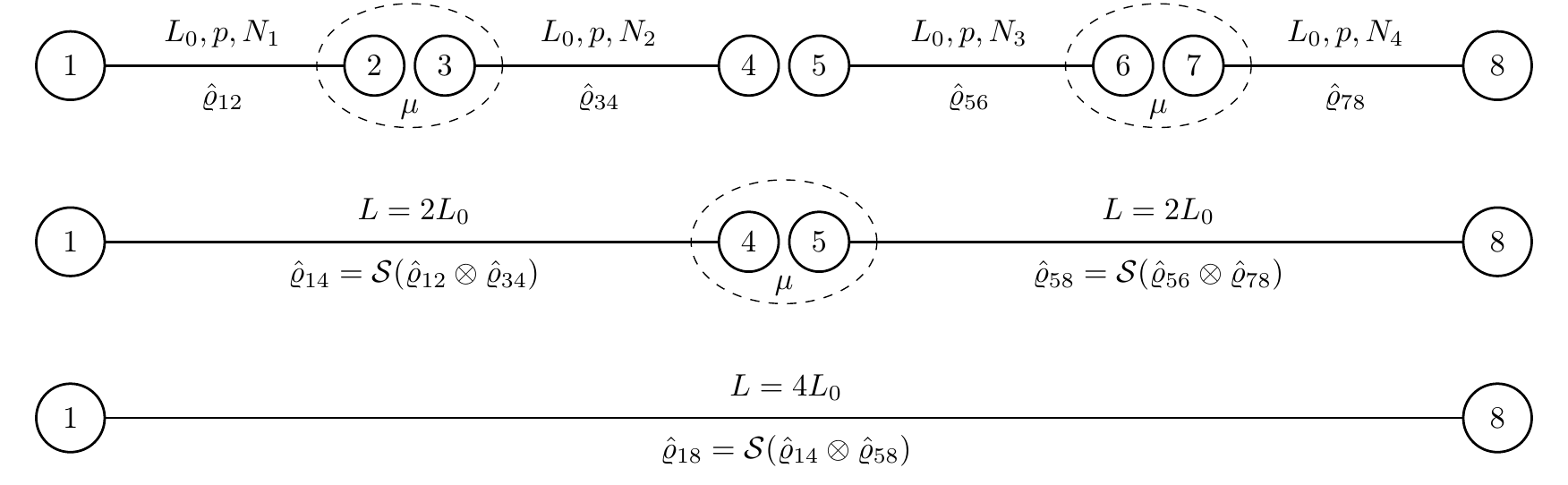}
	\caption{``Doubling'' swapping scheme for a four-segment quantum repeater. This is the most common swapping strategy which allows to systematically include entanglement distillation at each repeater ``nesting level". Without extra distillation, however, ``doubling'' is never optimal: combined with fast, parallel distributions it exhibits increased parallel storage times and hence memory dephasing (while combined with sequential distributions the repeater waiting times become suboptimal). Memory cut-off parameters are omitted in the illustration.}\label{fig:4segD}
\end{figure*}

\begin{figure*}
	\includegraphics{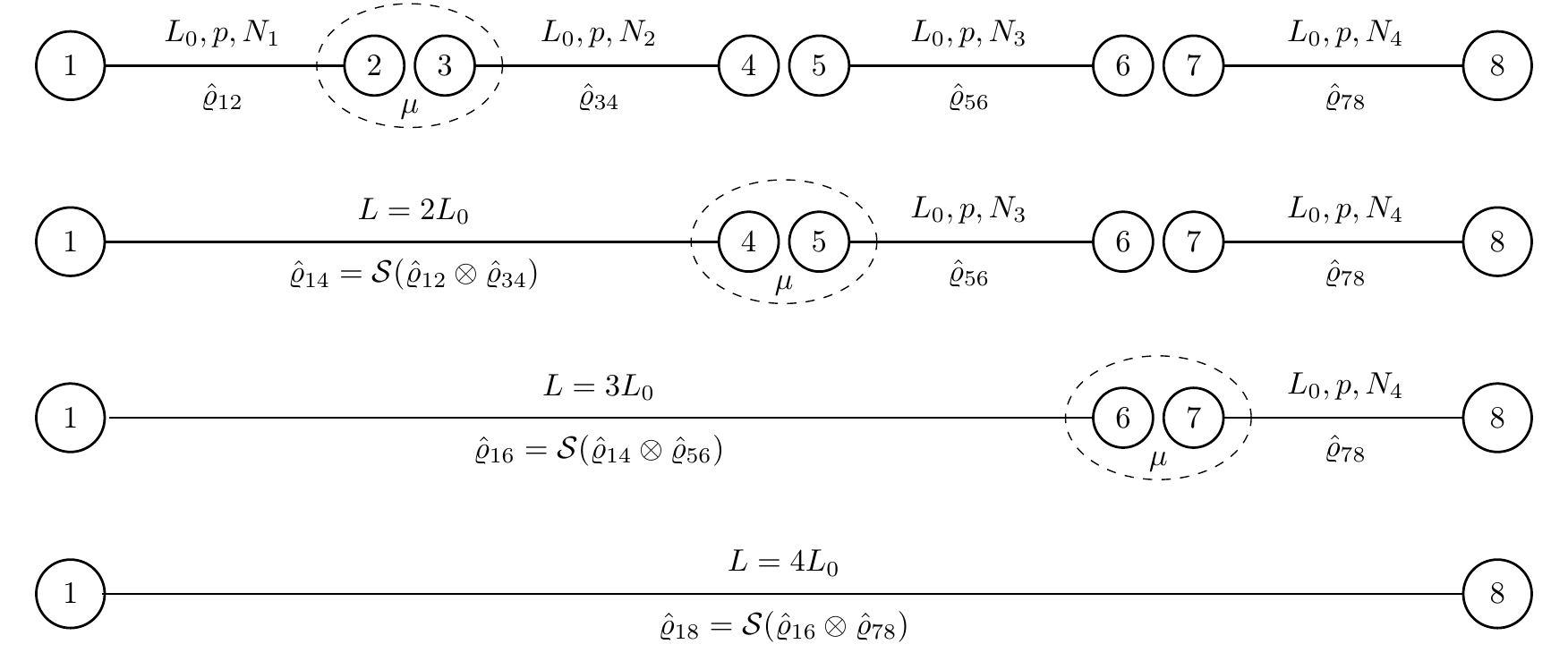}
	\caption{``Iterative'' swapping scheme for a four-segment quantum repeater. The swapping operations are performed step by step (here from left to right). Also this scheme, when executed with parallel distributions in each segment, leads to an increase of the total dephasing. However, if combined with sequential distributions, the accumulated dephasing times can be reduced (with always at most one spin or spin pair being subject to a long dephasing) at the expense of a growing repeater waiting time. Memory cut-off parameters are omitted in the illustration.}\label{fig:4segI}
\end{figure*}

Larger repeaters with more than two segments and one middle station can now be modeled in a way similar to the two-segment case
discussed above.
However, the extended, more general case is also more complex and there are both different ways to perform the initial
entanglement distributions in all elementary segments and different ways to connect the successfully distributed
segments via entanglement swapping. For the initial distributions we make a distinction between
sequential and parallel schemes, where the former refers to a scheme in which, according to a predetermined order,
the distributions are attempted step by step starting from e.g. the first segment. In a parallel scheme,
the distributions are attempted simultaneously in all segments, which obviously leads to a smaller total
repeater waiting time than for the sequential distribution schemes. Nonetheless, since the sequential schemes
do make use of the quantum memories, they do already offer the repeater-like scaling advantage over
point-to-point quantum communication links. Even for a two-segment quantum repeater, we may choose 
a sequential scheme, where we first only distribute e.g. the left segment and only once we succeeded there we attempt
to distribute the right segment. Experimentally, this can be of relevance for those realizations where only a single short-term
quantum memory is available at every station for the light-matter interface
and another quantum memory for the longer-term storage (e.g., respectively, an electronic and a nuclear spin in colour-center-based repeater nodes) \cite{WehnerNV,HansonNV}.
Theoretically and conceptually, there are at least two advantages of a (fully) sequential distribution approach \cite{tf_repeater}.

First, the two basic random variables of a quantum repeater are very simple and so the secret key rates are fairly easy to calculate. Second, always only at most one entangled qubit pair (or even only a single spin if e.g. Alice measures her qubit immediately) may be subject to memory dephasing during all distribution steps.

For the entanglement connections via entanglement swapping, the two-segment case is special,
as there is only one swapping to be performed at the end when pairs in both segments are available.
However, already with three segments and two repeater stations there is no unique swapping order anymore,
and we may either fix the order or ``dynamically'' choose where we swap as soon as swapping is possible
for two neighboring, successfully distributed segments. In a fixed scheme, two neighboring segments,
though ready, may have to wait before being connected.
Thus, the choice of the entanglement swapping scheme
has a significant impact on the totally accumulated dephasing time. In a worst-case scenario, we could wait until
all segments have been distributed and then do all the entanglement connections at the very end;
for deterministic entanglement swapping, like in our model, this would not affect the raw waiting times,
but it would lead to a maximal total dephasing.
In this case, a sequential distribution where entanglement swapping takes place immediately when a new, successfully bridged segment is available can lead to a higher secret key rate than a combination of parallel distribution and swapping at the end (where the rates of the latter scheme may still only be obtainable approximately) \cite{tf_repeater}.
The crucial innovation in our analytical treatment here is that we will be able to obtain the exact secret key rates for schemes that combine fast, parallel distributions with fast, immediate swappings (and hence a suppressed level of parallel storage). In other words, among all parallel-distribution schemes we will calculate the exact rates that are optimized with regards to the total repeater dephasing.

\subsection{Waiting times}

The average total waiting times in a quantum repeater or even the full statistics
of the waiting-time random variable can be, in principle, obtained
via the Markov chain formalism, even when the swapping is probabilistic \cite{PvL, Shchukin2021}.
More generally, the PGFs as introduced earlier contain the full statistical information,
and for deterministic swapping, we can obtain the PGF of $K_n$ through combinatorics.
In order to minimize the total waiting time, the distributions should occur in parallel.
However, there is no unique way to
perform the entanglement swapping, and so let us briefly consider this aspect in the context
of the waiting times. For example, for a four-segment repeater, two possible swapping strategies are shown in Figs.~\ref{fig:4segD} and
\ref{fig:4segI}. Both schemes are for a fixed swapping order, while we may distribute the
individual segments in parallel.
In the first scheme, typically referred to as ``doubling'', we swap the two halves of the repeater independently and only when both are ready, we
swap them too. In the second scheme, we swap the segments one after the other starting in one of the repeater's ends
(here the left end); we may refer to this scheme as ``iterative'' swapping.
Other schemes are possible, and the more segments the repeater has, the more possibilities for performing swappings
there are. The raw rate of a repeater is characterized by the number of steps, $K_n$, needed to successfully distribute
an entangled pair, and this random variable can be expressed in terms of the geometric random variables $N_i$ associated with each segment. For
example, for the swapping schemes shown in Figs.~\ref{fig:4segD} and \ref{fig:4segI}, when combined with parallel distributions, we have $K_4 = \max(N_1, N_2, N_3, N_4)$, so the two schemes have the same raw rate.
In general, the waiting times of all such schemes that distribute in parallel are of a similar form.
Those schemes that we later classify as ``optimal'' in terms of the whole secret key rate
are assumed to be parallel distribution schemes. Conversely, combining iterative swapping with sequential distribution can lead to a reduced accumulated dephasing time at the expense of an increased total repeater waiting time. We shall discuss the accumulated dephasing times next.

\subsection{Dephasing times}

In order to treat the total dephasing time in a quantum repeater with more than two segments,
we have to generalize the methods and the model that led to the result for the distributed state for two segments,
Eq.~\eqref{eq:rho14} and Eq.~\eqref{eq:Fd}, and the discussion below, to larger repeaters
with, in pinciple, an arbitrary number of segments $n$.
In fact, we did the two-segment derivations in such a way that
an $n$-segment extension is now straightforward.
We obtain the following expression for the final, distributed state in the general case:
\begin{equation}\label{eq:rhon}
\begin{split}
	\hat{\varrho}_n = \tilde{\Gamma}_{\mu_n}\Biggl[&\frac{1
	+(2F_0 - 1)^n e^{-\alpha D_n}}{2} \dyad{\Psi^+} \Biggr. \\	
	+\Biggl.&\frac{1 - (2F_0 - 1)^n e^{-\alpha D_n}}{2} \dyad{\Psi^-}\Biggr],
\end{split}
\end{equation}
where $\mu_n = \mu^{n-1} \mu^n_0$ and $D_n = D_n(N_1, \ldots, N_n)$ is a random variable describing the total number of
time units that contribute to the total dephasing in the final output state. For $n=2$, the expression for $D_2(N_1, N_2) = |N_1 - N_2|$ has been obtained before, for larger $n$ the value of $D_n$ now depends on the swapping scheme. As before, we omitted explicit factors depending on the number of memory qubits that are subject to dephasing in a single repeater segment (one or two spins in our model) which also depends on the application and the specific execution of the protocol. Such factors can always be absorbed into $\alpha$. 
The precise physical meaning of $\alpha$ will be discussed later when we calculate the memory-assisted secret key rates in a quantum repeater. 

The QBERs for the state in Eq.~\eqref{eq:rhon} are easy to compute,
\begin{equation}\label{eq:QBER}
\begin{split}
	e_z &= \frac{1}{2}(1 - \mu^{n-1}\mu^n_0), \\
	e_x &= \frac{1}{2}(1 - \mu^{n-1}\mu^n_0 (2F_0 - 1)^n e^{-\alpha D_n}).
\end{split}
\end{equation}
For one of the averages, we have $\overline{e_z} = e_z$, and in order to obtain the other average $\overline{e_x}$ we need to calculate the expectation value
$\mathbf{E}[e^{-\alpha D_n}]$. This average can be obtained with the help of Eq.~\eqref{eq:PGF_2} if we know the
PGF of $D_n$. Again, in principle, we can get the full statistics of $D_n$ (and functions of it) from this PGF. More specifically, according to Eq.~\eqref{eq:PGF_2}, for the random variable $e^{-\alpha D_n}$ we can easily obtain all statistical moments of order $k$, $\mathbf{E}[e^{-\alpha D_n k}]$. This may be useful for a rate analysis that includes keys of a finite length, though here in this work we shall focus on asymptotic keys.  
The PGF of $D_n$, however, is generally harder to obtain than that of $K_n$. For example, the PGF of $D_n$ is not obtainable
via the absorption time of a Markov chain (unlike that of $K_n$, which is obtainable even when the entanglement swapping is probabilistic)  \cite{PvL, Shchukin2021}.
Nonetheless, at least without considering the more complicated case including
a memory cut-off, we can calculate the relevant PGF of $D_n$ by analyzing all permutations of the basic variables
(there are also other, more elegant, but still not so efficient and well scalable methods to treat the statistics of $D_n$,
e.g. based on algebraic geometry).

We see that in order to compute the secret key rate of a quantum repeater we need to study the two integer-valued random variables
$K_n$ and $D_n$. The former describes the number of steps to successfully distribute entanglement and is responsible for
the repeater's raw rate. The latter describes the quality of the final state and strongly depends on the swapping
scheme. For example, for a four-segment repeater with a predetermined swapping order like the iterative scheme
in Fig.~\ref{fig:4segI}, we could actually also choose to adapt the initial entanglement distributions
to the swapping strategy and hence wait with every subsequent distribution step until the corresponding connection from the left
has been performed. Since this is no longer parallel distribution (it is ``sequential'' distribution), we would obtain an increased total waiting time.
However, the accumulated dephasing time may be reduced this way, as we discuss in the next subsection.

In general, we may also consider schemes with a memory cut-off, where we put a certain restriction of $m$ time units on the
maximum time a qubit can be kept in memory. So, in this case, we study four variables --- the total number of distribution steps and the total dephasing, both with and without cut-off. In order to maximize the secret key rate we need a scheme with small $\mathbf{E}[K_n]$ and large
$\mathbf{E}[e^{-\alpha D_n}]$. In the following subsections, we will introduce different schemes for performing
the entanglement swapping and, where
possible, compute the PGFs of the corresponding random variables. The PGF of $K_n$ is denoted as $G_n(t)$ and that of $D_n$ as
$\tilde{G}_n(t)$. For the corresponding quantities with cut-off $m$ we use the superscript $[m]$, e.g. $K^{[m]}_n$.
We will see and argue that there are three basic properties that a quantum repeater protocol (unassisted by additional
quantum error detection or correction) should exhibit: distribute the entangled states in each segment in parallel,
swap the initially distributed states as soon as possible, and avoid parallel storage of already distributed pairs
as much as possible. It is obvious that all these three ``rules'' cannot be fully obeyed at the same time.
In particular, parallel distribution will ultimately lead to some degree of parallel storage.

\subsection{Sequential distribution schemes}

In what we refer to as a sequential entanglement distribution scheme,
the initial, individual pairs are no longer distributed in parallel but
strictly sequentially according to a predetermined order.
If this order is chosen in a suitable way, it is possible that
at any time during the repeater protocol at most one entangled pair
is subject to dephasing (apart from small constant dephasing units
for single attempts), because once a new pair is available
an entanglement connection can be immediately performed and only then
another new segment starts distributing.
This may lead to a reduced accumulated dephasing time.
Moreover, from a secret key rate analysis point of view,
an appropriate sequential scheme can allow for a straightforward
calculation of the statistics of both random variables,
the total waiting and the accumulated dephasing times, even when a memory
cut-off is included.

Let us consider a simple, sequential distribution and swapping scheme where the
above discussion applies and the secret key rate can be computed exactly by means of elementary
combinatorics. In this scheme, we start by distributing entanglement in segment 1 (most left segment),
and only after a success we start to attempt distributions in segment 2.
As soon as we succeed there too, we immediately swap segments 1 and 2 and start to distribute entanglement in segment 3. As soon
as we succeed with the distribution in segment 3, we swap segment 3 with the first two, already connected segments, start distributing
in segment 4, and so on, repeating this process until entanglement has also been distributed in the most right segment
followed by a final entanglement swapping step.
This scheme, for $n=4$, is also illustrated by Fig.~\ref{fig:4segI}. The variables $K_n$ and $D_n$ for this scheme and general $n$
are thus defined as
\begin{equation}
	K^{\mathrm{seq}}_n = N_1 + \ldots + N_n, \quad D^{\mathrm{seq}}_n = N_2 + \ldots + N_n.
\end{equation}
The PGFs of these random variables are just powers of the PGF of the geometric distribution:
\begin{equation}
    G^{\mathrm{seq}}_n(t) = \left(\frac{pt}{1 - qt}\right)^n, \  \tilde{G}^{\mathrm{seq}}_n(t) = \left(\frac{pt}{1 - qt}\right)^{n-1}.
\end{equation}
In App.~\ref{app:SeqPGF} we derive the following expressions for the PGFs of the random variables with memory cut-off. We assume an accumulated, global cut-off where the total storage (dephasing) time across all segments must not exceed the value $m$. The
PGF of $K^{[m]}_n$ is given by
\begin{equation}\label{eq:Gmnt}
	G^{[m]}_n(t) = \frac{p^n t^n \sum^{m-n+1}_{j=0}\binom{j+n-2}{n-2}q^j t^j}{1-qt-p\sum^{n-2}_{i=0}\binom{m}{i}p^i q^{m-i} t^{m+1}},
\end{equation}
and the PGF of $D^{[m]}_n$ becomes
\begin{equation}
	\tilde{G}^{[m]}_n(t) = \frac{t^{n-1}\sum^{m-n+1}_{j=0} \binom{j+n-2}{n-2}q^j t^j}{\sum^{m-n+1}_{i=0}\binom{m}{i+n-1}p^i q^{m-n+1-i}}.
\end{equation}
Because it takes at least one time step for each segment to succeed, we have the inequalities $n \leqslant K^{[m]}_n$
and $n-1 \leqslant D^{[m]}_n \leqslant m$, which agree with the PGFs of these quantities presented above. Moreover, for
$m \to +\infty$ we have
\begin{equation}\label{eq:GGinf}
	G^{[+\infty]}_n(t) = G^{\mathrm{seq}}_n(t), \quad \tilde{G}^{[+\infty]}_n(t) = \tilde{G}^{\mathrm{seq}}_n(t).
\end{equation}
These relations are easy to prove, just note that
\begin{equation}
\begin{split}
	\sum^{m-n+1}_{i=0} &\binom{m}{i+n-1}p^i q^{m-n+1-i} \\
	&= \frac{1}{p^{n-1}} \left[1 - \sum^{n-2}_{i=0}\binom{m}{i}p^i q^{m-i}\right].
\end{split}
\end{equation}
The binomial coefficient $\binom{m}{i}$ is polynomial in $m$ of $i$-th degree, and thus $\binom{m}{i} q^m \to 0$ when $m
\to +\infty$ for all $i = 0, \ldots, n-2$, which proves the relations of Eq.~\eqref{eq:GGinf}.

There are also variations of the above sequential cutoff scheme.
In the previous scheme we only abort a round when we already waited $m$ time units. Now consider the case where we already waited $m/2$ time units, but only a small number of segments succeeded. Hence, it is highly unlikely that we will succeed in all segments within the $m$ time steps. Therefore, it is better not to waste time and already abort the current round to start from scratch. A very simple strategy following this idea makes use of an individual (local) cutoff in each segment. However, it is beneficial to use a different cutoff in every segment; one should choose a smaller cutoff in the first segments and then increase the cutoff for later segments. The rationale behind this is that in the first segments we have not invested much effort and can discard rather aggressively, whereas later we should discard less aggressively since we already consumed lots of resources.

The advanced protocol is uniquely defined by a vector of cutoffs $\vec{m}=(m_1,\dots,m_{n-1})$ and the random variables $K_n$ and $D_n$ for this protocol and general $n$ are given by
\begin{equation}
    K_{n}^{\mathrm{seq},\vec{m}}= \tilde{N}^{(m_{n-1})}+(T_{n-1}-1)m_{n-1}+\sum_{j=1}^{T_{n-1}} K_{n-1,j}^{\mathrm{seq},\vec{m}},
\end{equation}
where $K_1^{\mathrm{seq},\vec{m}}$ is geometrically distributed with parameter $p$, $\tilde{N}^{(m_{n-1})}$ follows a truncated geometric distribution with cutoff $m_{n-1}$, and $T_{n-1}$ is a geometric random variable with parameter $(1-q^{m_{n-1}})$ describing the number of starts of the protocol. For the dephasing we have
\begin{equation}
    D_{n}^{\mathrm{seq},\vec{m}}=\tilde{N}^{m_1}+\ldots+\tilde{N}^{m_{n-1}}.
\end{equation}

The PGF of $K_{n}^{\mathrm{seq},\vec{m}}$ is calculated in App.~\ref{app:SeqPGF} and given recursively by 
\begin{equation}
    G^{[\vec{m}]}_n(t)=\tilde{G}_2^{[m_{n-1}]}(t)t^{-m_{n-1}}P^{(m_{n-1})}\left(G_{n-1}^{[\vec{m}]}(t)t^{m_{n-1}}\right),
\end{equation}
where $P^{(m)}(t)=\frac{(1-q^m)t}{1-q^m t}$ and $G^{[\vec{m}]}_1=G^{seq}_1$.
The PGF of $D_{n}^{\mathrm{seq},\vec{m}}$ is simply given by 
\begin{equation}
   \tilde{G}^{[\vec{m}]}_{n}(t)=\prod_{j=1}^{n-1} \tilde{G}^{[m_j]}_2(t)\,, 
\end{equation}
since the sum of independent random variables translates to a product for PGFs.
As the state quality only depends on the total dephasing time, the best sequential protocol would count the total number of storage steps and would discard following a cutoff which is a function of the number of already succeeded segments, and one may also make use of the early aggressive discarding. 

\subsection{Parallel distribution schemes}

A more efficient class of schemes is constructed when we do not wait for some segments to finish before we start others.
In these schemes we start all segments independently and distribute in parallel. It follows that for these schemes without cut-off we
have
\begin{equation}
	K^{\mathrm{par}}_n = \max(N_1, \ldots, N_n),
\end{equation}
which means that all such schemes give the same raw rate. In App.~\ref{app:GKn} we derive the following expressions
for the PGF of $K_n$:
\begin{equation}\label{eq:GKn}
\begin{split}
	G^{\mathrm{par}}_n(t) &= t\sum^n_{i = 1}(-1)^{i+1} \binom{n}{i}\frac{1-q^i}{1 - q^i t} \\
    &= 1 + (1-t)\sum^n_{i=1} (-1)^i \binom{n}{i} \frac{1}{1-q^i t}.
\end{split}
\end{equation}
The two expressions are identical, since their difference reduces to $(1 - 1)^n = 0$. From the first expression it is clear that the values of $K_n$ start at 1, as it must be, because it takes at least one time unit to distribute
entanglement. In the other expression the necessary property of all PGFs becomes manifest, $G_n(1) = 1$. From the first relation
of Eqs.~\eqref{eq:PGF} we get the well-known expression for the average waiting time of a quantum repeater with parallel distribution and deterministic entanglement swapping (at any time when possible, e.g. at the very end)
\begin{equation}\label{eq:Knpar}
	\overline{K^{\mathrm{par}}_n} = \frac{\mathrm{d}}{\mathrm{d}t}G^{\mathrm{par}}_n(t) \Big\vert_{t=1} = \sum^n_{i=1} (-1)^{i+1} \binom{n}{i} \frac{1}{1 - q^i},
\end{equation}
which has been obtained in Ref.~\cite{PhysRevA.83.012323} (but the full waiting time probability distribution has not). Importantly, however, all other relevant expressions, the total number of
distribution steps including memory cut-off as well as the finally distributed quantum state including memory imperfections, both for the model with and without memory cut-off, depend on the particular swapping
strategy chosen (e.g. unnecessarily postponing some or even all entanglement swapping steps until the very end maximizes the amount of parallel storage and hence the total dephasing in the final state). For this, there is a growing number of choices for larger repeaters, and in the following
we shall derive an optimal swapping scheme that results in a minimal total dephasing time
(while sharing the high raw rates, i.e. the minimal total waiting times, with all parallel distribution schemes).

\subsubsection{Optimal swapping scheme}\label{sec:optimalswapscheme}

Because all schemes (without cut-off) considered in this subsection have equal raw rates, the best secret key rate is
determined by the optimal scheme with regards to the secret key fraction.
In this subsection we shall present this scheme. In contrast to the
schemes presented in Figs.~\ref{fig:4segD} and \ref{fig:4segI}, which are fixed, the optimal swapping scheme is
dynamic. In a fixed scheme the order of swappings is fixed at the beginning and does not depend on the order in which
the segments become ready. For example, for the ``doubling'' scheme as shown in Fig.~\ref{fig:4segD} for $n=4$, we never swap
segments 2 and 3, even if they are ready and segments 1 and 4 are not. We always wait for segments 1 and 2 or segments 3
and 4 to become ready, swap these pairs, and then swap the larger segments to finish the entanglement distribution over
the whole repeater. In a dynamical scheme we do not follow a prescribed order and can swap the segments based on their
state. Of course, we can freely mix and match fixed and dynamic behaviours. For example, for $n=8$, we can first swap four
pairs of segments in a fixed way and then swap the four new, larger segments dynamically. We now show that the fully
dynamic scheme, where we always swap the segments that are ready, is the optimal one.

To prove this statement, we give two characterizations of this fully dynamic scheme. One is the straightforward
translation of the description to the definition, but this definition is not explicitly optimal. The other one is
optimal by construction, but it is not fully dynamic explicitly. We then show that the two constructions coincide, which
will demonstrate the validity of our statement.

Swapping an earliest pair of segments means that we choose an index $i$ for which $\max(N_i, N_{i+1})$ is minimal (there
can be several such indices), swap the pair of segments $i$ and $i+1$, and recursively apply this procedure to the other
segments (if there are several such pairs, choose one of them arbitrarily). If we denote the dephasing random variable
of this scheme as $\tilde{D}_n$, then its formal definition reads as
\begin{widetext}
\begin{equation}\label{eq:Dtilde}
	\tilde{D}_n(N_1, \ldots, N_n) = |N_{i_0} - N_{i_0 + 1}|
	+\tilde{D}_{n-1}(N_1, \ldots, N_{i_0-1}, \max(N_{i_0}, N_{i_0+1}), N_{i_0+2}, \ldots, N_n),
\end{equation}
\end{widetext}
where $i_0 = \argmin_i \max(N_i, N_{i+1})$. This definition is a greedy, locally optimal scheme, which optimizes only
one step. As it is known from algorithm theory, greedy algorithms do not always produce globally optimal results. By
doing only locally optimal steps, we may miss an opportunity for a much better reward in the future if we make a
non-optimal step now. Fortunately, in this case the greedy, locally optimal scheme expressed by Eq.~\eqref{eq:Dtilde}
does give the globally optimal result, as we show below.

In any scheme, the first step will be to swap a pair of neighbouring segments, let us say segments $i$ and $i+1$. We do
this at the time moment $\max(N_i, N_{i+1})$, and the contribution of these segments to the total dephasing is $|N_i -
N_{i+1}|$. After this swapping, we are left with $n-1$ new segments, one of which is the combination of two original
ones. Any initial segment $j$, where $j \not= i, i+1$, generates an entangled state after $N_j$ time units, and the
combined segment ``generates'' entanglement after $\max(N_i, N_{i+1})$ time units. If we swap these $n-1$ segments in
any way in $D_{n-1}$ time units, then the total swapping takes $D_n = |N_i - N_{i+1}| + D_{n-1}$ time units. To find the
minimal dephasing we simply take the minimum over $i = 1, \ldots, n-1$ of this expression, and recursively apply it for
the new segments. If we denote the dephasing random variable corresponding to this scheme as $D^\star_n$, then this
description translates into the following definition:
\begin{equation}\label{eq:Dstar}
\begin{split}
	&D^\star_n(N_1, \ldots, N_n) = \min_{i = 1, \ldots, n-1}\Bigl[|N_i - N_{i+1}| \Bigr. \\
	\Bigl.&+ D^\star_{n-1}(N_1, \ldots, N_{i-1}, \max(N_i, N_{i+1}), N_{i+2}, \ldots, N_n) \Bigr].
\end{split}
\end{equation}
The base case of this recursive definition is $D^\star_2(N_1, N_2) \equiv D_2(N_1, N_2) = |N_1 - N_2|$. This definition
by construction gives the globally minimal number of dephasing time units required to distribute long-distance entanglement if it takes $N_i$ time units for segment $i$ to generate entanglement.

We now have two quantities, the locally optimal one, given by Eq.~\eqref{eq:Dtilde}, and the globally optimal one, given
by Eq.~\eqref{eq:Dstar}. The former has semantics of swapping the earliest, but may not be globally optimal. The latter
is optimal by construction, but does not necessarily correspond to the swapping earliest strategy. It turns out that the
two quantities coincide, at least for all $n = 2, \ldots, 8$. A straightforward way to check this is to consider all
possible inequality relations between $N_i$. There are $n!$ such relations, which correspond to the permutations of
$N_i$ in the following inequality
\begin{equation}\label{eq:N1Nn}
	N_1 \leqslant \ldots \leqslant N_n.
\end{equation}
For any given inequality relation between $N_i$ we can compute both quantities explicitly in terms of $N_i$. For
example, for the relation in Eq.~\eqref{eq:N1Nn} both quantities reduce to the same expression, $\tilde{D}_n = D^\star_n = N_n -
N_1$. For all other possible relations we have
\begin{equation}
	\tilde{D}_n(N_1, \ldots, N_n) = D^\star_n(N_1, \ldots, N_n),
\end{equation}
for all $n = 2, \ldots, 8$. This can be easily verified with the help of a computer algebra system. Our conjecture is
that the statement is valid for all $n \geqslant 2$, but in this work we consider repeaters with up to eight segments only,
and for such $n$ we have verified this statement directly.

In contrast to the sequential scheme introduced earlier, there is no compact expression for the PGF of the optimal
scheme here. Each case will be considered separately in the next subsections. Where possible, we present explicit expressions
of the PGFs of the quantities in question. The main difficulty is encountered for those schemes with memory cut-off, and hence when including a cut-off, even for smaller repeaters (but $n>2$) we only consider the fully sequential scheme, for which we have got the exact expressions.
In the following subsections, we discuss quantum repeaters for $n=2$, $3$, $4$, and $8$ segments.
Although the case $n=2$ is rather well known and there is no set of different swapping strategies to choose from in this case, it will 
be briefly reproduced based on the formalism introduced in this work.
The case $n=3$ is interesting, as it represents the simplest, nontrivial case beyond one middle station,
already requiring a choice regarding distribution and swapping strategies
(here, in the main text, the focus remains on schemes with an optimal dephasing for parallel distribution;
in App.~\ref{app:Optimality 3 segments}, we discuss the full secret key rate for $n=3$ including all possible distribution schemes).
Finally, the cases $n=4$ and $n=8$ are chosen, as they allow for a comparison with ``doubling''
(see Fig.~\ref{fig:4segD}). Larger quantum repeaters with $n>8$ become increasingly difficult to treat
(in terms of the optimized total dephasing). 
We will later also see that for $n=8$, without additional methods of quantum error detection or correction,
the necessary experimental parameter values in our model become already highly demanding.

\subsubsection{Two-segment repeater}

This is the simplest kind of a quantum repeater. The PGF $G_2(t)$ of $K_2 = \max(N_1, N_2)$ is given by
Eq.~\eqref{eq:GKn} with $n=2$ and in this case reads as
\begin{equation}
	G_2(t) = \frac{p^2 t (1 + qt)}{(1 - qt)(1 - q^2 t)}.
\end{equation}
As we noted before, there is only one choice for the dephasing variable, $D_2 = |N_1 - N_2|$ (parallel distribution). In Appendix~\ref{app:PGF Parallel schemes}, we derive
the following expression for the PGF of this variable:
\begin{equation}
	\tilde{G}_2(t) = \frac{p^2}{1 - q^2} \frac{1 + q t}{1 - q t}.
\end{equation}
There we also show that the PGFs of the variables with cut-offs are
\begin{equation}
\begin{split}
	G^{[m]}_2(t) &= \frac{p^2 t (1 + qt - 2(qt)^{m+1})}{(1 - qt)(1 - q^2 t - 2p (qt)^{m+1})}, \\
	\tilde{G}^{[m]}_2(t) &= \frac{p}{1 + q - 2q^{m+1}} \frac{1 + qt - 2(qt)^{m+1}}{1 - qt}.
\end{split}
\end{equation}
It is obvious that we have the same consistency relations as for the sequential distribution scheme:
\begin{equation}
	G^{[+\infty]}_2(t) = G_2(t), \quad \tilde{G}^{[+\infty]}_2(t) = \tilde{G}_2(t).
\end{equation}

\subsubsection{Three-segment repeater}\label{sssec:Par-distr: 3-segment repeater}
For three segments there are various ways how to distribute entanglement. One could use a fully sequential scheme, start at one end and distribute entanglement in concurrent segments. Alternatively, one could consider schemes where pairs of segments generate entanglement in parallel and the remaining segment goes last or, the other way around, it goes first. There are also combined distribution schemes with ``overlapping" parallel and sequential distributions. Finally, there are those schemes which attempt to generate entanglement in all segments at once and thereby use different swapping schemes. Among the latter here only the potentially optimal scheme is of interest, as it minimizes the accumulated dephasing, while having the same total waiting time as any other parallel distribution scheme.

However, it could still be the case that a scheme from the other, slower class of schemes performs better in terms of the full secret key rate. This is possible, because there is typically a trade-off between the raw rate and the dephasing or, more generally, the QBER. In particular, the fully sequential distribution scheme is interesting, since its total dephasing becomes minimal, as there is basically always only one segment waiting at every time step. On the other hand, for the fully parallel schemes the raw rate is optimal.

In App.~\ref{app:Optimality 3 segments} we present all possible schemes for $n=3$ and calculate the PGFs of their total waiting and dephasing times. Then we use these results to obtain the secret key rate for each scheme and to compare the different schemes. We also show in the appendix that the PGF of the optimal dephasing random variable, equivalently defined by Eqs.~\eqref{eq:Dtilde} and \eqref{eq:Dstar}, reads as
\begin{equation}
	\tilde{G}^\star_3(t) = \frac{p^3}{1-q^3} \frac{1 + (q+2q^2)t - (2q^2+q^3)t^3 - q^4 t^4}{(1-qt)(1-q^2t)(1-qt^2)}.
\end{equation}

It turns out that with regards to the full secret key rate the parallel-distribution optimal-dephasing scheme is indeed optimal in all relevant regimes and especially in the limit of improving hardware parameters, which can be seen in Fig.~\ref{fig:Comparison_3_segments_non tau=0.1} and Fig.~\ref{fig:Comparison_3_segments_non tau=10} for two different memory coherence times. In the same section one can also find a more detailed discussion of the figure. In addition, aiming at the most general treatment of the $n=3$ case, we also consider the scenario where Alice and Bob measure their qubits immediately, thus suppressing their memory dephasing, and we apply this to all possible schemes. The comparison of these ``immediate-measurement" schemes is shown in Fig.~\ref{fig:Comparison_3_segments_immediate tau=0.1} and Fig.~\ref{fig:Comparison_3_segments_immediate tau=10}, again for two different coherence times. The conclusion remains the same: overall ``optimal" is optimal.
However, note that the option with immediate measurements for Alice and Bob only exists when they operate the quantum repeater for the purpose of long-range QKD. More advanced quantum repeater applications may require quantum storage for the qubits at each end (user) node. In any case, the memory qubits at each intermediate repeater node are (jointly) measured as soon as possible when the two adjacent segments are filled with an entangled pair (or even later, depending on the particular swapping strategy, but in App.~\ref{app:Optimality 3 segments} we only consider swap-as-soon-as-possible schemes that minimize the dephasing).

The above discussion leads us to the conclusion that there are three basic properties that a quantum repeater protocol (unassisted by additional quantum error detection or correction) should exhibit: distribute the entangled states in each segment in parallel, swap the initially distributed states as soon as possible, and avoid parallel storage of already distributed pairs as much as possible. It is obvious that all these three``rules" cannot be fully obeyed at the same time. However, our optimal scheme has the optimal balance with regards to these rules for three segments. We conjecture that this also holds true for larger $n>3$-segment repeaters.

\subsubsection{Four-segment repeater}\label{sssec:Par-distr: 4-segment repeater}

Of particular interest to us is the case of a four-segment repeater which is commonly operated via ``doubling". Here we are now able to discuss more general schemes, especially those that would always swap as soon as possible, unlike doubling where the second and third segments may not be immediately connected even when they are both ready. Overall there are many more schemes than in the previous $n=3$ case, and here for $n=4$ we focus on the parallel-distribution schemes. All these schemes (without cut-off) have identical $K_4 = \max(N_1, N_2, N_3, N_4)$, whose PGF is given by Eq.~\eqref{eq:GKn}
for $n=4$. The dephasing variable $D_4$ and its PGF become different for different schemes. One such scheme, the common ``doubling", is illustrated in
Fig.~\ref{fig:4segD}, where we first swap the pairs of segments 1, 2 and 3, 4 independently and then swap the two larger
segments. Note that the swappings will typically take place at different
moments in time - one pair of segments will usually swap earlier than the other. The state of the faster pair that goes into the final swapping operation is the state of these segments after their connection and at the moment when the final swapping is done, and so the state has been subject to a corresponding memory dephasing. For example, if the swapping of segments
1 and 2 is done first, the state of the distributed state over segments 1 and 2 just after the swapping is
$\hat{\varrho}_{14} = \mathcal{S}(\hat{\varrho}_{12} \otimes \hat{\varrho}_{34})$. If $k$ time units later segments 3
and 4 swap, producing the state $\hat{\varrho}_{58} = \mathcal{S}(\hat{\varrho}_{56} \otimes \hat{\varrho}_{78})$, the
former state becomes $\Gamma_{k\alpha}(\hat{\varrho}_{14})$, and the state distributed over the whole repeater is
\begin{equation}\label{eq:Sk}
	\hat{\varrho}_{18} = \mathcal{S}(\Gamma_{k\alpha}(\mathcal{S}(\hat{\varrho}_{12} \otimes \hat{\varrho}_{34})) \otimes
	\mathcal{S}(\hat{\varrho}_{56} \otimes \hat{\varrho}_{78})),
\end{equation}
instead of just $\hat{\varrho}_{18} = \mathcal{S}(\mathcal{S}(\hat{\varrho}_{12} \otimes \hat{\varrho}_{34}) \otimes
\mathcal{S}(\hat{\varrho}_{56} \otimes \hat{\varrho}_{78}))$. Again, as before, we omitted any extra factors that depend on the number of spins subject to dephasing in a single repeater segment. So, Fig.~\ref{fig:4segD} shows just a workflow of swapping
operations, while the exact expressions should be adjusted according to the respective time differences. The dephasing variable $D_4$ in
this doubling scheme is defined as follows:
\begin{equation}
\begin{split}\label{eq:doublingvariable}
	D^{\mathrm{dbl}}_4 &= |N_1 - N_2| + |N_3 - N_4| \\
	&+ |\max(N_1, N_2) - \max(N_3, N_4)|.
\end{split}
\end{equation}
The first two terms are due to the possible time difference for generating entangled states within each pair of
segments. The last term is due to the time difference between the pairs (e.g. the difference of the two maxima is $k$ time steps in
Eq.~\eqref{eq:Sk}). Note that this particular form of $D^{\mathrm{dbl}}_4$ is consistent with the commonly used "doubling" where the initial distributions happen in parallel, but the swapping strategy is fixed and sometimes disallows to swap as soon as possible. In Appendix~\ref{app:PGF Parallel schemes}, we derive the PGF of this random dephasing variable,
\begin{equation}
	\tilde{G}^{\mathrm{dbl}}_4(t) = \frac{p^4}{1-q^4} \frac{P^{\mathrm{dbl}}_4(q, t)}{Q^{\mathrm{dbl}}_4(q, t)},
\end{equation}
where the numerator and denominator are given by
\begin{displaymath}
\begin{split}
	P^{\mathrm{dbl}}_4(q, t) &= 1 + (q^2+3q^3)t + (3q+3q^2-q^5)t^2 \\
	&- (q^3-q^5)t^3 + (q^3-3q^6-3q^7)t^4 \\
	&- (3q^5+q^6)t^5 - q^8t^6, \\
	Q^{\mathrm{dbl}}_4(q, t) &= (1-q^2t)(1-q^3t)(1-qt^2)(1-q^2t^2).
\end{split}
\end{displaymath}

The dephasing variable corresponding to the iterated scheme as shown in Fig.~\ref{fig:4segI} differs from that of the doubling
scheme. In the iterative scheme we first distribute entanglement over segments 1 and 2, then extend it over segment 3,
and finally over segment 4. Note that the figure can be understood to illustrate both sequential distribution and iterated swapping. In the
sequential distribution scheme, we would start to generate entanglement in each segment only when all previous segments (e.g. from left to right) have successfully generated entanglement. In the iterated swapping scheme, all segments may start simultaneously (parallel distribution), thus increasing the chances to swap sooner, but also the number of qubits potentially stored in parallel. The variable
$D^{\mathrm{itr}}_4$ for this scheme is
\begin{displaymath}
\begin{split}
	D^{\mathrm{itr}}_4(N_1, N_2, N_3, N_4) &= |N_1 - N_2| + |\max(N_1, N_2) - N_3| \\
	&+ |\max(N_1, N_2, N_3) - N_4|.
\end{split}
\end{displaymath}
The PGF of this random variable is rather large and reads as
\begin{equation}
	\tilde{G}^{\mathrm{itr}}_4(t) = \frac{p^4}{1-q^4} \frac{P^{\mathrm{itr}}_4(q, t)}{Q^{\mathrm{itr}}_4(q, t)},
\end{equation}
where the numerator and denominator are given by
\begin{displaymath}
\begin{split}
	P^{\mathrm{itr}}_4(q, t) &= 1+3q^3t+(4q^2-q^4-2q^5)t^2 \\
	&+(q-q^2-3q^3-6q^4+2q^5+q^6)t^3\\
	&+(-2q^2-5q^3+q^4+2q^5-q^6-3q^7)t^4\\
	&+(-2q^2+4q^4-4q^6+2q^8)t^5\\
	&+(3q^3+q^4-2q^5-q^6+5q^7+2q^8)t^6 \\
	&+(-q^4-2q^5+6q^6+3q^7+q^8-q^9)t^7\\
	&+(2q^5+q^6-4q^8)t^8-3q^7t^9-q^{10}t^{10}, \\
	Q^{\mathrm{itr}}_4(q, t) &= (1-qt)(1-q^2t)(1-q^3t)(1-qt^2)\\
	&\times (1-q^2t^2)(1-qt^3).
\end{split}
\end{displaymath}

We present an example for another, mixed swapping strategy
in App.~\ref{app:mixedstr}.

For the dephasing random variable $D^\star_4$, corresponding to the optimal swapping scheme given by
Eq.~\eqref{eq:Dstar} for $n=4$, we derive the following PGF:
\begin{equation}
	\tilde{G}^\star_4(t) = \frac{p^4}{1-q^4} \frac{P^\star_4(q, t)}{Q^\star_4(q, t)},
\end{equation}
where the numerator and denominator read as
\begin{displaymath}
\begin{split}
	P^\star_4(q, &t) = 1 + (q+2q^2+3q^3)t + (q+2q^2+q^4)t^2 \\
	&-(3q^2+4q^3+4q^4)t^3 - (4q^5+4q^6+3q^7)t^4 \\
	&+ (q^5+2q^7+q^8)t^5 + (3q^6+2q^7+q^8)t^6 + q^9t^7, \\
	Q^\star_4(q, &t) = (1-qt)(1-q^2t)(1-q^3t)(1-qt^2)(1-q^2t^2).
\end{split}
\end{displaymath}

\subsubsection{Eight-segment repeater}\label{sssec:Par-distr: 8-segment repeater}

As before, again all parallel-distribution schemes (without cut-off) have identical total waiting times, 
$K_8 = \max(N_1, \ldots, N_8)$, whose PGF is given by
Eq.~\eqref{eq:GKn} for $n=8$. For the dephasing variable there are many more possibilities now. We shall consider and compare five
different schemes -- the doubling and the optimal schemes, and three less important schemes, which nevertheless exhibit
an interesting behavior. The somewhat less important ones
are described and discussed in App.~\ref{app:mixedstr}.

The optimal dephasing $D^\star_8$ is defined equivalently by 
Eqs.~\eqref{eq:Dtilde}-\eqref{eq:Dstar} for $n=8$ and the doubling dephasing $D^{\mathrm{dbl}}_8$ is defined recursively
as
\begin{equation}
\begin{split}
	D^{\mathrm{dbl}}_8&(N_1, \ldots, N_8) = D^{\mathrm{dbl}}_4(N_1, \ldots, N_4) \\
	&+ D^{\mathrm{dbl}}_4(N_5, \ldots, N_8) \\
	&+ |\max(N_1, \ldots, N_4) - \max(N_5, \ldots, N_8)|,
\end{split}
\end{equation}
with $D^{\mathrm{dbl}}_4$ defined as in Eq.~\eqref{eq:doublingvariable}.
The comparison of the five different schemes can be found in App.~\ref{app:mixedstr}.
In this appendix, App.~\ref{app:mixedstr}, we present some figures
showing the ratios between the average dephasing
of the four sub-optimal schemes and the optimal scheme,
with and without exponentiation.   
We can then compare the relative positions of the curves in Fig.~\ref{fig:Ee} with those of the curves of the ratios
\begin{equation}\label{eq:r3}
	\frac{\mathbf{E}[D^{\mathrm{sch}}_8]}{\mathbf{E}[D^{\mathrm{opt}}_8]} =
	\frac{\tilde{G}^{\mathrm{sch}\prime}_8(1)}{\tilde{G}^{\mathrm{opt}\prime}_8(1)},
\end{equation}
which are shown in Fig.~\ref{fig:Ea}. Looking at the two figures, we see that
\begin{equation}
	\mathbf{E}[D^{\mathrm{dbl}}_8] > \mathbf{E}[D^{44}_8], \quad
	\mathbf{E}[e^{-\alpha D^{\mathrm{dbl}}_8}] < \mathbf{E}[e^{-\alpha D^{44}_8}].
\end{equation}
This behavior is in full agreement with the properties of the exponential function: if $x > y \geqslant 0$ and $\alpha >
0$, then $e^{-\alpha x} < e^{-\alpha y}$. But for the other pair of schemes we have
\begin{equation}\label{eq:DD}
	\mathbf{E}[D^{242}_8] > \mathbf{E}[D^{2222}_8], \quad
	\mathbf{E}[e^{-\alpha D^{242}_8}] > \mathbf{E}[e^{-\alpha D^{2222}_8}].
\end{equation}
Nonetheless there is no contradiction here. This is a known property of nonlinear functions of random variables. This property can be
observed even in the simplest case of random variables $X$ and $Y$ each taking two values only. One can easily construct
an example such that $\mathbf{E}[X] > \mathbf{E}[Y]$ and $\mathbf{E}[e^{-\alpha X}] > \mathbf{E}[e^{-\alpha Y}]$. However,
the inequalities \eqref{eq:DD} show that it is not necessary to consider artificial constructions. This property can be
observed for simple and natural schemes.

The important conclusion is that the optimal scheme by construction minimizes $\mathbf{E}[D]$, but to have the highest fidelity of the
distributed state we need to maximize $\mathbf{E}[e^{-\alpha D}]$. For an ordinary nonnegative function $f(x)$ and a
positive parameter $\alpha > 0$ the minimum of $f(x)$ is the maximum of $e^{-\alpha f(x)}$ and vice versa, but for
random variables this is not necessarily true. Strictly speaking, in general, we know only the scheme that minimizes
$\mathbf{E}[D]$, but not the scheme that maximizes $\mathbf{E}[e^{-\alpha D}]$. The two schemes seem to be identical, but
there is no strict proof of this statement. We have to rely on evidence based on computing the properties of some schemes
explicitly and comparing them. For the examples for $n=8$ given in this section
and in the appendix, we see that dividing the exponentiated dephasing of 
all other schemes by that of the optimal scheme gives a number smaller than one,
whereas the same ratios without exponentiation give a number greater than one.
Thus, minimal dephasing corresponds to minimal dephasing errors,
and the optimal dephasing scheme exhibits the smallest fraction of dephasing errors. 

To summarize, our optimization of the secret key rates obtainable 
with different distribution and swapping strategies is based
on three steps. First, we can rely upon the proof of the minimal dephasing variable for up to $n=8$ segments given in Sec.~\ref{sec:optimalswapscheme}
assuming parallel initial distributions
(it is already non-trivial to extend this proof to larger $n>8$).
Second, in order to compare the average dephasing errors in the final density operators, we need to consider the average dephasing exponentials for the different schemes. Finally, in order to assess the optimality of the secret key rate over all possible schemes, we have to take into account also those schemes where the initial distributions no longer occur in parallel which generally leads to smaller raw rates, but at the same time can result in a smaller dephasing by (partially) avoiding parallel storage. For the first non-trivial case beyond a single middle station, we have explicitly gone through all these three steps, namely for the case of a three-segment repeater with two intermediate stations (App.~\ref{app:Optimality 3 segments}), and found that ``optimal" is optimal. For larger repeaters beyond eight segments, $n>8$, we conjecture that our ``optimal" scheme also gives the best secret key rate. This includes conjecturing that our minimized dephasing is minimal also for $n>8$, that it minimizes the dephasing errors in the final density operator, and that overall the dephasing-optimized parallel-distribution approach is superior to any partially or fully sequential distribution scheme. Especially the last point cannot be taken for granted. In App.~\ref{app:8segmentsimmediate} we present some rate calculations for $n=8$ where, beyond a certain distance, ``optimal" can be beaten by a sequential scheme. However, there we allow for immediate measurements at an end node only for the sequential scheme (for which this is easy to include), but not for ``optimal"; a comparison which is slightly unfair and also only relevant for QKD applications. In the case of non-immediate-measurement schemes including potential beyond-QKD applications, ``optimal" remains optimal.

\section{Secret key rate analysis}\label{sec:Secret Key Rate}
A useful and practically relevant figure of merit for quantifying a quantum repeater's performance is its secret key rate in long-range QKD, which determines the amount of secret key generated in bits per channel use or second. As briefly reviewed in Sec.~\ref{sec:skr}, the secret key rate consists of two parts: the raw rate or yield and the secret key fraction. The former quantifies how long it takes to send a raw quantum bit or to (effectively) generate entanglement, independent of the quality of the final state; the latter then determines the average amount of secret key that can be extracted from a single raw bit, depending on the particular QKD protocol chosen and including the corresponding procedures for the classical post-processing.  

Here we will focus on the asymptotic BB84 secret key rate $S=Rr=r/T$ with one-way post-processing. In the most general scenario of long-range memory-assisted QKD, i.e.
including a finite swapping probability $a$ and a memory cut-off parameter $m$,
this secret key rate is given by
\begin{equation}\label{eq:secret key rate}
    S(p,a,m)=\frac{1-h(\overline{e_x}(p,a,m))-h(\overline{e_z}(p,a,m))}{T(p,a,m)},
\end{equation}
where \(h\) is the binary entropy function, \(T\) the average number of steps needed to successfully distribute long-distance entanglement, and \(e_x\), \(e_z\) are the QBERs of Eq.~\eqref{eq:QBER}. 
The probability of successful entanglement generation in a single attempt in a single elementary segment is $p$, as introduced in Sec.~\ref{sec:rawrate}.
The denominator of $S$, $T = \mathbf{E}[K]$, is basically the total raw waiting time of the repeater which generally depends on $p$ and $a$ where $a$ is a finite success probability of the entanglement swapping using the same notation
as in Refs.~\cite{PvL,Shchukin2021} (where it was shown how to compute \cite{PvL} and optimize \cite{Shchukin2021}
$T=\mathbf{E}[K]$ for arbitrary $a$).
The dependency on the cut-off parameter $m$ means: the smaller $m$ becomes, the longer it takes to distribute an entangled state. The numerator of $S$, $r$, generally also depends on $p$, $a$, and $m$ through the QBERs. 
Recall that we have to take the averages here, $\overline{e_z} = e_z$ and $\overline{e_x}$ obtainable via $\mathbf{E}[e^{-\alpha D_n}]$. 
A smaller $m$ can lead to a higher state quality with a smaller total dephasing and thus to a larger secret key fraction $r$. It is generally hard to optimize $S$ over general $p$, $a$, and $m$. Our approach here is based on the simplifying (and experimentally still relevant) assumption $a=1$ (deterministic entanglement swapping) and the idea that the highest secret key rates will be obtainable with the fastest schemes (parallel distributions minimizing the total waiting time) and, among these, with those that swap entanglement as soon as possible (minimizing the total dephasing time, see Sec.~\ref{sec:optimalswapscheme}). While for a two-segment repeater the cases of deterministic and non-deterministic swapping can be treated similarly, for repeater chains with more than a single middle station ($n>2$) our results for optimizing distribution and swapping strategies only hold for the deterministic swapping case. Using the results of all previous sections the secret key rate can then be calculated. Therefore, in what follows we always have $a=1$. 

The above secret key rate $S$ is expressed in terms of bits per channel use. For a rate per second, the average total number of distribution attempts $T$ must by multiplied with the duration of a single attempt in seconds, i.e. the elementary time unit $\tau = L_0/c_f$. Note that a single attempt or channel use is uniquely defined only for direct channel transmission in a point-to-point link, whereas the channel in a quantum repeater is used directly only between neighboring memory stations. Since our model always assumes that the interfaces at each station connect a single channel (to the left or to the right) with a single memory qubit (unit memory ``buffer"), those channel segments that belong to already successfully distributed pairs remain unused until new attempts in these segments will be started (e.g. when the memory cut-off has been exceeded or when a long-distance pair has been finally created). Nonetheless, at every attempt, we shall always count a full channel use over the entire distance despite the growing number of unused channel segments during memory-assisted long-distance entanglement distribution. Thus, strictly speaking, we underestimate the secret key rate per channel use and one could continue distributing pairs in all channel segments provided sufficient memory qubits are available.

The parameter values as given in Tab.~\ref{tab:constants} have been used to obtain the quantitative results discussed in this section.  
Most parameters there have been introduced in the previous sections in the context of our physical model.
The resulting probability to distribute entanglement over one link in terms of the parameters of Tab.~\ref{tab:constants} now includes a
zero-distance link-coupling efficiency 
\begin{equation}
p(L_0)=p_{\mathrm{link}}\cdot e^{-\frac{L_0}{L_{\mathrm{att}}}},
\end{equation}
with $p(0) = p_{\mathrm{link}}$ and where $p_{\mathrm{link}} = \eta_\mathrm{c} \cdot \eta_\mathrm{d} \cdot \eta_\mathrm{p}$
incorporates various efficiencies of the experimental hardware independent of the channel transmission itself, especially wavelength conversion, fiber coupling, preparation, and detector efficiencies.

\begin{table*}
	\begin{tabular}{c|c|c|c}
		Constant & Meaning & Current value & Improved value \\
		\hline
		\hline
		$a$ & swapping probability & $1$ & $1$\\
		$\tau_{\mathrm{coh}}$ & coherence time & $\unit[0.1]{s}$ & $\unit[10]{s}$ \\
		$\mu$ & gate depolarisation (Bell measurement) & $0.97$ & $1$ \\
		$\mu_0$ & initial state depolarisation & $0.97$ & $1$ \\
		$F_0$ & initial state fidelity (dephasing) &  $1$ & $1$ \\
		$L_{\mathrm{att}}$ & attenuation length & $\unit[22]{km}$ & $\unit[22]{km}$ \\
		$n_\mathrm{r}$ & index of refraction & $1.44$ & $1.44$ \\
		$\eta_\mathrm{p}$ & preparation efficiency & * & * \\
		$\eta_\mathrm{c}$ & \begin{tabular}{@{}c@{}}
			photon-fibre coupling efficiency $\times$\\
			wavelength conversion\\
		\end{tabular}  & * & *\\
		$\eta_\mathrm{d}$ & detector efficiency & * & * \\
		\hline
		$p_{\mathrm{link}}:=\eta_\mathrm{c} \cdot \eta_\mathrm{d} \cdot \eta_\mathrm{p}$ & total efficiency & $0.05$ & $0.7$
	\end{tabular}
	\caption{Experimental parameter values used to calculate secret key rates. The star symbols * allow for various choices. The exact choices vary for each experimental platform. Some of the ``improved values" are the ideal values which allow to consider idealized, fundamental scenarios such as ``channel-loss-only" or ``channel-loss-and-memory-dephasing-only" (for which we may also set $p_{\mathrm{link}}=1$).}
	\label{tab:constants}
\end{table*}

In the context of our statistical and physical model
the memory coherence time \(\tau_{\mathrm{coh}}\) in Tab.~\ref{tab:constants},
an experimentally determined parameter that describes the average speed of the memory dephasing, can be converted into a (dimensionless) effective coherence time in units of the repeater's elementary time unit, $\tau_{\mathrm{coh}}/\tau$. Equivalently, we can say that the (number of) dephasing time (steps) $D_n$ is to be multiplied with an elementary time $\tau$ before it can be divided by $\tau_{\mathrm{coh}}$ in $\mathbf{E}[e^{-D_n \tau/\tau_{\mathrm{coh}}}]$. In any case, we absorb both $\tau$ and $\tau_{\mathrm{coh}}$ in our dimensionless $\alpha$ dephasing parameter,
\begin{equation}
\alpha(L_0)=\frac{\tau}{\tau_{\mathrm{coh}}}=\frac{L_0}{c_f \tau_{\mathrm{coh}}}.
\end{equation}
Thus, $\alpha$ can be referred to as an inverse effective coherence time.
Note that in order to count the dephasing times appropriately in a specific protocol, we may have to add an extra factor of 2 (depending on the number of spins dephasing at each time step in a certain elementary or extended segment) and a constant dephasing term $\sim 2n$ that takes into account memory dephasing that occurs even when the first distribution attempt in a segment succeeds.
Any missing factors in the dephasing can be reinterpreted in terms of $\alpha$ or $\tau_{\mathrm{coh}}$, e.g. a missing factor of 2 corresponds to a coherence time twice as big. 

In Tab.~\ref{tab:constants}, two sets of current and improved parameter values are listed, which specifically refer to $\tau_{\mathrm{coh}}$ and $p_{\mathrm{link}}$ for which we choose 0.1s or 10s and 0.05 or 0.7, respectively. The other state and gate fidelity parameters will be either set to unity or close to but below one
(in some of the following plots we will also treat them as a free parameter).
We will see that in memory-assisted QKD without additional quantum error detection or correction, the fidelity parameters must always be above a certain threshold value which (obviously) grows with the number of stations (and which generally depends on the particular QKD protocol and the classical post-processing method).

To compare the performance of each repeater protocol with a direct point-to-point link over the total distance $L$, we will use the PLOB bound \cite{PLOB}, which is given by
\begin{equation}
 S^{\mathrm{PLOB}}(L)=-\log_2(1-e^{-\frac{L}{L_{\mathrm{att}}}}).
\end{equation}
It represents an upper bound on the number of secret bits that can be shared per channel use. For example, for $e^{-\frac{L}{L_{\mathrm{att}}}}=1/2$ corresponding to $L=15$km, we have $S^{\mathrm{PLOB}} = 1$, and so at most one secret bit can be distributed per channel use (per mode) independent of the optical encoding. 
It will also be useful to consider an upper bound on the number of secret bits that can be shared with the help of a quantum repeater \cite{PLOB_QR},

\begin{equation}
 S^{\mathrm{PLOB,QR}}(L_0)=-\log_2(1-e^{-\frac{L_0}{L_{\mathrm{att}}}}),
\end{equation}
corresponding to the PLOB bound for one segment (in the case of equal segment lengths $L_0$). For a point-to-point link, $n=1$ with $L=L_0$, we thus use the notation $S^{\mathrm{PLOB}}=S^{\mathrm{PLOB,QR}}$.
The rates we will focus on first in the following are to be understood as secret key rates per channel use. Later we shall also discuss secret key rates per second.

\subsection{Two-segment repeater}\label{sec:Two-Segment Repeater}

Let us start with the rates for the simplest case: a two-segment quantum repeater with one middle station. We shall only consider one scheme, the ``optimal" scheme, with and without a memory cut-off.
First, we address the question whether and when it is possible to overcome the PLOB bound with a two-segment repeater given the (current and improved) parameter values from Tab.~\ref{tab:constants}. We stick to \(F_0=1\) and, for illustrative clarity, we set \(\mu=\mu_0\) (while, first, $\mu$ is not fixed). Physically, this means that the repeater states when initially distributed in each segment and then manipulated at the middle station for the Bell measurement are subject to the same depolarizing error channels (and there is no extra initial dephasing). The cut-off parameter \(m\) is chosen most appropriately such that the final secret key rate is close to optimal over the entire range. 

\begin{figure*}[ht]
	\centering
	\subfloat[a][$\tau_{\mathrm{coh}}={\unit[0.1]{s}}$, $p_{\mathrm{link}}=0.05$, $m=10$]{\includegraphics[width=0.5\linewidth]{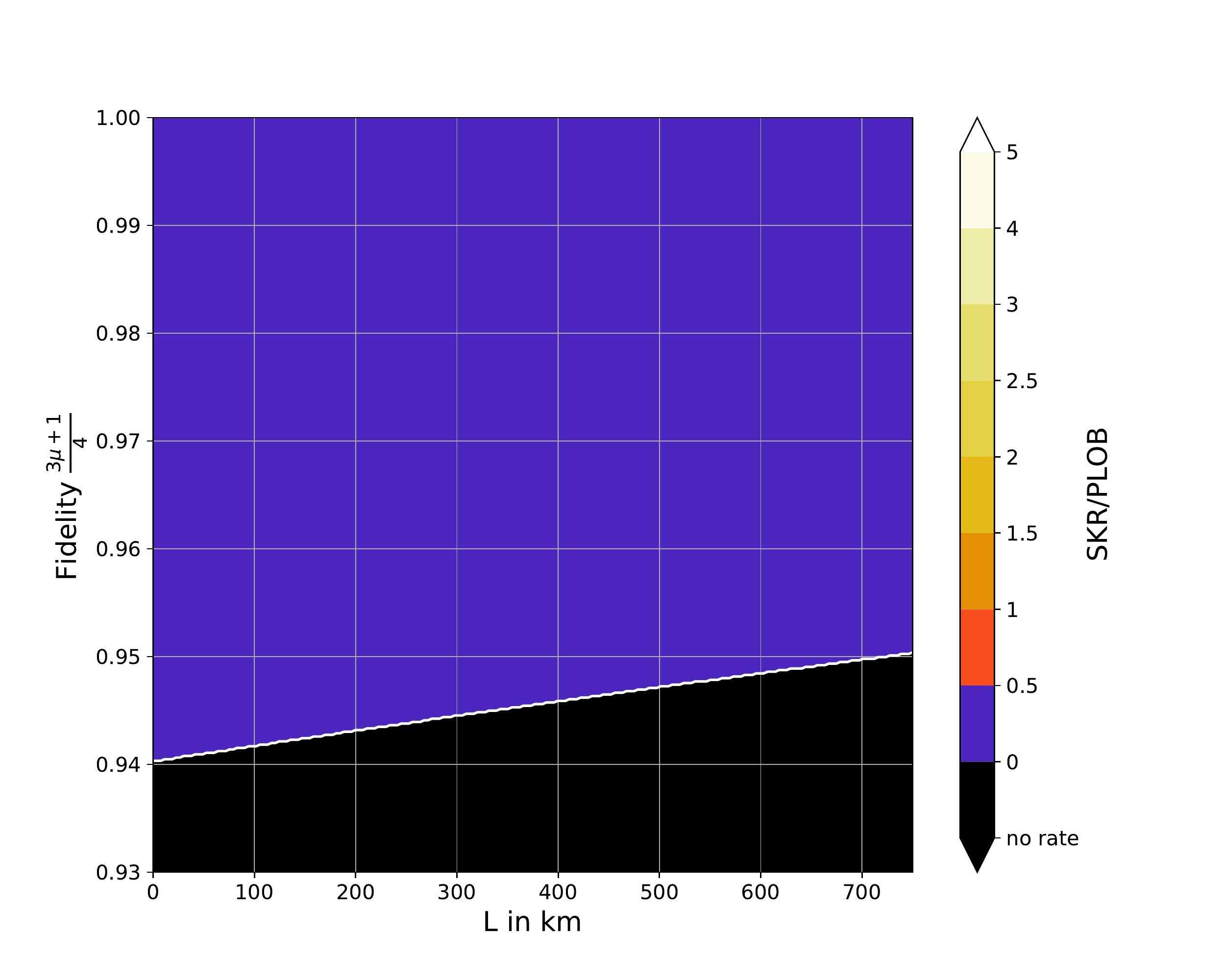}}
	\subfloat[][$\tau_{\mathrm{coh}}={\unit[0.1]{s}}$, $p_{\mathrm{link}}=0.7$, $m=50$]{\includegraphics[width=0.5\linewidth]{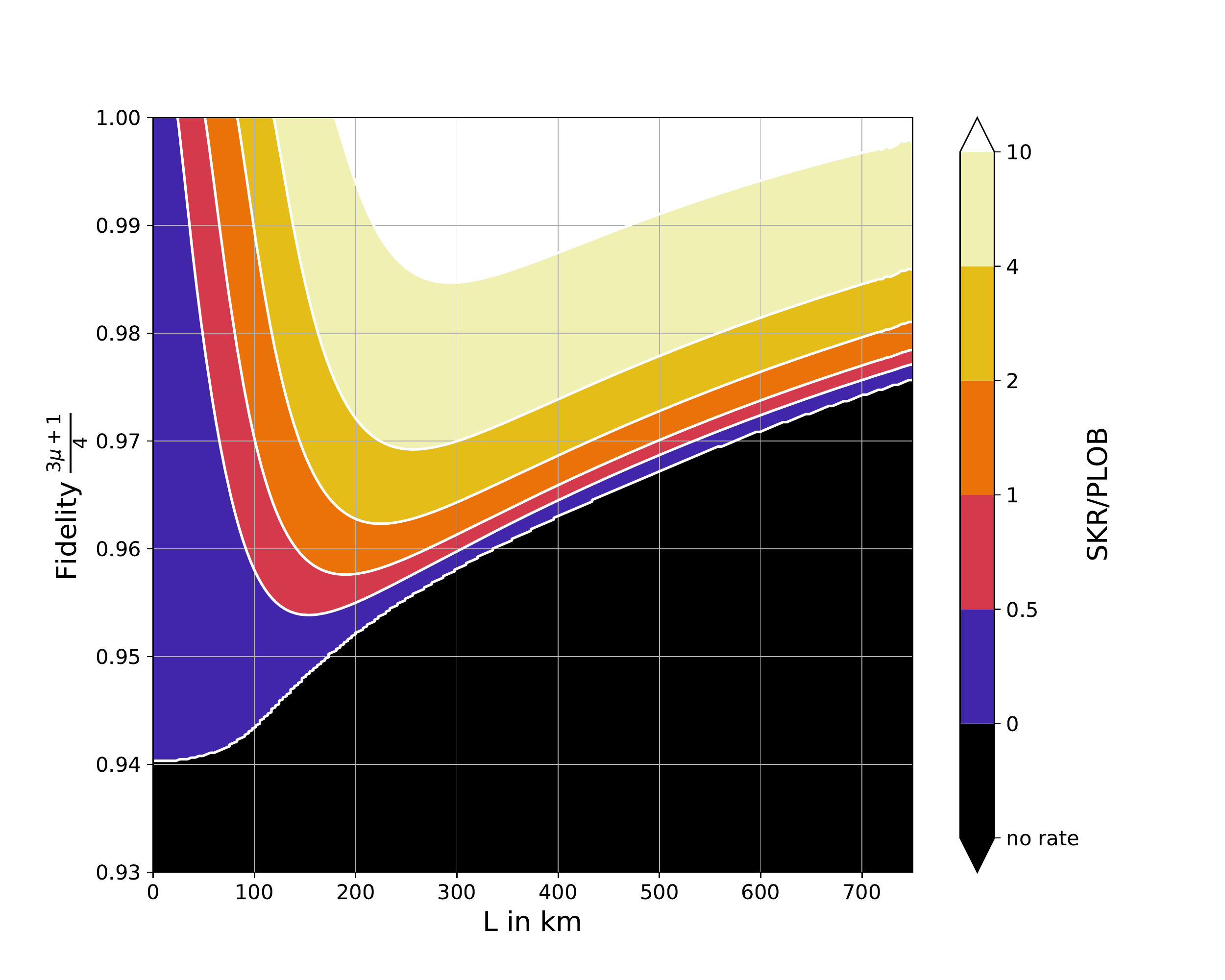}} \\
	\subfloat[][$\tau_{\mathrm{coh}}={\unit[10]{s}}$, $p_{\mathrm{link}}=0.05$, $m=3000$]{\includegraphics[width=0.5\linewidth]{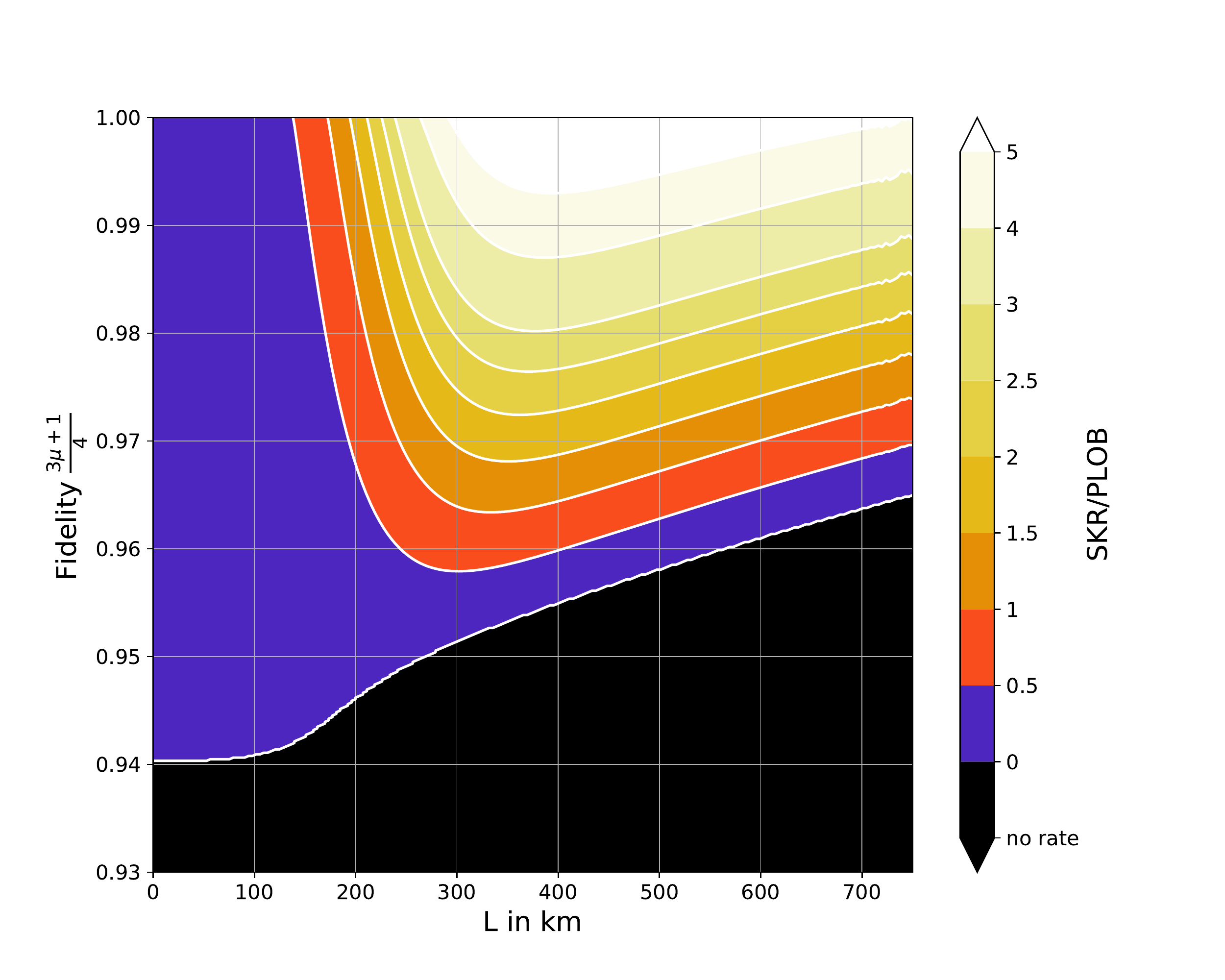}} 
	\subfloat[][$\tau_{\mathrm{coh}}={\unit[10]{s}}$, $p_{\mathrm{link}}=0.7$, $m=5000$]{\includegraphics[width=0.5\linewidth]{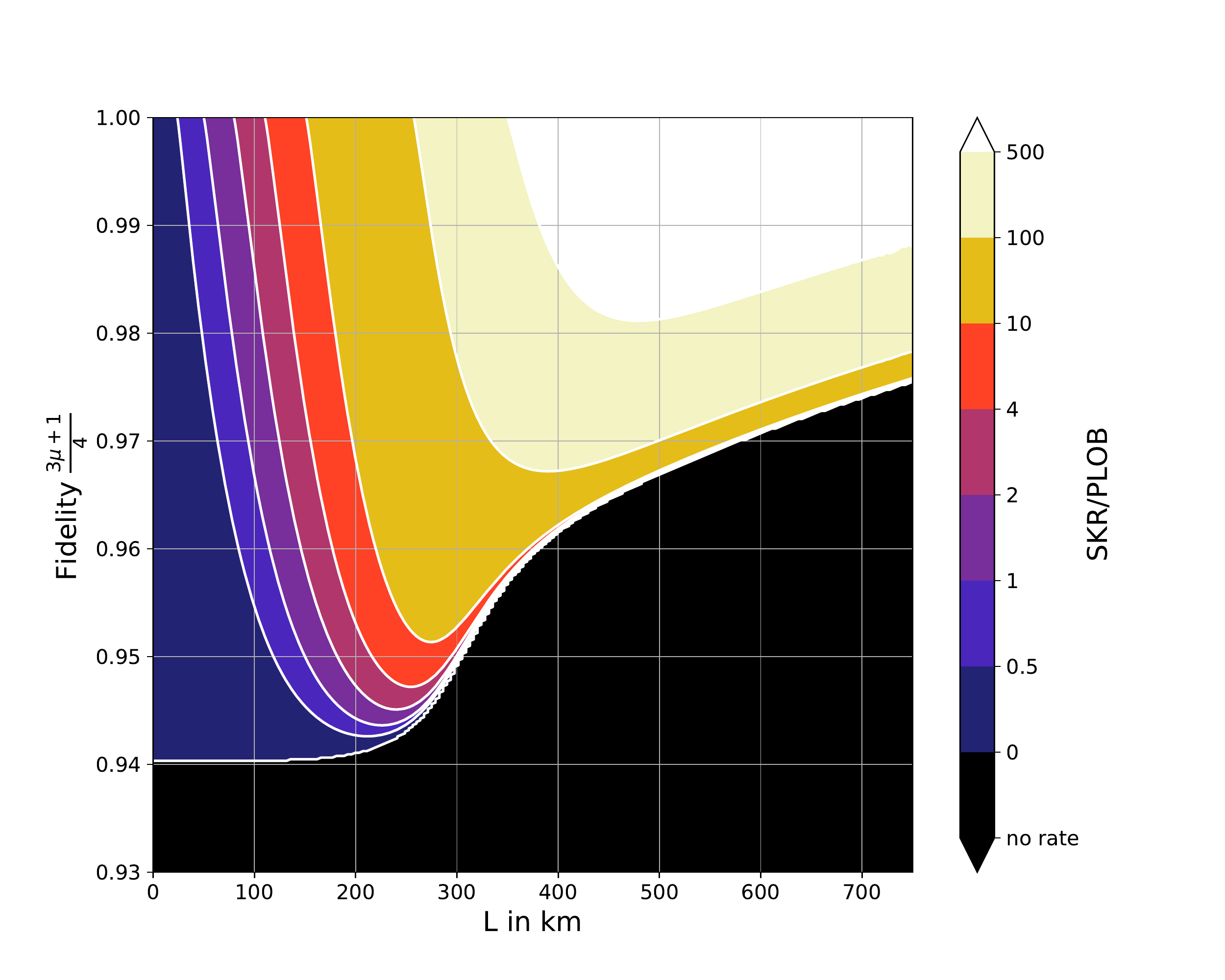}}
	\caption{Contour plots illustrating the minimal fidelity requirements to overcome the PLOB bound by a two-segment repeater for different parameter sets. In all contour plots, \(\mu = \mu_0\) and \(F_0=1\) has been used.}
	\label{fig:Contour_2_segments}
\end{figure*}

In Fig.~\ref{fig:Contour_2_segments} one can see various contour plots of the secret key rate. For convenience, we translated the error parameter \(\mu\) into a fidelity, $F = (3\mu + 1)/4$. The plots clearly indicate the minimal fidelity values below which the rates drop below the PLOB bound or even to zero rates, for different total repeater distances \(L\). The resulting contours are color-coded such that a particular color represents the secret key rate to be e.g. twice the rate of the PLOB bound. Thus, one can see that in certain parameter regimes it becomes impossible to beat the PLOB bound with a two-segment repeater. However, if both the memory coherence time $\tau_{\mathrm{coh}}$ and the link efficiency \(p_{\mathrm{link}}\) take on their improved values, it is possible to reach secret key rates as high as \(500\)-times the rate of the PLOB bound, and beyond, in a certain distance regime.

\begin{figure*}[ht]
	\centering
	\subfloat[a][$\tau_{\mathrm{coh}}={\unit[0.1]{s}}$, $p_{\mathrm{link}}=0.05$, $\mu = \mu_0=0.97$]{\includegraphics[width=0.33\linewidth]{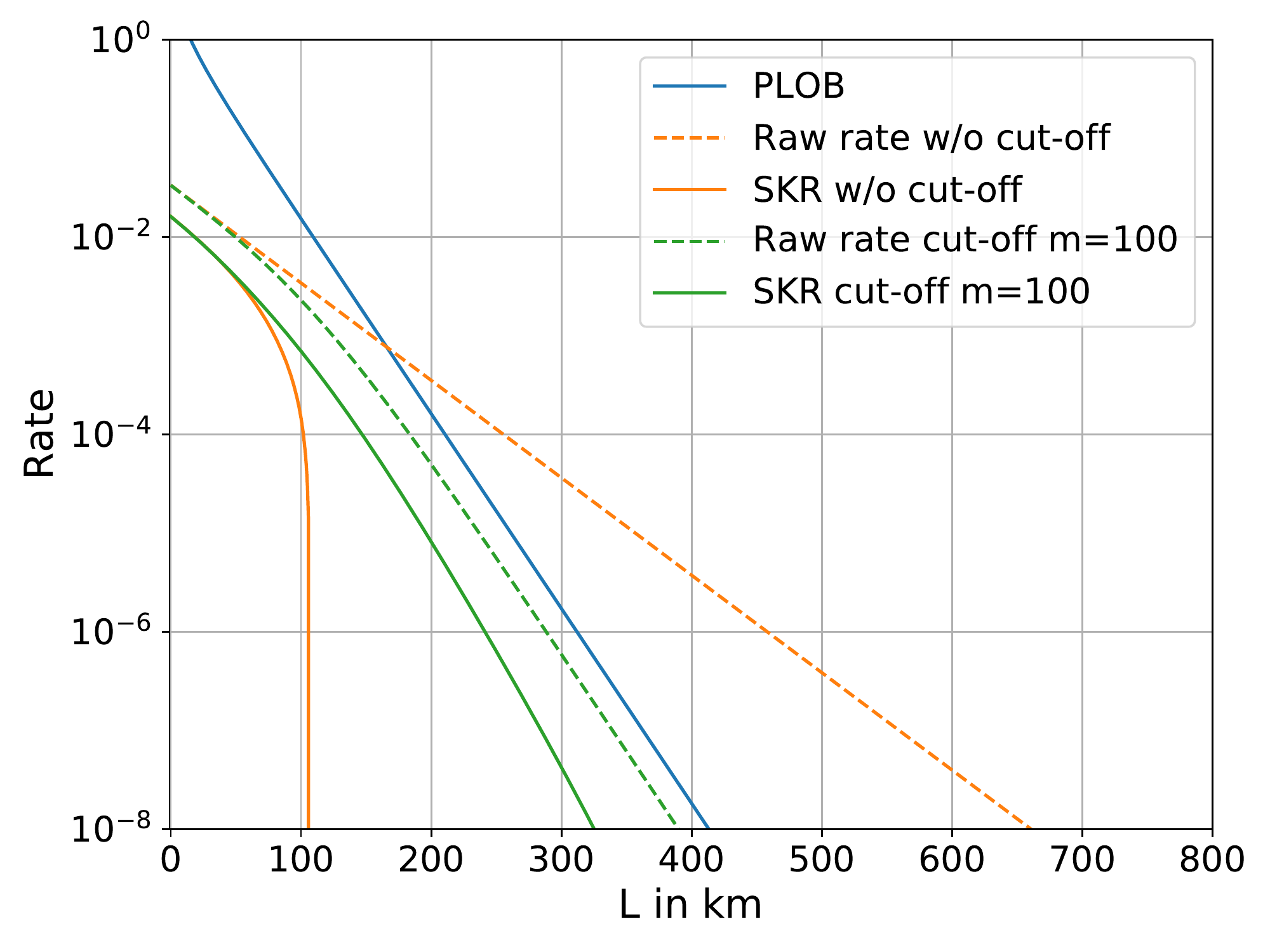}}
	\subfloat[][$\tau_{\mathrm{coh}}={\unit[0.1]{s}}$, $p_{\mathrm{link}}=0.05$, $\mu = \mu_0=1$]{\includegraphics[width=0.33\linewidth]{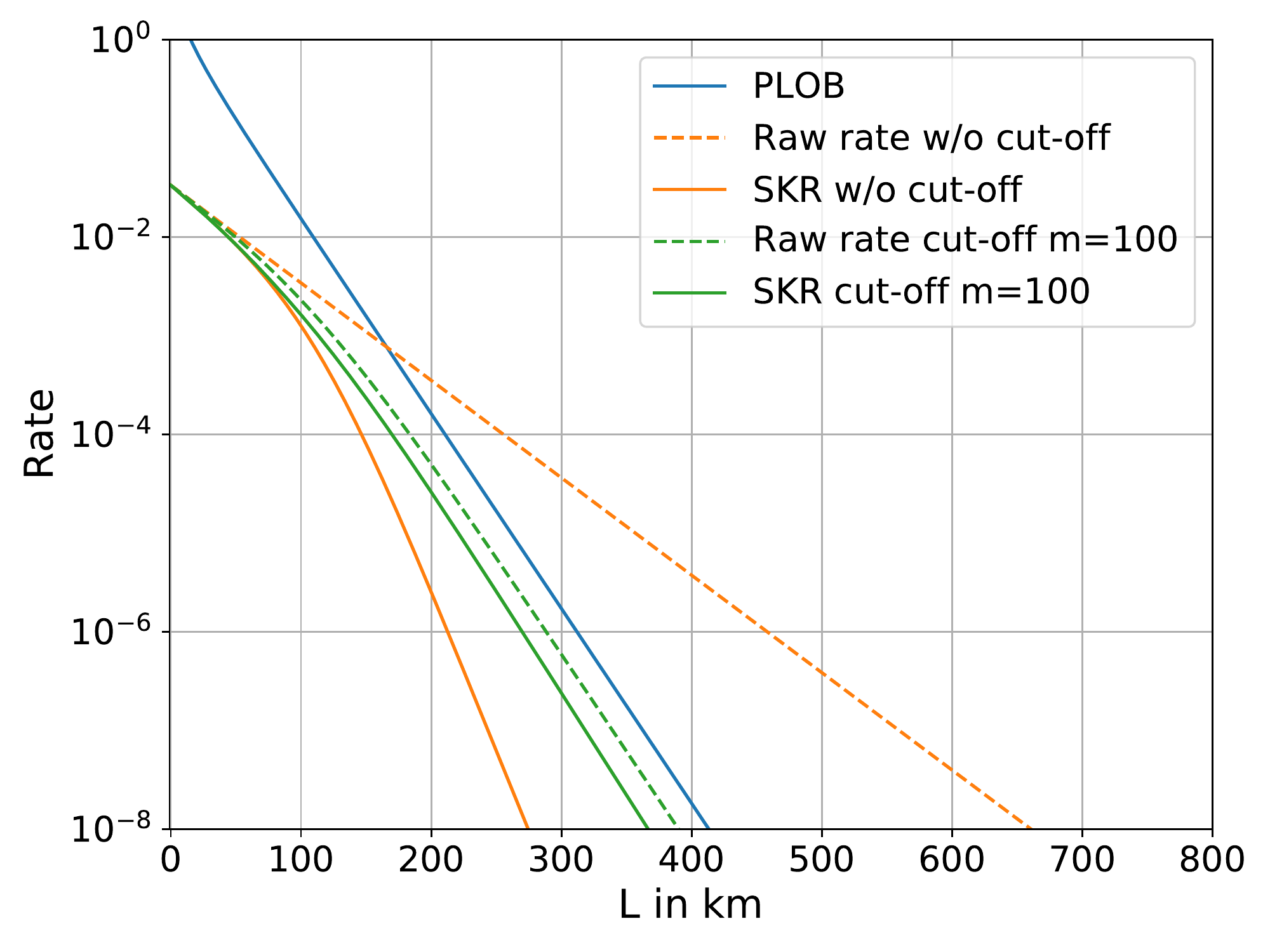}}
	\subfloat[][$\tau_{\mathrm{coh}}={\unit[0.1]{s}}$, $p_{\mathrm{link}}=0.7$, $\mu = \mu_0=0.97$]{\includegraphics[width=0.33\linewidth]{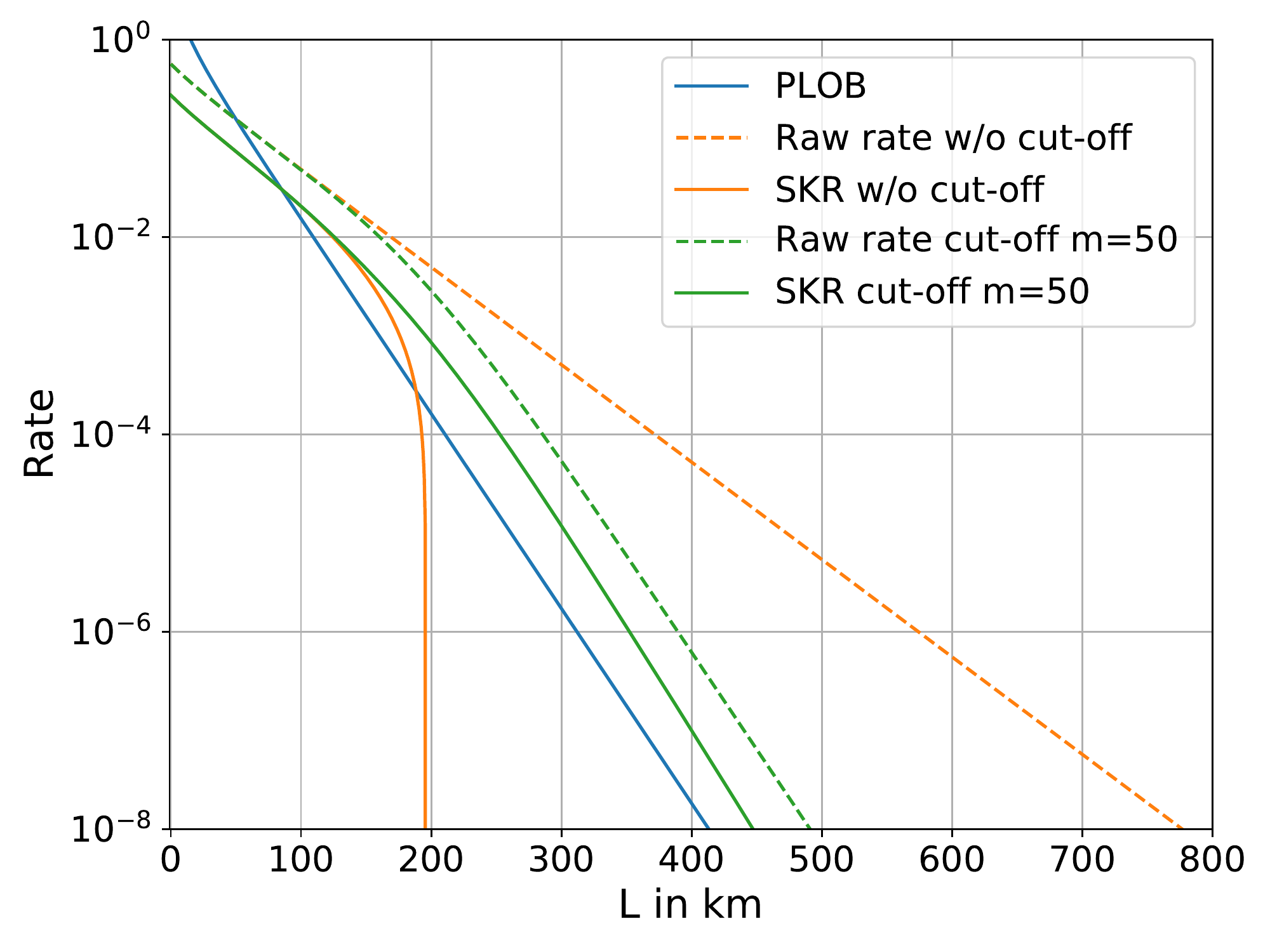}} \\
	\subfloat[][$\tau_{\mathrm{coh}}={\unit[0.1]{s}}$, $p_{\mathrm{link}}=0.7$, $\mu = \mu_0=1$]{\includegraphics[width=0.33\linewidth]{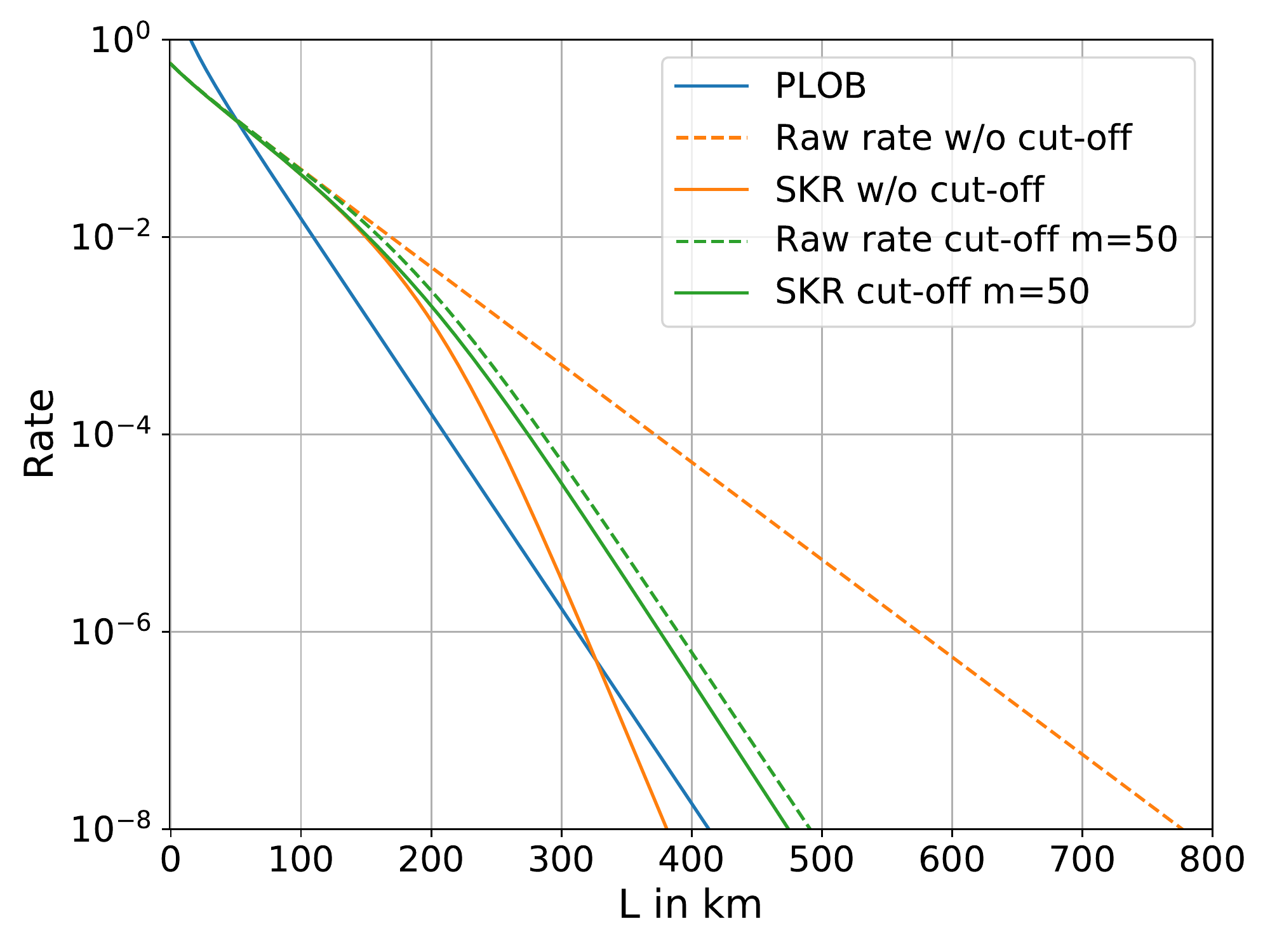}}
	\subfloat[a][$\tau_{\mathrm{coh}}={\unit[10]{s}}$, $p_{\mathrm{link}}=0.05$, $\mu = \mu_0=0.97$]{\includegraphics[width=0.33\linewidth]{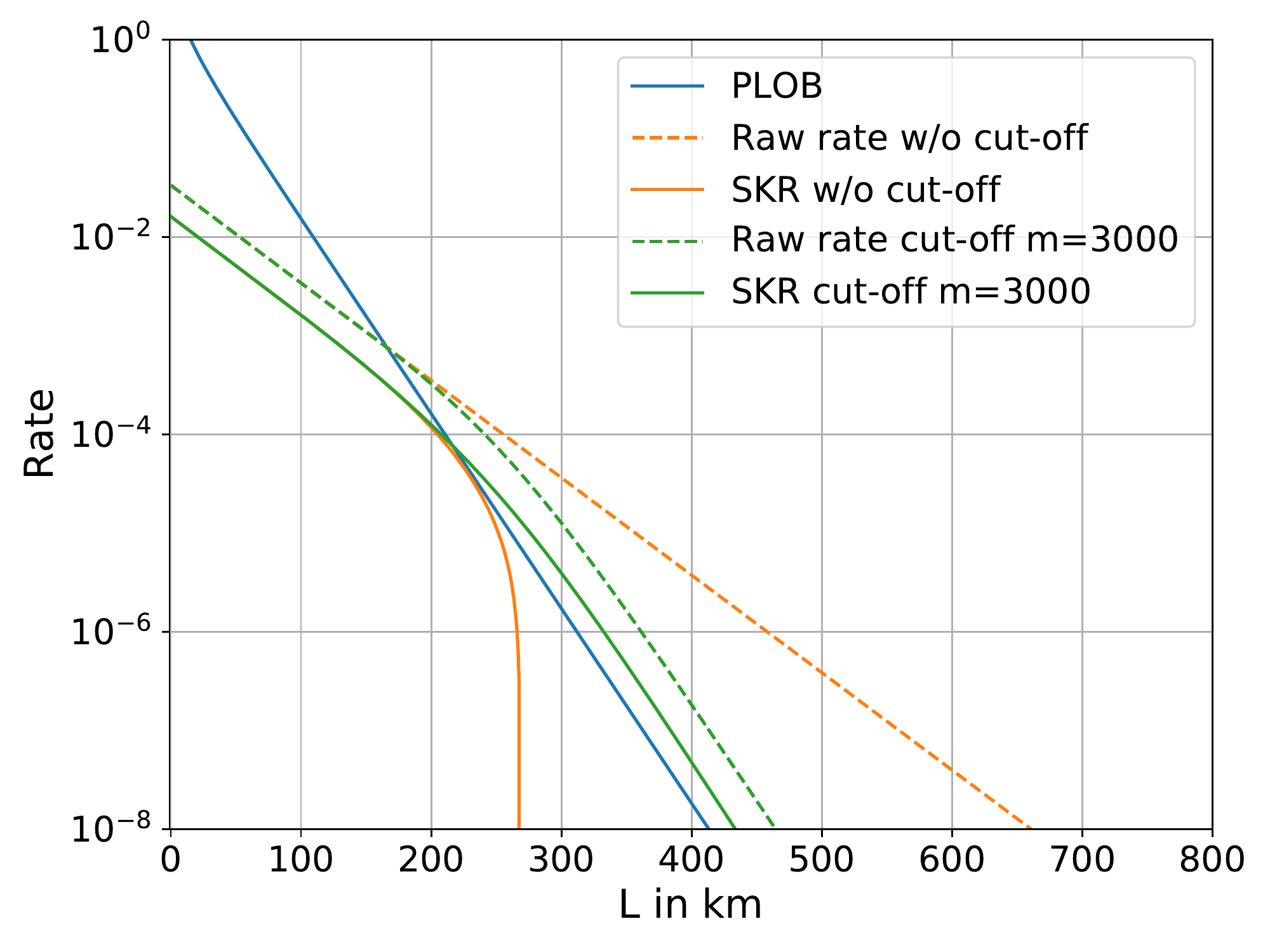}}
	\subfloat[][$\tau_{\mathrm{coh}}={\unit[10]{s}}$, $p_{\mathrm{link}}=0.05$, $\mu = \mu_0=1$]{\includegraphics[width=0.33\linewidth]{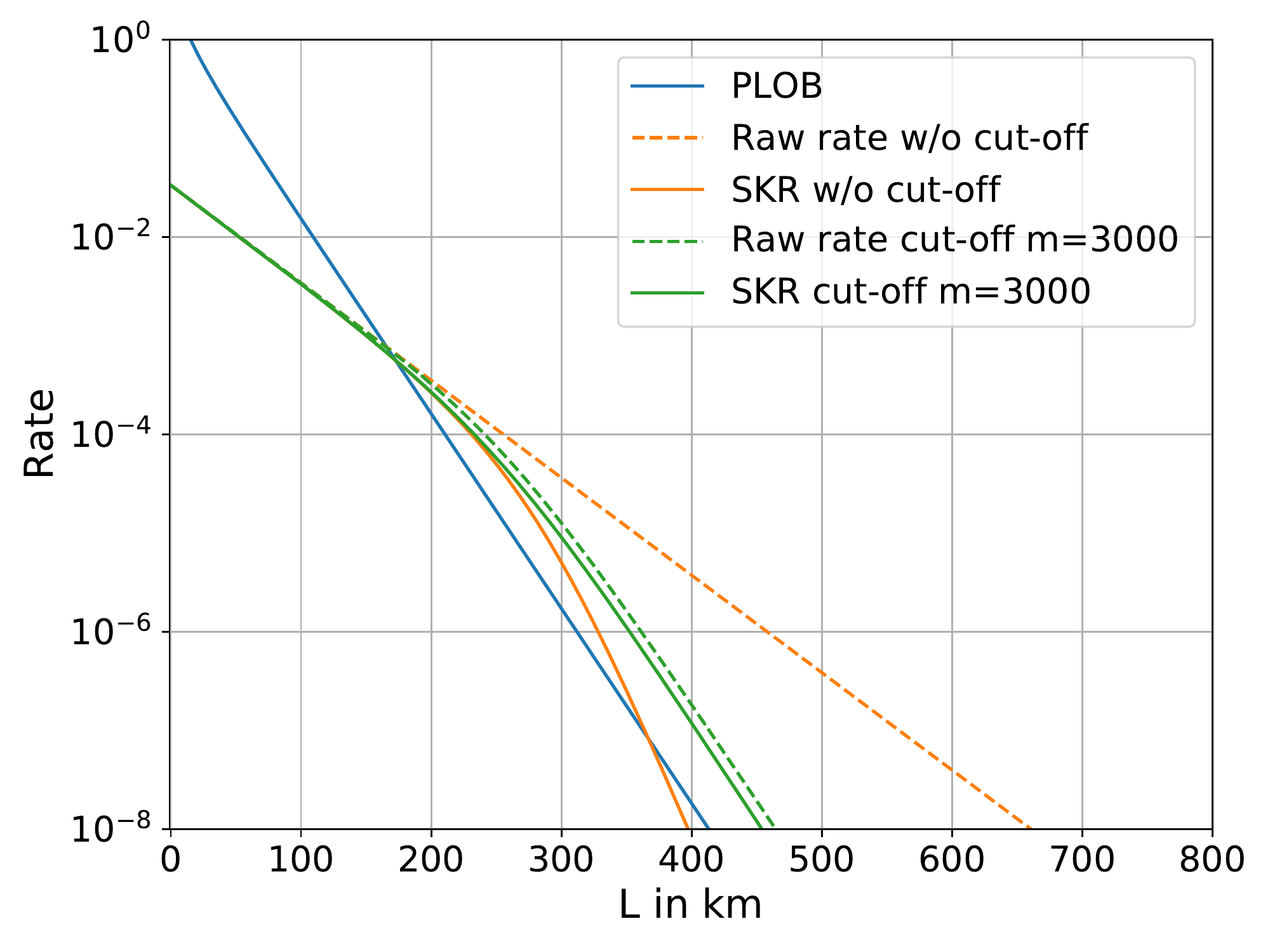}} \\
	\centering
	\subfloat[][$\tau_{\mathrm{coh}}={\unit[10]{s}}$, $p_{\mathrm{link}}=0.7$, $\mu = \mu_0=0.97$]{\includegraphics[width=0.33\linewidth]{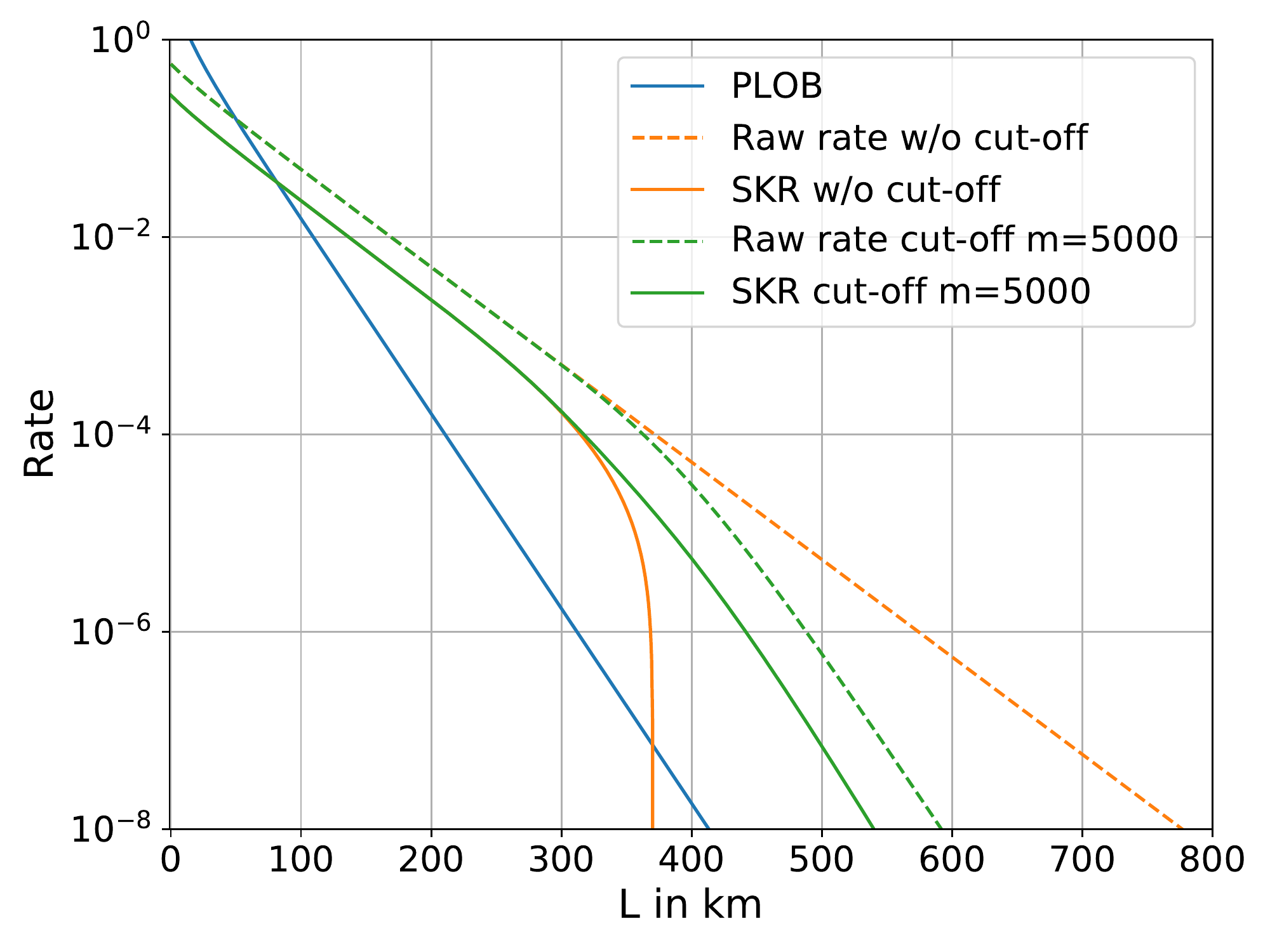}} 
	\subfloat[][$\tau_{\mathrm{coh}}={\unit[10]{s}}$, $p_{\mathrm{link}}=0.7$, $\mu = \mu_0=1$]{\includegraphics[width=0.33\linewidth]{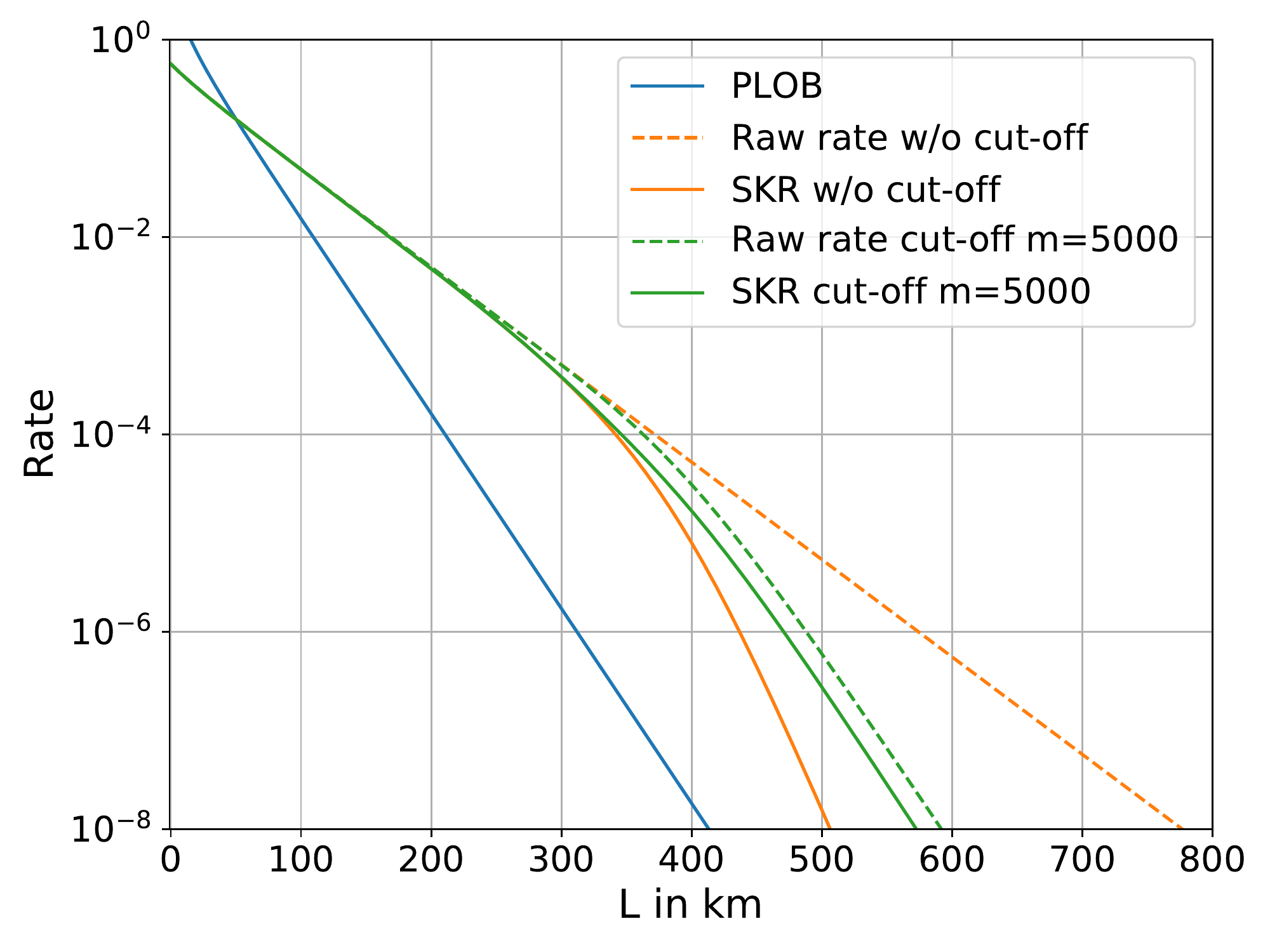}}
	\caption{Rates (secret key or raw) for a two-segment repeater over distance \(L\) for different experimental parameters.}
	\label{fig:SKR_2_segments}
\end{figure*}

In Fig.~\ref{fig:SKR_2_segments}, we show the resulting secret key rates for the experimental parameters from Tab.~\ref{tab:constants}, for both the scheme with and without a memory cut-off. This time the error parameter \(\mu=\mu_0\) is fixed, and it either takes on its ``current" or its ``improved" (ideal) value. For comparison, as a reference, we also included the raw rates in each case. The loss scaling of the rates in all schemes is, as expected, proportional to $p_{\mathrm{link}} \,e^{-\frac{L}{2 L_{\mathrm{att}}}}=p_{\mathrm{link}}\sqrt{e^{-\frac{L}{L_{\mathrm{att}}}}}$ (corresponding to a linear decrease with distance due to the log scale representation). The effect of the different experimental parameter values is clearly visible. The choice of $p_{\mathrm{link}}=0.05$ or $p_{\mathrm{link}}=0.7$ determines the offset along the $y$-axis (rate axis) at zero distance. A higher $p_{\mathrm{link}}$ allows to cross the PLOB bound at a smaller distance. Note that the PLOB bound itself can arbitrarily exceed the value of one secret bit towards zero distance; in our schemes we always distribute qubits and so one secret bit per channel use is the maximum (and depending on the number of modes to encode the photonic qubits there could be extra factors, ``per mode"). The choice of $\tau_{\mathrm{coh}}=\unit[0.1]{s}$ or $\tau_{\mathrm{coh}}=\unit[10]{s}$ determines when (at which distance) the (negative) slope of the secret key rate increases such that the repeater switches from a $\sqrt{e^{-\frac{L}{L_{\mathrm{att}}}}}$ to a $e^{-\frac{L}{L_{\mathrm{att}}}}$ (PLOB-like) scaling, or even worse. This effect is an effect of the memory dephasing that occurs even when \(\mu=\mu_0=1\). If, in addition, \(\mu=\mu_0=0.97<1\), the secret key rates can drop abruptly down to zero, since then the QBERs have nonzero contributions both in $e_z$ and $e_x$, see Eq.~\eqref{eq:QBER}. Note that this effect happens also when either of the two parameters, $\mu$ or $\mu_0$, drop below one, i.e. when either the gates or the initial states become imperfect. Also note that non-unit $\mu$ or $\mu_0$ in addition lead to an increased $y$-axis offset which will become more apparent for larger repeaters with larger $n$.

However, a memory cut-off can significantly change the picture, and it can  increase the achievable distance compared to the scheme without a cut-off (compare the solid yellow with the solid green curves in Fig.~\ref{fig:SKR_2_segments}). More specifically,
beyond distances when the rates of the no cut-off scheme drop dramatically, the cut-off scheme still scales proportional to the PLOB bound.
Note that for the scheme with cut-off, even the raw rates (dashed green curves) can switch from an $L/2$ to an $L$ scaling (like PLOB), because a finite cut-off value ``simulates" an imperfect memory in the raw rate (whose loss scaling  resembles the scaling without a quantum memory, i.e. that of the PLOB bound, in the limit of $m=1$) \cite{CollinsPrl}.

Again, one can also see that with ``current" parameter values, see Fig.~\ref{fig:SKR_2_segments}(a), it is impossible to beat the PLOB bound
(here even when \(\mu=\mu_0=1\), see Fig.~\ref{fig:SKR_2_segments}(b)), but with improving values for the coherence time and the link efficiency, it becomes possible. This holds even when only one of the two parameters, $p_{\mathrm{link}}$ or $\tau_{\mathrm{coh}}$, is improved, as long as we can cross PLOB at a sufficiently small distance or maintain the repeater's slope for sufficiently long, respectively.

In the next section we will turn to a four-segment repeater
(a three-segment repeater is discussed in great detail in App.~\ref{app:Optimality 3 segments}). 

\subsection{Four-segment repeater}\label{sec:Four-Segment Repeater}

\begin{figure*}[ht]
	\centering
	\subfloat[a][$\tau_{\mathrm{coh}}={\unit[0.1]{s}}$, $p_{\mathrm{link}}=0.05$]{\includegraphics[width=0.5\linewidth]{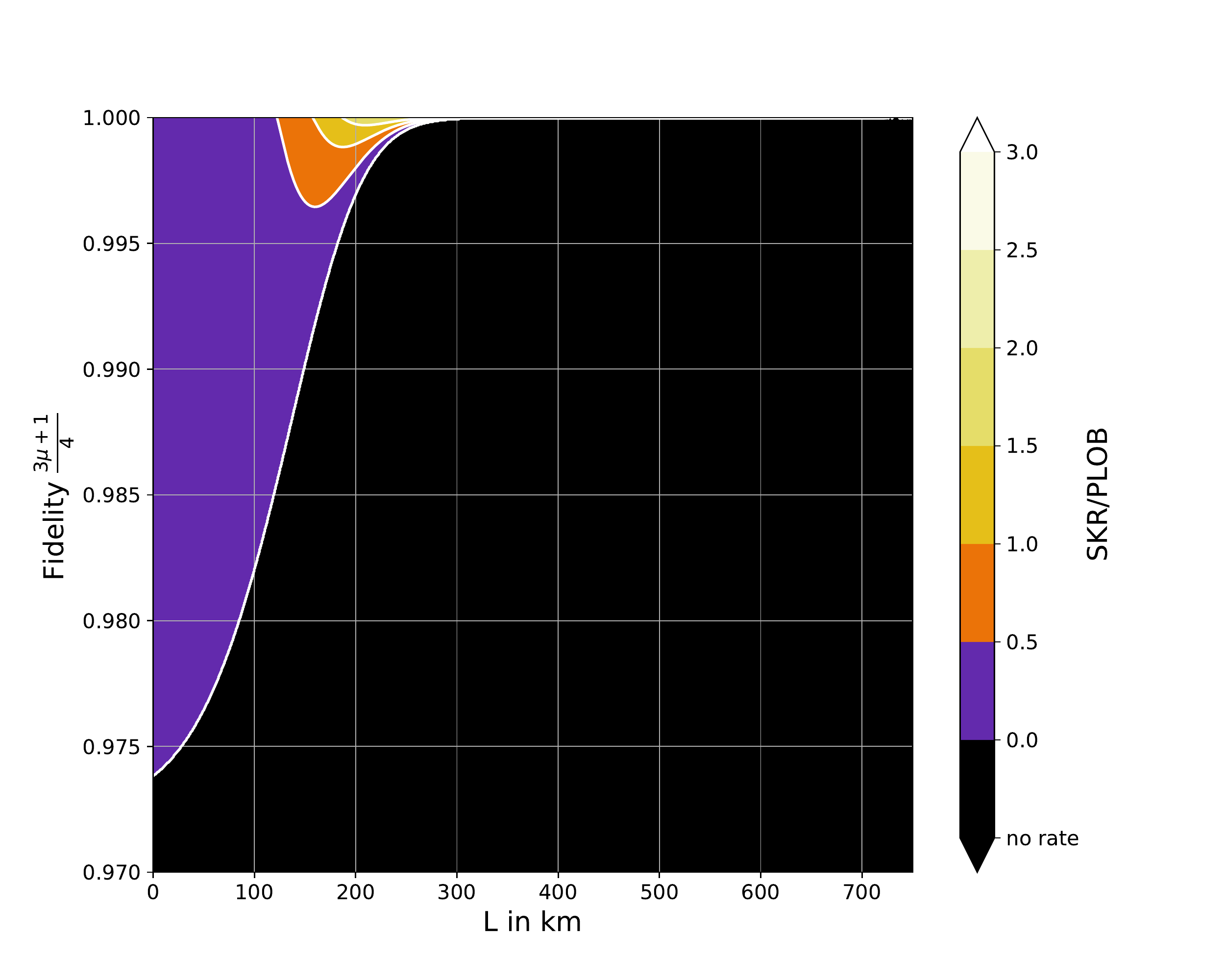}}
	\subfloat[][$\tau_{\mathrm{coh}}={\unit[0.1]{s}}$, $p_{\mathrm{link}}=0.7$]{\includegraphics[width=0.5\linewidth]{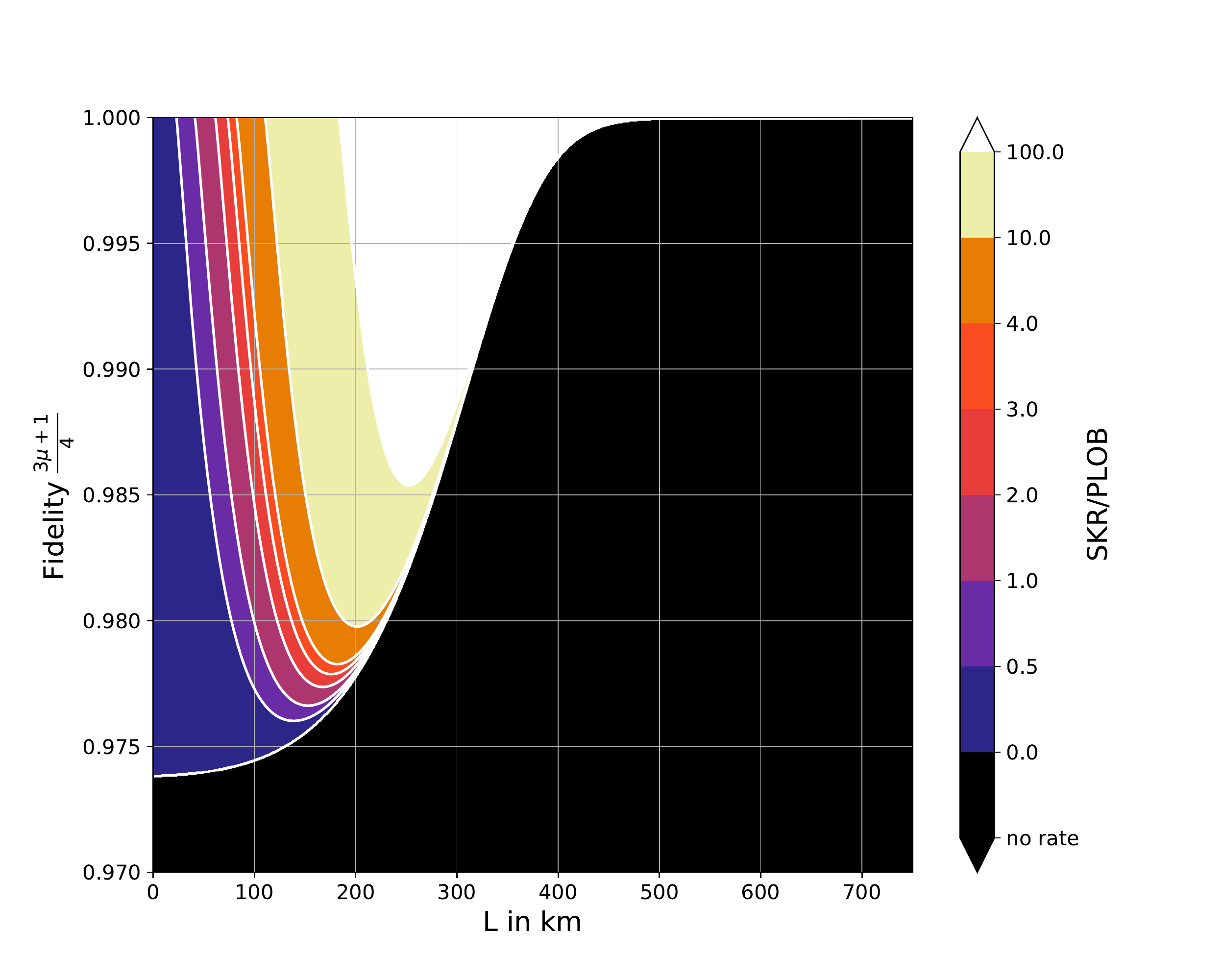}} \\
	\subfloat[][$\tau_{\mathrm{coh}}={\unit[10]{s}}$, $p_{\mathrm{link}}=0.05$]{\includegraphics[width=0.5\linewidth]{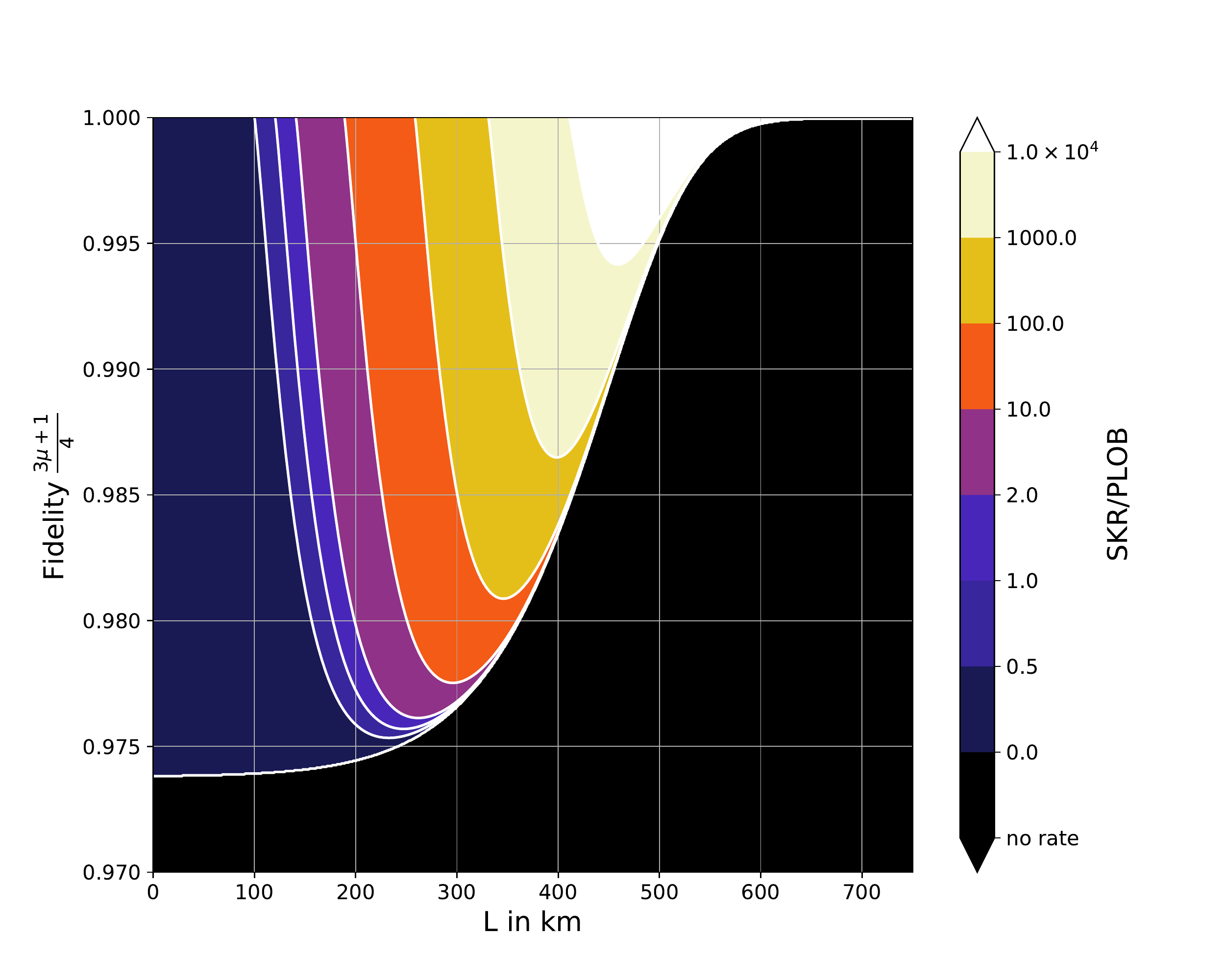}} 
	\subfloat[][$\tau_{\mathrm{coh}}={\unit[10]{s}}$, $p_{\mathrm{link}}=0.7$]{\includegraphics[width=0.5\linewidth]{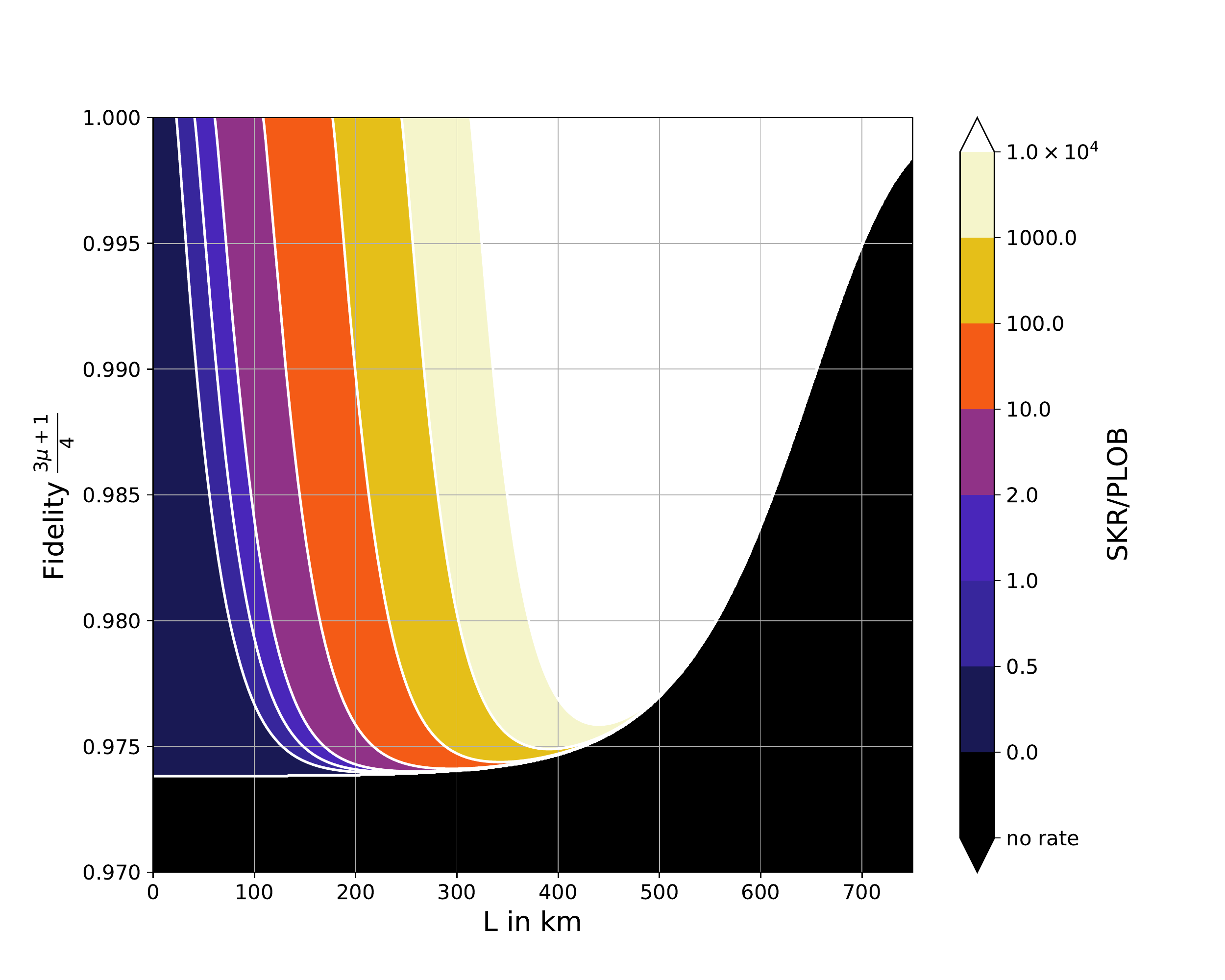}}
	\caption{Contour plots illustrating the minimal fidelity requirements to overcome the PLOB bound by a four-segment repeater for different parameter sets. In all contour plots, \(\mu = \mu_0\) and \(F_0=1\) has been used.}
	\label{fig:Contour_4_segments}
\end{figure*}

As we have seen in Sec.~\ref{sssec:Par-distr: 4-segment repeater}, there are various swapping strategies possible for a four-segment repeater in contrast to a simple two-segment repeater. Our conjecture is (see also App.~\ref{app:Optimality 3 segments} for the case $n=3$) that the ``optimal" scheme is optimal in the regimes of increasingly good hardware parameters. Thus, let us first again focus on the minimal fidelities to overcome the PLOB bound for this scheme, similar to our analysis for two segments, but now without cut-off only. The results are shown in Fig.~\ref{fig:Contour_4_segments}. It becomes apparent that now a much higher fidelity or equivalently \(\mu\) is needed, but in turn also much higher secret key rates, \(10^4\)-times the PLOB rate and beyond, are possible. Since we have $n=4$ now, non-unit $\mu$ values have a stronger impact on the QBERs, see Eq.~\eqref{eq:QBER}. At the same time, however, the loss scaling becomes proportional to $p_{\mathrm{link}} \,e^{-\frac{L}{4 L_{\mathrm{att}}}}=p_{\mathrm{link}}\sqrt[4]{e^{-\frac{L}{L_{\mathrm{att}}}}}$. Furthermore, note that a different scaling of the contours is observable. This effect is due to the lack of a memory cut-off.

\begin{figure*}[ht]
	\centering
	\subfloat[a][$\tau_{\mathrm{coh}}={\unit[0.1]{s}}$, $p_{\mathrm{link}}=0.05$, $\mu = \mu_0=0.97$]{\includegraphics[width=0.33\linewidth]{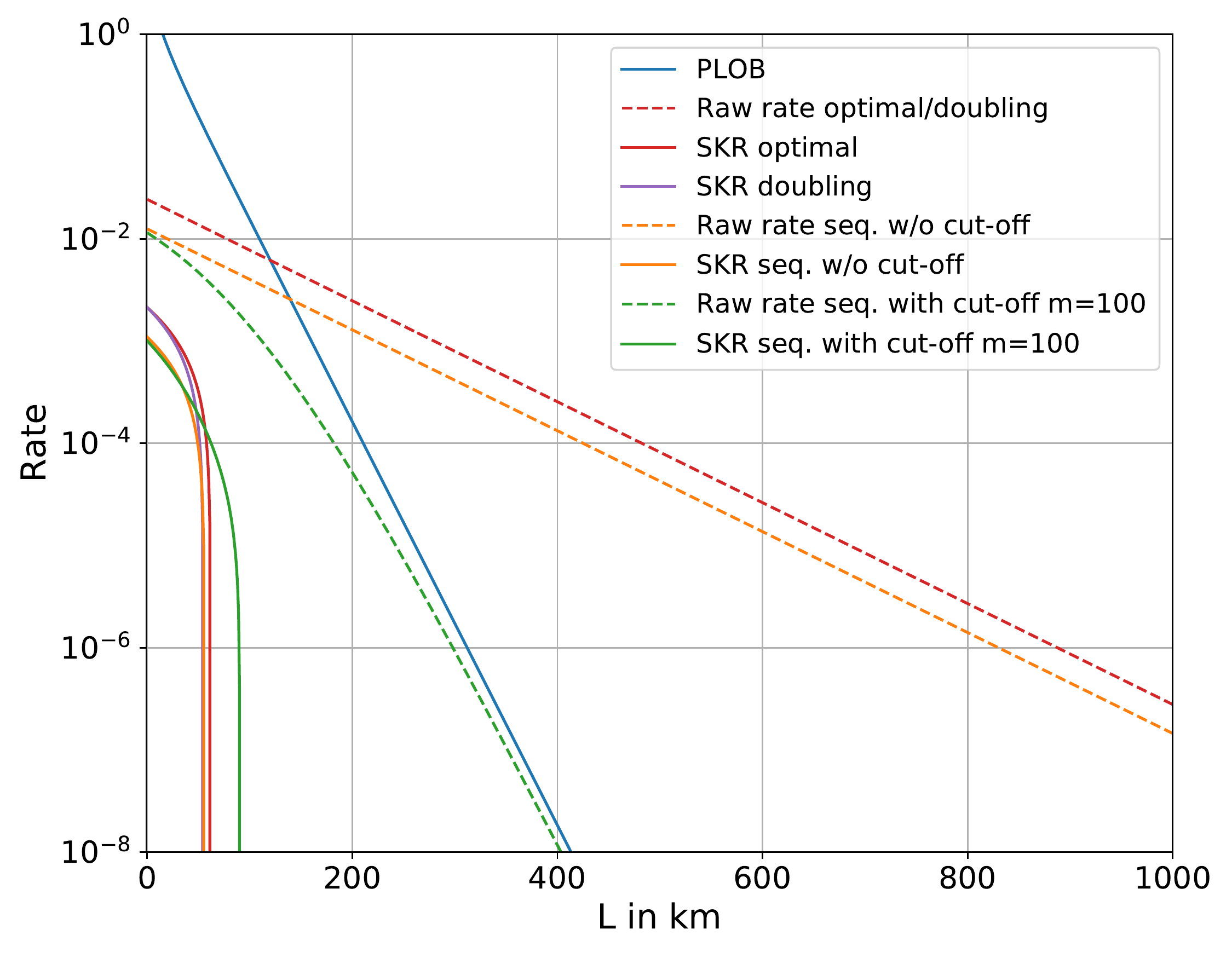}}
	\subfloat[][$\tau_{\mathrm{coh}}={\unit[0.1]{s}}$, $p_{\mathrm{link}}=0.05$, $\mu = \mu_0=1$]{\includegraphics[width=0.33\linewidth]{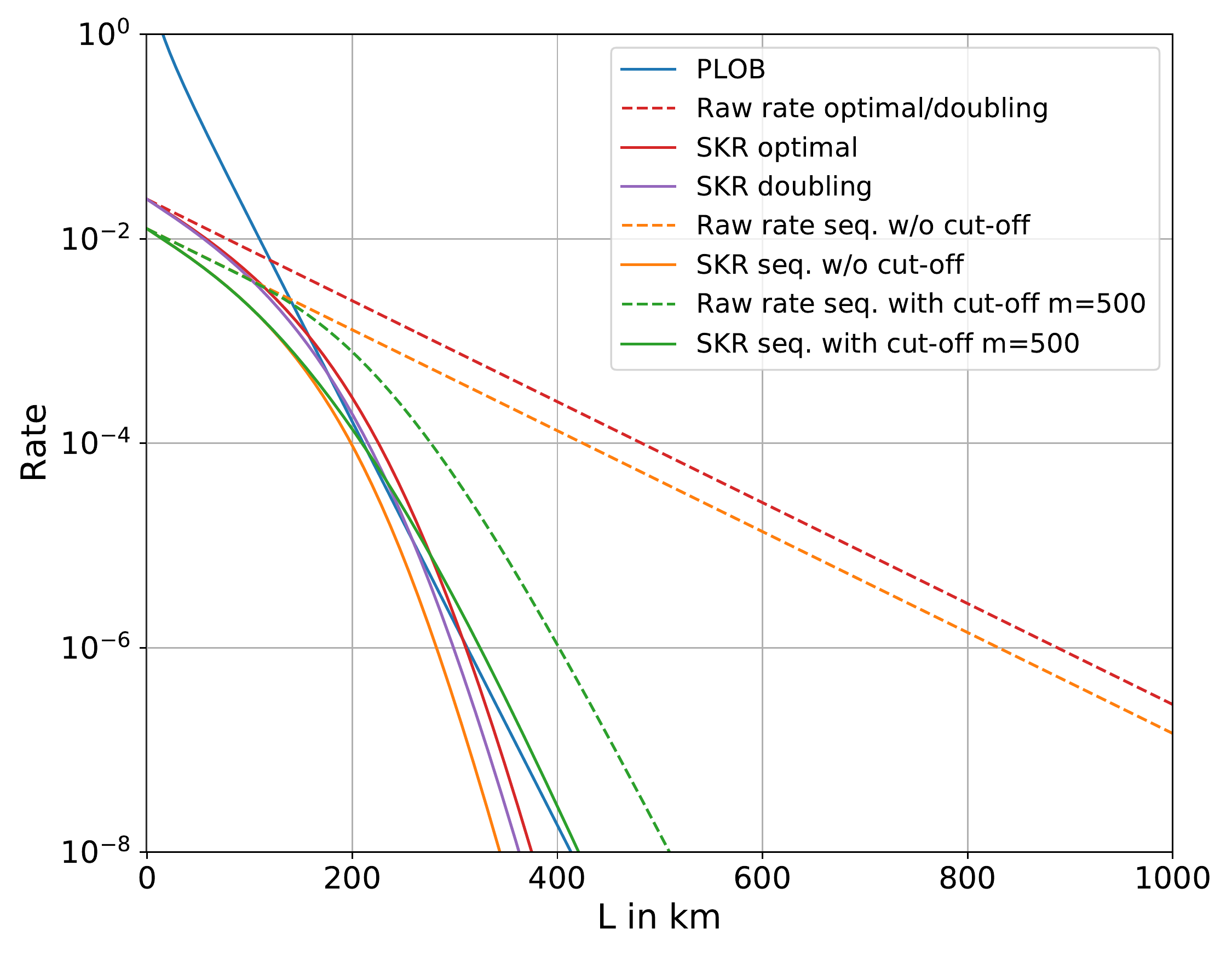}}
	\subfloat[][$\tau_{\mathrm{coh}}={\unit[0.1]{s}}$, $p_{\mathrm{link}}=0.7$, $\mu = \mu_0=0.97$]{\includegraphics[width=0.33\linewidth]{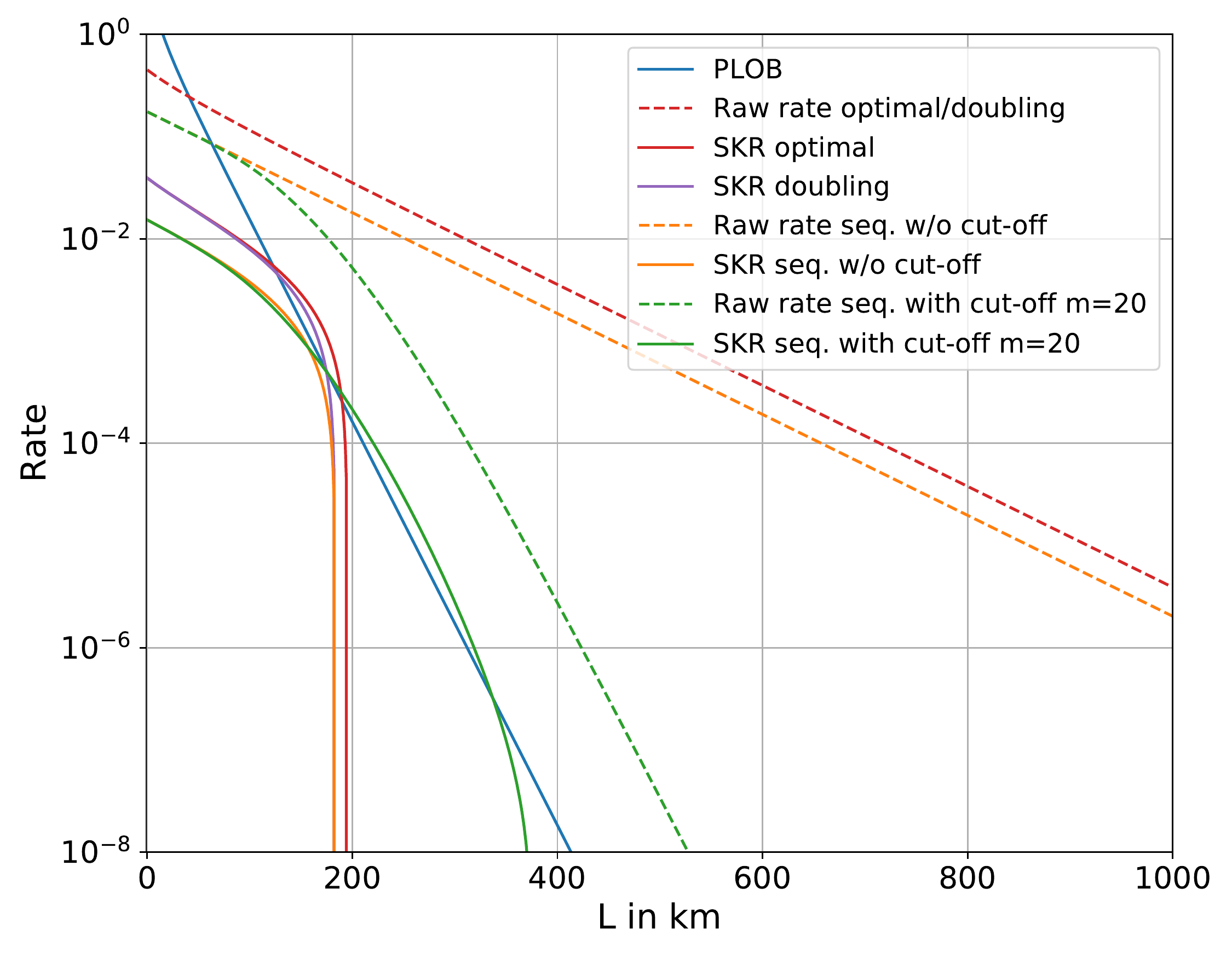}} \\
	\subfloat[][$\tau_{\mathrm{coh}}={\unit[0.1]{s}}$, $p_{\mathrm{link}}=0.7$, $\mu = \mu_0=1$]{\includegraphics[width=0.33\linewidth]{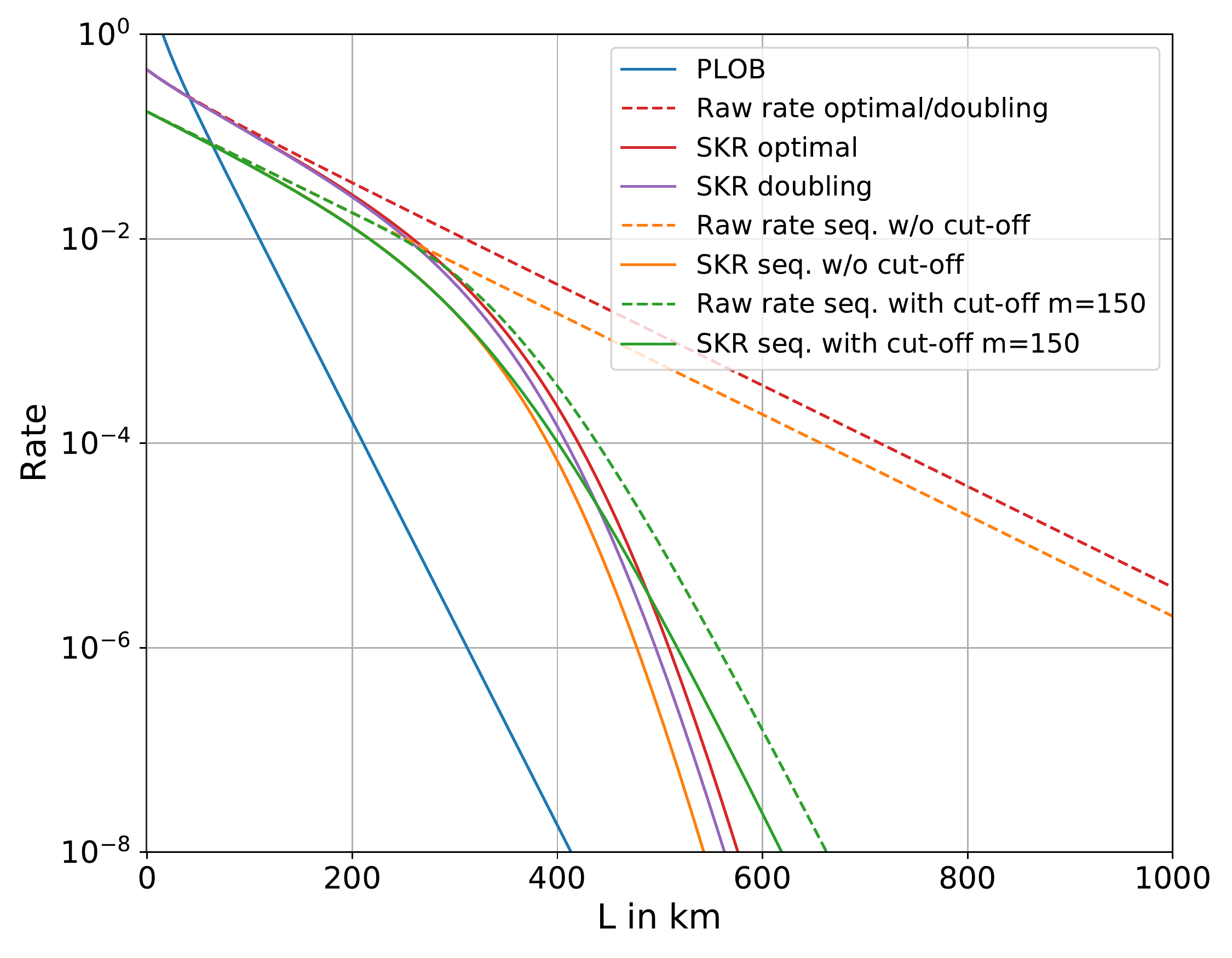}}
	\subfloat[a][$\tau_{\mathrm{coh}}={\unit[10]{s}}$, $p_{\mathrm{link}}=0.05$, $\mu = \mu_0=0.97$]{\includegraphics[width=0.33\linewidth]{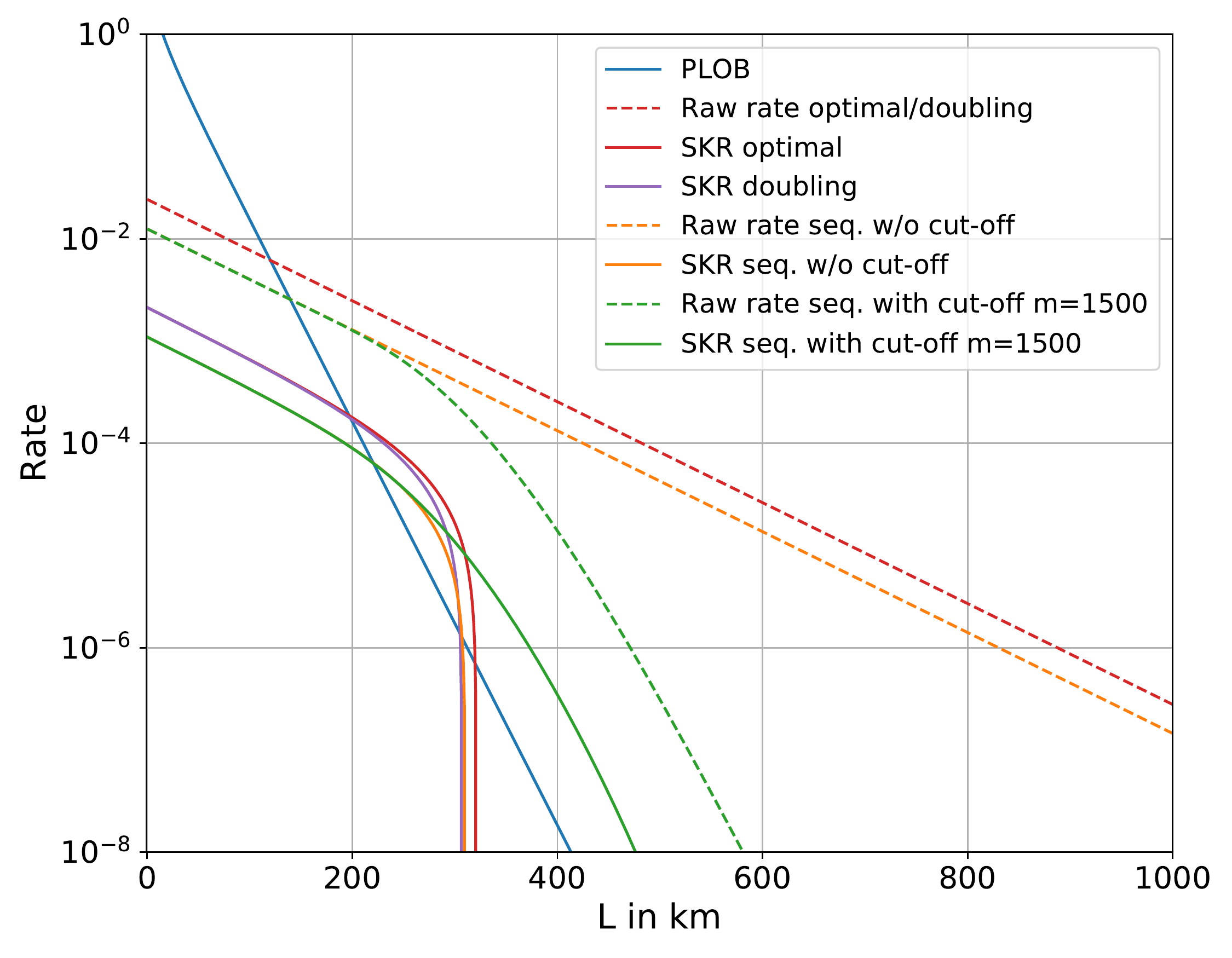}}
	\subfloat[][$\tau_{\mathrm{coh}}={\unit[10]{s}}$, $p_{\mathrm{link}}=0.05$, $\mu = \mu_0=1$]{\includegraphics[width=0.33\linewidth]{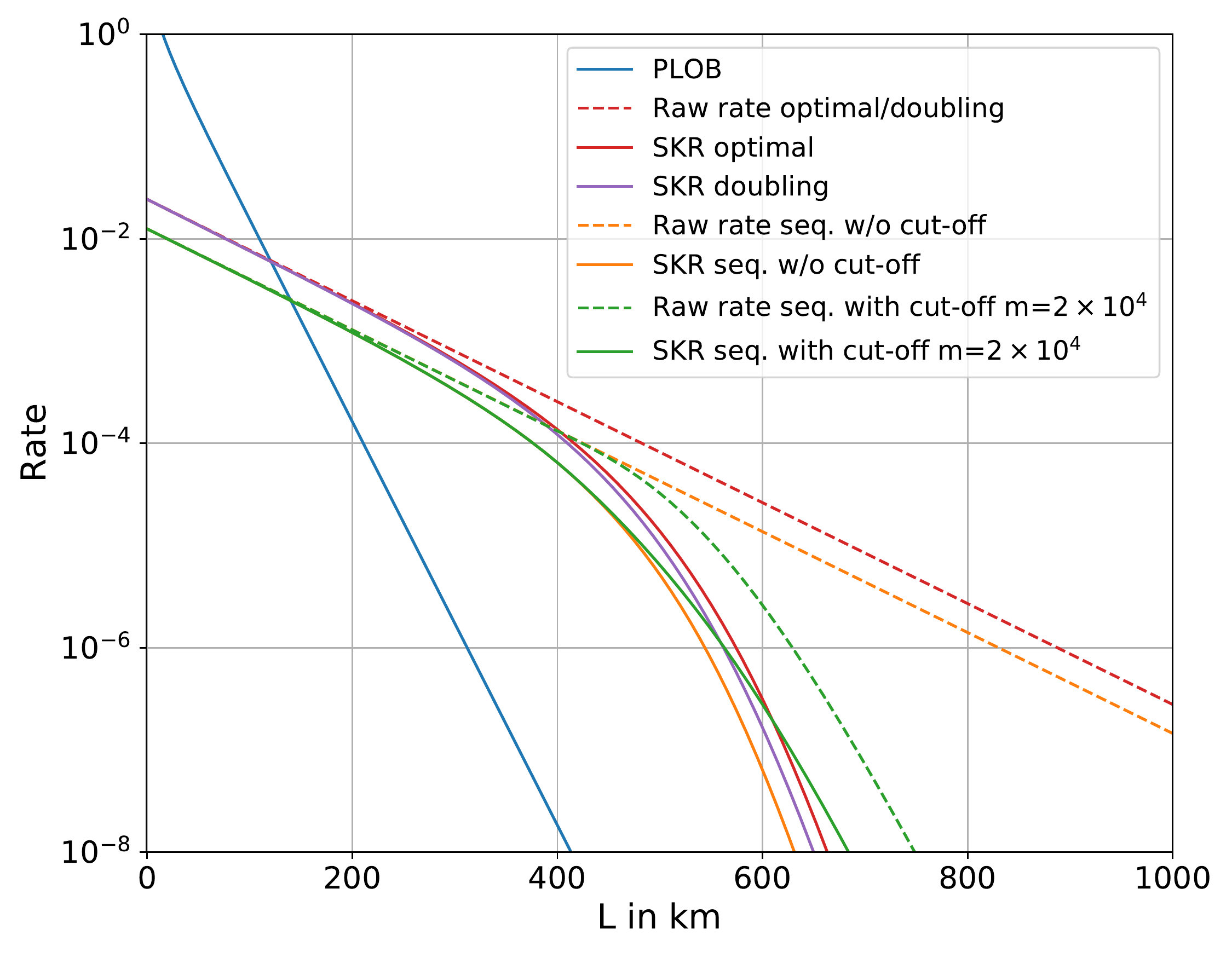}} \\
	\subfloat[][$\tau_{\mathrm{coh}}={\unit[10]{s}}$, $p_{\mathrm{link}}=0.7$, $\mu = \mu_0=0.97$]{\includegraphics[width=0.33\linewidth]{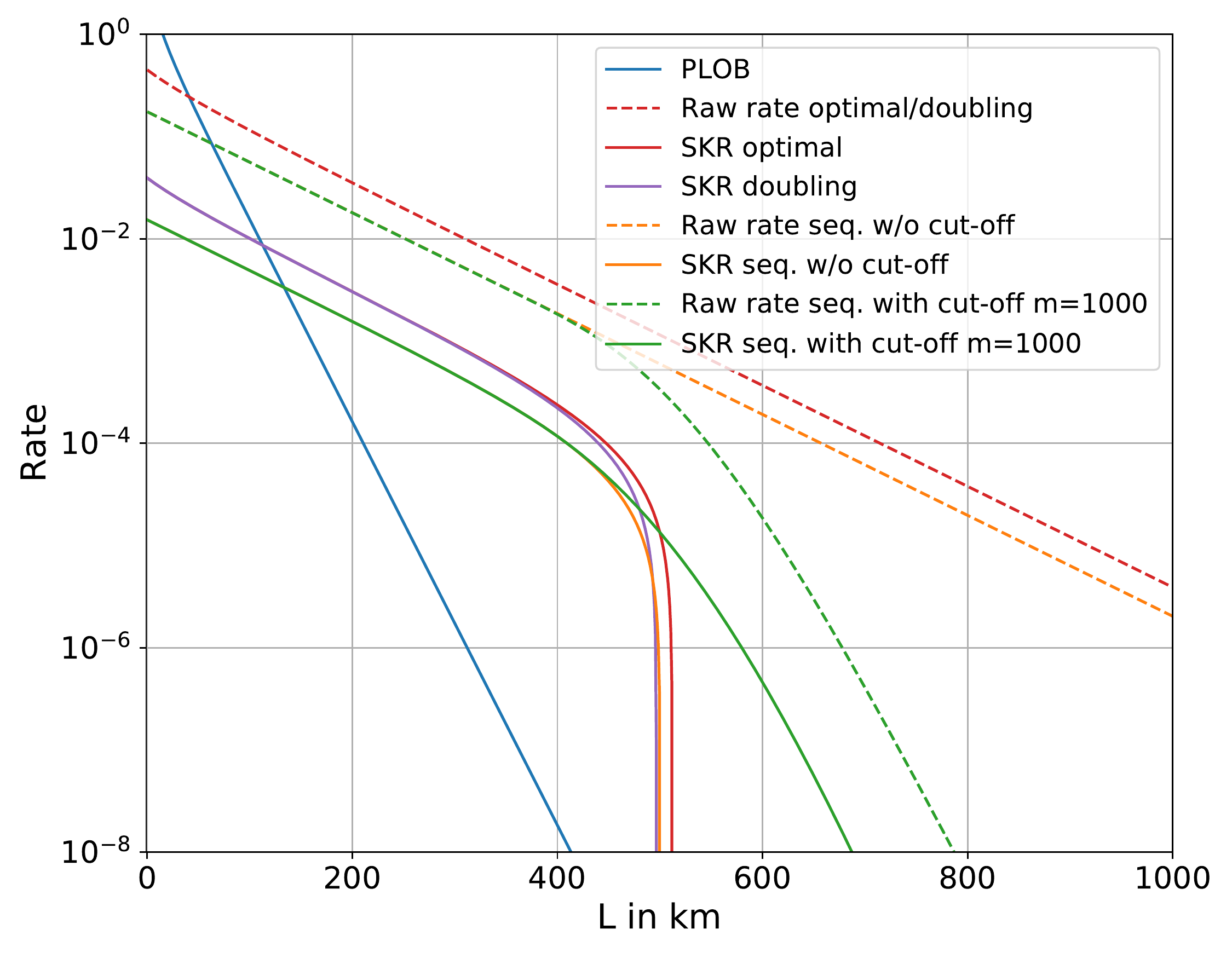}} 
	\subfloat[][$\tau_{\mathrm{coh}}={\unit[10]{s}}$, $p_{\mathrm{link}}=0.7$, $\mu = \mu_0=1$]{\includegraphics[width=0.33\linewidth]{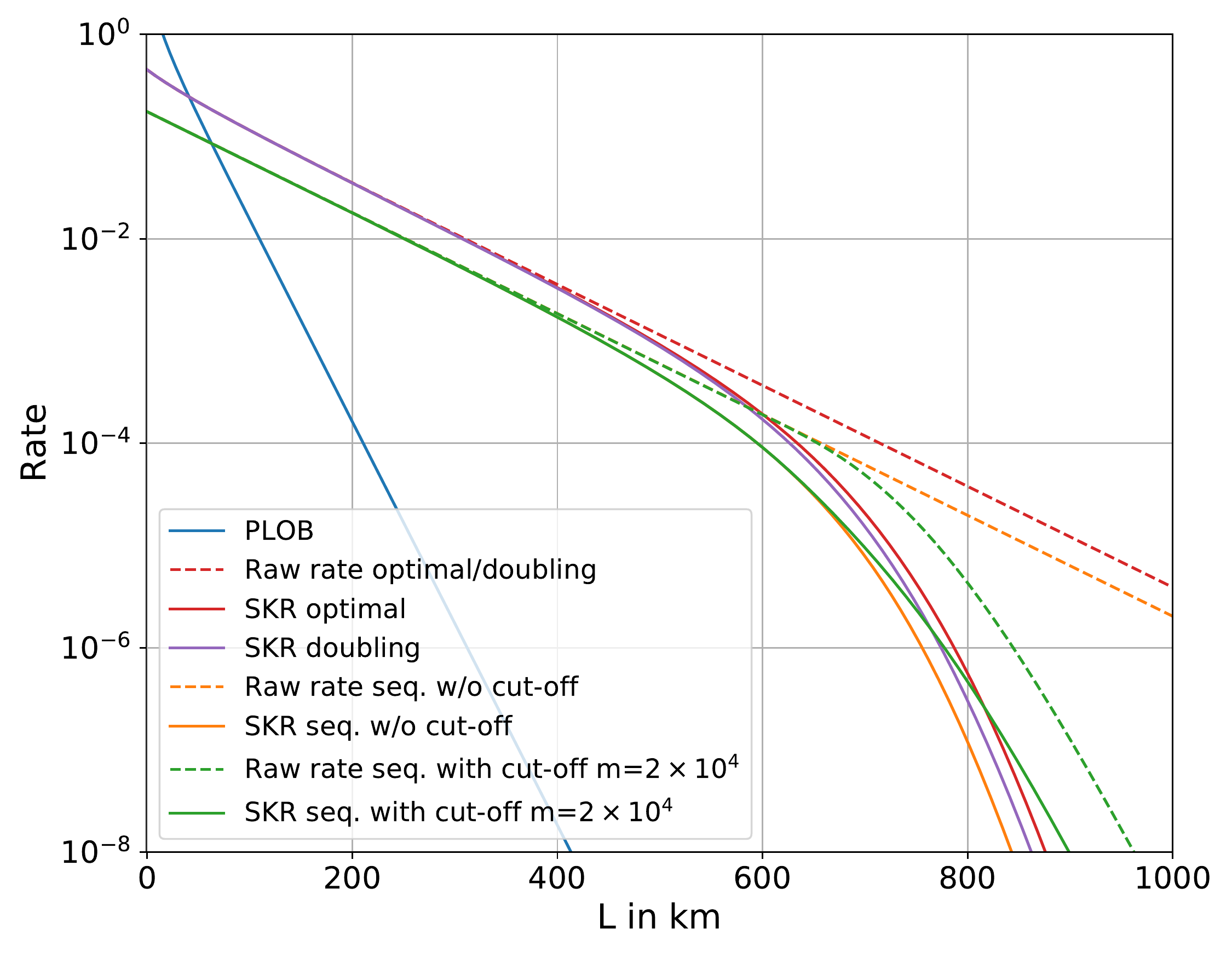}}
	\caption{Rates (secret key or raw) for a four-segment repeater over distance \(L\) for different experimental parameters.}
	\label{fig:SKR_4_segments}
\end{figure*}

Next, we consider the secret key rates for a particular choice of the experimental parameters including $\mu = \mu_0$ according to Tab.~\ref{tab:constants}. Besides the ``optimal" scheme, now we also include the sequential and the doubling schemes in the rate analysis (sequential/iterative swapping together with sequential distributions and doubling with parallel distributions). In Fig.~\ref{fig:SKR_4_segments}, one can see the
PLOB bound and the secret key rates for the sequential scheme with and without a cut-off, for the doubling scheme and for the optimal scheme (both without a cut-off). In addition, again the raw rates are shown as a reference, and the corresponding three dashed curves are the raw rates for (equivalently) doubling and ``optimal", and for the sequential scheme with and without cut-off. Compared to the previous two-segment repeater, it is now easier to overcome the PLOB bound, but the crossing happens at longer distances, since the four-segment repeater starts with a lower rate at \(L=\unit[0]{km}\).

\subsection{Eight-segment repeater}\label{sec:Eight-Segment Repeater}

\begin{figure*}[ht]
	\centering
	\subfloat[a][$\tau_{\mathrm{coh}}={\unit[0.1]{s}}$, $p_{\mathrm{link}}=0.05$]{\includegraphics[width=0.5\linewidth]{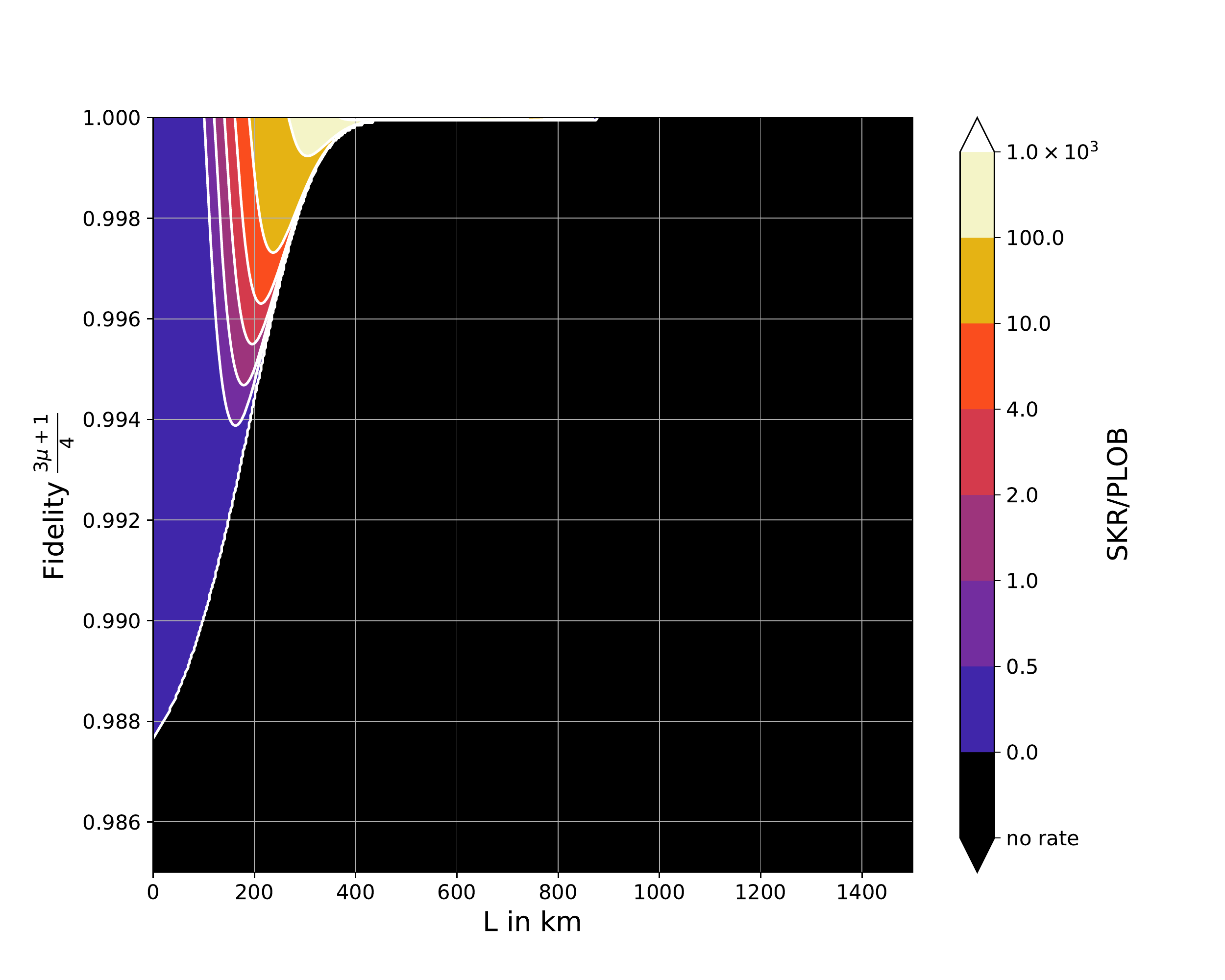}}
	\subfloat[][$\tau_{\mathrm{coh}}={\unit[0.1]{s}}$, $p_{\mathrm{link}}=0.7$]{\includegraphics[width=0.5\linewidth]{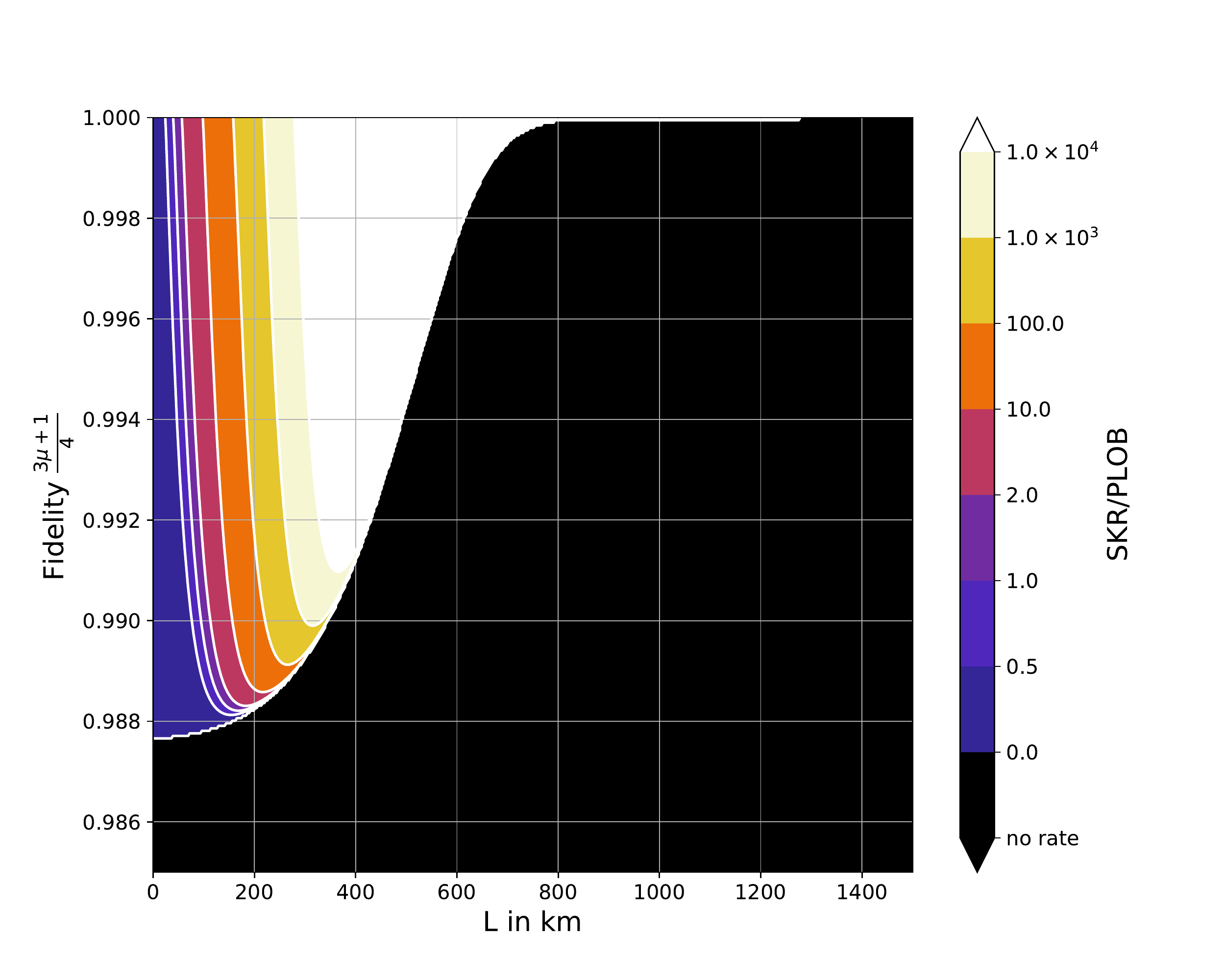}} \\
	\subfloat[][$\tau_{\mathrm{coh}}={\unit[10]{s}}$, $p_{\mathrm{link}}=0.05$]{\includegraphics[width=0.5\linewidth]{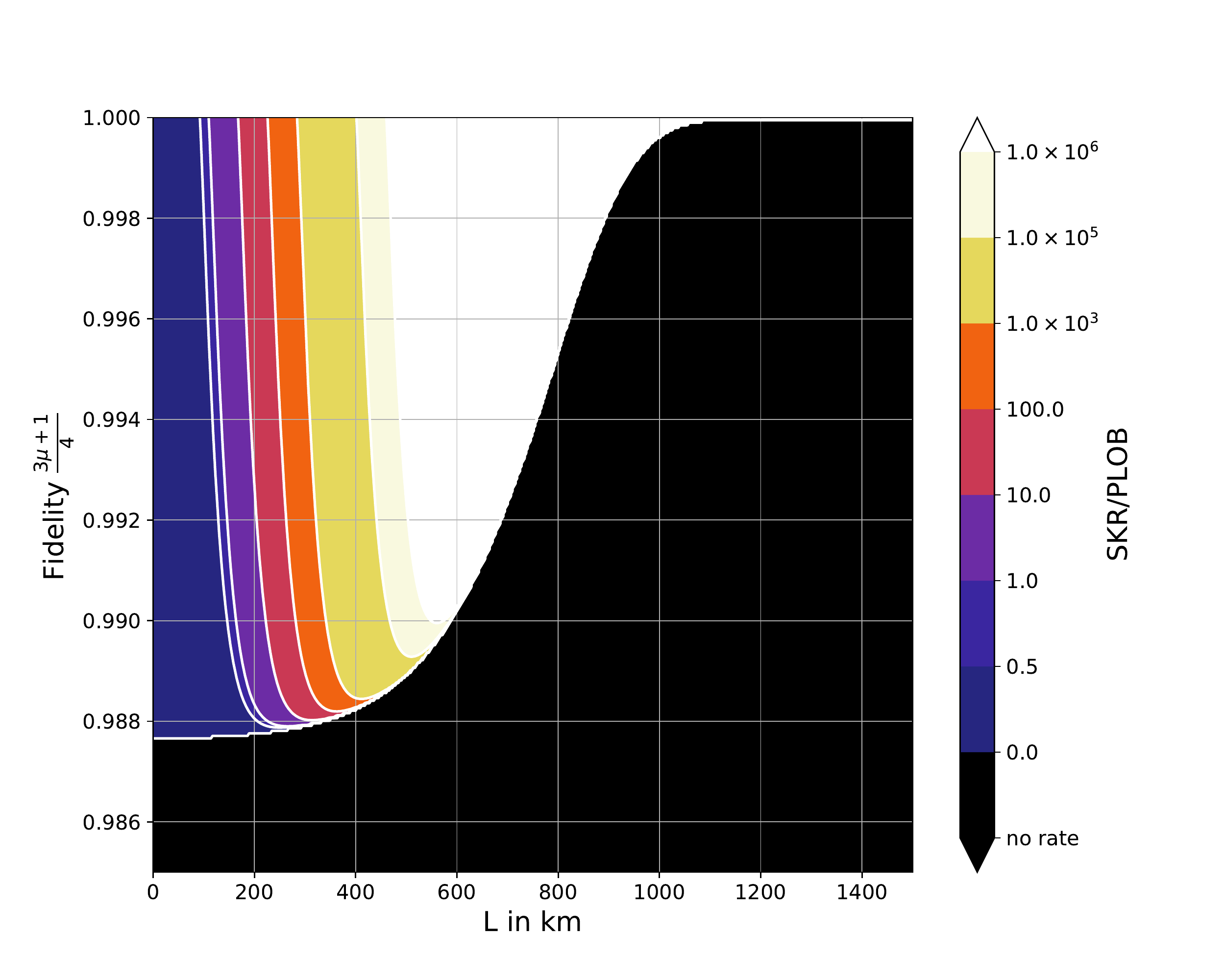}} 
	\subfloat[][$\tau_{\mathrm{coh}}={\unit[10]{s}}$, $p_{\mathrm{link}}=0.7$]{\includegraphics[width=0.5\linewidth]{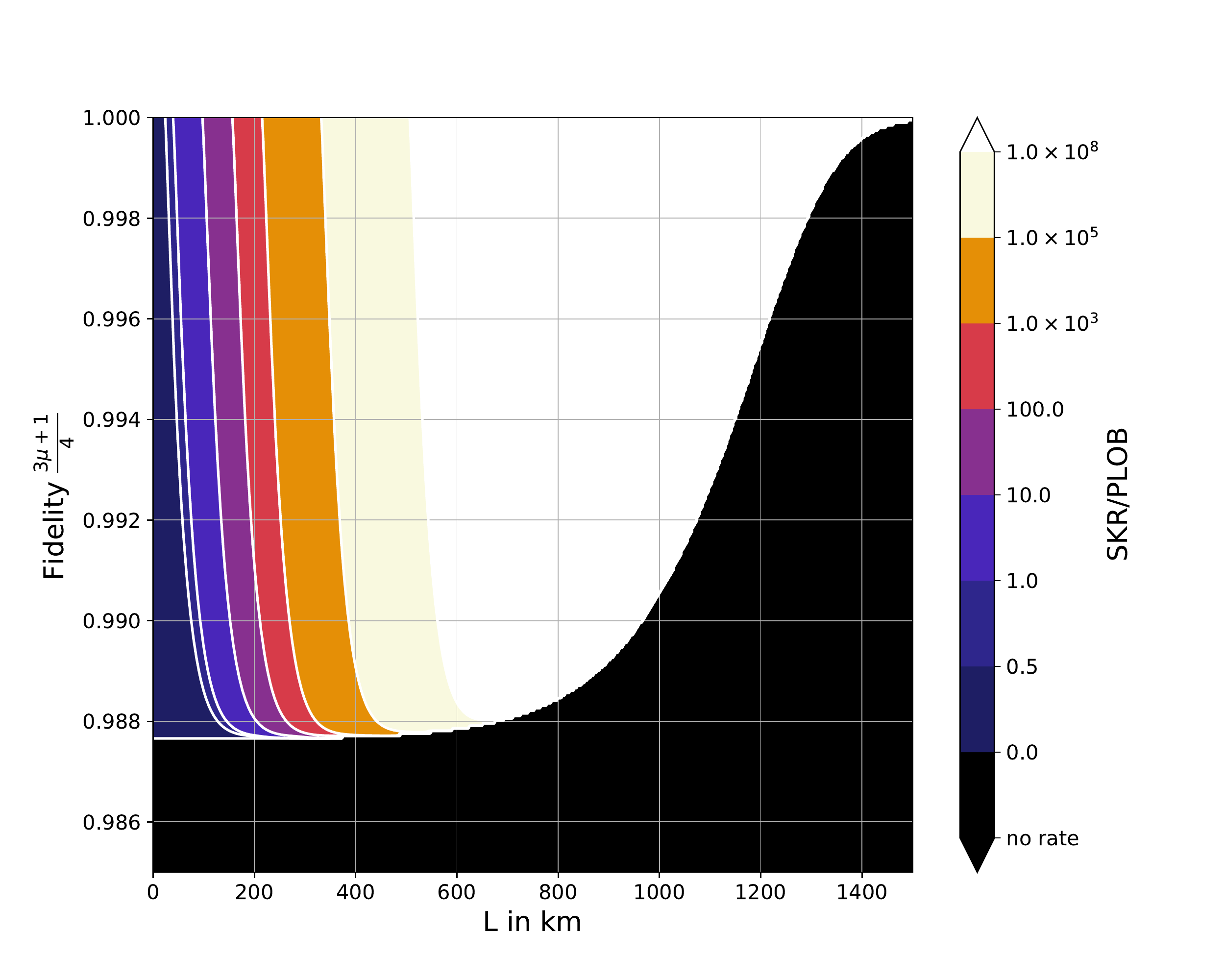}}
	\caption{Contour plots illustrating the minimal fidelity requirements to overcome the PLOB bound by an eight-segment repeater for different parameter sets. In all contour plots, \(\mu = \mu_0\) and \(F_0=1\) has been used.}
	\label{fig:Contour_8_segments}
\end{figure*}

In comparison with the usual treatment of quantum repeaters via doubling the links at each repeater level, the next logical step is to consider an eight-segment repeater. For eight segments, there is an increasing number of possible distribution and swapping strategies, and for the swapping we have discussed this in more detail in Sec.~\ref{sssec:Par-distr: 8-segment repeater}. Here we will only consider the sequential, the doubling, and the optimal schemes (the former one with sequential distributions, the latter two with parallel distributions). Again, in Fig.~\ref{fig:Contour_8_segments}, we present limitations on the error parameter \(\mu\) to overcome the PLOB rate at different distances. The regions are color-coded as before. Compared to the limits observed for a two-segment repeater they exhibit a different behaviour now, but this is again due to the fact that we do not consider a cut-off scheme here. The requirements for the fidelity or \(\mu\) are higher, but this was expected, since the secret key fraction includes terms \(\propto \mu^{2n-1} \), again setting \(\mu_0=\mu\). Nevertheless, for sufficiently high fidelities, the attainable secret key rates are much higher than for any of the previously considered repeater schemes, becoming as high as \(10^8\)-times the rate of the PLOB bound, and beyond.

\begin{figure*}[ht]
	\centering
	\subfloat[a][$\tau_{\mathrm{coh}}={\unit[0.1]{s}}$, $p_{\mathrm{link}}=0.05$, $\mu = \mu_0=0.99$]{\includegraphics[width=0.33\linewidth]{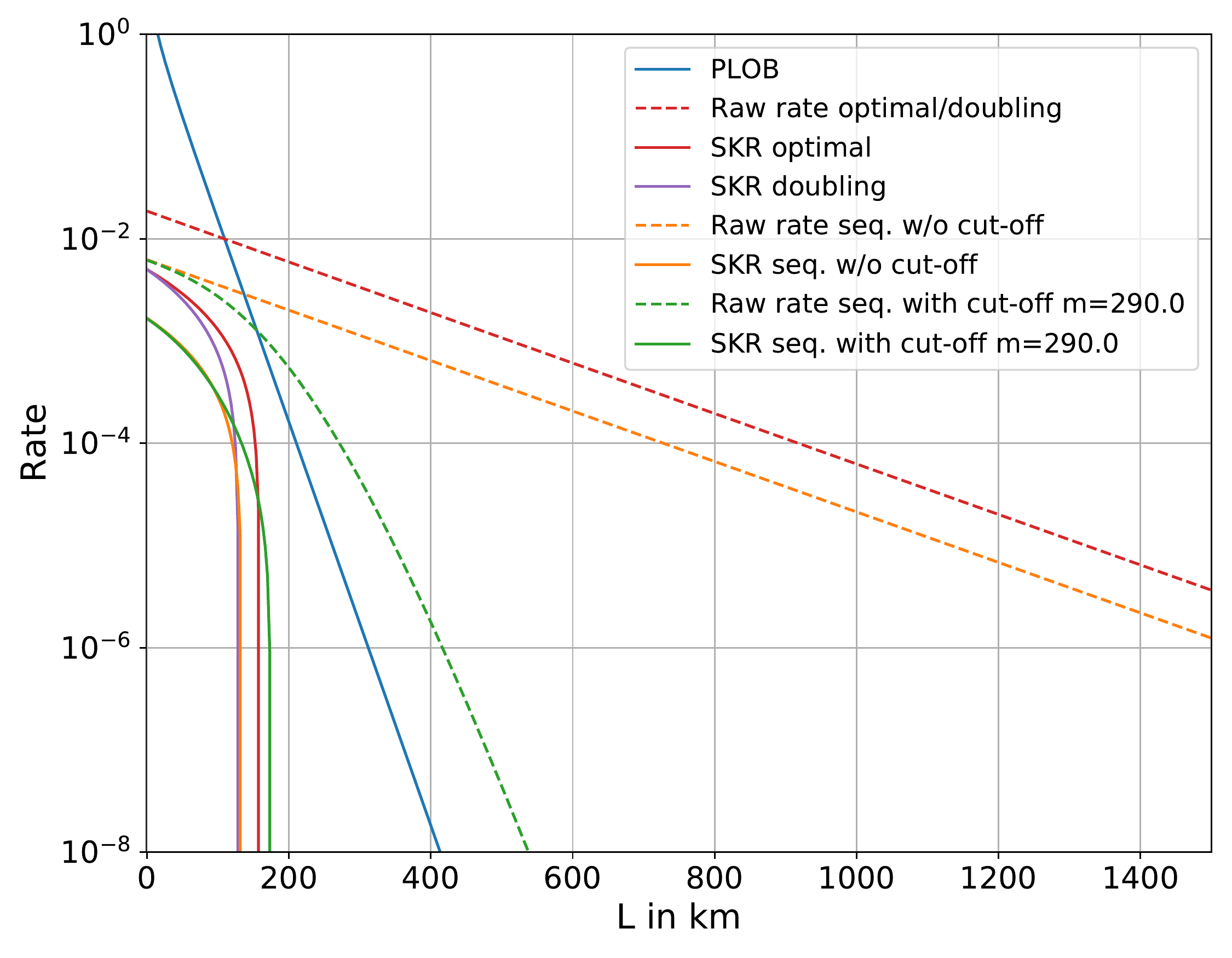}}
	\subfloat[][$\tau_{\mathrm{coh}}={\unit[0.1]{s}}$, $p_{\mathrm{link}}=0.05$, $\mu = \mu_0=1$]{\includegraphics[width=0.33\linewidth]{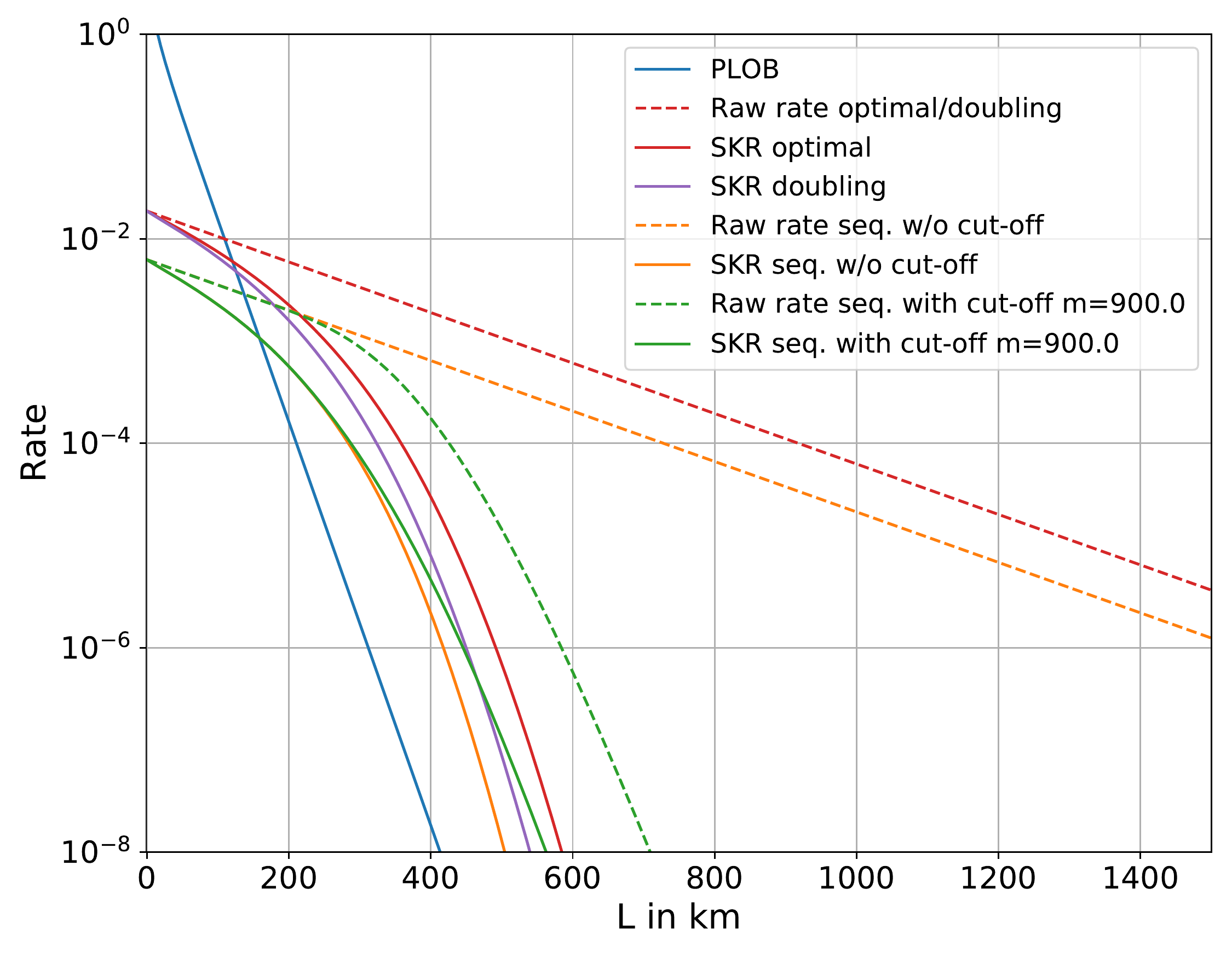}}
	\subfloat[][$\tau_{\mathrm{coh}}={\unit[0.1]{s}}$, $p_{\mathrm{link}}=0.7$, $\mu = \mu_0=0.99$]{\includegraphics[width=0.33\linewidth]{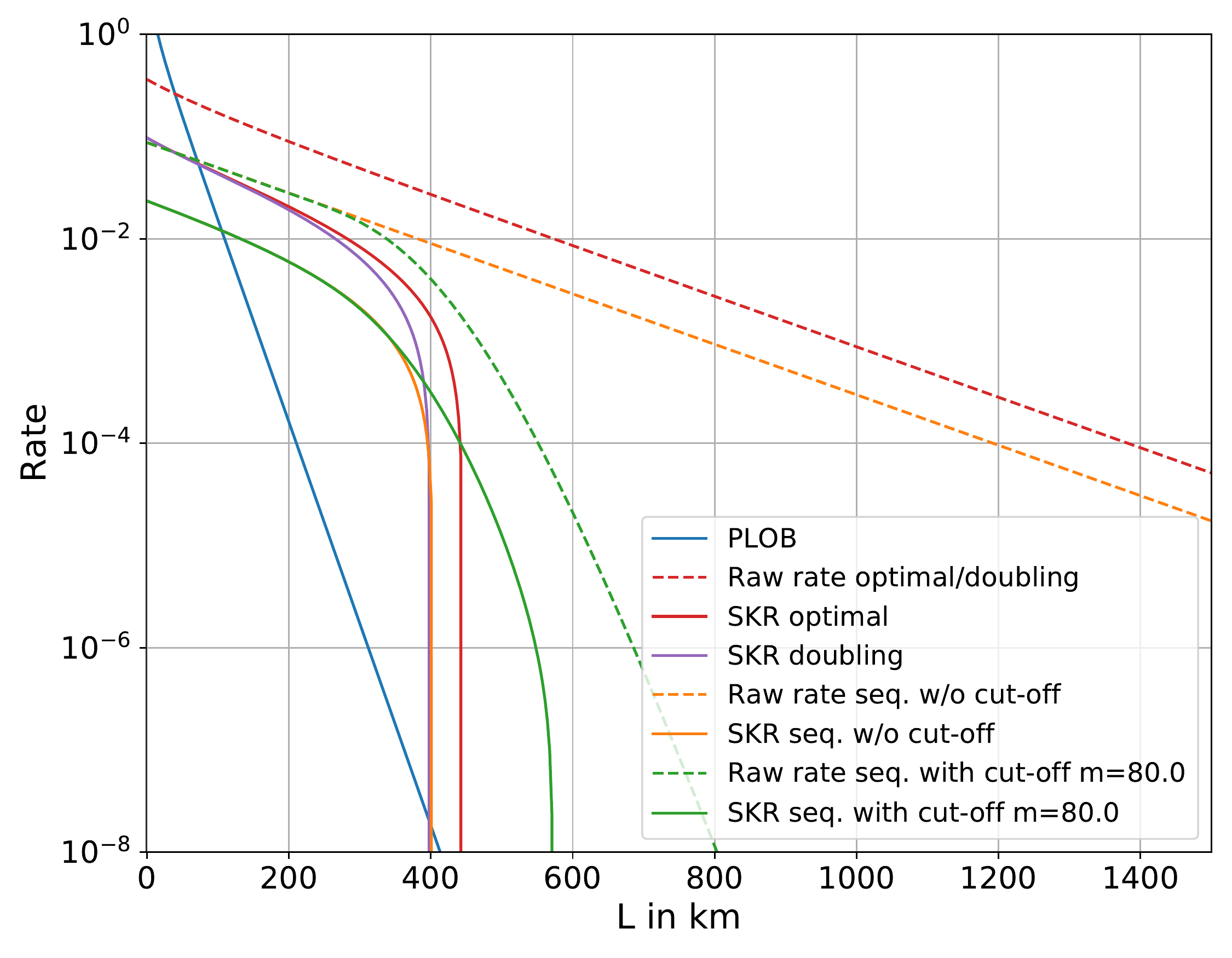}}\\
	\subfloat[][$\tau_{\mathrm{coh}}={\unit[0.1]{s}}$, $p_{\mathrm{link}}=0.7$, $\mu = \mu_0=1$]{\includegraphics[width=0.33\linewidth]{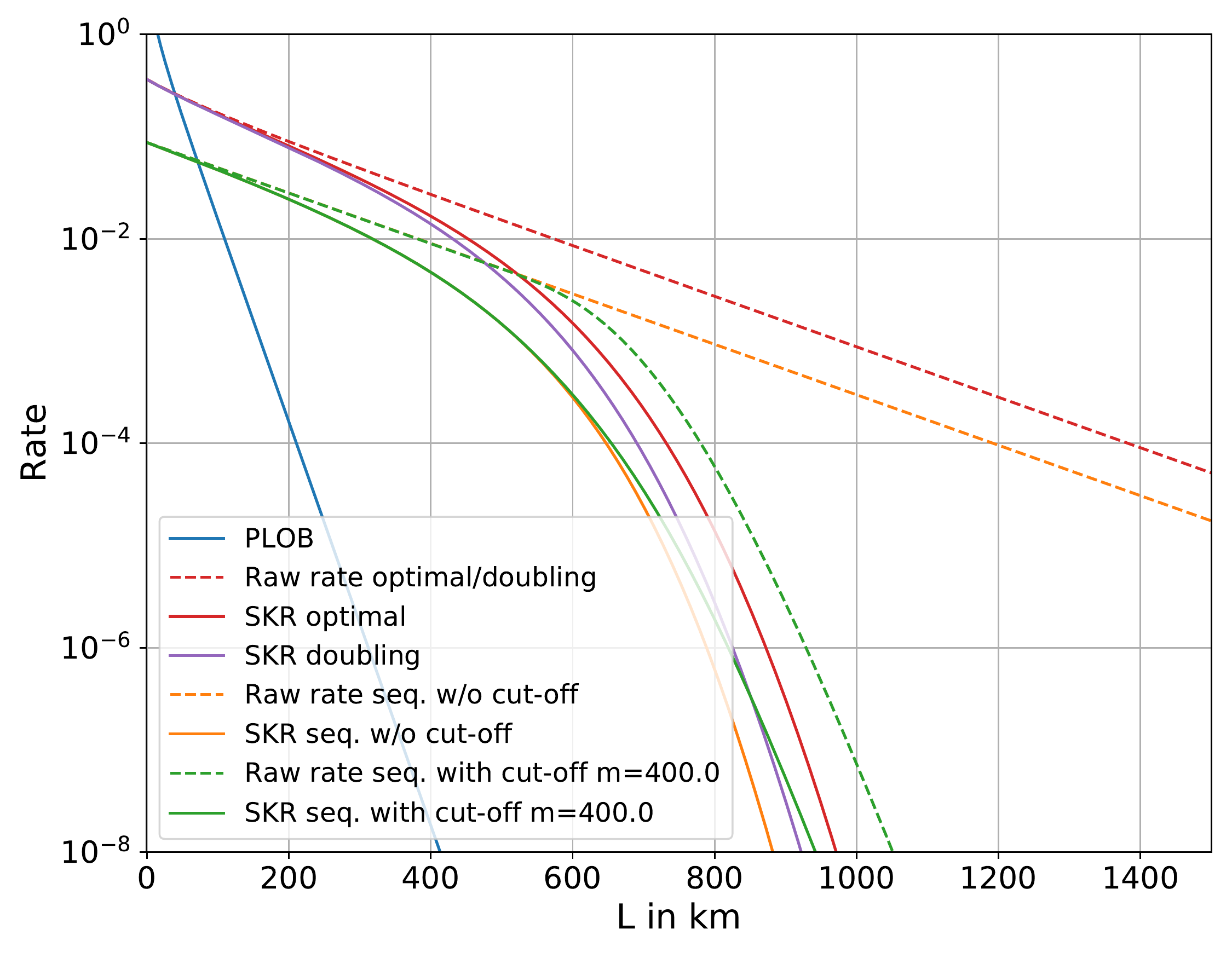}}
	\subfloat[][$\tau_{\mathrm{coh}}={\unit[10]{s}}$, $p_{\mathrm{link}}=0.05$, $\mu = \mu_0=0.99$]{\includegraphics[width=0.33\linewidth]{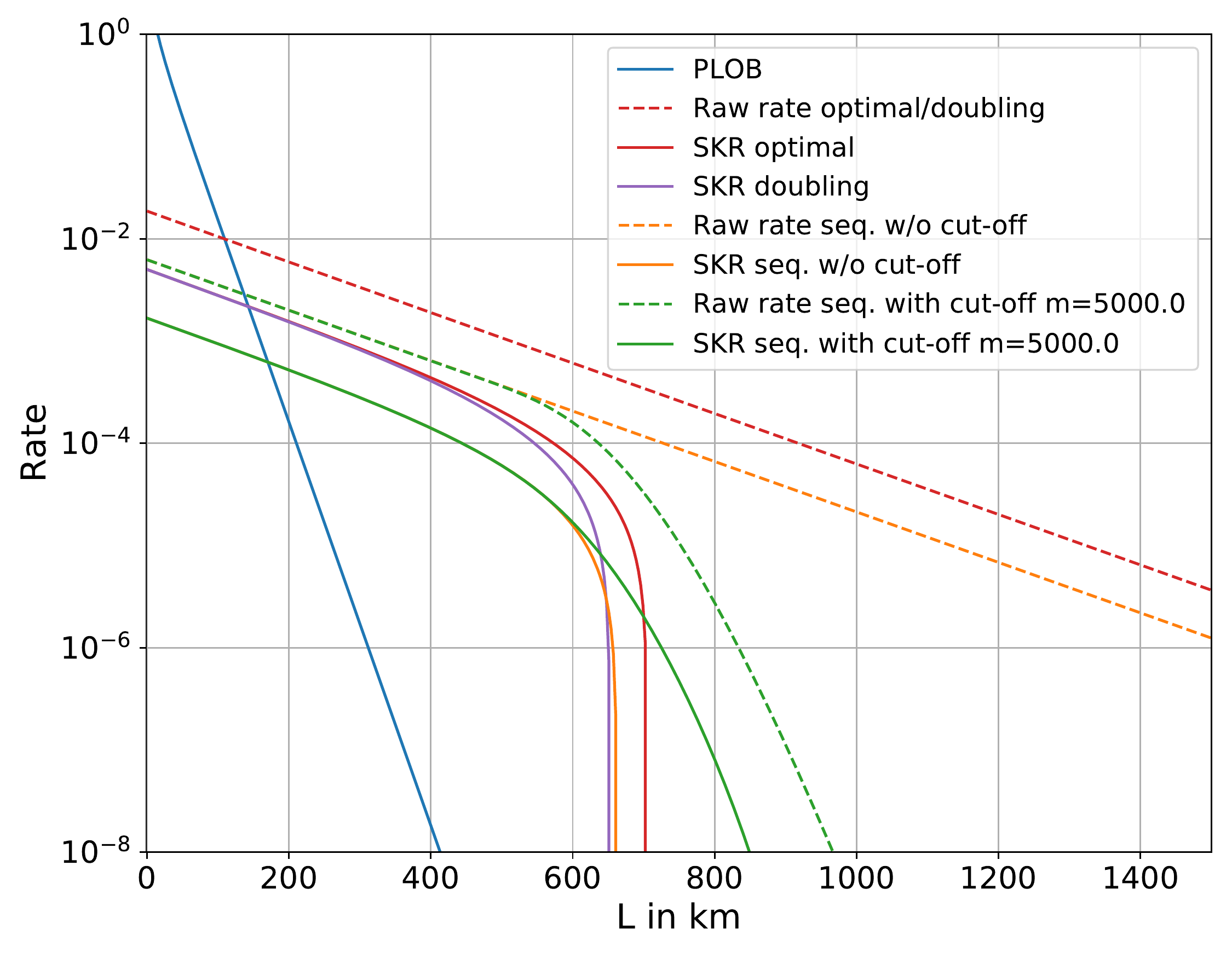}}
	\subfloat[][$\tau_{\mathrm{coh}}={\unit[10]{s}}$, $p_{\mathrm{link}}=0.05$, $\mu = \mu_0=1$]{\includegraphics[width=0.33\linewidth]{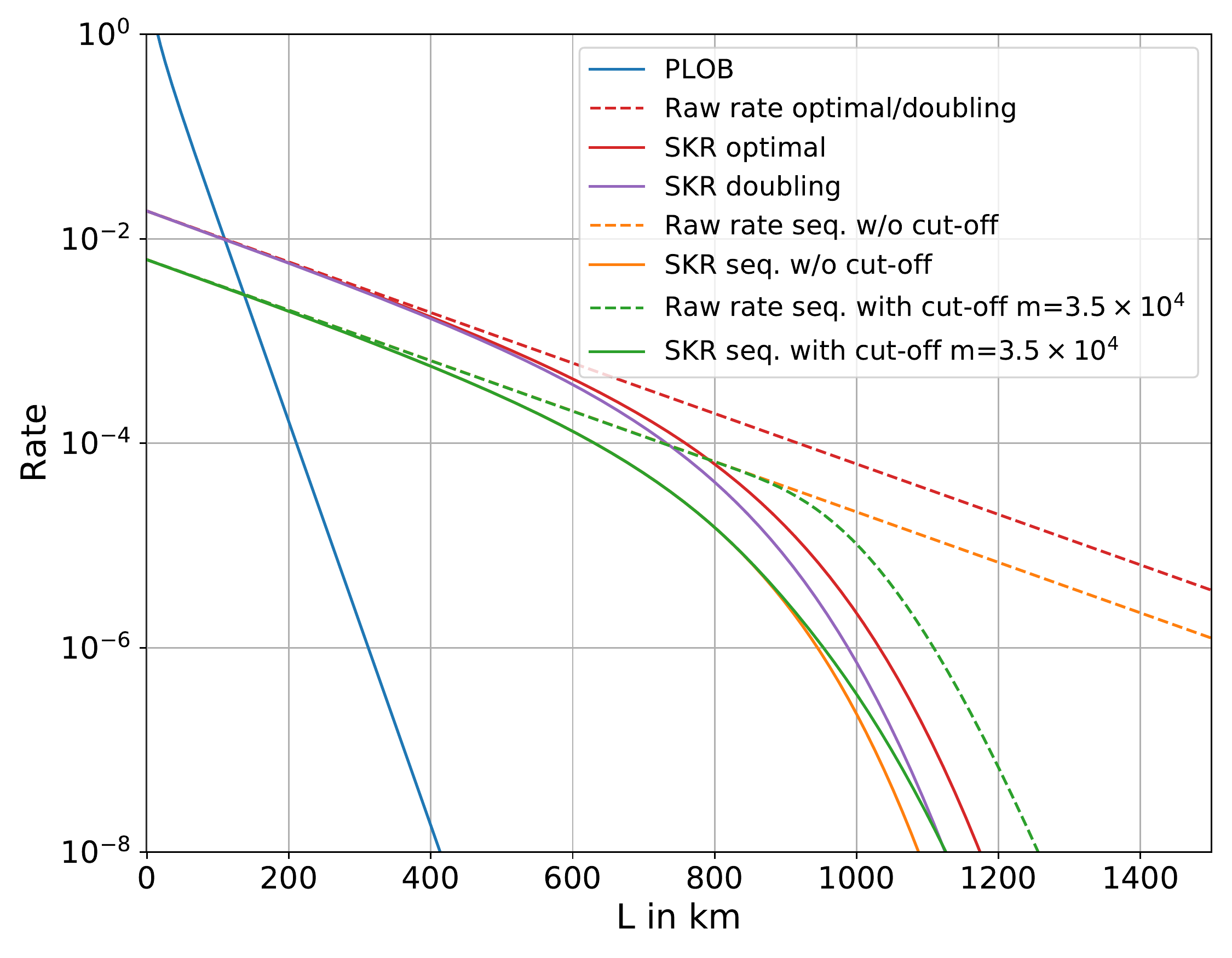}} \\
	\subfloat[][$\tau_{\mathrm{coh}}={\unit[10]{s}}$, $p_{\mathrm{link}}=0.7$, $\mu = \mu_0=0.99$]{\includegraphics[width=0.33\linewidth]{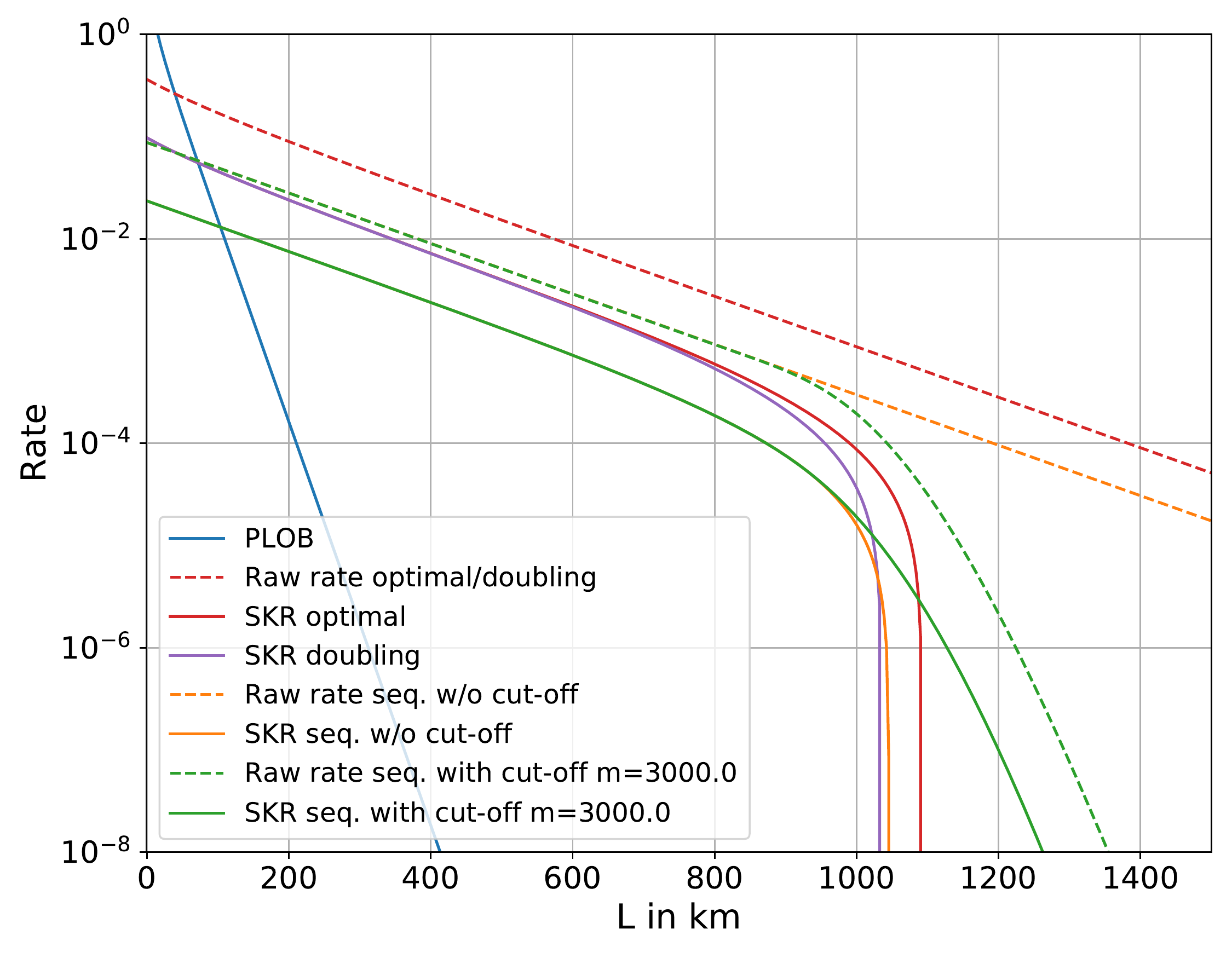}} 
	\subfloat[][$\tau_{\mathrm{coh}}={\unit[10]{s}}$, $p_{\mathrm{link}}=0.7$, $\mu = \mu_0=1$]{\includegraphics[width=0.33\linewidth]{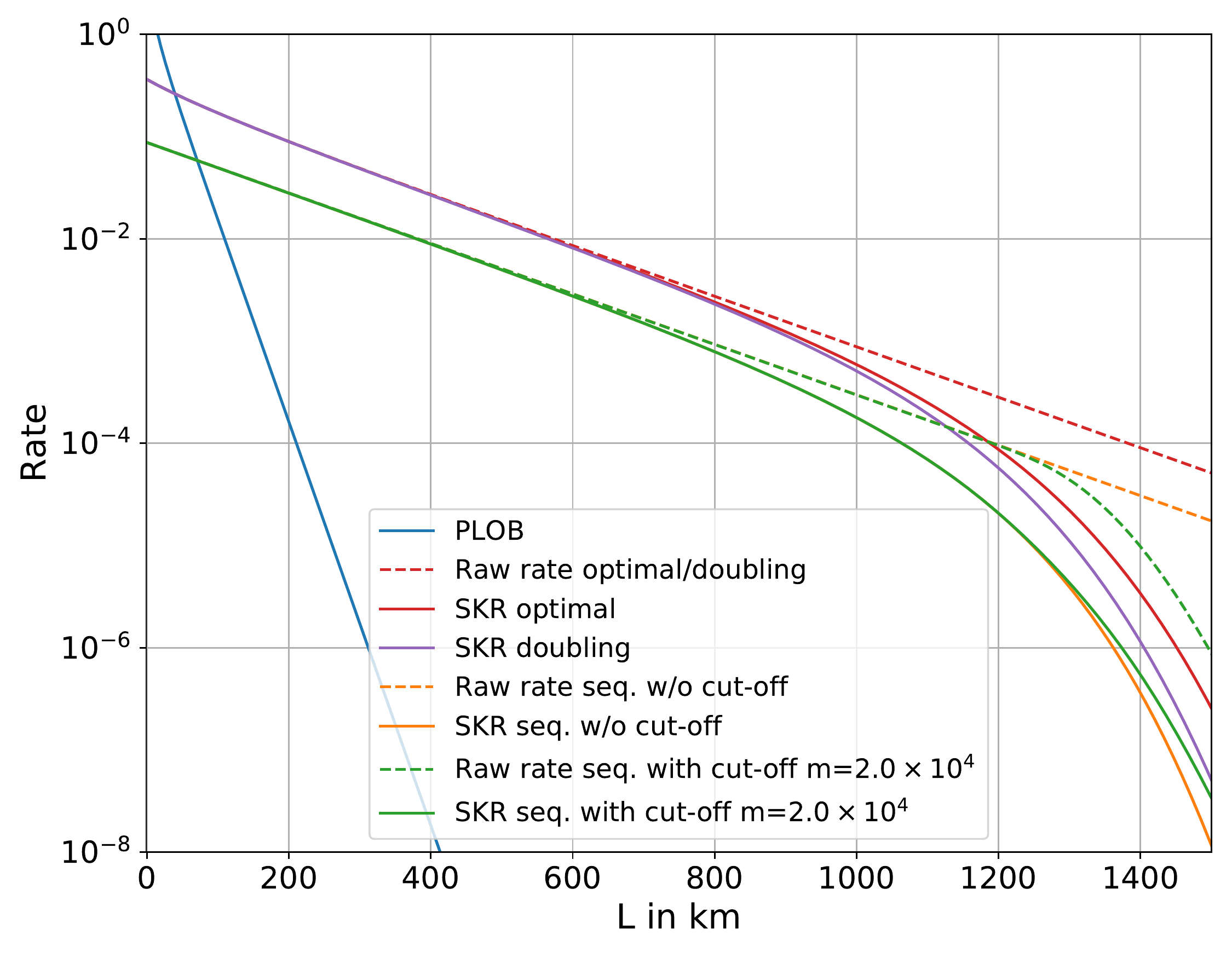}}
	\caption{Rates (secret key/raw) for an eight-segment repeater over distance \(L\) for different experimental parameters.}
	\label{fig:SKR_8_segments}
\end{figure*}

Finally, we have also evaluated the performance of an eight-segment repeater for our experimental parameter set. Now caution is required when these plots are compared directly with the previous ones, as we had to improve the ``current", non-unit value of \(\mu\) to \(\mu=0.99\). Without this fidelity adjustment, it would be impossible to achieve a non-zero secret key rate for an eight-segment repeater (see next section). The $\mu$-scaling with $n$ in the QBERs prohibits to scale up a realistic quantum repeater to arbitrarily large distances and $n$ values, as long as no extra elements for quantum error detection or correction are included. For example, in a 2nd-generation quantum repeater, the effective $\mu_0$ and $\mu$ values could be kept close to one, at the expense of extra resources for quantum error correction and a typically decreasing initial distribution efficiency $p$ (for instance, due to an extra step of entanglement distillation for the distributed, encoded memory qubits). In principle, our formalism could be also applied to such a more sophisticated scenario by considering the effective changes of $\mu$, $\mu_0$, and $p$ (and possibly $\alpha$ too).
Nevertheless, our plots presented in Fig.~\ref{fig:SKR_8_segments} show that an eight-segment quantum repeater in a memory-assisted QKD scheme is, in principle, already able to cover large distances by reaching usable rates up to
\(\unit[1000]{km}\) or even \(\unit[1200]{km}\), provided that \(\mu=0.99\) or \(\mu\rightarrow 1\), respectively. Apart from this, the behaviour of an eight-segment repeater is very similar to that of the previous four-segment repeater.

\subsection{Minimal $\mu$ values}

We have already seen that the secret key rate of memory-assisted QKD is highly sensitive to the depolarizing errors that we use to model the imperfect gates and the imperfect initial states in the quantum repeater. Here let us explicitly give some minimal values for the error parameter $\mu$ which at least have to be achieved in order to obtain a non-zero secret key fraction for QKD protocols restricted to one-way post-processing (see Tab.~\ref{tab_minimalmu}). More generally, in principle, much higher error rates can be tolerated by allowing for two-way post-processing in the QKD protocols \cite{twowayqkd}. However, in this work, we primarily utilize the secret key rate as a practical and useful quantitative figure of merit to assess a quantum repeater's performance. Nonetheless, the quantum repeater schemes that we consider may also be employed for other, more general quantum information and communication tasks. Thus, we decided not to include schemes with two-way post-processing, as this would certainly lead to a narrower specialization towards QKD applications. Clearly, in the context of long-range QKD, we believe that considering schemes with two-way post-processing will be very valuable, since potential, future large-scale quantum repeaters will be rather noisy and therefore protocols which still work for large error rates are very useful. Such a further optimization of our schemes with a special focus on long-range QKD is possible and we leave this option for future work.

It is easy to check that the concatenation of two depolarizing channels with parameters $\mu_1$ and $\mu_2$ is equivalent to a single depolarizing channel with parameter $\mu_1\mu_2$. Thus, for an $n$-segment repeater, we would expect  a total depolarizing channel with parameter $\mu_n=\mu_0^n\mu^{n-1}$. We have carefully and systematically checked and confirmed this in the first part of the paper including other parameters too, such as constant initial and time-dependent memory dephasing.

For the BB84 and the six-state protocols, the amount of tolerable noise, such that a secret key can still be obtained with one-way post-processing, has been extensively studied. For BB84 the error threshold lies at $Q=11.0\%$ and for the six-state protocol it is $Q=12.6\%$ \cite[App. A]{RevModPhys.81.1301}. Since a maximally mixed state results in an error rate of $50\%$, this gives us a constraint on the minimal values of $\mu_n\geq1-2Q$.

More specifically, the BB84 secret key fraction of Eq.~\eqref{eq:skf} on which we focus here vanishes when the two QBERs both exceed $Q=11\%$. This is the case for $\mu_n < 1-2Q$ even when all other elements are perfect, i.e. even when there is no memory dephasing at all ($\alpha \rightarrow 0$). In this case, the two QBERs as described by Eq.~\eqref{eq:QBER} coincide (also assuming zero initial dephasing $F_0=0$) and neither includes a random variable. These two constant QBERs then express the sole faultiness of the repeater elements without any time-dependent quantum storage (i.e., only the initial states and the gates) which can suffice to prevent Alice and Bob from finally sharing a non-zero secret key.     

\begin{table}[]
\begin{tabular}{l|l|l|l|l}
$n$ & \begin{tabular}{@{}c@{}}\quad $\mu_0=1$, \quad \\\quad BB84 \quad \end{tabular}  & \begin{tabular}{@{}c@{}} \quad $\mu_0=\mu$, \quad \\ \quad BB84 \quad \end{tabular} & \begin{tabular}{@{}c@{}} \quad $\mu_0=1$, \quad \\ \quad 6-state \quad \end{tabular} & \begin{tabular}{@{}c@{}} \quad $\mu_0=\mu$, \quad \\ \quad 6-state \quad \end{tabular} \\\hline
2   & 0.780           & 0.920            & 0.748             & 0.908              \\\hline
4   & 0.920           & 0.965            & 0.908             & 0.959              \\\hline
8   & 0.965           & 0.984            & 0.959             & 0.981             
\end{tabular}
\caption{Minimal values of $\mu$ required for a non-zero secret key rate in one-way post-processing protocols.}
\label{tab_minimalmu}
\end{table}

\subsection{Comparisons}

\subsubsection{Sequential vs. doubling vs. optimal schemes}\label{sec:Comparison: Sequential vs. Doubling vs. Optimal scheme}

In the previous sections (together with the appendix) we have presented our results for the obtainable secret key rates of two-, three-, four- and eight-segment quantum repeaters based on various entanglement distribution and swapping strategies. While it is generally straightforward to include a memory cut-off for the case of two segments, for more than two segments, we have achieved this only for the fully sequential scheme.
This was depicted in green in the (non-contour) plots for four and eight segments. 
The memory cut-off allows to maintain a scaling proportional to the PLOB bound even beyond the distance where the scheme without cut-off drops more quickly. As a consequence, the cut-off can significantly increase the achievable distance. However, it is hard to obtain an exact result for the secret key rate for the more complicated swapping strategies. Nonetheless, for larger distances, one could extrapolate the behaviour of the doubling and optimal schemes including a cut-off by simply continuing the curves with lines parallel to the PLOB bound after the drops. Alternatively, inferring from our plots, at larger distances one can rely on a continuation of the curves that behaves exactly like the sequential scheme with memory cut-off. Both approaches give us a fairly good picture of the behaviour of the doubling and optimal schemes including the cut-off.

Nevertheless, the optimal scheme outperforms all other schemes without a cut-off before each one drops completely. The doubling scheme achieves almost similar rates, although it starts earlier to decline. The secret key rates are similar thanks to the equivalent, high raw rates of the doubling and optimal schemes (both being based upon parallel entanglement distributions), and due to our general assumption of deterministic entanglement swapping with $a=1$ \cite{Shchukin2021} \footnote{for $a<1$, regimes exist where in terms of the raw rates ``doubling'' performs strictly worse than ``swap as soon as possible'' \cite{Shchukin2021}, similar to regimes here for the full secret key rates with $a=1$ when the dephasing becomes dominant.}.

Thus, for the doubling scheme one could additionally incorporate nested entanglement distillations in the usual, well-known way, which would allow to reduce the QBERs at the expense of the effective raw rates and with the need of extra physical resources.
While the differences between the doubling and optimal schemes may not be
so large for the repeater sizes mainly considered here ($n\leq 8$),
our exact statistical treatment enabled us to determine the optimal swapping scheme (optimizing the dephasing) and thus allows for a rigorous, quantitative comparison with the non-optimal doubling and possible other (including ``mixed") schemes.
The fully sequential scheme, based on sequential entanglement distributions, leads to the lowest raw rate. The longer total waiting times of this scheme also contribute to an increased accumulated dephasing. On the other hand, the dephasing of the fully sequential scheme remains limited, as only one segment is waiting at any time step. Thus, although theoretically the sequential scheme is the
easiest to calculate, experimentally it would typically result in the lowest secret key rate. Nonetheless, the fully sequential scheme is conceptually special and serves as a very useful reference for comparison with the other schemes.

\subsubsection{Two- vs. four- vs. eight-segment repeaters}\label{sec:Comparison: 2 vs. 4 vs. 8 segment repeaters}

In this section, let us finally address one of the main questions that motivates the exact secret key rate analysis that we have presented: is there an actual benefit of additional (memory) stations and repeater segments compared with schemes that work entirely without quantum memories (such as point-to-point links or twin-field QKD) or compared to schemes with a smaller number of memory stations? More specifically, is it useful to replace a simple two-segment repeater by a four- or eight-segment repeater in a realistic setting, i.e. even when the extra quantum memories are subject to additional preparation and operational errors and contribute to an increased accumulated memory dephasing?
In the preceding section with Tab.~\ref{tab_minimalmu} we saw that the sole faultiness of the memory qubit initial states and gates, even with no time- and distance-dependent memory dephasing, can make the secret key rate completely vanish, and this effect grows with the segment number $n$. 
In the last section of the paper, we shall also look at schemes that minimize the actual number of memory stations by combining the twin-field QKD and repeater memory concepts, for instance, in a four-segment scheme with only one of the three intermediate stations being equipped with memory qubits.

Now here we only consider the ``optimal" scheme (generally and rigorously only without memory cut-off, as discussed before), since this ensures we always consider the highest possible secret key rates. By adding extra repeater stations the requirements on the initial state preparations and the Bell measurements become much higher, where the corresponding terms scale as \(\propto \mu^{n-1} \mu_0^n \) in the QBERs. We stress again that in order to achieve a non-zero secret key rate for the eight-segment repeater, we had to alter the non-ideal value of \(\mu\) of Tab.~\ref{tab:constants} to a sufficiently large value, \(\mu=0.99\), see also Tab.~\ref{tab_minimalmu}. 
For a fair comparison, this value is then also used here to obtain the curves of the two- and four-segment repeaters.

\begin{figure*}[ht]
	\centering
	\subfloat[a][$\tau_{\mathrm{coh}}={\unit[0.1]{s}}$, $p_{\mathrm{link}}=0.05$, $\mu = \mu_0=0.99$]{\includegraphics[width=0.33\linewidth]{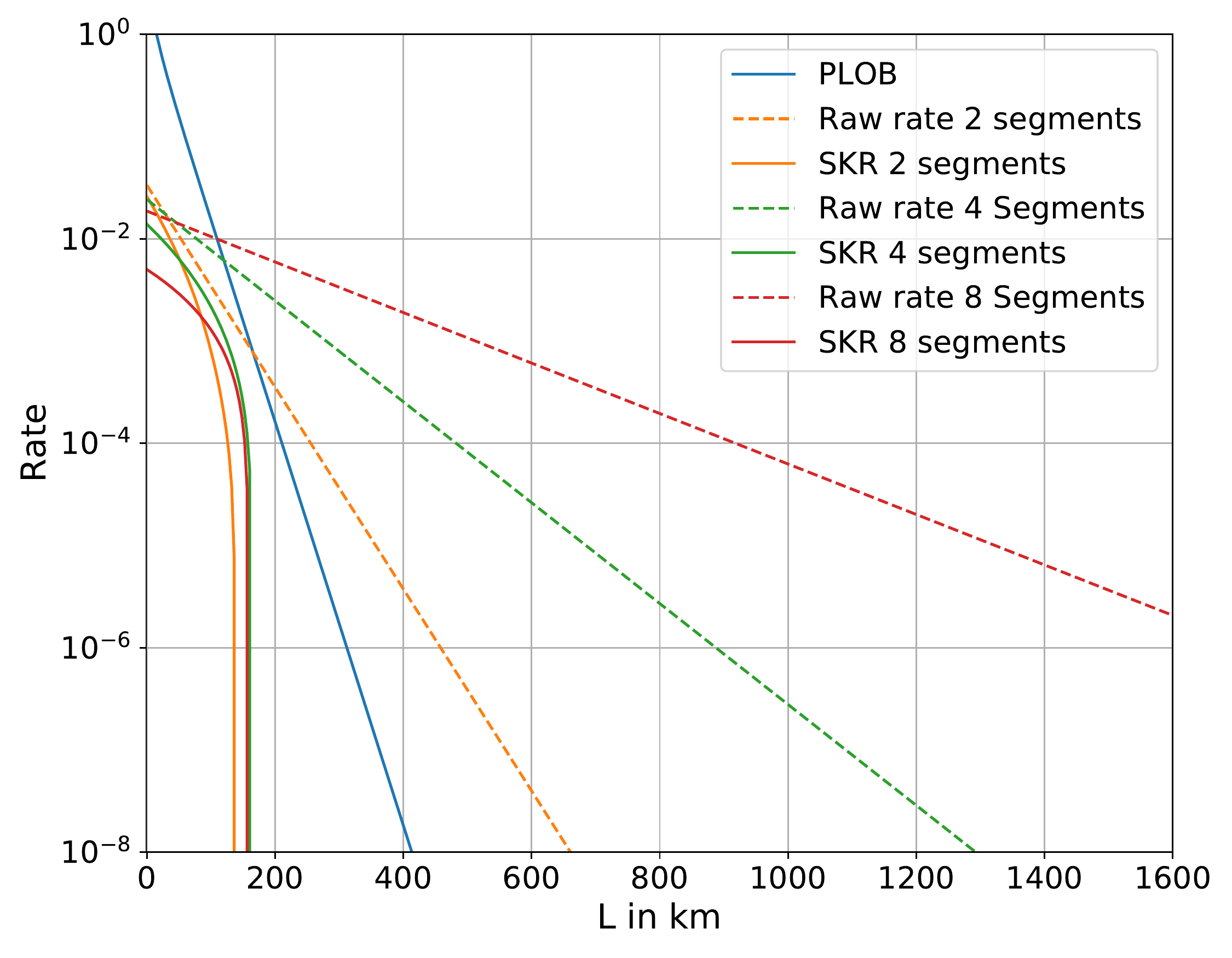}}
	\subfloat[][$\tau_{\mathrm{coh}}={\unit[0.1]{s}}$, $p_{\mathrm{link}}=0.05$, $\mu = \mu_0=1$]{\includegraphics[width=0.33\linewidth]{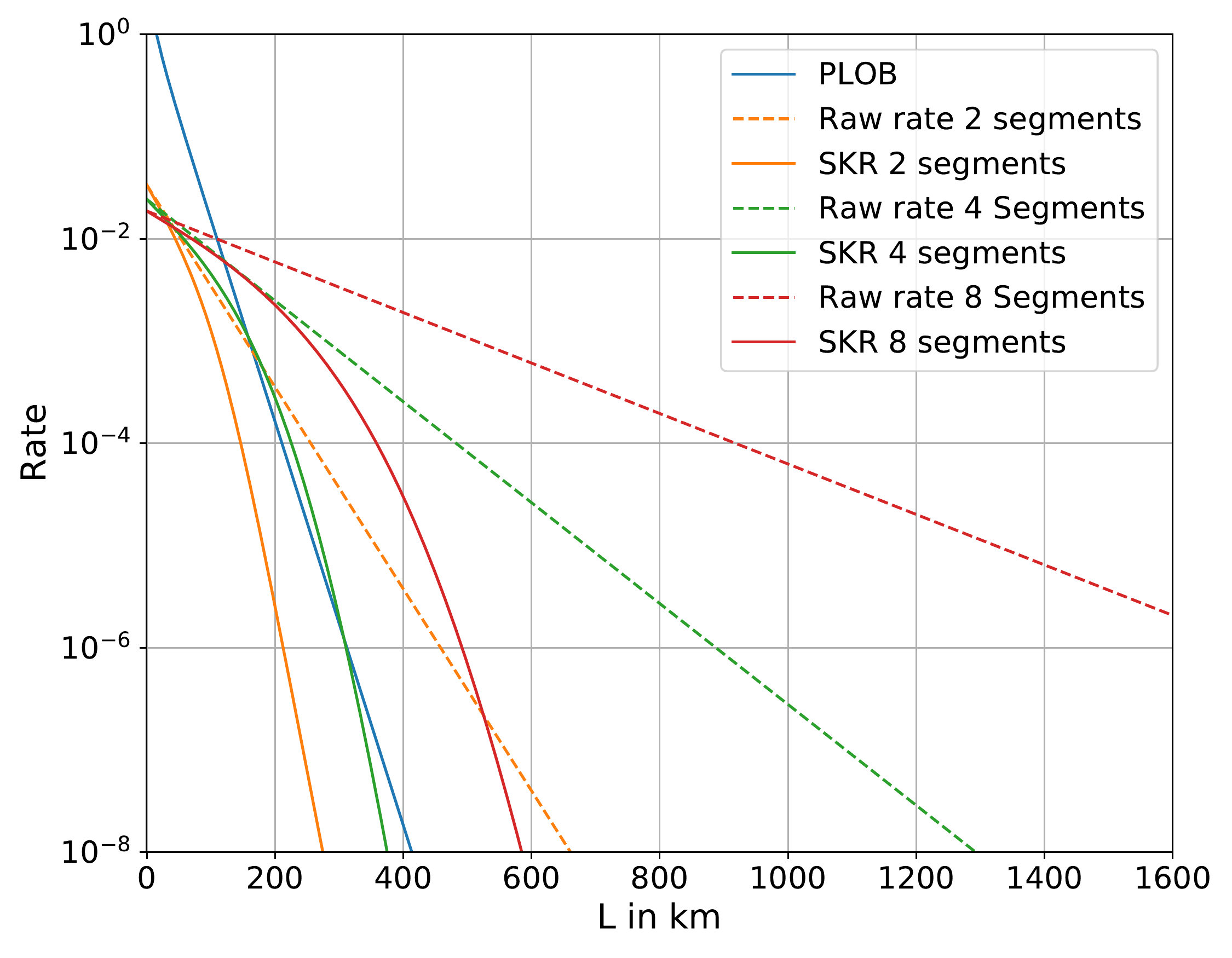}}
	\subfloat[][$\tau_{\mathrm{coh}}={\unit[0.1]{s}}$, $p_{\mathrm{link}}=0.7$, $\mu = \mu_0=0.99$]{\includegraphics[width=0.33\linewidth]{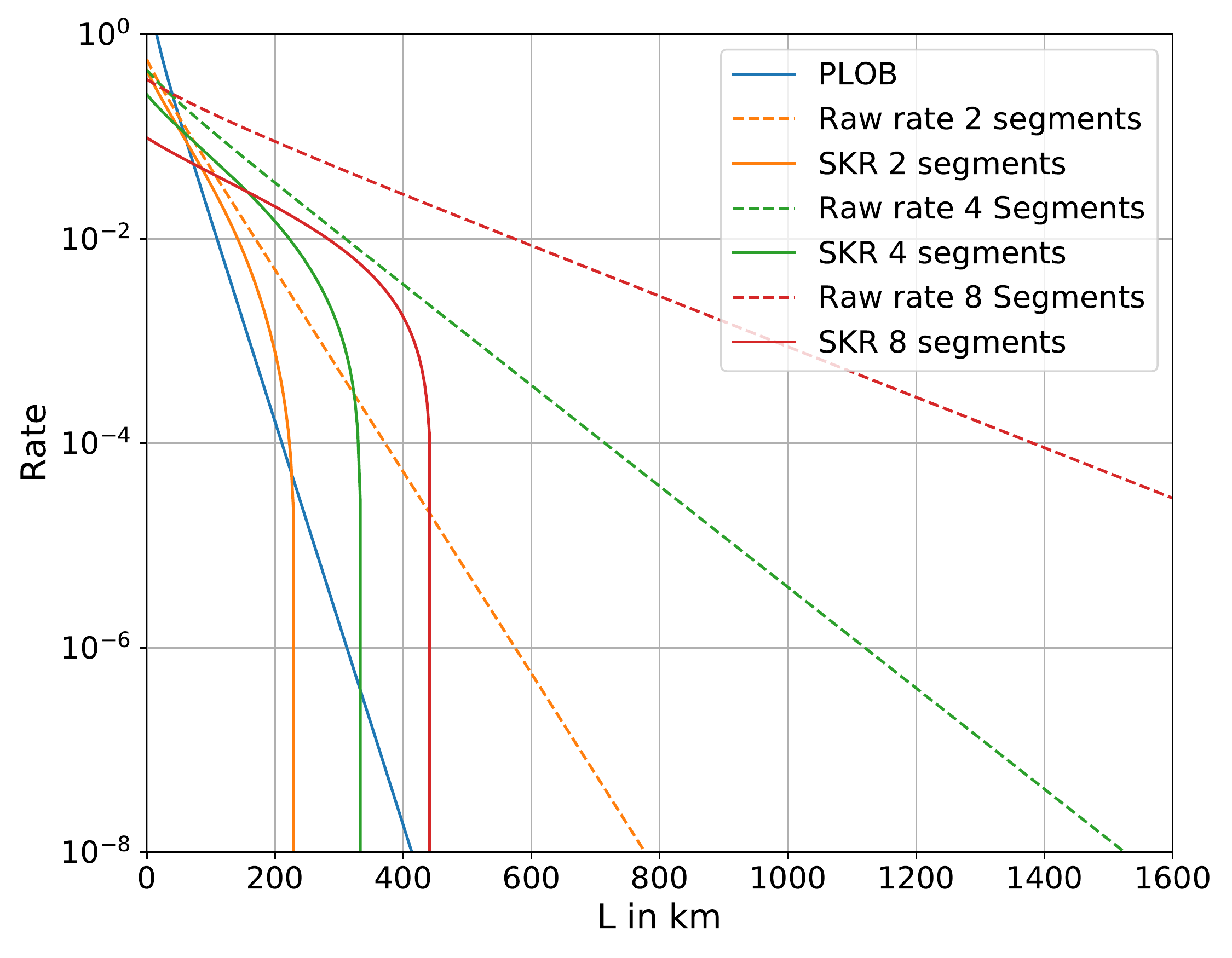}} \\
	\subfloat[][$\tau_{\mathrm{coh}}={\unit[0.1]{s}}$, $p_{\mathrm{link}}=0.7$, $\mu = \mu_0=1$]{\includegraphics[width=0.33\linewidth]{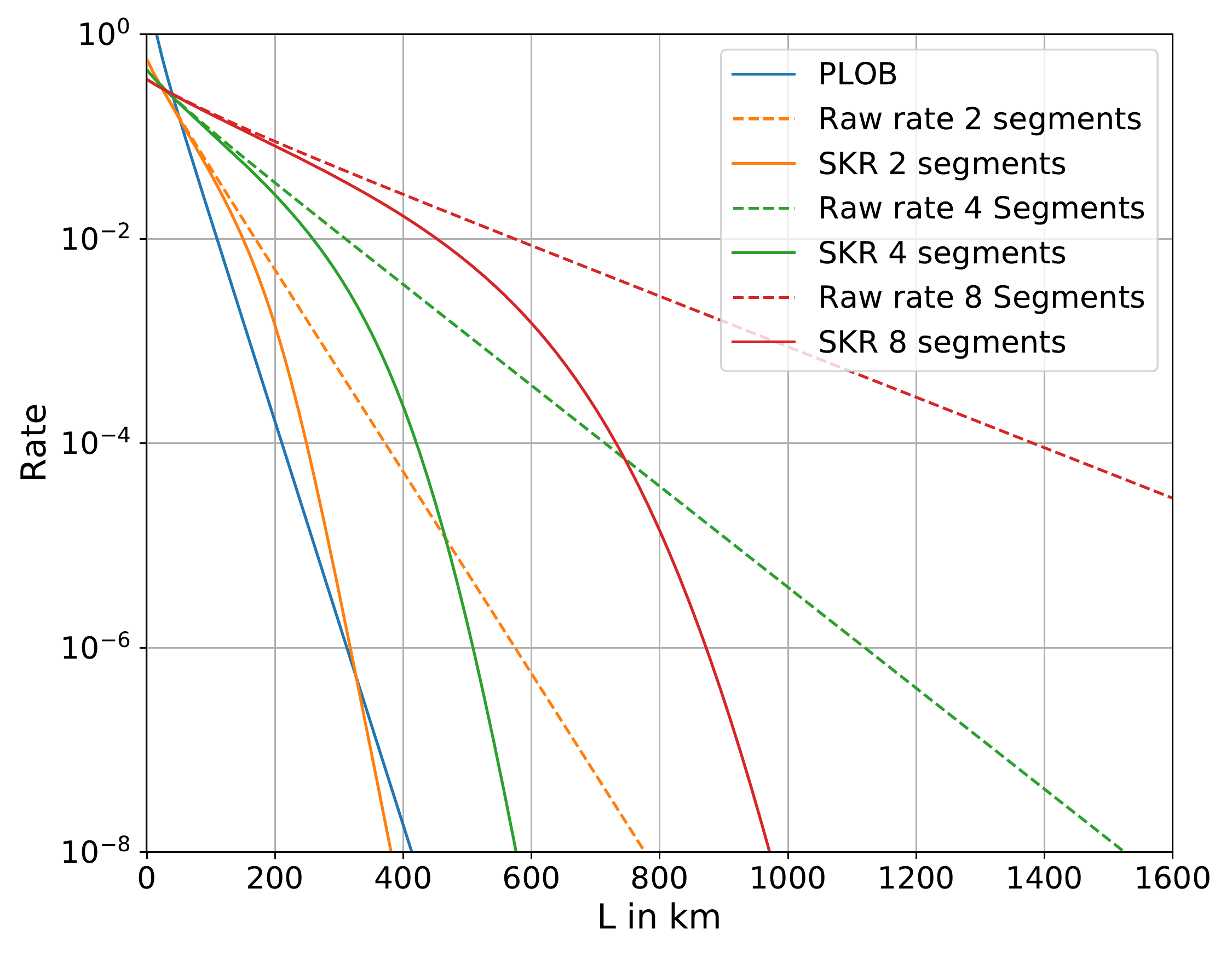}}
	\subfloat[a][$\tau_{\mathrm{coh}}={\unit[10]{s}}$, $p_{\mathrm{link}}=0.05$, $\mu = \mu_0=0.99$]{\includegraphics[width=0.33\linewidth]{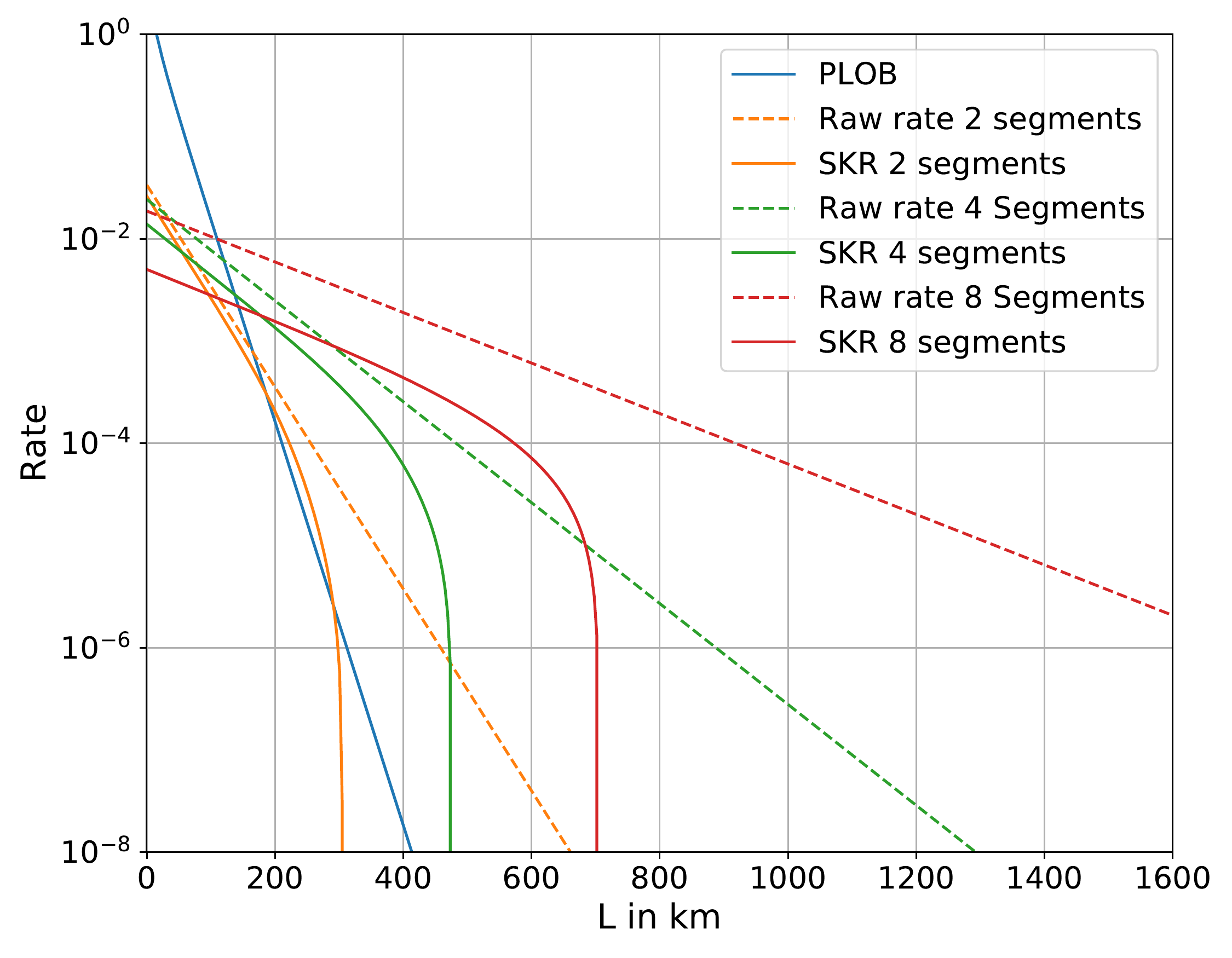}}
	\subfloat[][$\tau_{\mathrm{coh}}={\unit[10]{s}}$, $p_{\mathrm{link}}=0.05$, $\mu = \mu_0=1$]{\includegraphics[width=0.33\linewidth]{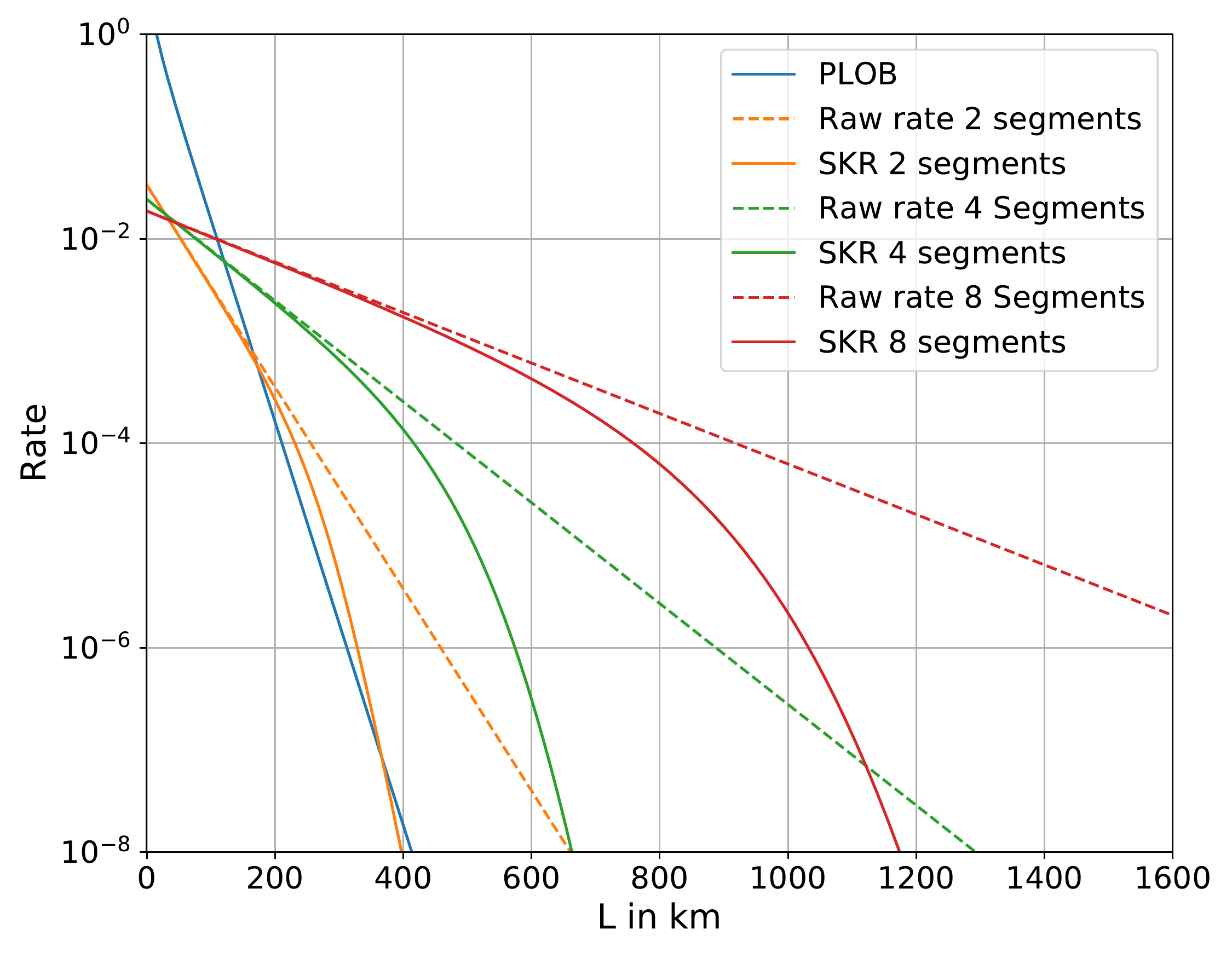}} \\
	\subfloat[][$\tau_{\mathrm{coh}}={\unit[10]{s}}$, $p_{\mathrm{link}}=0.7$, $\mu = \mu_0=0.99$]{\includegraphics[width=0.33\linewidth]{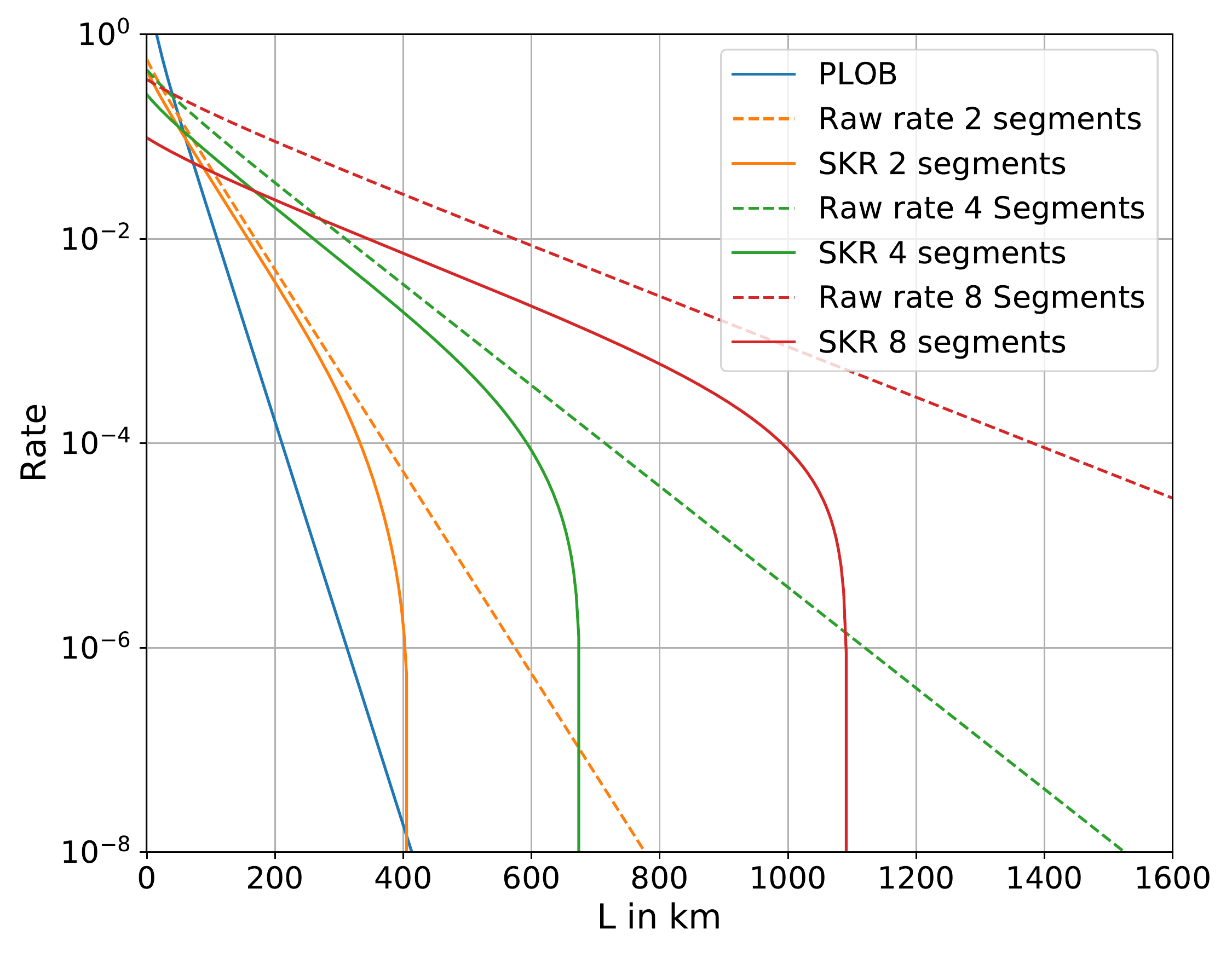}} 
	\subfloat[][$\tau_{\mathrm{coh}}={\unit[10]{s}}$, $p_{\mathrm{link}}=0.7$, $\mu = \mu_0=1$]{\includegraphics[width=0.33\linewidth]{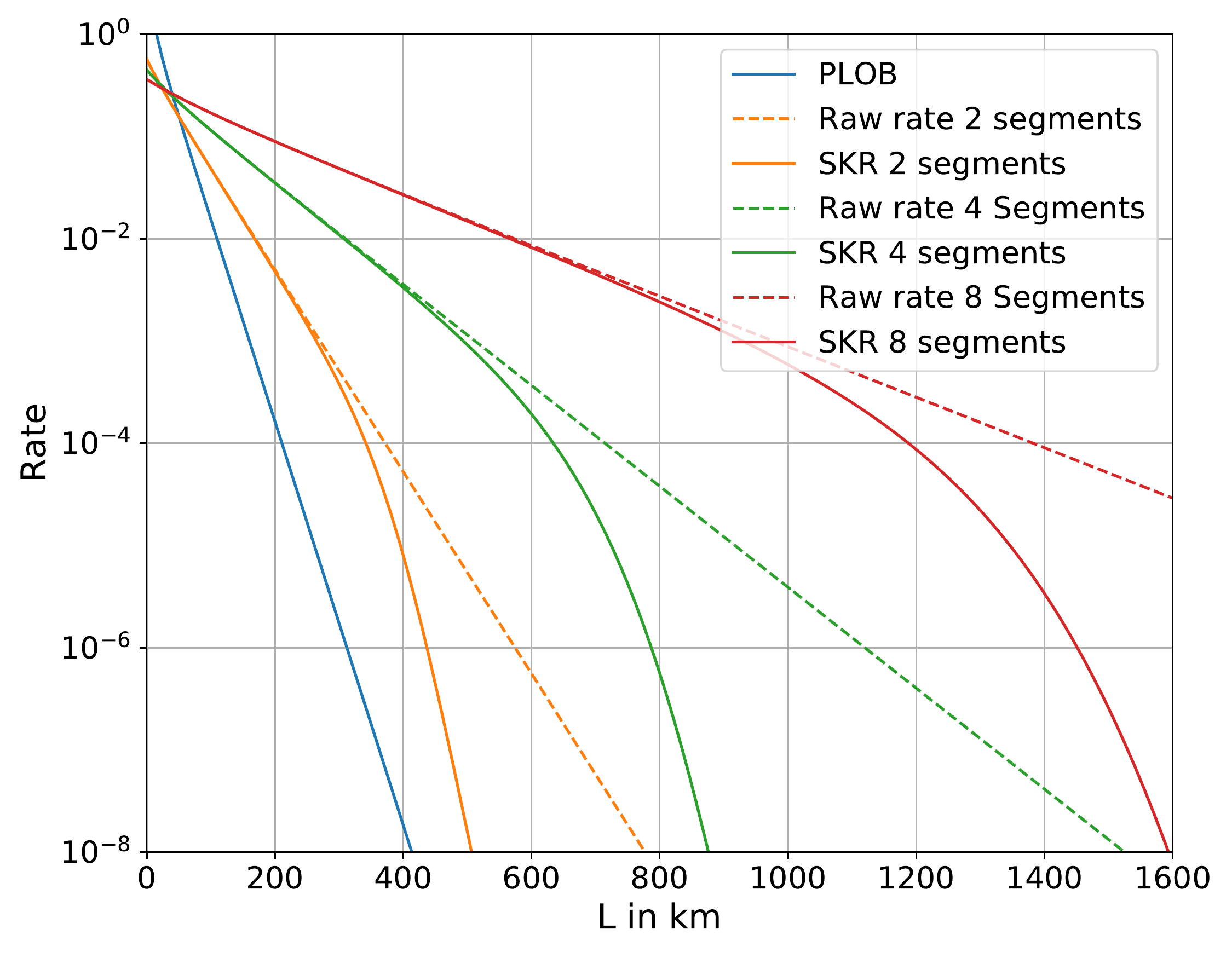}}
	\caption{Comparison of secret key rates of the two-, four-, and eight-segment repeaters at total distances \(L\) for different experimental parameters.}
	\label{fig:SKR_comparison}
\end{figure*}

The resulting secret key rates can be seen in Fig.~\ref{fig:SKR_comparison}. As one would expect, for example, the scaling
changes from \(\sqrt{e^{-\frac{L}{L_{\mathrm{att}}}}} \) to \(\sqrt[8]{e^{-\frac{L}{L_{\mathrm{att}}}}} \)
when the transition from a two-segment to an eight-segment repeater is considered. 
However, the rate at \(L=\unit[0]{km} \) decreases when increasing the number of segments. 
This effect occurs for the raw rates (and the secret key rates assuming $\mu=1$), but it becomes more apparent for $\mu=0.99$.
Still, at long distances, eight segments are superior to a smaller number of segments. 
Therefore, acknowledging that the necessary $\mu$ requirements are extremely demanding but not entirely impossible to achieve in practice,  we conclude that it is indeed beneficial to add repeater stations. In particular, the effect of the memory dephasing alone (besides channel loss), for possible coherence times like those in Tab.~\ref{tab:constants} and used throughout the plots, will not prevent the benefit of adding more stations.
Even when both $p_{\mathrm{link}}$ and $\tau_{\mathrm{coh}}$ take on their lowest of the two considered values as shown in Fig.~\ref{fig:SKR_comparison}(b), by placing seven memory stations along the channel it is in principle still possible to exceed the PLOB bound significantly. However, realistically, when $\mu<1$ like in Fig.~\ref{fig:SKR_comparison}(a), all secret key rates stay below the PLOB bound.
In this case it becomes crucial that either $p_{\mathrm{link}}$ (Fig.~\ref{fig:SKR_comparison}(c)) or $\tau_{\mathrm{coh}}$ (Fig.~\ref{fig:SKR_comparison}(e)) is sufficiently large such that the curves can cross PLOB at a sufficiently small distance (thanks to the small $y$-axis offset) or they can maintain their repeater loss scaling for sufficiently long distances, respectively. Recall that all rates shown and discussed here are per channel use. 
Further it should be stressed here that we did not explicitly include time-dependent memory loss (assuming that the memory imperfections are dominated by the time-dependent memory dephasing), which can additionally jeopardise the benefits of adding more, in this case lossy memory stations \cite{PirEisert}. (If this loss is detectable it may lead to a non-deterministic entanglement swapping like in the ``DLCZ" quantum repeater, which is harder to accurately analyze and optimize even for a constant swapping probability \cite{Shchukin2021}; if the loss remains partially undetected at each station, it can lead to a reduced final state fidelity and thus an increased QBER.)

Let us discuss the comparison of repeaters with different segment numbers in a little more detail.
It is indeed quite subtle and for this we shall also take into account larger repeater systems, far beyond the $n=8$ case. For the general discussion, it is helpful to first consider the fully sequential scheme, as in this case we have access to all relevant (physical and statistical) quantities even for large repeaters, see Tab.~\ref{tab:seqperchanneluse}. If we only consider channel loss or, equivalently, if we only look at the raw rates, there is an optimal number of segments for a given total distance.
In Tab.~\ref{tab:seqperchanneluse}, among the possibilities considered there, this is $n=80$ for $L=800$km, and so we should put stations every $L_0=10$km.
If we include the memory dephasing (``channel-loss-and-memory-dephasing-only case"), we observe that not only the average (number of) waiting time (steps) $\mathbf{E}[K_n]$, but also the average (number of) dephasing time (steps) $\mathbf{E}[D_n]$ is minimized for $n=80$ when $L=800$km. In fact, these two averages, $n/p$ and $(n-1)/p$, respectively, become identical for larger $n$, and both grow in the two limits of many and very few segments, $L_0\rightarrow 0$ ($n\rightarrow \infty$) and $L_0\rightarrow L/2$ ($n\rightarrow 2$), respectively. However, when changing the segment length $L_0$, also the inverse effective coherence time $\alpha=L_0/(c_f \tau_{\mathrm{coh}})$ will change, where now $\alpha$ is simply maximal at $L_0=L/2$ and it steadily becomes smaller when $L_0\rightarrow 0$ at fixed $\tau_{\mathrm{coh}}$. Note that below a certain $L_0$ value the repeater's elementary time unit is no longer dominated by the classical communication times and instead the maximal local processing times must go into $\alpha$ which we refer to as $\alpha^{\mathrm{loc}}$. This effect implies that in order to maximize the effective coherence time $\tau_{\mathrm{coh}}/\tau$, one should simply use as many stations as possible, eventually approaching the limitation given by the local processing times at each station. For these we may typically assume 
$\alpha^{\mathrm{loc}}_1=\tau/\tau_{\mathrm{coh}}={\rm MHz}^{-1}/0.1{\rm s}=0.00001$ and 
$\alpha^{\mathrm{loc}}_2=\tau/\tau_{\mathrm{coh}}={\rm MHz}^{-1}/10{\rm s}=0.0000001$.  

However, the first really relevant quantity to assess the effect of the memory dephasing is the effective average dephasing time $\alpha \mathbf{E}[D_n]$ that is related to the memory dephasing channel evolution. Interestingly, for the fully sequential scheme, this quantity, $\alpha \mathbf{E}[D_n]=(L/n)(n-1)/(c_f \tau_{\mathrm{coh}} p)$, converges for growing $n$ (small $L_0$) to $L/(c_f \tau_{\mathrm{coh}} p)$ with $p\rightarrow 1$. For example, in Tab.~\ref{tab:seqperchanneluse}, for $L=800$km, we have $L/(c_f \tau_{\mathrm{coh}} p)=0.0374$ for $\tau_{\mathrm{coh}}=0.1$s and $L/(c_f \tau_{\mathrm{coh}} p)=0.0004$ for $\tau_{\mathrm{coh}}=10$s. These limits are attainable for about $n=8000$ and for $n=800$, respectively. With $\tau_{\mathrm{coh}}=10$s the limit is also almost attainable for $n=80$, so again $L_0=10$km, and there is no further benefit by further increasing $n$. However, we also have $\alpha^{\mathrm{loc}}_1 \mathbf{E}[D_n]=0.00001\times (n-1)/p=0.0804$ for $n=8000$ and $\alpha^{\mathrm{loc}}_2 \mathbf{E}[D_n]=0.0000001\times (n-1)/p=0.0001$ for $n=800$.

\begin{table*}
	\begin{tabular}{c|c|c|c|c|c|c|c}
        $n$ & 1 & 2 & 4 & 8 & 80 & 800 & 8000 \\
        \hline
        \hline
        $L_0$[km] & 800 & 400 & 200 & 100 & 10 & 1 & 0 \\
        \hline
        $\mathbf{E}[K_n]$ & $\sim 10^{16}$ & $\sim 10^{8}$ & 35497 & 754 & 126 & 837 & 8036 \\
        \hline
        $R$ & $\sim 10^{-16}$ & $\sim 10^{-8}$ & $\sim 10^{-5}$ & 0.0013 & 0.0079 & 0.0012 & 0.0001 \\
        \hline
        $\mathbf{E}[D_n]$ & - & $\sim 10^{8}$ & 26623  & 659  & 124  & 836  & 8035  \\
        \hline
        $\alpha_1$ & - & 0.0192  & 0.0096  & 0.0048  & 0.0005  & $\sim 10^{-5}$ & $\sim 10^{-6}$ \\
        \hline
        $\alpha_1 \mathbf{E}[D_n]$ & - & $\sim 10^{6}$ & 256 & 3.1674 & 0.0598 & 0.0402 & 0.0386 \\
        \hline
        $\alpha_2$ & - & 0.0002  & 0.0001  & $\sim 10^{-5}$ & $\sim 10^{-6}$ & $\sim 10^{-7}$ & $\sim 10^{-8}$ \\
        \hline
        $\alpha_2 \mathbf{E}[D_n]$ & - & 15131  & 2.5576 & 0.0317 & 0.0006 & 0.0004 & 0.0004 \\
        \hline
        $\mathbf{E}[e^{-\alpha_1 D_n}]$ & - & $\sim 10^{-7}$ & $\sim 10^{-6}$ & $0.0729$ & $0.9420$ & $0.9606$ & $0.9621$ \\
        \hline
        $\mathbf{E}[e^{-\alpha_2 D_n}]$ & - & $ \sim 10^{-6}$ & $0.1573$ & $0.9689$ & $0.9994$ & $0.9996$ & $0.9996$ \\
        \hline
        $r_1(\mu=1)$ & - & $\sim 10^{-13}$ & $\sim 10^{-12}$ & $0.0038$ & $0.8106$ & $0.8603$ & $0.8646$ \\
        \hline
        $r_2(\mu=1)$ & - & $\sim 10^{-9}$ & $0.0179$ & $0.8843$ & $0.9961$ & $0.9972$ & $0.9973$ \\
        \hline
        $r_1(\mu=0.99)$ & - & $0$ & $0$ & $0$ & $0$ & $0$ & $0$ \\
        \hline
        $r_2(\mu=0.99)$ & - & $0$ & $0$ & $0.2203$ & $0$ & $0$ & $0$ \\
        \hline
        $S_1(\mu=1)$ & - & $\sim 10^{-21}$ & $\sim 10^{-17}$ & $\sim 10^{-6}$ & $0.0064$ & $0.0010$ & $0.0001$ \\
        \hline
        $S_2(\mu=1)$ & - & $\sim 10^{-17}$ & $\sim 10^{-7}$ & $0.0012$ & $0.0079$ & $0.0012$ & $0.0001$ \\
        \hline
        $S_1(\mu=0.99)$ & - & $0$ & $0$ & $0$ & $0$ & $0$ & $0$ \\
        \hline
        $S_2(\mu=0.99)$ & - & $0$ & $0$ & $0.0003 $ & $0$ & $0$ & $0$ \\
        \hline
		$S^{\mathrm{PLOB,QR}}(L_0)$
		& $\sim 10^{-16}$ & $\sim 10^{-8}$ & 0.0002 & 0.0154 & 1.4530 & 4.4921 & 7.7846 
		\end{tabular}
	\caption{Overview of the relevant quantities for the {\it fully sequential scheme}: segment number $n$, segment length $L_0$[km], average (number of) waiting time (steps) $\mathbf{E}[K_n]$, raw rate $R$, average (number of) dephasing time (steps) $\mathbf{E}[D_n]$, inverse effective coherence time $\alpha_1=L_0/(c_f 0.1{\rm s})$, effective average dephasing time $\alpha_1 \mathbf{E}[D_n]$, inverse effective coherence time $\alpha_2=L_0/(c_f 10{\rm s})$, effective average dephasing time $\alpha_2 \mathbf{E}[D_n]$, average dephasing fractions $\mathbf{E}[e^{-\alpha_1 D_n}]$ and $\mathbf{E}[e^{-\alpha_2 D_n}]$, secret key fractions and rates, $r$ and $S$, for different $\mu=\mu_0$ (subscript corresponds to the choice of $\alpha_1$ or $\alpha_2$, $\mu=1$ is the channel-loss-and-memory-dephasing-only case), and the (repeater-assisted) capacity bound $S^{\mathrm{PLOB,QR}}(L_0)$.
	We further assumed $p_{\mathrm{link}}=F_0=1$ for the link coupling efficiency and the initial state dephasing.}
	\label{tab:seqperchanneluse}
\end{table*}
		
\begin{table*}
	\begin{tabular}{c|c|c|c|c|c|c|c}
		$n$ &1 & 2 & 4 & 8 & 80 & 800 & 8000\\
        \hline
        \hline
        $L_0$[km] &800 & 400 & 200 & 100 & 10 & 1 & 0.1\\
        \hline
        $ \mathbf{E}[K_n] $ & $ \sim 10^{16} $ & $ \sim 10^{8} $ & $ 18487$ & $ 255$ & $ 5.4 $ & $ 2.9 $ & $ 2.2 $\\
        \hline
        $R $ & $ \sim 10^{-16} $ & $ \sim 10^{-8} $ & $ \sim 10^{-5} $ & $ 0.0039$ & $ 0.1841 $ & $ 0.3490 $ & $ 0.4646 $\\
        \hline
        $\mathbf{E}[D_n]$ &  - & $ \sim 10^{8} $ & $ 22923$ & $ 488$ & $<124$ & $<836$ & $<8035$\\
        \hline
        $\alpha_1$ & - & $ 0.0192$ & $ 0.0096$ & $ 0.0048$ & $ 0.0005$ & $ \sim 10^{-5} $ & $ \sim 10^{-6} $\\
        \hline
        $\alpha_1 \mathbf{E}[D_n]$ &- & $ \sim 10^{6} $ & $ 220$ & $ 2.3484$ & $ <0.0582$ & $ <0.0391$ & $ <0.0376$ \\
        \hline
        $\alpha_2$ & - & $ 0.0002$ & $ 0.0001 $ & $ \sim 10^{-5} $ & $ \sim 10^{-6} $ & $ \sim 10^{-7} $ & $ \sim 10^{-8} $\\
        \hline
        $\alpha_2 \mathbf{E}[D_n]$ &- & $ 15131$ & $ 2.2022$ & $ 0.0235$ & $<0.0006$ & $<0.0004$ & $<0.0004$\\
        \hline
        $\mathbf{E}[e^{-\alpha_1 D_n}]$ &- &$ \sim 10^{-6} $ & $ \sim 10^{-5} $ & $ 0.1552$ & $ >0.9420$ & $ >0.9606$ & $ >0.9621$ \\
        \hline
        $\mathbf{E}[e^{-\alpha_2 D_n}]$ &- &$ \sim 10^{-4} $ & $ 0.2215$ & $ 0.9769$ & $>0.9994$ & $>0.9996$ & $>0.9996$ \\
        \hline
        $r_1(\mu=1)$ &- &$ \sim 10^{-13} $ & $ \sim 10^{-11} $ & $ 0.0174$ & $>0.8106$ & $>0.8603$ & $>0.8646$ \\
        \hline
        $r_2(\mu=1)$ &- &$ \sim 10^{-9} $ & $ 0.0357$ & $ 0.9090$ & $>0.9961$ & $>0.9972$ & $>0.9973$ \\
        \hline
        $r_1(\mu=0.99)$ &- &$ 0$ & $ 0$ & $ 0$ & $ 0$ & $ 0$ & $ 0$\\
        \hline
        $r_2(\mu=0.99)$ &- &$ 0$ & $ 0$ & $ 0.2323$ & $ 0$ & $ 0$ & $ 0$\\
        \hline
        $S_1(\mu=1)$ &- &$ \sim 10^{-21} $ & $ \sim 10^{-15} $ & $ 0.0001 $ & $>0.0064$ & $>0.0010$ & $>0.0001$ \\
        \hline
        $S_2(\mu=1)$ &- &$ \sim 10^{-17} $ & $ \sim 10^{-6} $ & $ 0.0036$ & $>0.0079$ & $>0.0012$ & $>0.0001$ \\
        \hline
        $S_1(\mu=0.99)$ &- &$ 0$ & $ 0$ & $ 0$ & $ 0$ & $ 0$ & $ 0$\\
        \hline
        $S_2(\mu=0.99)$ &- &$ 0$ & $ 0$ & $ 0.0009$ & $ 0$ & $ 0$ & $ 0$\\
        \hline
		$S^{\mathrm{PLOB,QR}}(L_0)$
		& $\sim 10^{-16}$ & $\sim 10^{-8}$ & 0.0002 & 0.0154 & 1.4530 & 4.4921 & 7.7846 
		\end{tabular}
	\caption{Overview of the relevant quantities for the {\it optimal scheme}: segment number $n$, segment length $L_0$[km], average (number of) waiting time (steps) $\mathbf{E}[K_n]$, raw rate $R$, average (number of) dephasing time (steps) $\mathbf{E}[D_n]$, inverse effective coherence time $\alpha_1=L_0/(c_f 0.1{\rm s})$, effective average dephasing time $\alpha_1 \mathbf{E}[D_n]$, inverse effective coherence time $\alpha_2=L_0/(c_f 10{\rm s})$, effective average dephasing time $\alpha_2 \mathbf{E}[D_n]$, average dephasing fractions $\mathbf{E}[e^{-\alpha_1 D_n}]$ and $\mathbf{E}[e^{-\alpha_2 D_n}]$, secret key fractions and rates, $r$ and $S$, for different $\mu=\mu_0$ (subscript corresponds to the choice of $\alpha_1$ or $\alpha_2$, $\mu=1$ is the channel-loss-and-memory-dephasing-only case), and the (repeater-assisted) capacity bound $S^{\mathrm{PLOB,QR}}(L_0)$. For the cases $n>8$, not all exact values are available and hence we inserted approximate values or (lower or upper) bounds.
	We assumed $p_{\mathrm{link}}=F_0=1$ for the link coupling efficiency and the initial state dephasing.}
	\label{tab:optperchanneluse}
\end{table*}

Next let us consider the relevant quantities for the optimal scheme as presented in Tab.~\ref{tab:optperchanneluse}. In this case we no longer have access
to all exact values for larger repeaters $n>8$. However, there is a distinction between the waiting times $K_n$ and the dephasing times $D_n$. For the total waiting times or the raw rates $R$ we can calculate the numbers for small and also for larger $n$ according to the exact analytical expression in Eq.~\eqref{eq:Knpar}.
There are also good approximations for both small $n$ (small $p$) and larger $n$ ($p$ closer to one) which may be easier to calculate \cite{PvL, Elkouss2021, Eisenberg}.
Importantly, unlike the case of the fully sequential scheme, the raw rate $R$ now grows with all $n$ (though slowly for larger $n$) thanks to the fast, parallel distributions in all segments together with the loss scaling that improves with $n$. 
This behaviour even matches that of the repeater-assisted capacity bounds for increasing $n$, as given in the last row of Tab.~\ref{tab:optperchanneluse}.
However, recall that for our qubit-based quantum repeaters the raw rate can never exceed one secret bit per channel use, whereas $S^{\mathrm{PLOB,QR}}(L_0)$ can, for decreasing $L_0$.

For the average total dephasing we can calculate the exact values up to $n=8$. Comparing these values in Tabs.~\ref{tab:seqperchanneluse} and \ref{tab:optperchanneluse}, we see that the optimal scheme accumulates less dephasing than the fully sequential scheme when $n=4, 8$. The two competing effects in the fully sequential scheme, long total waiting time versus minimal number of simultaneously stored memory qubits per elementary time unit, overall result in a larger dephasing rate in comparison with our optimal scheme for $n\leq 8$. We extrapolate this relative behaviour to larger $n$ and therefore assume that the dephasing values of the fully sequential scheme may serve as upper bounds on those for the optimal scheme when $n>8$ in Tab.~\ref{tab:optperchanneluse}. We make the same assumption for the other dephasing-dependent quantities, in particular, the secret key fractions, for which the fully sequential values then serve as lower bounds. Looking at the entries of Tab.~\ref{tab:optperchanneluse} for the optimal scheme, as a final result, we conclude that while for $\mu=1$ (``channel-loss-and-memory-dephasing-only'' case) it may be best to choose as many segments as $n=80$ (i.e. stations are placed at every 10km), similar to what is best for the fully sequential scheme (Tab.~\ref{tab:seqperchanneluse}), for $\mu=0.99<1$ we must not go to segment numbers higher than $n=8$. In fact, for $\mu=0.99$, both for the sequential and the optimal schemes, effectively the only non-zero secret key rate is obtainable for $n=8$ and the larger of the two coherence times considered, with a factor-three enhancement for the optimal scheme over the sequential one. If $n>8$, the faulty states and gates make $S$ vanish, if $n<8$ the small raw rates and the high effective average dephasing times do not permit practically usable secret key rates. Note that the entire discussion here in the context of Tabs.~\ref{tab:seqperchanneluse} and \ref{tab:optperchanneluse} is for a total distance of $L=800$km. We may infer that an elementary segment length of $L_0 \sim 100$km is not only highly compatible with existing classical repeater and fiber network architectures, but also seems to offer a good balance between an improved memory-assisted loss scaling and an only limited addition of extra faulty elements. This conclusion here holds for our repeater setting based upon heralded loss-tolerant entanglement distribution, deterministic entanglement swapping, and a memory dephasing model. Similar elementary lengths have been used before for schemes with probabilistic entanglement swapping and memory loss \cite{DLCZ, Sangouard}.
For schemes with deterministic entanglement swapping, but a less loss-tolerant entanglement distribution mechanism, \cite{HybridPRL} smaller segment lengths may be preferable.
We will include such schemes, exhibiting an intrinsic channel-loss-dependent dephasing, into the discussion in a later section. Let us now consider a simple form of multiplexing in order to improve the repeater performance, provided sufficient extra resources are available.

\subsection{Multiplexing}

Operating $M$ repeater chains in parallel automatically leads to an enhancement of the overall rates by a factor of $M$. However, since in this case the corresponding number of channels grows as well by a factor of $M$, the rates per channel use remain unchanged. The situation becomes different though when the chains can ``interact" with each other. In particular, the loss scaling of heralded entanglement distributions can be improved, at least for small systems in an MDI QKD setting (even without the use of quantum memories but with the need for a nondestructive heralding) \cite{Azuma2015}.
For memory-based quantum repeaters, memory imperfections may be compensated via multiplexing techniques \cite{CollinsPrl,MunroNatPhot,LutPRA,RazLut}.

Experimentally, multiplexing can be realized through various degrees of freedom.
Apart from spatial multiplexing with additional memory qubits at each station that can be coupled to additional fiber channels, this can be forms of temporal or spectral multiplexing where a single fiber may be employed sequentially at a high clock rate \cite{Cody_Jones} or at the same time with multiple wavelengths, respectively.

In this section, we shall incorporate a simple form of multiplexing into our formalism and our repeater models and systems. We have seen that either high total efficiencies or sufficiently long coherence times are needed to achieve usable secret key rates at long distances. We will now see that multiplexing can be understood as a means to effectively enhance the memory coherence time. In the following we will describe in more detail which kind of multiplexing we consider and why it indeed effectively increases the coherence time.

\begin{figure}
	\centering
	\includegraphics[width=\linewidth]{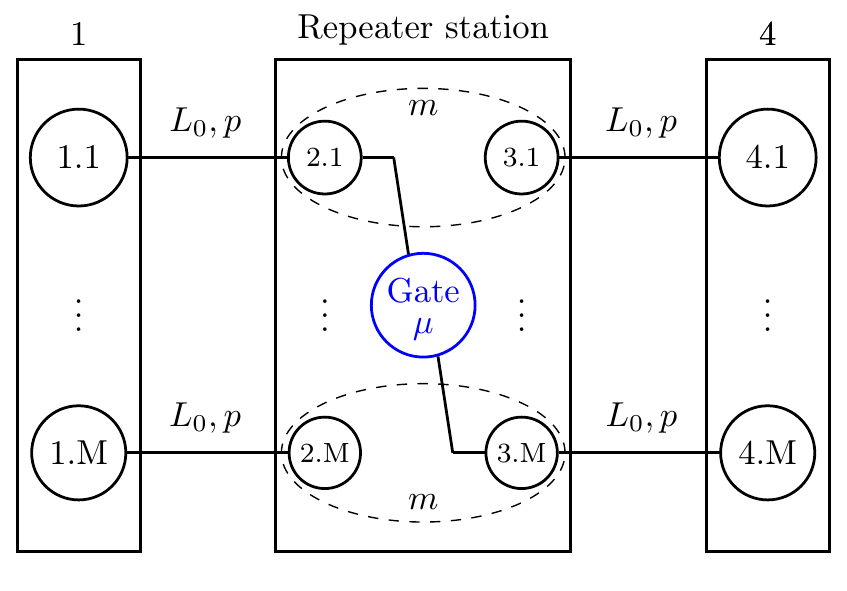}
	\caption{Multiplexing in a two-segment repeater.}
	\label{fig:figure-multiplexing}
\end{figure}

The simplest way to include multiplexing in our repeater models is by using \(M\) memories simultaneously to generate entanglement. These memories can either be connected to the same fiber by a switch or they may each be coupled to their own fiber channel. For simplicity, we consider the switch to be perfect such that both approaches become equivalent (and where the additional channel uses take place either in time or in space). A lossy switch could be easily incorporated into our model by using an additional parameter which is included in \(p_{\mathrm{link}}\) (note that the loss from the switch is time-independent and so 
always the same). A possible setup for a two-segment repeater with multiplexing is shown in Fig.~\ref{fig:figure-multiplexing}. Here all entanglement distribution attempts happen simultaneously. Since we have \(M\) replica of all  memories and channels, this setup acts as if \( p \mapsto 1-(1-p)^M\), provided that memory qubits from different chains can talk to each other in the middle station so that we may again swap as soon as possible.

For an $M$-multiplexing let us thus define the effective distribution probability $p_{\mathrm{eff}}=1-(1-p)^M$. For small $p$, only keeping linear terms, we have $p_{\mathrm{eff}}\approx M p$.
As the expected waiting time in a single segment is then given by $\frac{1}{Mp}$, we can already gain insight on the possibility that multiplexing increases the effective coherence time by a factor of $M$.
More specifically, for example, for the fully sequential scheme the expectation value of $D_n$ is $(n-1)/p$, thus the transition \( p \mapsto p_{\mathrm{eff}}\approx M p\) reduces the number of dephasing steps, on average, by a factor of $1/M$. 
This is equivalent to an increase of the coherence time by a factor $M$.
In the following, let us be more precise and show what `small' $p$ really means in terms of the corresponding segment length $L_0$. 
In fact, including multiplexing, the secret key rates in dependence of the repeater distance behave in a more complicated way and one can see that for small distances the rate is nearly constant and only for larger distances the rates  behave as we would expect from the non-multiplexed schemes.

\begin{figure}
    \centering
    \subfloat[]{\includegraphics[width=\linewidth]{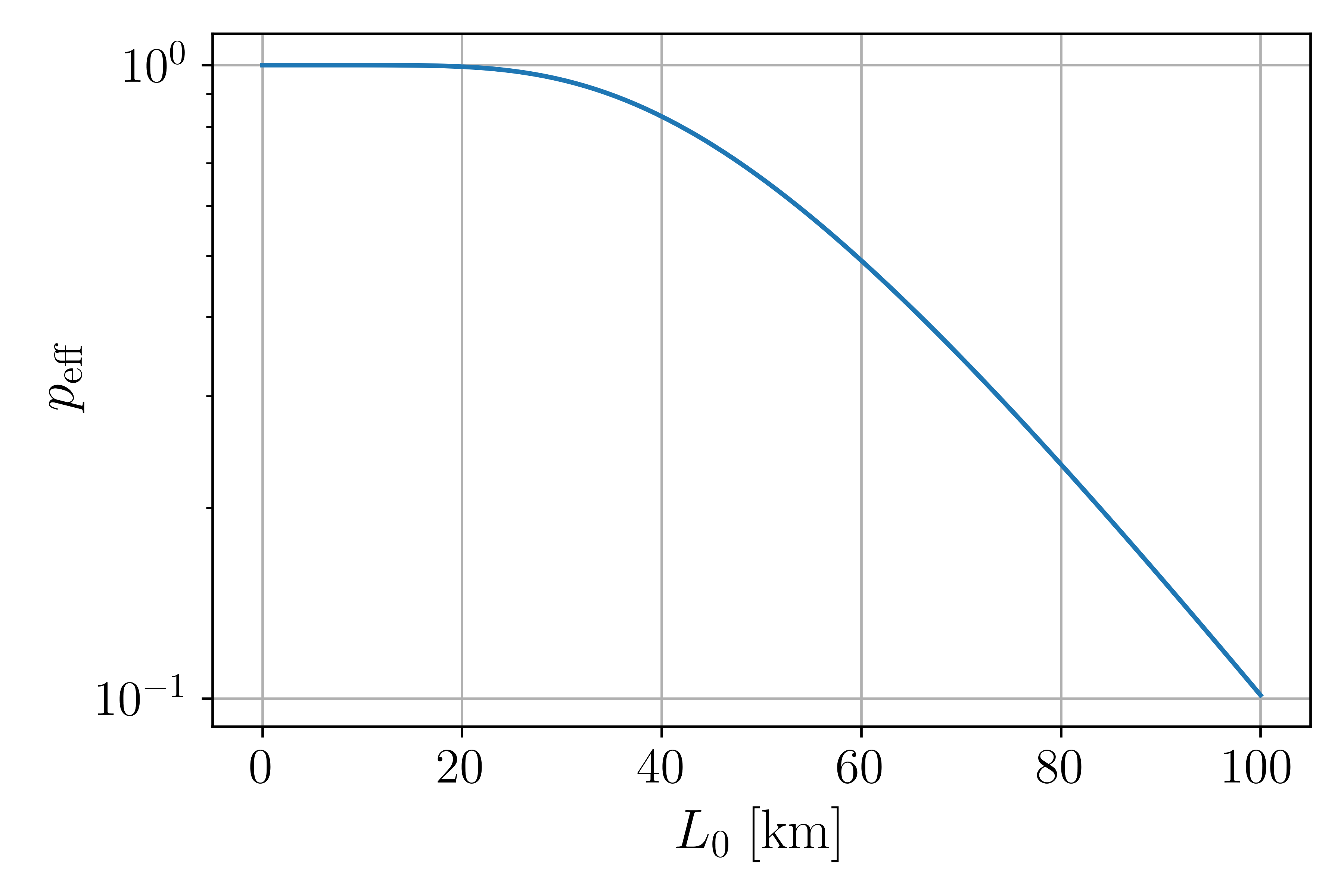} }\\
    \subfloat[]{\includegraphics[width=\linewidth]{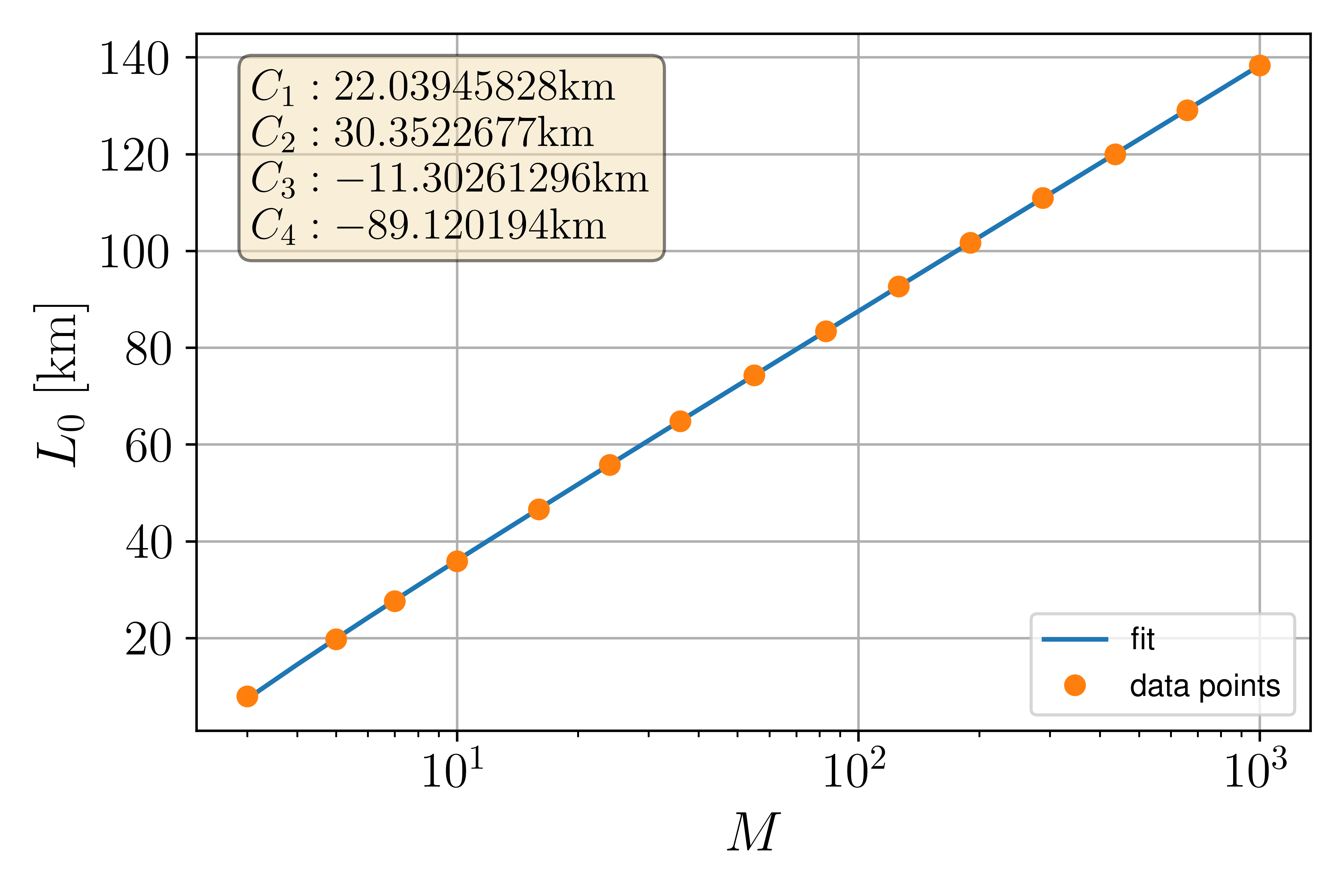}}
    \caption{(a) $p_{\mathrm{eff}}$ for $M=10$ (b) rule of thumb: the orange points show the numerical minimization for different $M$ and the blue line shows the fitted function. It was obtained by fitting the numerical function for all values in the interval (3,1000)\footnote{For $M=2$ our algorithm has convergence problems.}. However, it also works well for larger $M$ like e.g. $10^4$ up to some small deviations at high $M$ probably due to the numerical precision. For the meaning of the fitting parameters, see main text. As always we assumed $L_{\mathrm{att}}=\unit[22]{km}$.}
    \label{fig:ruleofthumb}
\end{figure}

In the general, exact model using $p_{\mathrm{eff}}=1-(1-p)^M$, it becomes clear that the above-mentioned behaviour originates from this general expression for $p_{\mathrm{eff}}$.
In Fig.~\ref{fig:ruleofthumb}(a) one can see that $p_{\mathrm{eff}}$ can be divided in three regimes. In the first regime of small $L_0$, $p_{\mathrm{eff}}$ is a constant. In the second regime of large $L_0$, $p_{\mathrm{eff}}$ is a simple exponential decay, while in between it has a more complicated form interpolating both regimes.
In the first regime, the effective probability is nearly constant, because in our simple multiplexing protocol we only make use of a single `entanglement excitation' in each segment of the parallelized repeater chains, but for small $L_0$ we would typically have multiple excitations in each segment.
Thus, increasing $L_0$ decreases the number of excitations, but as we anyway only make use of a single one, this barely matters
(making use of more excitations and keeping the `residual entanglement' could potentially further enhance the rates \cite{RazaviProc}; 
however, here our focus is on a simple and clear interpretation of the impact of the multiplexing on the coherence time and the memory dephasing in our statistical model).
In the second regime of rather large $L_0$, the contributions of multiple excitations can be neglected and therefore the rates behave exactly like in the $M=1$ case.
Hence, this regime two is exactly that where we can increase the effective coherence time by a factor of $M$ with the help of multiplexing.
We can give a rough rule of thumb for the minimal length of $L_0$ when one may use the simple approximation of increasing the coherence time by a factor of $M$.
For this we assume $p=\exp(-\frac{L_0}{L_{\mathrm{att}}})$ \footnote{When considering $p_{\mathrm{link}}<1$ one can incorporate this as an additional length of $-\ln(p_{\mathrm{link}})L_{\mathrm{att}}$ regarding $L_0$.} and take the minimizing argument of $\frac{\partial^2\ln\left(p_{\mathrm{eff}}\right)}{\partial L_0^2}$ for a given $M$ in order to estimate the midpoint of the interpolating regime.
For general $M$ this expression can be nicely fitted to an expression of the form $c_1 \ln\left(c_2 M+c_3\right)+c_4$, as one can see in Fig.~\ref{fig:ruleofthumb}(b).
One should then consider $L_0$ to be slightly larger for the approximation to hold.

\begin{figure*}
	\centering
	\subfloat[a][Two segments with $\tau_{\mathrm{coh}}={\unit[0.1]{s}}$, $p_{\mathrm{link}}=0.7$, $\mu=\mu_0=0.97$ \hspace{10pt} and $M=10$ for multiplexing.]{\includegraphics[width=0.5\linewidth]{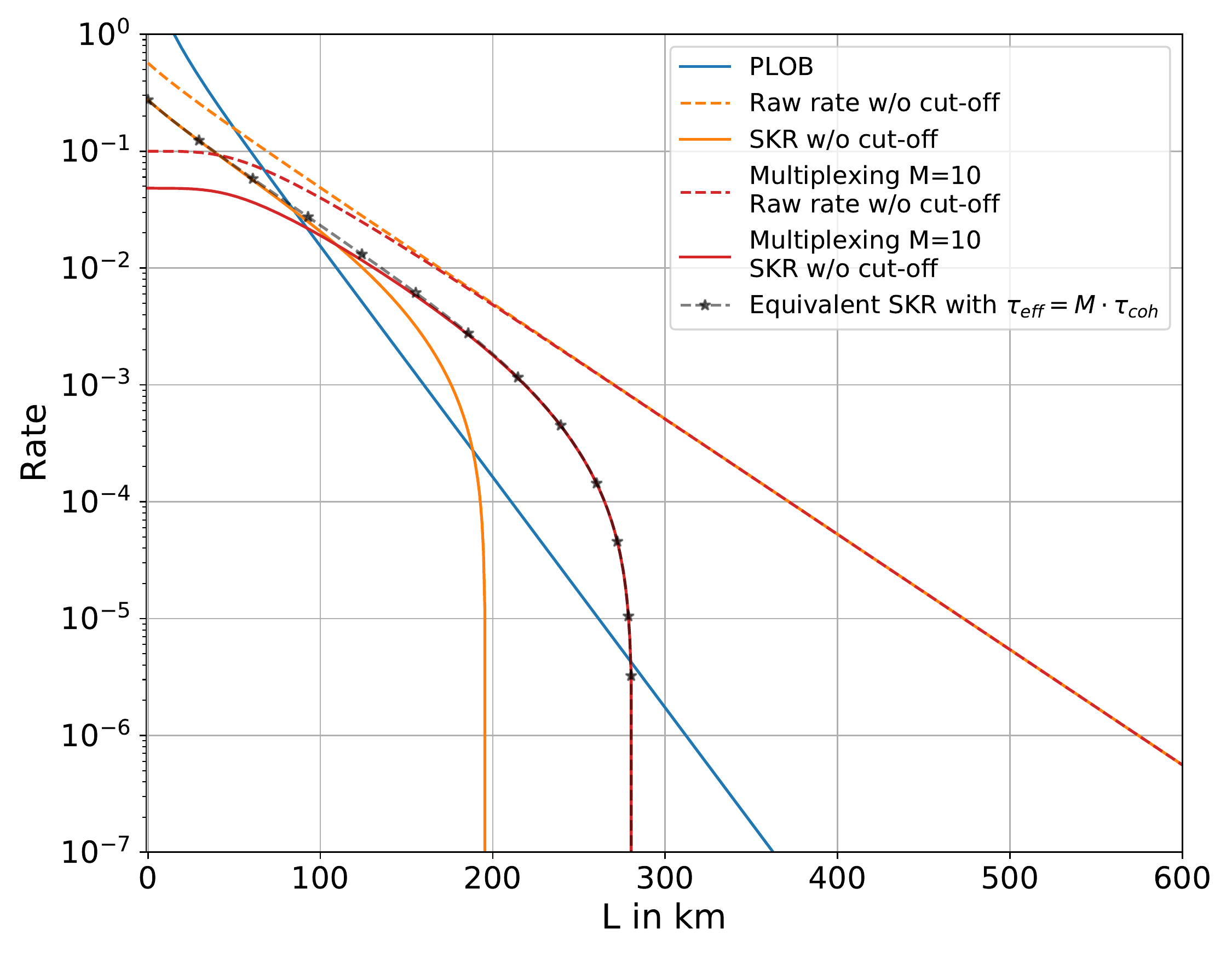}}
	\subfloat[][Two segments with $\tau_{\mathrm{coh}}={\unit[10]{s}}$, $p_{\mathrm{link}}=0.05$, $\mu=\mu_0=0.97$ \hspace{10pt} and $M=10$ for multiplexing.]{\includegraphics[width=0.5\linewidth]{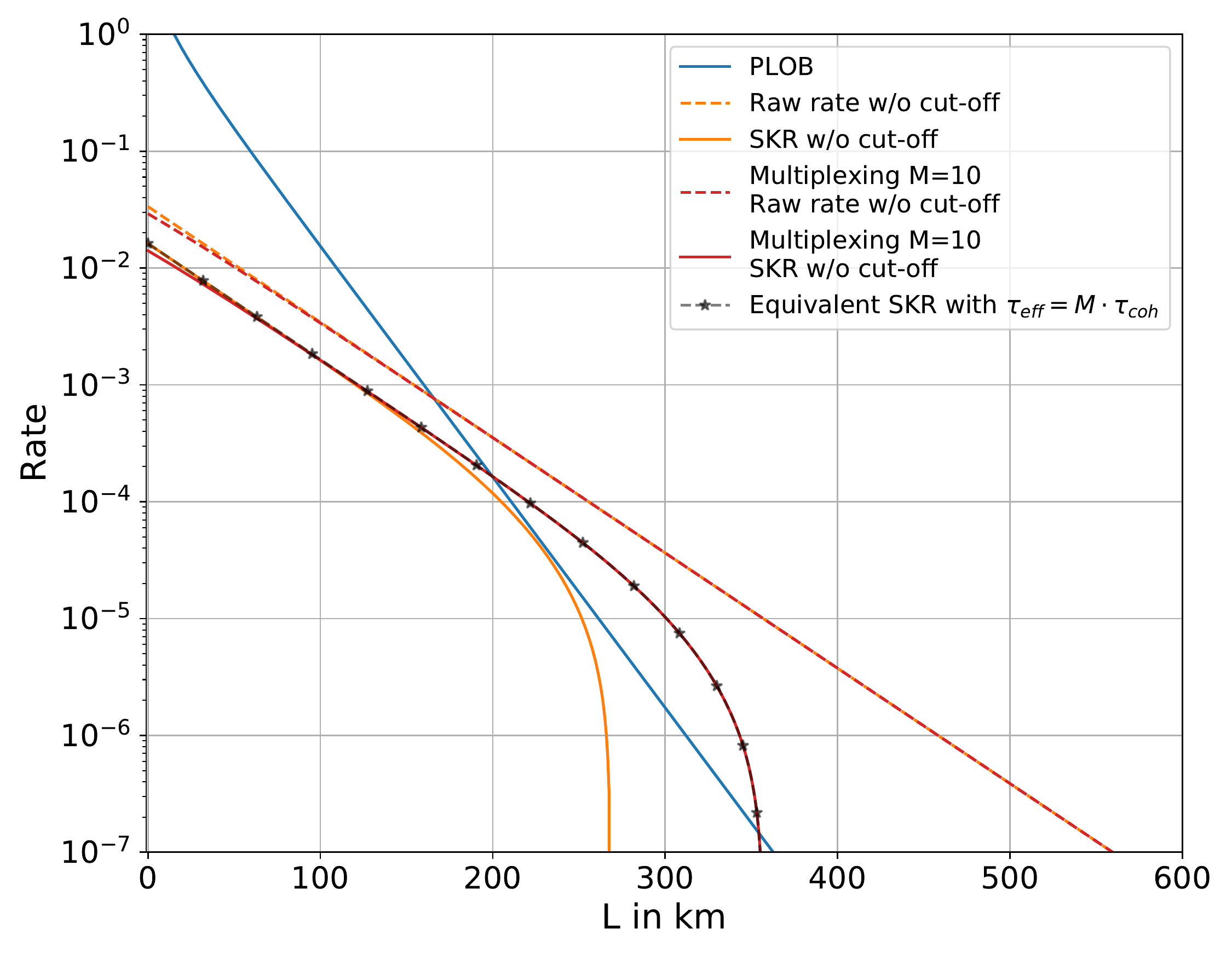}} \\
	\subfloat[][Four segments with $\tau_{\mathrm{coh}}={\unit[0.1]{s}}$, $p_{\mathrm{link}}=0.7$, $\mu=\mu_0=0.97$ \hspace{10pt} and $M=10$ for multiplexing.]{\includegraphics[width=0.5\linewidth]{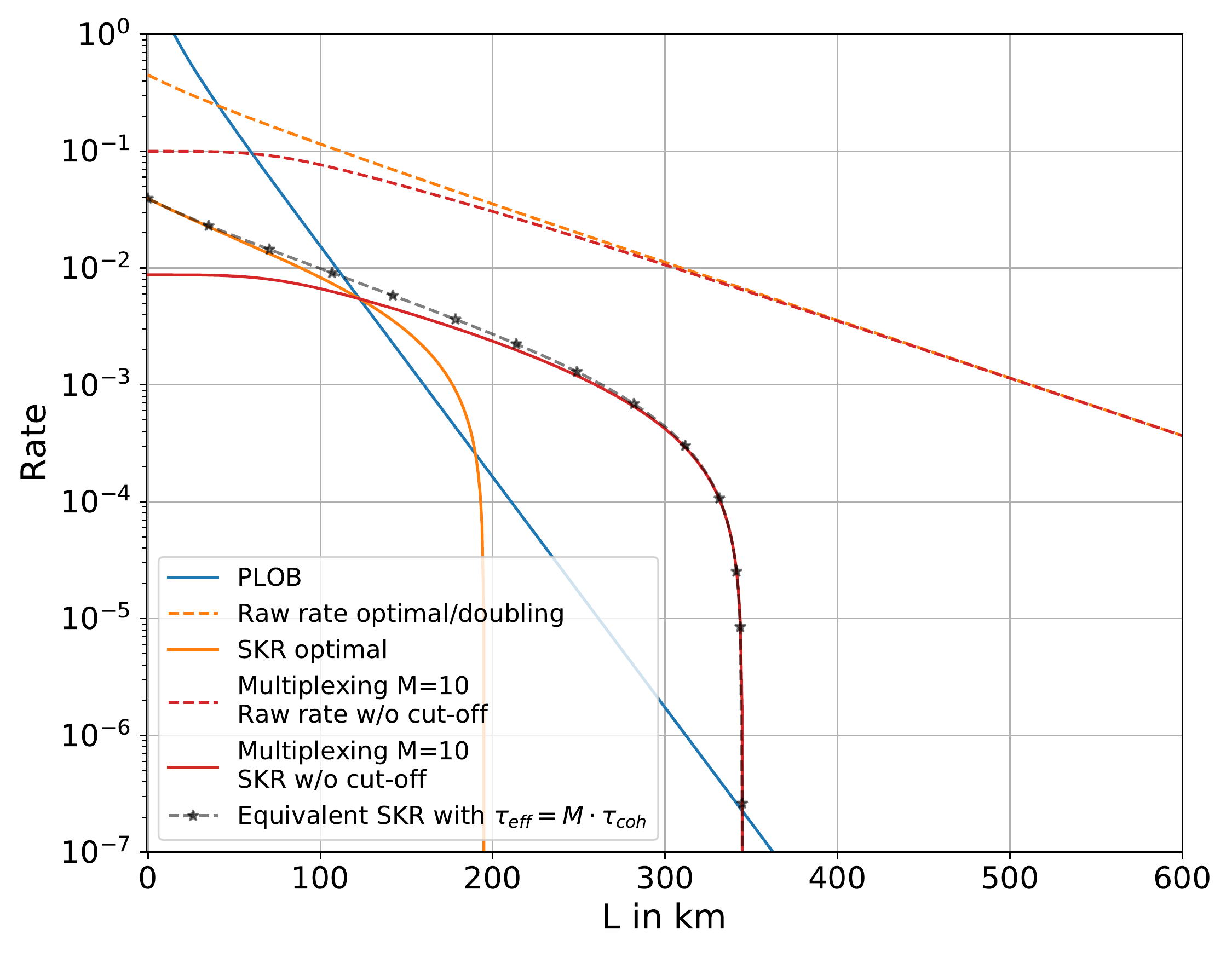}}
	\subfloat[][Four segments with $\tau_{\mathrm{coh}}={\unit[10]{s}}$, $p_{\mathrm{link}}=0.05$, $\mu=\mu_0=0.97$ \hspace{10pt} and $M=10$ for multiplexing.]{\includegraphics[width=0.5\linewidth]{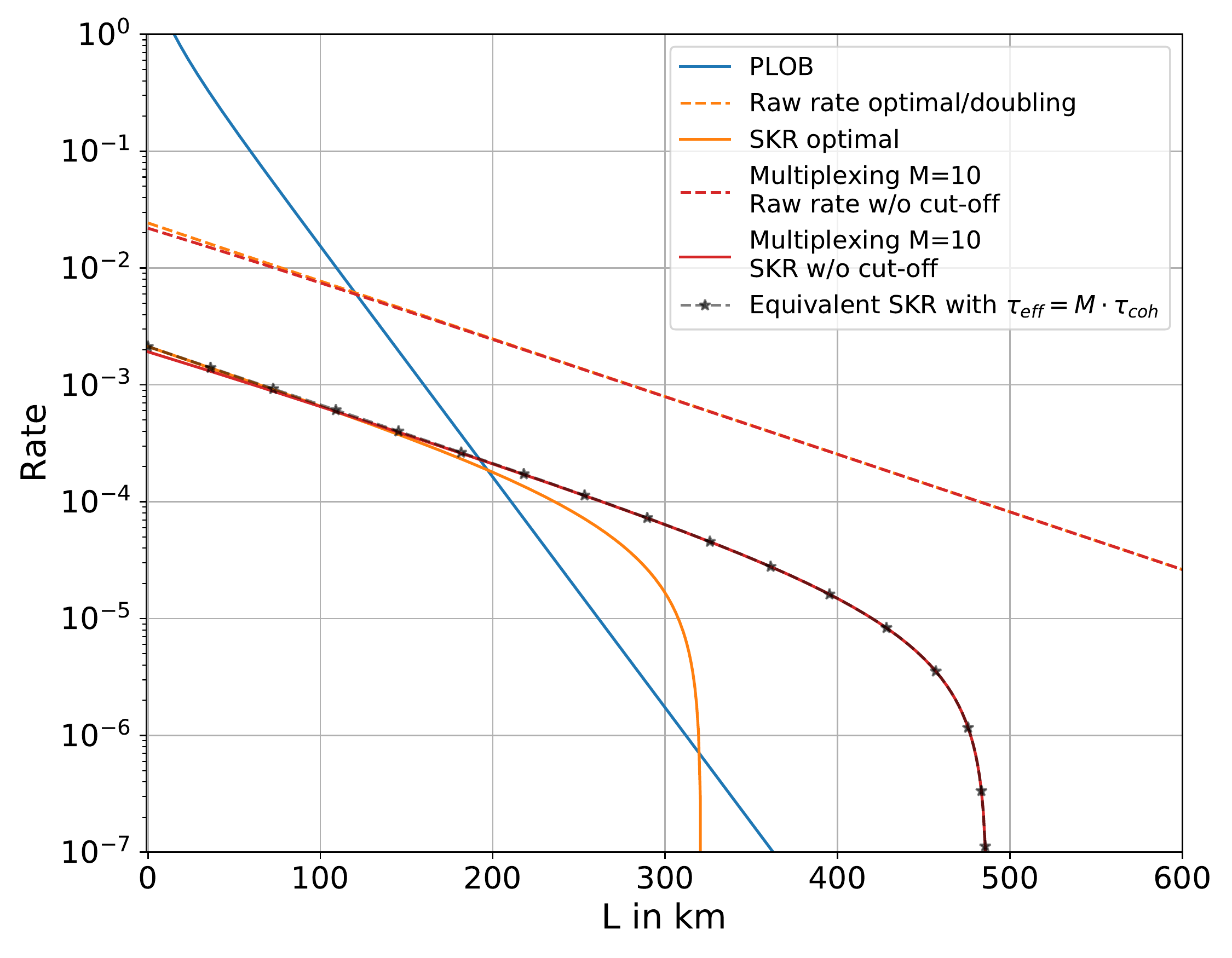}}
	\caption{Rates (secret key/raw) of (a,b) two- and (c,d) four-segment repeaters using multiplexing $M=10$ at distances \(L\) for different experimental parameters. The rate of a repeater without multiplexing, but with the same coherence time is shown in orange, whereas the rate of a repeater using multiplexing is shown in red. Additionally, a repeater without multiplexing, but with an equivalent effective coherence time is presented in dashed black. All rates are expressed per channel use and hence include a division by $M$.}
	\label{fig:SKR_multiplexing}
\end{figure*}

Let us give another, more rigorous derivation of the effective coherence time in the presence of multiplexing. 
The coherence time primarily characterises the increasing decline of the secret key rate with distance. However, a massive drop actually happens when the secret key fraction \(r\) reaches zero, which is possible when $e_z>0$, i.e. when $\mu<1$ or $\mu_0<1$. Thus, let us determine the probability at which \(r=0\) holds with multiplexing and from that deduce an equivalent coherence time without multiplexing. Since the QBER \(e_z\) is constant ($e_z=\overline{e_z})$, we have to solve for the expectation value of \(\overline{e_x}\) such that
\begin{equation}
	1-h(e_z) \overset{!}{=} h(\overline{e_x}).
\end{equation}	
In order to find the probability \(p\) or equivalently the distance at which the drop happens, let us use the Taylor series of the binary entropy function at \(x=\frac{1}{2}\), 
\begin{align}
	h(x)= 1- \frac{1}{2 \ln(2)}\sum_{n=1}^{\infty} \frac{\left(1-2x\right)^{2n}}{n\left(2n-1\right)},\; \forall\; 0<x<1.
\end{align}
Then one finds for \(\overline{e_x}\) up to first order:
\begin{equation}
	\overline{e_x}= \frac{1}{2} - \sqrt{\frac{\ln(2)h(e_z)}{2}},
\end{equation}
where only the negative root is possible, as \(0\leq e_x \leq \frac{1}{2}\). Inserting \(\overline{e_x}\) and solving for $\mathbf{E}[e^{-\alpha D_n}]$ gives
\begin{equation}
	\mathbf{E}[e^{-\alpha D_n}] = \frac{\sqrt{2\ln(2)h(e_z)}}{\mu^{n-1} \mu_0^{n} \left(2 F_0-1\right)^n}.
\end{equation}
If $\mu=\mu_0=1$, including especially the channel-loss-and-memory-dephasing-only case (for which also $F_0=1$), we have $h(e_z)=0$ and so the requirement becomes $\mathbf{E}[e^{-\alpha D_n}]=0$, which is impossible.
However, as soon as $e_z>0$, i.e. $\mu<1$ or $\mu_0<1$, a sufficiently small non-zero (average) dephasing fraction $\mathbf{E}[e^{-\alpha D_n}]$ leads to a zero secret key fraction.
As we can always calculate this expectation value by our previously derived PGFs, we now have an accurate and systematic way to derive the probability $p$ (or the total distance $L=n L_0$) at which the drop takes place for given values of $n$, $\tau_{\mathrm{coh}}$, $\mu$, $\mu_0$, and $F_0$. Recall that the inverse effective coherence time $\alpha=L_0/(c_f \tau_{\mathrm{coh}})$ typically also depends on $L_0$. 
On the other hand, we may use the above relation to determine an (inverse) effective coherence time by calculating the drop for a repeater with multiplexing and then the equivalent \(\alpha\), which would be needed to achieve the same distance without any multiplexing. From this \(\alpha\) one can recover the coherence time \(\tau_{\mathrm{coh}}\) and finds the approximate relation
\begin{equation}\label{eq:MPrelation}
	\tau_{\mathrm{coh}} \mapsto M \cdot \tau_{\mathrm{coh}},
\end{equation}
when a multiplexing of \(M\) is used and the remaining setup is kept the same. Thus, one can achieve an \(M\)-times longer effective coherence time with the help of multiplexing.

In Fig.~\ref{fig:SKR_multiplexing}, we show the rates of two- and four-segment repeaters using a multiplexing of \(M=10\) in red. Note that because we use the SKR per channel use, the rates are obtained including a division by $M$. The rates of the same repeaters without multiplexing are presented in orange. Furthermore, a repeater without multiplexing, but with the equivalent `effective' coherence time of \(\tau_{\mathrm{eff}} = M \tau_{\mathrm{coh}}\) is shown in dashed black. One can see that for small distances, i.e large probabilities, the multiplexed repeater does not quite behave like its non-multiplexed counterpart with an effectively increased  coherence time. A clear splitting between the red and black curves is visible. However, for larger distances, especially after crossing the PLOB bound, the multiplexed repeater behaves exactly the same as if simply memories with an effectively longer coherence time were used. For smaller link efficiencies, the splitting becomes much less pronounced, as can be seen in the plots on the right of Fig.~\ref{fig:SKR_multiplexing}. All this holds for both two and four segments, according to Fig.~\ref{fig:SKR_multiplexing}.
In particular, for small link efficiencies, the secret key rate of an equivalent repeater with \(\tau_{\mathrm{eff}} = M \tau_{\mathrm{coh}}\) is
almost indistinguishable from a repeater with multiplexing. This is in agreement with the above discussion on the occurrence of single versus multiple `entanglement excitations' in each segment where the latter are then highly suppressed even at short distances due to the small value of $p_{\mathrm{link}}$. 
Thus, for practical purposes, in all our discussions, we may treat several cases equivalently, for instance, a repeater with $\tau_{\mathrm{coh}}=10$s and $M=1$ would be equivalent to a repeater with $\tau_{\mathrm{coh}}=1$s and $M=10$.   

\subsection{Secret key rate per second}\label{sec:Secret Key Rate per Second}

In a real-world application, the important figure of merit is not the rate per channel use, it will be the rate per second. In particular, a memory-asissted QKD system or generally a memory-based quantum repeater, as typically based upon light-matter interactions and classical communication at least between neighboring stations, has a limited `clock rate'.
Classical communication is needed to declare successful
transmission of photons for the entanglement distribution. In general, also extra communication would be needed to signal any successful entanglement swapping, but as we assumed deterministic swapping no such communication is needed in our repeater models. 

As we already discussed frequently throughout the paper, a repeater's performance generally depends on an elementary time unit $\tau$, which is contained in the inverse effective coherence time $\alpha=\tau/\tau_{\mathrm{coh}}$, where generally $\tau=\tau_{\mathrm{clock}}+L_0/c_f$ including the experimental local processing time $\tau_{\mathrm{clock}}$. We have mostly argued that in the relevant distance regimes, this quantity is dominated by the (quantum and classical) communication times between neighboring stations, thus $\tau = L_0/c_f$ and $\alpha=L_0/(c_f \tau_{\mathrm{coh}})$. 
Already with segment
lengths above \(\unit[10]{km}\), one can neglect the local clock rates, since these are much higher than the rates given
by the transmission times. 
An extra factor of two could be included in $\tau$ for some protocols due to the $L_0$-transmission of a photon entangled with a memory qubit and the classical answer (sent back over $L_0$) heralding its successful transmission. However, this would depend on the specific protocol and so we have chosen the simplest, minimal form $\tau = L_0/c_f$.
Only for very short segment lengths do we have $\alpha\approx \alpha^{\mathrm{loc}}=\tau_{\mathrm{clock}}/\tau_{\mathrm{coh}}={\rm MHz}^{-1}/\tau_{\mathrm{coh}}$ assuming experimental clock rates $\tau_{\mathrm{clock}}^{-1}$ typically of the order of MHz.

However, there are repeater schemes that are independent of additional classical communication and the decision to keep or reinitialize a memory state can be made at the memory station. These schemes may be referred to as ``node receives photons" (NRP) as opposed to the class of schemes with ``node sends photons" (NSP) \cite{White}.
An NRP protocol and application that circumvents the need of extra signal waiting times can be realized with two ``segments'' and a middle station in memory-assisted MDI QKD \cite{White}.

Such a scheme, when treating it as an elementary quantum repeater unit or module
many of which a large-scale repeater can be made of, may be referred to as a ``quantum repeater cell'', actually composed of two half-segments \cite[Fig. 6b]{White}.
In this case, even for large (half-)segment length $L_0$, we have $\alpha = \alpha^{\mathrm{loc}}=\tau_{\mathrm{clock}}/\tau_{\mathrm{coh}}$. 
For completeness, we show the rates of such an NRP-based two-segment scheme
in the form of contour plots in App.~\ref{app:nrp}.
By circumventing the need for extra classical communication
and thus significantly reducing the effective memory dephasing,
the minimal state and gate fidelity values can even be kept constant 
over large distance regimes.
However, as soon as the NRP concept is applied to repeaters beyond a single middle station effectively connecting complete repeater segments, \cite[Fig. 6a]{White}
the need for extra classical communication to initiate an entanglement swapping operation can no longer be entirely avoided
(though there are ideas to still partially benefit from the NRP concept) \cite{Cody_Jones}.
A quantum repeater cell can also be considered employing the NSP protocol \cite{NL} and one such cell (two half-segments) or the corresponding complete segment can then be used as an elementary quantum repeater unit \cite[Fig. 4]{White}. 
For the NSP concept, the extra signal waiting time is generally required at every distribution attempt. 
In any case or protocol, the repeater's elementary time unit $\tau$ determines the effective coherence time $\tau_{\mathrm{coh}}/\tau$ and as such, even when the rates per channel use are considered, it determines how many distribution attempts are possible within a given $\tau_{\mathrm{coh}}$ and hence how big the effective dephasing time $\alpha D_n$ becomes. 

Compared with memory-assisted quantum communication schemes, a big asset of an all-optical point-to-point quantum communication link is that it can operate at a very high clock rate, typically of the order of GHz, only limited by the speed of Alice's laser (quantum state) source and Bob's (quantum state) detector. 
For such a direct state transmission, no extra classical communication is required as for heralding the successful transfer of entangled photons between repeater links. Thus, the rate per second is simply given by the two local clock rates, especially the time it takes to generate the photonic qubit states
or any other quantum states in QKD based on different types of encoding
(however, thanks to the known linear bounds on the key distribution via a long and lossy point-to-point quantum communication channel \cite{PLOB, TGW}, it is clear that the rate scaling of qubit-based QKD cannot be beaten by any form of non-qubit encoding).

Other all-optical schemes such as MDI QKD or twin-field QKD, which are no longer point-to-point and do include a middle station between Alice and Bob, also benefit from such high clock rates. The remarkable feature of twin-field QKD is that it shares both advantages: the high clock rate with point-to-point quantum communication and the $L \rightarrow L/2$ loss scaling gain with memory-based two-segment quantum repeaters. In order to assess whether there is a real benefit of employing a two-segment quantum repeater or even adding extra repeater stations, we must eventually consider the rates per second and take into account the corresponding clock rates in all schemes. As a consequence, comparing clock rates of MHz with those of GHz (of memory-based versus all-optical quantum communication), there is a penalty of a factor of about 1000 from the start for the memory-based approach. In the regime where $\alpha\approx L_0/(c_f \tau_{\mathrm{coh}})$, this penalty even gets worse. In this case, when $\tau\approx L_0/c_f$, there are at least two disadvantages of $\tau$ growing with $L_0$: a reduced effective coherence time $\tau_{\mathrm{coh}}/\tau$ and a reduced raw rate per second $R/\tau$. Beating the PLOB bound for the rates per channel use is only a necessary criterion that a quantum repeater can be beneficial. In order to confirm a real benefit, we have to consider the secret key rates per second $S/\tau=r R/\tau$. Thus, even with perfect memories $\tau_{\mathrm{coh}}\rightarrow\infty$, the different $\tau$ values matter. The situation is similar to throwing two or more dices at once at a fast rate. To get all dices showing six eyes this may still be faster than throwing them very slowly while being allowed to only continue with the unsuccessful dices in each round. The final raw and secret key rates per second obtainable with our two most prominent and mostly discussed repeater schemes, the fully sequential scheme and the optimal scheme, are given in Tabs.~\ref{tab:seqpersecond} and \ref{tab:optpersecond}, respectively.

\begin{table*}
	\begin{tabular}{c|c|c|c|c|c|c|c}
        $n$ &1 & 2 & 4 & 8 & 80 & 800 & 8000\\
        \hline
        \hline
        $L_0$[km] &800 & 400 & 200 & 100 & 10 & 1 & 0.1\\
        \hline
        $R /\tau $ & $\unit[\sim 10^{-14}]{Hz}$ & $\unit[\sim 10^{-6}]{Hz}$ & $\unit[0.0293]{Hz}$ & $\unit[2.8]{Hz}$ & 
        $\unit[165.2]{Hz}$ & $\unit[248.7]{Hz}$ & $\unit[259.1]{Hz}$\\
        \hline
        $S_1(\mu=1)/\tau $ &- &$\unit[\sim 10^{-18}]{Hz}$ & $\unit[\sim 10^{-14}]{Hz}$ & $\unit[0.0106]{Hz}$ & $\unit[133.9]{Hz}$ & $\unit[213.9]{Hz}$ & $\unit[224.0]{Hz}$\\
        \hline
        $S_2(\mu=1)/\tau $ &- &$\unit[\sim 10^{-14}]{Hz}$ & $\unit[0.0005]{Hz}$ & $\unit[2.4]{Hz}$ & $\unit[164.5]{Hz}$ & $\unit[248.0]{Hz}$ & $\unit[258.4]{Hz}$\\
        \hline
        $S_1(\mu=0.99)/\tau $ &- &$\unit[0]{Hz}$ & $\unit[0]{Hz}$ & $\unit[0]{Hz}$ & $\unit[0]{Hz}$ & $\unit[0]{Hz}$ & $\unit[0]{Hz}$\\
        \hline
        $S_2(\mu=0.99)/\tau $ &- &$\unit[0]{Hz}$ & $\unit[0]{Hz}$ & $\unit[0.6086]{Hz}$ & $\unit[0]{Hz}$ & $\unit[0]{Hz}$ & $\unit[0]{Hz}$\\
        \hline
        $S^{\mathrm{PLOB,QR}}(L_0)/\tau $ &$\unit[\sim 10^{-7}]{Hz}$ & $\unit[18.3]{Hz}$ & $\unit[0.2]{MHz}$ & $\unit[15.5]{MHz}$ & $\unit[1.5]{GHz}$ & $\unit[4.5]{GHz}$ & $\unit[7.8]{GHz}$
		\end{tabular}
	\caption{Overview of the relevant quantities for the {\it fully sequential scheme} of Tab.~\ref{tab:seqperchanneluse} calculated per second (shown are only those entries that change, but again with segment number $n$, segment length $L_0$[km]): raw rate $R/\tau$, secret key rate $S/\tau$ for different $\mu=\mu_0$ (again subscript corresponds to the choice of $\alpha_1$ or $\alpha_2$, $\mu=1$ is the channel-loss-and-memory-dephasing-only case), and the (repeater-assisted) capacity bound per elementary time unit $S^{\mathrm{PLOB,QR}}(L_0)/\tau$ where we choose $\tau={\rm GHz}^{-1}$ for the cases $n=1,2$, i.e. the bounds, expressed per second, on all-optical point-to-point and twin-field QKD. Note that for realistic but still GHz-clock-rate twin-field QKD we rather have $S/\tau\sim1$Hz. In any of the other, memory-based scenarios, we choose $\tau=\tau_{\mathrm{clock}}+L_0/c_f$ with $\tau_{\mathrm{clock}}={\rm MHz}^{-1}$.
	We again assumed $p_{\mathrm{link}}=F_0=1$ for the link coupling efficiency and the initial state dephasing.}
	\label{tab:seqpersecond}
\end{table*}

\begin{table*}
	\begin{tabular}{c|c|c|c|c|c|c|c}
		$n$ &1 & 2 & 4 & 8 & 80 & 800 & 8000\\
        \hline
        \hline
        $L_0$[km] &800 & 400 & 200 & 100 & 10 & 1 & 0.1\\
        \hline
        $R /\tau $ & $\unit[\sim 10^{-14}]{Hz}$ & $\unit[\sim
        10^{-6}]{Hz}$ & $\unit[0.0563]{Hz}$ & $\unit[8.2]{Hz}$ & $\unit[3.8]{kHz}$ & $\unit[72.7]{kHz}$ & $\unit[967.2]{kHz}$\\
        \hline
        $S_1(\mu=1)/\tau $ &- &$\unit[\sim 10^{-18}]{Hz}$ & $\unit[\sim 10^{-12}]{Hz}$ & $\unit[0.1423]{Hz}$ & $>\unit[3.1]{kHz}$ & $>\unit[62.5]{kHz}$ & $>\unit[832.1]{kHz}$\\
        \hline
        $S_2(\mu=1)/\tau $ &- &$\unit[\sim 10^{-14}]{Hz}$ & $\unit[0.0020]{Hz}$ & $\unit[7.4]{Hz}$ & $>\unit[3.8]{kHz}$ & $>\unit[72.4]{kHz}$ & $>\unit[964.5]{kHz}$\\
        \hline
        $S_1(\mu=0.99)/\tau $ &- &$\unit[0]{Hz}$ & $\unit[0]{Hz}$ & $\unit[0]{Hz}$ & $\unit[0]{Hz}$ & $\unit[0]{Hz}$ & $\unit[0]{Hz}$\\
        \hline
        $S_2(\mu=0.99)/\tau $ &- &$\unit[0]{Hz}$ & $\unit[0]{Hz}$ & $\unit[1.9]{Hz}$ & $\unit[0]{Hz}$ & $\unit[0]{Hz}$ & $\unit[0]{Hz}$\\
        \hline
		$S^{\mathrm{PLOB,QR}}(L_0)/\tau $ &$\unit[\sim 10^{-7}]{Hz}$ & $\unit[18.3]{Hz}$ & $\unit[0.2]{MHz}$ & $\unit[15.5]{MHz}$ & $\unit[1.5]{GHz}$ & $\unit[4.5]{GHz}$ & $\unit[7.8]{GHz}$
		\end{tabular}
	\caption{Overview of the relevant quantities for the {\it optimal scheme} of Tab.~\ref{tab:optperchanneluse} calculated per second (shown are only those entries that change, but again with segment number $n$, segment length $L_0$[km]): raw rate $R/\tau$, secret key rate $S/\tau$ for different $\mu=\mu_0$ (again subscript corresponds to the choice of $\alpha_1$ or $\alpha_2$, $\mu=1$ is the channel-loss-and-memory-dephasing-only case), and the (repeater-assisted) capacity bound per elementary time unit $S^{\mathrm{PLOB,QR}}(L_0)/\tau$ where we choose $\tau={\rm GHz}^{-1}$ for the cases $n=1,2$, i.e. the bounds, expressed per second, on all-optical point-to-point and twin-field QKD. Note that for realistic but still GHz-clock-rate twin-field QKD we rather have $S/\tau\sim1$Hz. In any of the other, memory-based scenarios, we choose $\tau=\tau_{\mathrm{clock}}+L_0/c_f$ with $\tau_{\mathrm{clock}}={\rm MHz}^{-1}$.
	We again assumed $p_{\mathrm{link}}=F_0=1$ for the link coupling efficiency and the initial state dephasing.}
	\label{tab:optpersecond}
\end{table*}

\subsection{Application and comparison of protocols}

Let us now consider various quantum repeater protocols
based on different types of the optical encoding 
and calculate their corresponding secret key rates per second
using the methods developed in the preceding sections.
We shall look at (i) a kind of standard scheme employing 
two-mode (dual-rail, DR) photonic qubits distributed through the optical-fiber channels (either emitted from a central source of entangled photon pairs
and written into the spin memory qubits or emitted from the repeater nodes
employing spin-photon entangled states and utilizing two-photon interference in the middle of each segment), \cite{White}
(ii) a scheme based upon spin-photon (spin-light-mode) entanglement and one-photon interference with an encoding similar to that 
introduced by Cabrillo et al. \cite{cabrillo} effectively using one-mode (single-rail, SR) photonic qubits,
(iii) a scheme that extends the concepts of twin-field QKD with coherent states 
to a specific variant of memory-assisted QKD, i.e. a kind of twin-field quantum repeater \cite{tf_repeater}.
We refer to scheme (ii) as the Cabrillo scheme and discuss it in more detail in App.~\ref{app:cabrillo}.
For all three schemes we consider a quantum repeater with $n=1,2,3,4,8$ segments
matching the size of the repeater systems that we have formally/theoretically treated in great detail in the first parts of this paper. 
We always use the previously derived ``optimal'' quantum repeater protocol
that belongs to the fastest schemes and gives the smallest dephasing among all fast schemes.

The two schemes (ii) and (iii) share the potential benefit that 
for quantum repeaters with $n$ segments and $n-1$ intermediate memory stations
(not counting the memories at Alice and Bob or assuming immediate measurements there) they lead to an improved loss scaling with a $2n$-times bigger
effective attenuation distance compared with a point-to-point link
(unlike the standard scheme (i) that only achieves an $n$-times bigger
effective attenuation distance), but a final state fidelity parameter 
still decreasing as the power of $2n-1$ (assuming equal gate and initial state 
error rates) like the standard scheme (i).
However, scheme (ii) has an intrinsic error 
during the distribution step due to the initial two-photon terms 
in combination with channel loss.
Similarly, scheme (iii) is more sensitive to channel loss
exhibiting an intrinsic loss-dependent dehasing error,
because the optical state is a phase-sensitive continuous-variable state \cite{HybridPRL}.
The two models of channel-loss-induced errors for schemes (ii) and (iii)
thus slightly differ, while the transmission loss scaling is identical.
As a consequence, for both (ii) and (iii), we have the constraint that 
the excitation amplitudes (the weights of the non-vacuum terms)
must not become too large.
Despite the above-mentioned benefits compared with scheme (i) it will turn out that the intrinsic errors of schemes (ii) and (iii)
represent an essential complication that prevents to fully exploit the improved 
scaling of the basic parameters in comparison with the standard repeater protocols. 

For a fair comparison, assuming similar types of initial state imperfections in all three schemes, we set $\mu_0=1$ with $F_0=0.99,0.98$ and so replace the initial depolarizing error for scheme (i) by an initial dephasing error.  
Thus, in the expressions of the QBERs as given by 
Eq.~\eqref{eq:QBER}, the contribution of $\mu_0^n$ to the initial error scaling
from the analysis of the preceding sections (where $F_0=1)$ is now replaced by a corresponding scaling with $F_0<1$. The gate error scaling with $\mu^{n-1}$ remains unchanged in all schemes. Of course, our formalism also allows to focus on specific schemes including initial state errors with $\mu_0<1$.
In this case, the specific contributions of the different elements in each elementary repeater unit (segments, half-segments, ``cells'') \cite{White} to the link coupling efficiency $p_{\mathrm{link}}$ and the initial state error parameters $\mu_0$ or $F_0$ depend of the particular protocol \cite{White}.

For example, zooming in on an NSP segment, \cite{White} we have a squared contribution from the two spin-photon entangled states on the left and on the right, $\mu_{\mathrm{sp,ph}}^2$, and another possible gate error factor, $\mu_{\mathrm{OBM}}$, coming from the optical Bell measurement in the middle of the segment.
In this scenario, already in a single segment, we effectively have one imperfect entanglement swapping operation (acting on the two photons in the middle of the segment) connecting two initially distributed, depolarized entangled states (the two spin-photon states), to which our physical model directly applies replacing our initial $\mu_0$ for one segment according to $\mu_0 \rightarrow \mu_{\mathrm{sp,ph}}^2 \mu_{\mathrm{OBM}}$. 
This overall initial distribution error will most likely be dominated by the 
imperfect spin-photon states, assuming near-error-free (though probabilistic)
photonic Bell measurements, thus $\mu_0 \sim \mu_{\mathrm{sp,ph}}^2$.

In a full NRP segment, the memory write-in may be realized via quantum teleportation using a locally prepared spin-photon state and an optical Bell measurement on the photon that arrives from the fiber channel and the local photon. In this scenario, already in a single complete segment, we may effectively have three initial entangled states (two local spin-photon states on the left and on the right together with one distributed entangled photon pair emitted from a source in the middle of the segment) and two optical Bell measurements, \cite[Fig. 6a]{White} with our model resulting in a $\mu_0 \sim \mu_{\mathrm{ph,ph}} \mu_{\mathrm{sp,ph}}^2 \mu_{\mathrm{OBM}}^2$ scaling of the initial error parameter for one segment (i.e., similar to the effective final scaling of a three-segment repeater in our more abstract model,
with $\mu_0 \rightarrow \mu_{\mathrm{sp,ph}}$ and $\mu \rightarrow \mu_{\mathrm{OBM}}$, and setting for this simplifying analogy, quite unrealistically, $\mu_{\mathrm{sp,ph}}=\mu_{\mathrm{ph,ph}}$). Assuming near-error-free Bell measurements, and near-perfect (though possibly only probabilistically created) photon pairs, we would again arrive at an overall scaling of $\mu_0 \sim \mu_{\mathrm{sp,ph}}^2$ for the initial error parameter. In case of an entangled photon pair source that deterministically produces imperfect photon-photon states (such as a quantum dot source), we would have $\mu_0 \sim \mu_{\mathrm{ph,ph}}\mu_{\mathrm{sp,ph}}^2$ instead. There is also the option of a heralded memory write-in that no longer relies on the generation of local spin-photon states and optical Bell measurements \cite{Rempe}.
In this case, our physical model has to be slightly adapted to such a scenario and a decomposition of the different error channels, including an imperfect memory write-in operation, into one effective initial error channel should be considered.

Thus, zooming in on our general initial-state error parameters $\mu_0$ or $F_0$
for a specific implementation is straightforwardly possible, but it will eventually lead to even stronger fidelity requirements for the individual experimental components that contribute to $\mu_0$ or $F_0$. 
The different contributions to the link coupling efficiencies $p_{\mathrm{link}}$
can be similarly decomposed into the different experimental elements,
also including some differences for the different types of quantum repeater units and protocols \cite{White}.
However, note that for our comparison in this section, especially assuming that two photonic states are combined in the middle of each segment (i.e. in a kind of NSP scenario), the two-photon interference of scheme (i) results in 
a quadratic disadvantage not only for the channel transmission
but also in terms of the link coupling efficiency $p_{\mathrm{link}}$ in comparison
with the protocols based on one-photon interference
(schemes (ii) and (iii)), $p_{\mathrm{link,(i)}}=p_{\mathrm{link,(ii)}}^2=p_{\mathrm{link,(iii)}}^2$.
For this let us write in short $p_{\mathrm{link,DR}}=p_{\mathrm{link,TF}}^2$, given the similarity of schemes (ii) and (iii).

\begin{figure*}[ht]
	\centering
	\subfloat[a][$\mu=0.999$, $F_0=0.99$]{\includegraphics[width=0.33\linewidth]{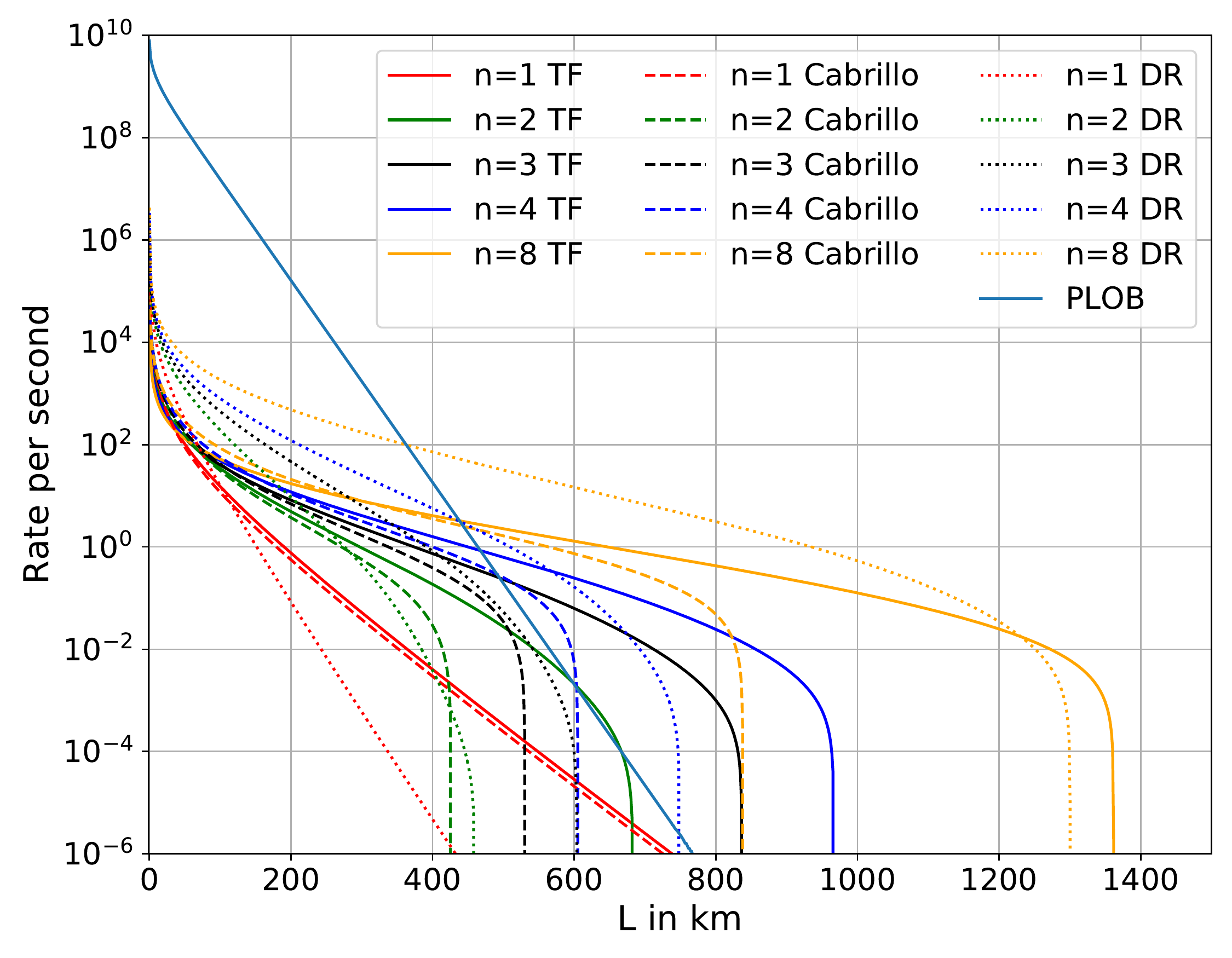}}
	\subfloat[][$\mu=0.999$, $F_0=0.98$]{\includegraphics[width=0.33\linewidth]{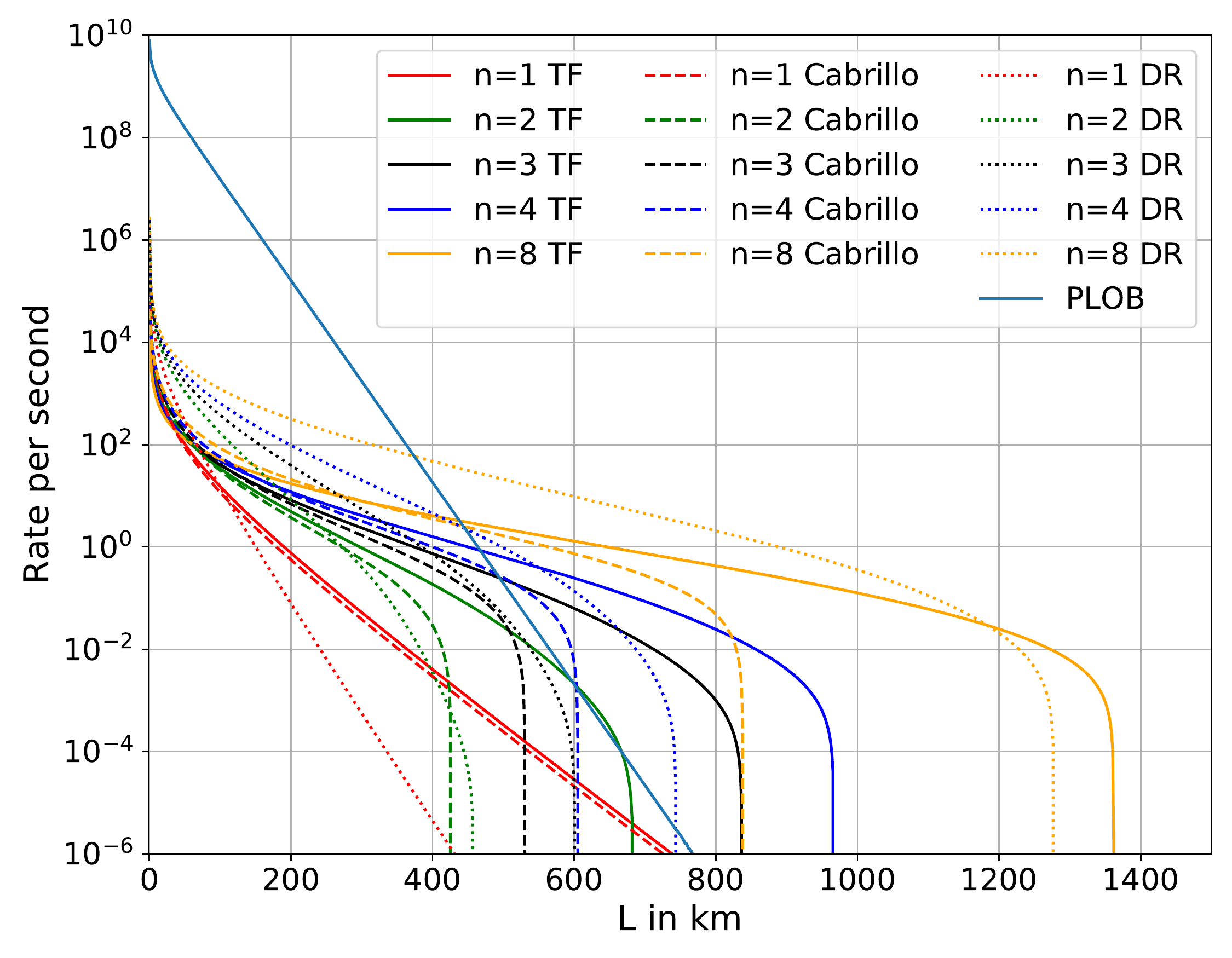}}
	\subfloat[][$\mu=0.99$, $F_0=0.99$]{\includegraphics[width=0.33\linewidth]{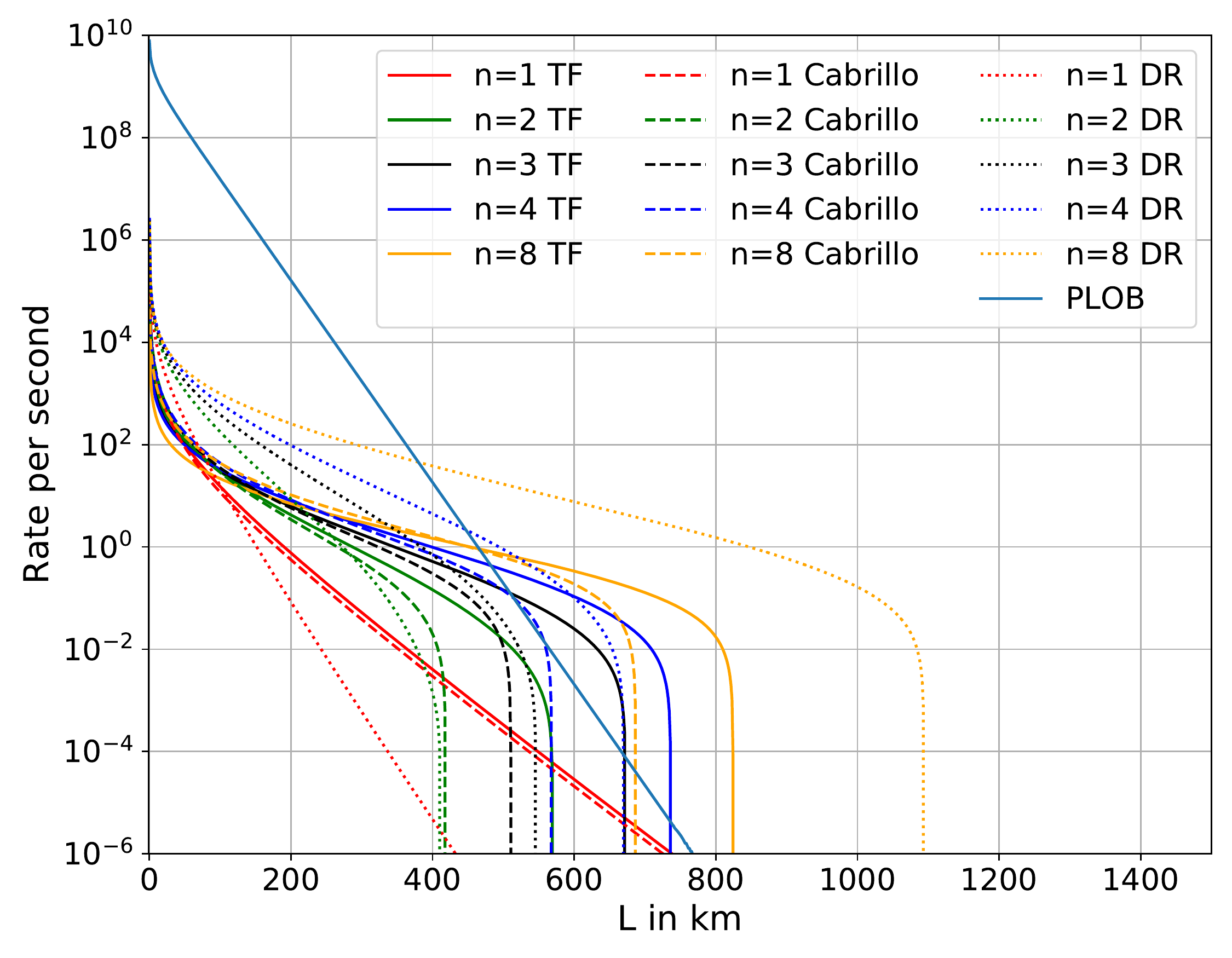}} \\
	\subfloat[][$\mu=0.99$, $F_0=0.98$]{\includegraphics[width=0.33\linewidth]{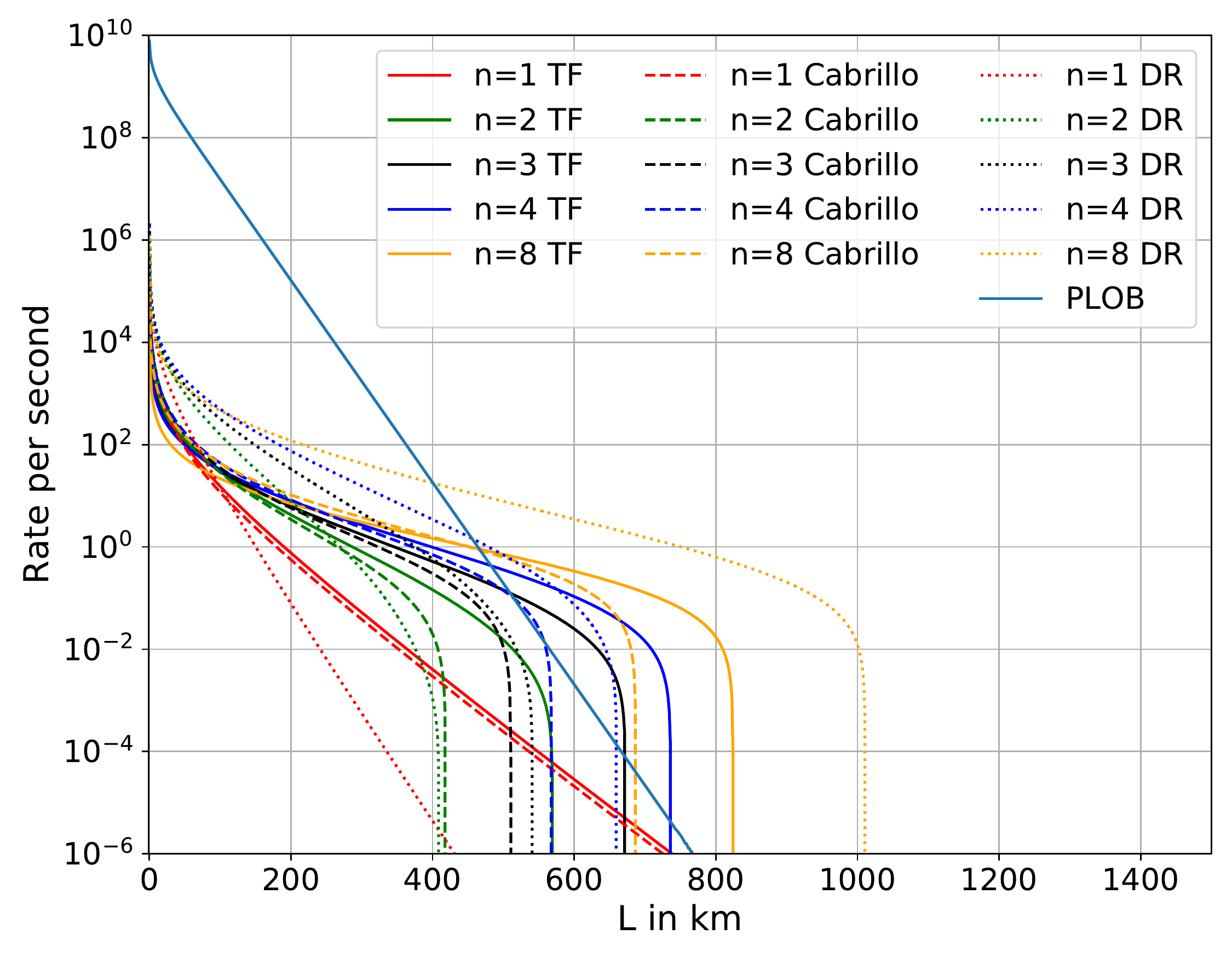}} 
	\subfloat[][$\mu=0.98$, $F_0=0.99$]{\includegraphics[width=0.33\linewidth]{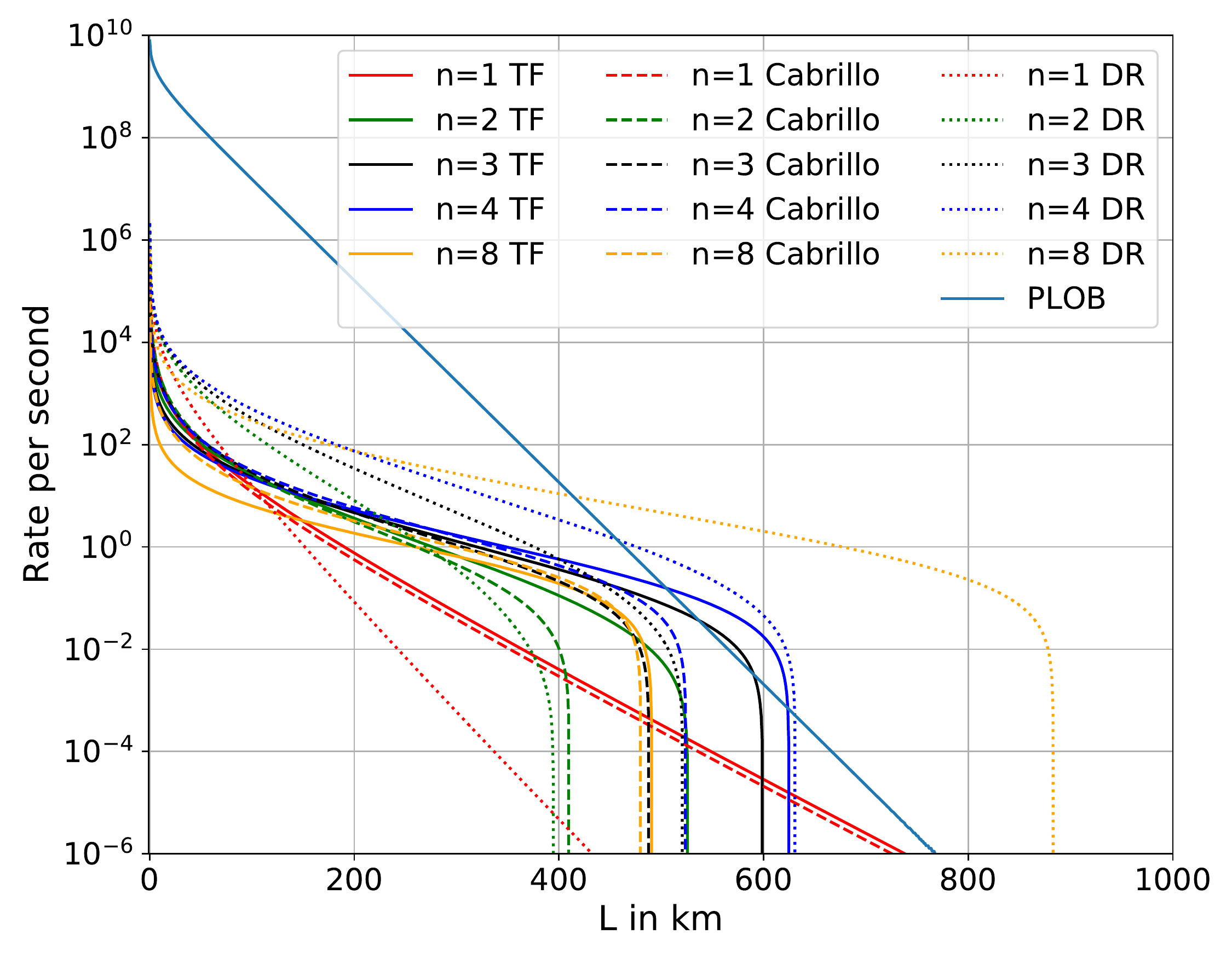}} \\
	\subfloat[][$\mu=0.98$, $F_0=0.98$]{\includegraphics[width=0.33\linewidth]{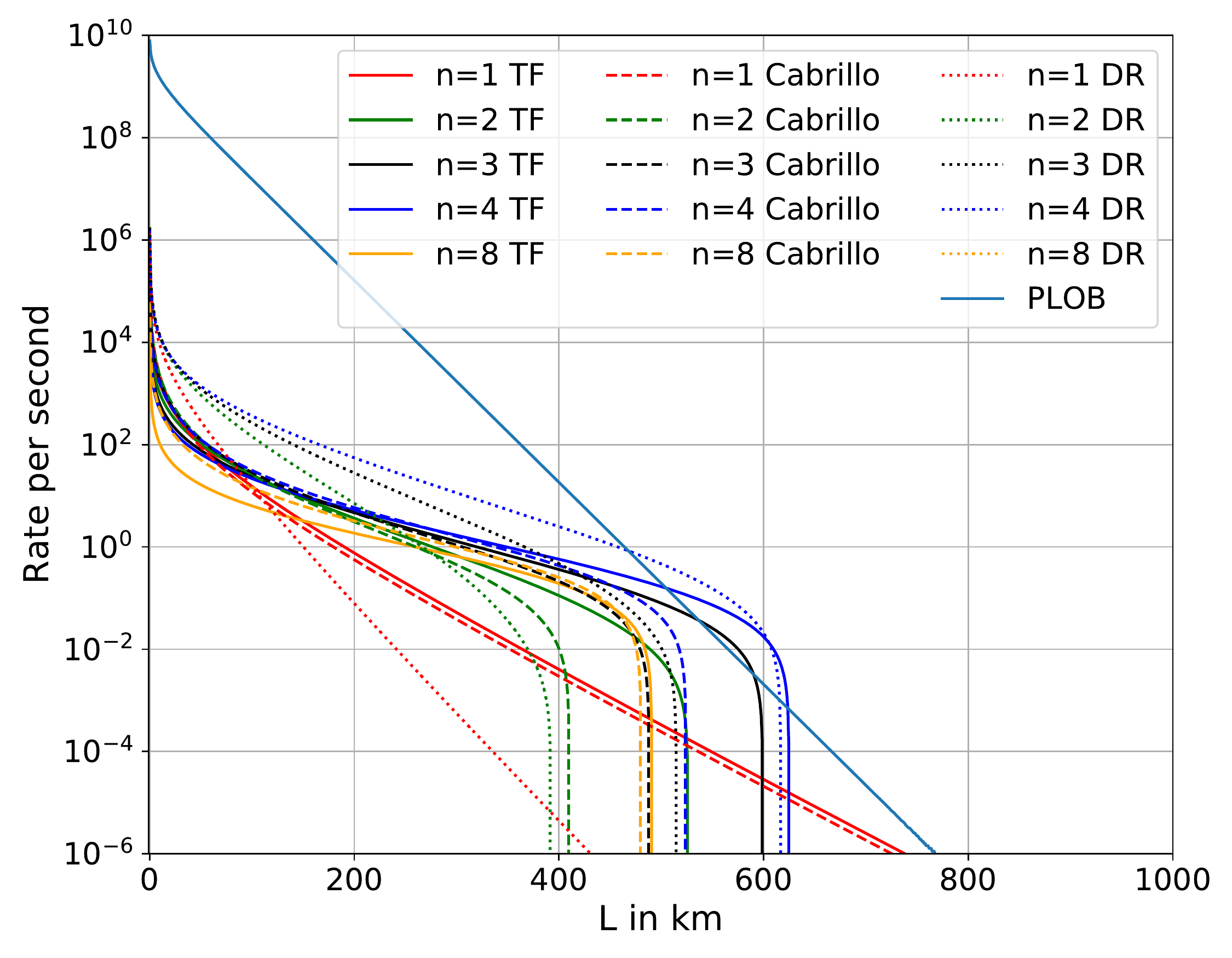}}
	\subfloat[][$\mu=0.97$, $F_0=0.99$]{\includegraphics[width=0.33\linewidth]{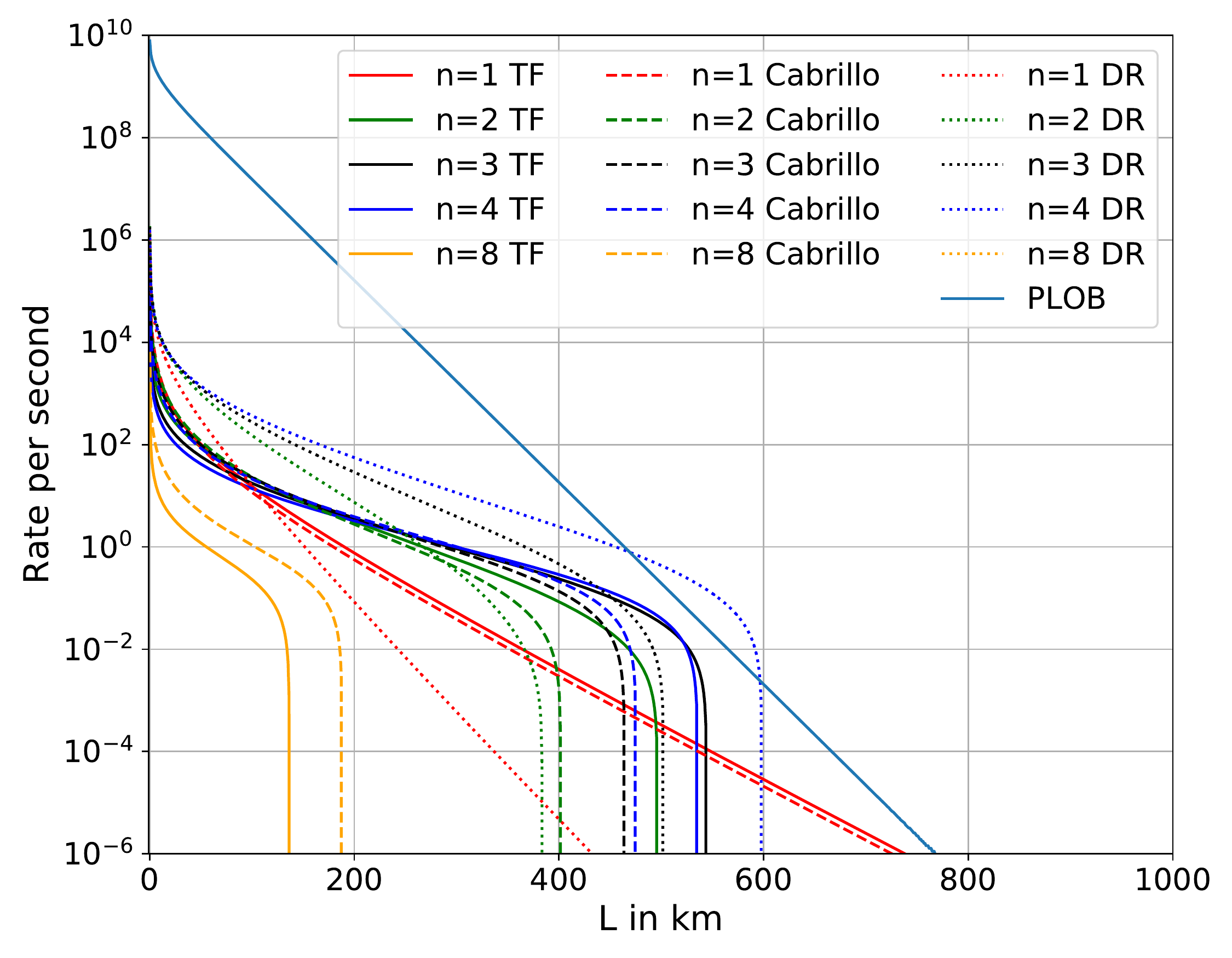}} 
	\caption{Secret key rates per second. We always assume a coherence time $\tau_{\mathrm{coh}}=\unit[10]{s}$, $p_{\mathrm{link,TF}}=0.9$, and $M=1$.}
	\label{fig:SKR_per_sec_tf}
\end{figure*}

In Fig. \ref{fig:SKR_per_sec_tf} we compare the secret key rates for the 
dual-rail scheme (i) (DR), the Cabrillo scheme (ii), and the twin-field repeater (iii) (TF).
The two twin-field-type schemes include a free parameter describing the number of excitations. More excitations lead to a higher transmission rate at the expense of a lower state quality. In the plots we optimize this parameter for each data point to obtain the maximal secret key rate.
Recall, for the DR scheme, we introduce a small dephasing via the parameter $F_0<1$ in order to avoid comparing perfect initial entangled states with noisy ones.
When comparing schemes (ii) and (iii) one can see that for $\mu\approx1$ (iii) performs better while for lower $\mu$ (ii) is the better performing scheme.
This is because the probability of an error is smaller for the Cabrillo scheme, but the error would affect both QBERs of the BB84 protocol, significantly reducing the secret key rate. For the TF scheme (iii) we have an effect on only one of the two error rates.
When $\mu$ gets smaller, all schemes have a non-vanishing error rate in both bases and therefore the lower error rate of the Cabrillo scheme is helpful.

Figure~\ref{fig:SKR_per_sec_tf} shows that, although the DR scheme has a scaling disadvantage in comparison to both other schemes, it is often highly competitive, since both twin-field-type schemes suffer from their low initial probabilities of success when only weak excitations can be used to avoid introducing too much noise from the loss channel. Considering a memory coherence time of 10 seconds, a gate error parameter $\mu\geq0.97$, and coupling efficiencies as $p_{\mathrm{link,TF}}=0.9$, one can already overcome the PLOB bound with only three memory stations using either the DR scheme (i) or the TF protocol (iii).
For this comparison in terms of secret bits per second, we assume a
source repetition rate of \unit[1]{GHz} for an ideal point-to-point link
as associated with the PLOB bound per channel use.
Note that we do not include an extra factor of $1/2$ for the final rates which would strictly be needed in the DR-based scheme in comparison with the PLOB bound for a single-mode loss channel. Here the parallel transmission of the two modes
for a DR qubit does not change the rates per second and this optical encoding does not cause an extra experimental resource overhead (in fact, it even simplifies the optical transmission circumventing the need for long-distance phase stabilization as for the TF-type schemes). Moreover, an optical point-to-point direct transmission would most likely be based on DR qubit transmission as well. The other, previously mentioned factor $2$ that occurs in front of the effective inverse coherence time $\alpha$ when the two spins of a two-qubit spin pair simultaneously dephase while waiting in one segment has now been included here for each segment
(i.e. a small improvement would be possible when Alice and Bob measure their spins immediately).

In Fig.~\ref{fig:SKR_per_sec_tf}, we always assume a coherence time $\tau_{\mathrm{coh}}=\unit[10]{s}$, $p_{\mathrm{link,TF}}=0.9$, and $M=1$.
Recall from our discussions of the possibility of multiplexing that
we may equivalently consider schemes for which, for instance, 
$\tau_{\mathrm{coh}}=\unit[1]{s}$ and $M=10$ according to Eq.~\eqref{eq:MPrelation}.
The plots lead to the following observations. 
The two TF-type schemes (ii) and (iii) more heavily rely upon
sufficiently good error parameters than the DR scheme (i).
In Figs.~\ref{fig:SKR_per_sec_tf}(a) and (b) for two different initial dephasing
fidelities (which is only relevant for DR), we see that only the TF scheme (iii)
performs as good as DR with a gate error as low as $\mu=0.999$.
In this case, for the given parameters, TF even allows to reach slightly larger distances compared with DR, both going well above $L=1200$ km
giving more than a hundredth of a secret bit per second at such distances.
Note that in order to achieve this, the TF scheme requires a loss scaling
with a $16$-times bigger
effective attenuation distance compared with a point-to-point link,
whereas the DR scheme only has to exhibit an $8$-times bigger
effective attenuation distance (``$n=8$ TF'' vs. ``$n=8$ DR'').
The number of memory stations is the same for both, namely seven (not counting those at Alice and Bob).

With increasing gate errors $\mu\leq 0.99$, as shown in Figs.~\ref{fig:SKR_per_sec_tf}(c)-(g), only the DR scheme allows to reach distances above or near $L=1000$km. If both error parameters, that for the gates, $\mu$, and that for the initial states, $F_0$, are no longer sufficiently good (both or in combination), also the DR scheme ceases to reach large distances and barely beats the PLOB bound (see Figs.~\ref{fig:SKR_per_sec_tf}(f) and (g)).
For the two TF-type schemes (ii) and (iii), we generally checked both types of detectors, on-off as well as photon-number-resolving
(Fig.14 shows the results for on-off detections), 
and we did not see a significant difference in the logarithmic plots
of the secret key rates for both schemes. The reason is that for larger distances the two-photon events at either of the two detectors (detectable via PNRDs) get increasingly unlikely compared with one-photon detection events coming from the two-photon terms in combination with the loss of one photon during transmission
(causing errors which remain undetectable via PNRDs).

The practically most relevant situation is shown in Figs.~\ref{fig:SKR_per_sec_tf}(c)-(e). In particular, for the numbers chosen there, i.e. state and gate errors of the order of 1-2\%, the DR scheme reaches 
a distance of $L=800$km with about one secret bit per second, and even beyond with a lower rate. The link coupling efficiency for this scenario, like in all others, is $p_{\mathrm{link,DR}}=p_{\mathrm{link,TF}}^2=0.81$; the coherence time is $\tau_{\mathrm{coh}}=\unit[10]{s}$. The number of segments is $n=8$ (``$n=8$ DR'', dotted yellow curve) corresponding to a memory station placed at every $L_0=100$km. 
The result for this scheme is consistent with the results obtained for $S_2(\mu=0.99)$ and especially $S_2(\mu=0.99)/\tau$ in Tabs.~\ref{tab:optperchanneluse} and \ref{tab:optpersecond}, respectively, for $n=8$. However, note that for the values in Tabs.~\ref{tab:optperchanneluse} and \ref{tab:optpersecond} we chose $p_{\mathrm{link}}=F_0=1$ and $\mu=\mu_0$, slightly different from the parameter choice for Fig.~\ref{fig:SKR_per_sec_tf}(c)
where $\mu_0=1$ and $F_0=0.99$ playing the role of an imperfect state parameter instead of $\mu_0$ (in addition, we have $p_{\mathrm{link}}=0.81$ for DR, and also two spins dephasing at any time step included).
Reiterating the previous discussions in Secs.~\ref{sec:Comparison: 2 vs. 4 vs. 8 segment repeaters}, the choice of $L_0 \sim 100$km seems not only highly compatible with existing classical repeater and fiber network architectures, but also offers a good balance between an improved memory-assisted loss scaling and an only limited addition of extra faulty elements.
Here now we found, in particular, that the standard DR scheme (i) provides another good choice in order to really benefit from these well balanced parameters.

Finally, we also considered the six-state QKD protocol \cite{sixstate} instead of BB84, but this only improved the final rates marginally. In the case of $\mu=0.98$ and $\mu_0=1$, the rate could be, in principle, improved significantly for $n=8$, but for these parameters, in practice, it is easier to use BB84 and $n=4$ instead. When considering sufficiently good error parameter values like $\mu=0.99$, such that $n=8$ outperforms $n=4$, then again there is only a minimal improvement by employing the six-state QKD protocol.

\section{Conclusion}\label{sec:Conclusion}

We presented a statistical model based on two random variables 
and their probability-generating functions (PGFs)
in order to describe, in principle, the full statistics
of the rates obtainable in a memory-based quantum repeater chain.
The physical repeater model assumes a heralded initial 
entanglement distribution with a certain elementary probability
for each repeater segment
(including fiber channel transmission and all link coupling efficiencies),
deterministic entanglement swapping to connect the segments,
and single-spin quantum memories at each repeater station that are subject
to time-dependent memory dephasing.
No active quantum error correction is performed on any of the repeater
``levels'', while our model does not even rely upon the basic assumption of any nested repeater level structure.
The two basic statistical variables associated with this
physical repeater model are the total repeater waiting time and the total, accumulated dephasing time.

In the context of an application in long-range quantum cryptography, our model corresponds to a form of memory-assisted quantum key distribution, for which we calculated the (asymptotic, primarily BB84-type) secret key rates as a figure of merit to assess the repeater performance against known benchmarks and all-optical quantum communication schemes.
Apart from the theoretical complexity that grows with the size of the repeater
(i.e., the number of repeater segments), it was clear from the start that experimentally the memory-assisted schemes of our model cannot go arbitrarily far while still producing a non-zero secret key rate. One motivation and goal of our work was to quantify this intuition and to provide an answer to the question whether it is actually beneficial, in a real setting, to add faulty memory stations to a quantum communication line. Existing works had their focus on the smallest repeaters with only two segments and one middle station. So, the aim was to further explore these smallest repeaters and then extend them to repeaters of a larger scale, answering the above question.

Within this framework, we determined an optimal repeater scheme
that belongs to the class of the fastest schemes (minimizing the average total waiting time and hence maximizing the long-distance entanglement distribution ``raw rate'') and, in addition, minimizes the average accumulated memory dephasing within the class of the fastest schemes. We have achieved this optimization for medium-size quantum repeaters with up to eight segments.
In particular, for the minimal dephasing, this led us to a scheme to ``swap as soon as possible''. The technically most challenging element of our treatment is to determine an explicit analytical expression for the random dephasing variable of the fast schemes and its PGF.  
In order to confirm the correspondence of the minimum of the dephasing variable with the minimal QKD quantum bit error rate (for the variable related to memory dephasing), we calculated the relevant expectation values and compared the optimal scheme with schemes based on other, different swapping strategies. More generally, our formalism enables one to also consider mixed strategies in which different types of entanglement distribution and swapping can be combined, including the traditionally used doubling strategy that allows to systematically incorporate methods for quantum error detection (entanglement distillation).

Our new results especially apply to quantum repeaters beyond one middle station for which an optimization of the distribution and swapping strategies is no longer obvious. For the special case of three repeater segments, assuming only channel loss and memory dephasing, we showed that our optimal scheme gives the highest secret key rate among not only all the fastest schemes but among all schemes including overall slower schemes that may still potentially lead to a smaller accumulated dephasing. We conjecture that our optimal scheme also gives the highest secret key rate for more than three segments under the same physical assumptions. A rigorous proof of this is non-trivial, because the number of distinct swapping and distribution strategies grows fast with the number of repeater segments. Moreover, in a long-range QKD application, some of the spin qubits may be measured immediately which is generally hard to include in the statistical analysis and the optimization for all possible schemes; for three segments though we did include this additional complexity of the protocols. Towards applications beyond QKD, this extra variation may no longer be relevant. 

We identified three criteria that should be satisfied by an optimal repeater scheme: distribute entanglement in parallel as fast as possible, store entanglement in parallel as little as possible, and swap entanglement as soon as possible. It is not always possible to satisfy these conditions at the same time, and we discussed specific schemes that are particularly good or bad with regards to some of the criteria. For example, a fully sequential repeater scheme is particularly slow, but avoids parallel storage of many spin qubits. Nonetheless, since it is overall slow, the fully sequential scheme can still accumulate more dephasing. We presented a detailed analysis comparing such different repeater protocols and approaches.          

With regards to a more realistic quantum repeater modelling, we considered additional tools and parameters such as memory cut-offs, multiplexing,
initial state and swapping gate fidelities in order to identify 
potential regimes in memory-assisted quantum key distribution beyond one middle station where, exploiting our optimized swapping strategy, it becomes useful to add further memory stations along the communication line and connect them via two-qubit swapping operations. Importantly, we found that the initial state and gate fidelities must exceed certain minimal values (generally depending on the specific QKD protocol including post-processing), as otherwise the sole faultiness of the spin-qubit preparations and operations prevents to obtain a non-zero secret key rate even when no imperfect quantum storage (no memory dephasing) at all takes place and independent of the finite channel transmission.
This effect becomes stronger with an increasing number of repeater nodes, scaling with the power of $2n-1$ for the error parameters in the QKD secret key rate. Once this minimal state and gate fidelity criterion is fulfilled and when the other experimental imperfections are included too, especially the time-dependent memory dephasing, it is essential to consider the exact secret key rates obtainable in optimized repeater protocols in order to conclude whether a genuine quantum repeater advantage over direct transmission schemes is possible or not.
This is what our work aimed at and achieved based on the standard notion of asymptotic QKD figures of merit. 

By quantifying the influence of (within our physical model) basically all relevant experimental parameters on the final long-range QKD rate, we were able to determine the scaling and trade-offs of these parameters and analytically calculate exact, optimal rates. A quantum repeater of $n=L/L_0$ segments is thereby characterized by the parameter set $(p,a,\alpha)$ where $p$ is the entanglement distribution probability per segment (including the $n$-dependent channel transmission and zero-distance link coupling efficiency per segment), $a$ is the entanglement swapping success probability, and $\alpha$ is the inverse effective memory coherence time which, in most protocols, depends on $n$ via the quantum and classical communication times per distribution attempt (we also considered small-scale two-segment protocols without this dependence and ideas exist to minimize the impact of the inevitable signal waiting times for the elementary units of larger repeaters in combination with high experimental source and processing clock rates \cite{Cody_Jones}).
In addition, we have introduced a set of initial state and gate parameters $(\mu_0/F_0, \mu)$ where $\mu_0$ and $F_0$ can be adapted to the specific protocols. Additional memory parameters can be collected as $(m,M,B)$ where $m$ is the memory cut-off (maximal time at which any spin qubit is stored), $M$ is the number of simultaneously employed memory qubits in a simple multiplexing scenario with $M$ repeater chains used in parallel, and $B$ is the ``memory buffer'' (the number of memory qubits per half station in a single repeater chain).
In our work, we focussed on schemes with $a=1$ and $B=1$. The use of $B>1$ memories at each station would allow to continue the optical quantum state transfer even in segments that already possess successfully distributed states
and to potentially replace the earlier distributed lower-quality pairs (subject to memory dephasing) by the later distributed pairs.
We also did not put the main emphasis on the use and optimization of $m$, though we did include this option in some schemes. We found that $M>1$ leads to an effective improvement of the memory coherence time by a factor of $M$. 

In this setting, the three essential experimental parameters that have to be sufficiently good are the link coupling efficiency (via $p$), the memory coherence time (via $\alpha$), and the state/gate error parameter $\mu_0$/$\mu$. While the latter must not go below the above-mentioned limits, generally two of these three parameters should be sufficiently good as a rule of thumb in order to exceed the repeaterless bound and obtain practically meaningful rates. If this is the case, or even better, if all three are of high quality, memory-assisted quantum key distribution based on heralded entanglement distribution and swapping without additional quantum error correction or detection is possible to allow Alice and Bob to share a secret key at a rate orders of magnitude faster than in all-optical quantum state transmission schemes. For instance, for a total distance of 800km and experimental parameter values that are highly demanding but not impossible (up to 10s coherence time, about 80\% link coupling, and state or gate infidelities in the regime of 1-2\%), one secret bit can be shared per second with repeater stations placed at every 100km, providing the best balance 
between a minimal number of extra faulty repeater elements and a sufficient number of repeater stations for an improved loss scaling.

{\it Acknowledgement:}
We thank the BMBF in Germany for support via Q.Link.X/QR.X and
the BMBF/EU for support via QuantERA/ShoQC.

\appendix

\section{Derivation of Eq.~\eqref{eq:GKn}}\label{app:GKn}

In this section we derive the PGF $G_n(t)$ of the random variable $K_n$ defined via
\begin{equation}
	K_n = \max(N_1, \ldots, N_n),
\end{equation}
where $N_i$ are the geometrically distributed random variables with parameter $p$. We have
\begin{equation}\label{eq:app:GKn}
\begin{split}
	G_n(t) &= \sum^{+\infty}_{k_1, \ldots, k_n = 1} p q^{k_1 - 1} \ldots p q^{k_n - 1} t^{\max(k_1, \ldots, k_n)} \\
	&= p^n t F_n(q, t),
\end{split}
\end{equation}
where the function $F_n(x, t)$ is defined as
\begin{equation}
	F_n(x, t) = \sum^{+\infty}_{k_1, \ldots, k_n = 0} x^{k_1 + \ldots + k_n} t^{\max(k_1, \ldots, k_n)}.
\end{equation}
The series on the right-hand side of this definition converges for all $|x|<1$ and $|t| \leqslant 1$, since we have
\begin{equation}
	|F_n(x, t)| \leqslant \sum^{+\infty}_{k_1, \ldots, k_n = 0} |x|^{k_1 + \ldots + k_n} = \frac{1}{(1-|x|)^n}.
\end{equation}
The function $F_n(x, t)$ can be written in a compact form, having only a finite number of terms. We have
\begin{equation}
\begin{split}
	&\frac{F_n(x, t)}{1-t} = \sum^{+\infty}_{k_1, \ldots, k_n = 0} \sum^{+\infty}_{k = \max(k_1, \ldots, k_n)}
	x^{k_1 + \ldots + k_n} t^k \\
	&= \sum^{+\infty}_{k = 0} t^k \sum^k_{k_1, \ldots, k_n = 0} x^{k_1 + \ldots + k_n}
	= \sum^{+\infty}_{k = 0} t^k \left(\frac{1 - x^{k+1}}{1-x}\right)^n.
\end{split}
\end{equation}
Expanding the $n$-th power on the right-hand side and applying simple algebraic transformations, we obtain the following
compact expression:
\begin{equation}
	F_n(x, t) = \frac{1 - t}{(1 - x)^n t} \sum^n_{i = 0} (-1)^i \binom{n}{i} \frac{1}{1 - x^i t}.
\end{equation}
From Eq.~\eqref{eq:app:GKn} we derive the following expression for the PGF of $K_n$:
\begin{equation}
\begin{split}
	G_n(t) &= (1 - t) \sum^n_{i = 0} (-1)^i \binom{n}{i} \frac{1}{1 - q^i t} \\
	&= 1 + (1-t)\sum^n_{i = 1} (-1)^i \binom{n}{i} \frac{1}{1 - q^i t},
\end{split}
\end{equation}
which is exactly the expression presented in the main text.

\section{Trace identities}\label{app:Trace Identities}

We have
\begin{equation}\label{eq:Trid}
\begin{split}
	{}_{23}\langle&\Psi^+| \tilde{\Gamma}_{\mu, 23}(\hat{\varrho}_{1234})|\Psi^+\rangle_{23} \\
	&= \mu \cdot {}_{23}\langle\Psi^+| \hat{\varrho}_{1234}|\Psi^+\rangle_{23}
	+ \frac{1 - \mu}{4} \Tr_{23}(\hat{\varrho}_{1234}).
\end{split}
\end{equation}
Here we show how to compute the quantities on the right-hand side of this equality. A
simple way is to work with density matrices. We use the order of basis elements
induced by the tensor product. From the one-qubit basis $(|0\rangle, |1\rangle)^T$
we obtain the two-qubit basis
\begin{equation}\label{eq:B2}
	\begin{pmatrix}
		|0\rangle \\
		|1\rangle
	\end{pmatrix}
	\otimes
	\begin{pmatrix}
		|0\rangle \\
		|1\rangle
	\end{pmatrix}
	=
	\begin{pmatrix}
		|00\rangle \\
		|01\rangle \\
		|10\rangle \\
		|11\rangle
	\end{pmatrix}.
\end{equation}
Taking the tensor product once again, we obtain the ordering of four-qubit basis vectors $|0000\rangle$, $|0001\rangle$,
$|0010\rangle$, $|0011\rangle$, $|0100\rangle$, $|0101\rangle$, $|0110\rangle$, $|0111\rangle$, $|1000\rangle$,
$|1001\rangle$, $|1010\rangle$, $|1011\rangle$, $|1100\rangle$, $|1101\rangle$, $|1110\rangle$, $|1111\rangle$. If a
four-qubit state is described by a density operator $\hat{\varrho}_{1234}$ which has a $16 \times 16$
density matrix $\varrho$ in the standard basis ordered as described above, then two-qubit partial diagonal states have the
following matrices in the basis \eqref{eq:B2}:
\begin{equation}\label{eq:D23}
\begin{split}
	{}_{23}\langle 00|\hat{\varrho}_{1234}|00\rangle_{23} &= \rho[1, 2, 9, 10] \\
	{}_{23}\langle 01|\hat{\varrho}_{1234}|01\rangle_{23} &= \rho[3, 4, 11, 12] \\
	{}_{23}\langle 10|\hat{\varrho}_{1234}|10\rangle_{23} &= \rho[5, 6, 13, 14] \\
	{}_{23}\langle 11|\hat{\varrho}_{1234}|11\rangle_{23} &= \rho[7, 8, 15, 16],
\end{split}
\end{equation}
where $\varrho[I]$, $I$ being a set of 1-based indices, is the submatrix of $\varrho$ with row and column indices in
$I$. For the off-diagonal states we have
\begin{equation}\label{eq:O23}
\begin{split}
	{}_{23}\langle 01|\hat{\varrho}_{1234}|10\rangle_{23} &= \rho[3, 4, 11, 12 | 5, 6, 13, 14] \\
	{}_{23}\langle 10|\hat{\varrho}_{1234}|01\rangle_{23} &= \rho[5, 6, 13, 14 | 3, 4, 11, 12],
\end{split}
\end{equation}
where $\varrho[I|J]$ is the submatrix of $\varrho$ with row indices in $I$ and column indices in $J$.

The state of the form given by Eq.~\eqref{eq:Drho}
\begin{equation}
	\hat{\varrho} = \tilde{\Gamma}_{\mu}\bigl(F|\Psi^+\rangle\langle\Psi^+| + (1 - F)|\Psi^-\rangle\langle\Psi^-|\bigr)
\end{equation}
has the following density matrix in the basis \eqref{eq:B2}:
\begin{equation}
	\varrho =
	\frac{1}{4}
	\begin{pmatrix}
		1 - \mu & 0 & 0 & 0 \\
		0 & 1 + \mu & 2\mu(2F - 1) & 0 \\
		0 & 2\mu(2F - 1) & 1 + \mu & 0 \\
		0 & 0 & 0 & 1 - \mu
    \end{pmatrix}.
\end{equation}
Taking the Kronecker product of two states of this form, Eq.~\eqref{eq:Trid} together with the relations
Eqs.~\eqref{eq:D23}-\eqref{eq:O23} lead to the final form of the distributed state given by Eq.~\eqref{eq:rho14}.

\section{Computing PGFs of the sequential scheme}\label{app:SeqPGF}

\begin{figure*}
	\subfloat[The general structure of failure periods (if any) and the success period.]{\includegraphics{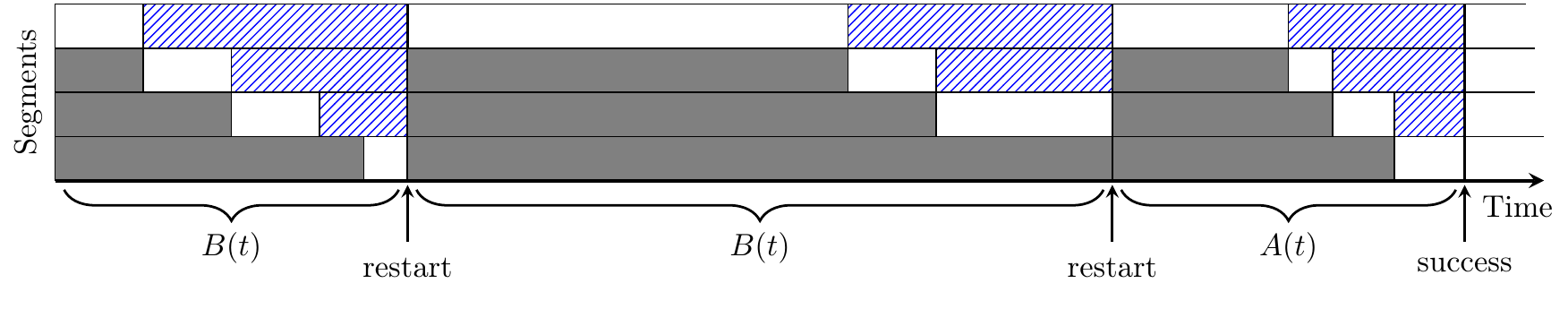}} \\
	\subfloat[A detailed view of the failure part generating function $B(t)$.]{\includegraphics{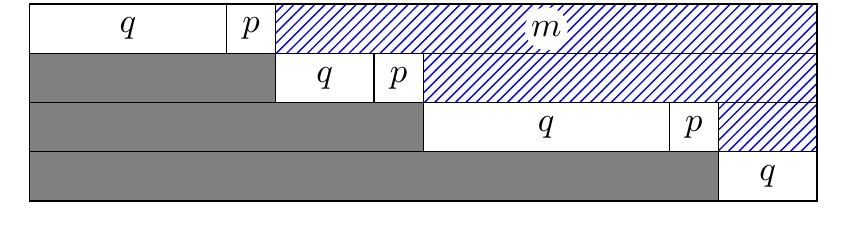}}
	\subfloat[A detailed view of the success part generating function $A(t)$.]{\includegraphics{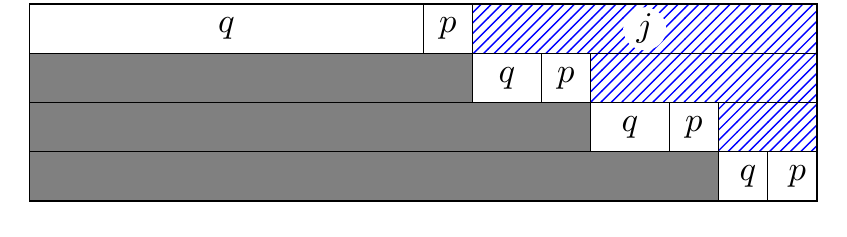}}
\caption{A visualization of the entanglement distribution process with the sequential scheme for $n=4$.}\label{fig:seq1}
\end{figure*}

In the sequential scheme the number of steps $K_n$ and the dephasing $D_n$ are given by
\begin{equation}
	K_n = N_1 + \ldots + N_n, \quad D_n = N_2 + \ldots + N_n.
\end{equation}
Their PGFs are thus the $n$-th and $(n-1)$-th power of the single-segment PGF:
\begin{equation}
	G_n(t) = \left(\frac{p t}{1 - q t}\right)^n, \quad \tilde{G}_n(t) = \left(\frac{p t}{1 - q t}\right)^{n-1}.
\end{equation}
In the case of cutoff, the process of entanglement distribution is visualized in Fig.~\ref{fig:seq1}. There are zero or
more failure parts, with number of steps generating function $B^{[m]}_n(t)$, and one and only one success part, with
generating function $A^{[m]}_n(t)$. The total PGF $G^{[m]}_n(t)$ of the number of steps $K^{[m]}_n$ is thus given by
\begin{equation}
	G^{[m]}_n(t) = \frac{A^{[m]}_n(t)}{1 - B^{[m]}_n(t)}.
\end{equation}
We start with the derivation of the failure parts's PGF. The PGF of the top line is clearly
\begin{equation}
    G_0(t) = \frac{pt}{1-qt}.
\end{equation}
Among the rest $n-1$ lines there are $i$ lines that succeed, where $0 \leqslant i \leqslant n-2$, so we have to put $i$ $p$'s into $m$ places and the rest $m-i$ places will be taken by $q$'s. We thus have
\begin{equation}
    B^{[m]}_n(t) = G_0(t) \sum^{n-2}_{i=0} \binom{m}{i}p^iq^{m-i}t^m.
\end{equation}
For the success part's PGF we have
\begin{equation}
    A^{[m]}_n(t) = G_0(t) \sum^m_{j=n-1} \binom{j-1}{n-2}p^{n-1}q^{j-n+1} t^j,
\end{equation}
since the length of the success part can vary from $n-1$ to $m$ (we need to put at least $n-1$ $p$'s there). The position of the last $p$ is fixed, so we need to place $n-2$ $p$'s into $j-1$ places and the rest $j-n+1$ will be taken by $q$'s. Making substitution $j \to j-n+1$, we arrive to the expression \eqref{eq:Gmnt} of the main text.

The random variable for the waiting time of the scheme involving multiple cutoffs is given by
\begin{equation}
    K_{n}^{\mathrm{seq},\vec{m}}= \tilde{N}^{(m_{n-1})}-m_{n-1}+\sum_{j=1}^{T_{n-1}} \left(K_{n-1,j}+m_{n-1}\right)\,.
\end{equation}
Exploiting that sums of independent random variables correspond to products of their PGFs and using \cite[Satz 3.8]{Klenke2020} for the sum one immediately obtains the result in the main text.

\section{Computing dephasing PGFs for parallel schemes}\label{app:PGF Parallel schemes}

In this section we derive explicit expressions for the PGFs of the dephasing random variables $D_n$ for different
schemes considered in the main text. All these schemes have the same property --- if the order of $N_i$'s is known then
one can obtain an analytical expression for the corresponding random variable $D_n$ explicitly. Having an explicit
expression for $D_n$, we can compute a part of its PGF corresponding to a given order of arguments. Combining these
parts for all possible ordering of arguments, we get the expression for PGF of $D_n$.

More formally, the space $\Omega = \mathbb{N}^n$ of elementary events consists of all $n$-vectors $\vec{N} = (N_1,
\ldots, N_n)$ of positive integers. The components $N_i$ are independent identically distributed (i.i.d.) random
variables with geometric distribution with success probability $p$, so $N_i$ is the number of attempts (including the
last successful one) of the $i$-th segment to distribute entanglement. The failure probability we denote $q = 1-p$. To
every point $\vec{N} = (N_1, \ldots, N_n) \in \Omega$ we assign the probability 
\begin{equation}
    \mathbf{P}(\vec{N}) = pq^{N_1 - 1} \ldots pq^{N_n-1} = p^n q^{N_1 + \ldots + N_n - n}.
\end{equation}
The sum of these probabilities is obviously 1, so we have a valid probability space $(\Omega, \mathbf{P})$.

The PGF of every component $N_i$ is given by the following simple expression:
\begin{equation}
    g_{N_i}(t) = \frac{pt}{1-qt}.
\end{equation}
To find PGFs of more complicated random variables involving several components, we appropriately partition $\Omega$,
compute the partial PGF on each part and then combine these partial results into the full expression. For every permutation
$\pi \in S_n$ we define a subset of $\Omega$ which is determined by the corresponding relations between $n$ arguments.
For $n=2$ we have two permutations $(12)$ and $(21)$ with corresponding relations $N_1 \leqslant N_2$ and $N_2 < N_1$.
For $n=3$ we have six permutations and six corresponding relations
\begin{equation}
\begin{split}
    &N_1 \leqslant N_2 \leqslant N_3 \quad N_1 \leqslant N_3 < N_2 \quad N_2 < N_1 \leqslant N_3 \\
    &N_2 \leqslant N_3 < N_1 \quad N_3 < N_1 \leqslant N_2 \quad N_3 < N_2 < N_1.
\end{split}
\end{equation}
To make all these subsets non-overlapping, we use strict inequality between an inversion and non-strict inequality in
other positions between numbers in permutations. We thus have the following decomposition:
\begin{equation}\label{eq:omega}
    \Omega = \bigsqcup_{\pi \in S_n} \Omega_\pi,
\end{equation}
where $\Omega_\pi$ is the subset determined by the relations corresponding to $\pi$. For any point $\vec{N} \in
\Omega_\pi$ we can obtain an explicit expression for $D_n$ for any scheme. In Table~\ref{tbl:1} we show all possible
relations between four arguments and the expression corresponding to the optimal and doubling schemes in the case of
$n=4$. Expressions corresponding to different $\pi$ might be the same, as can be seen for the doubling scheme.

The PGF of $D_n$ is defined as
\begin{equation}
    \tilde{G}_n(t) = \sum^{+\infty}_{d=0} \mathbf{P}(D_n=d) t^d = \sum_{\vec{N} \in \Omega} \mathbf{P}(\vec{N}) t^{D_n(\vec{N})}.
\end{equation}
Using the decomposition in Eq.~\eqref{eq:omega}, we introduce the partial PGFs via
\begin{equation}
    \tilde{G}_n(\pi|t) = \sum_{\vec{N} \in \Omega_\pi} p^nq^{N_1 + \ldots + N_n - n} t^{D_n(N_1, \ldots, N_n)},
\end{equation}
where $D_n(N_1, \ldots, N_n)$ is given explicitly as an appropriate linear combination of $N_i$'s. The total PGF $\tilde{G}_n(t)$ is then just the sum of all of these partial PGFs:
\begin{equation}
    \tilde{G}_n(t) = \sum_{\pi \in S_n} \tilde{G}_n(\pi|t).
\end{equation}
We demonstrate computing these sums by an example for $n=4$. We have the correspondence
\begin{equation}
    \pi = (2134) \to N_2 < N_1 \leqslant N_3 \leqslant N_4
\end{equation}
and the explicit expressions 
\begin{equation}
\begin{split}
    D^\star_4(N_1, N_2, N_3, N_4) &= N_4 - N_2, \\
    D^{\mathrm{dbl}}_4(N_1, N_2, N_3, N_4) &= 2N_4 - N_2 - N_3.
\end{split}
\end{equation}
For the partial PGFs we have
\begin{widetext}
\begin{equation}
\begin{split}
    \tilde{G}^\star_4(\pi|t) &= \sum^{+\infty}_{N_2=1}\sum^{+\infty}_{N_1=N_2+1}\sum^{+\infty}_{N_3=N_1}\sum^{+\infty}_{N_4=N_3} p^4 q^{N_1+N_2+N_3+N_4-4} t^{N_4 - N_2} = \frac{p^4}{1-q^4}\frac{q^3 t}{(1-qt)(1-q^2t)(1-q^3t)}, \\
    \tilde{G}^{\mathrm{dbl}}_4(\pi|t) &= \sum^{+\infty}_{N_2=1}\sum^{+\infty}_{N_1=N_2+1}\sum^{+\infty}_{N_3=N_1}\sum^{+\infty}_{N_4=N_3} p^4 q^{N_1+N_2+N_3+N_4-4} t^{2N_4 - N_2- N_3} = \frac{p^4}{1-q^4}\frac{q^3 t}{(1-q^2t)(1-q^3t)(1-qt^2)}.
\end{split}
\end{equation}
\end{widetext}
Summing up the expression for all $\pi \in S_4$, we obtain the expressions for $\tilde{G}^\star_4(t)$ and
$\tilde{G}^{\mathrm{dbl}}_4(t)$ presented in the main text. For completeness, we also give the optimal PGFs for $n=2$ and $n=3$:
\begin{displaymath}
\begin{split}
    \tilde{G}^\star_2(t) &= \frac{p^2}{1-q^2} \frac{1+qt}{1-qt}, \\
    \tilde{G}^\star_3(t) &= \frac{p^3}{1-q^3} \frac{1+(q+2q^2)t-(2q^2+q^3)t^3-q^4t^4}{(1-qt)(1-q^2t)(1-qt^2)}.
\end{split}
\end{displaymath}
The size of the expressions grows rather quickly with $n$, so we do not present them explicitly for $n>4$.

We see that obtaining $\tilde{G}_n(t)$ reduces to computing sums of many geometrical series, which is a rather trivial
task. The only nontrivial part of this algorithm is its superexponential $n!$-complexity. So, this algorithm is applicable
only for small $n$; we used it up to $n=8$, which is of practical relevance.

\begin{table}[ht]
\begin{tabular}{|l|l|l|}
    \hline
    \hfil Permutation & \hfil $D^\star_4(\vec{N})$ & \hfil $D^{\mathrm{dbl}}_4(\vec{N})$ \\ 
    \hline
    $N_1 \leqslant  N_2 \leqslant  N_3 \leqslant  N_4$ & $N_4-N_1$         & $2N_4-N_1-N_3$ \\
    $N_1 \leqslant  N_2 \leqslant  N_4 <  N_3$         & $2N_3-N_1-N_4$    & $2N_3-N_1-N_4$ \\
    $N_1 \leqslant  N_3 <  N_2 \leqslant  N_4$         & $N_2+N_4-N_1-N_3$ & $2N_4-N_1-N_3$ \\
    $N_1 \leqslant  N_3 \leqslant  N_4 <  N_2$         & $2N_2-N_1-N_3$    & $2N_2-N_1-N_3$ \\
    $N_1 \leqslant  N_4 <  N_2 \leqslant  N_3$         & $2N_3-N_1-N_4$    & $2N_3-N_1-N_4$ \\
    $N_1 \leqslant  N_4 <  N_3 <  N_2$                 & $2N_2-N_1-N_4$    & $2N_2-N_1-N_4$ \\
    $N_2 <  N_1 \leqslant  N_3 \leqslant  N_4$         & $N_4-N_2$         & $2N_4-N_2-N_3$ \\
    $N_2 <  N_1 \leqslant  N_4 <  N_3$                 & $2N_3-N_2-N_4$    & $2N_3-N_2-N_4$ \\
    $N_2 \leqslant  N_3 <  N_1 \leqslant  N_4$         & $N_4-N_2$         & $2N_4-N_2-N_3$ \\
    $N_2 \leqslant  N_3 \leqslant  N_4 <  N_1$         & $N_1-N_2$         & $2N_1-N_2-N_3$ \\
    $N_2 \leqslant  N_4 <  N_1 \leqslant  N_3$         & $2N_3-N_2-N_4$    & $2N_3-N_2-N_4$ \\
    $N_2 \leqslant  N_4 <  N_3 <  N_1$                 & $N_1+N_3-N_2-N_4$ & $2N_1-N_2-N_4$ \\
    $N_3 <  N_1 \leqslant  N_2 \leqslant  N_4$         & $N_2+N_4-N_1-N_3$ & $2N_4-N_1-N_3$ \\
    $N_3 <  N_1 \leqslant  N_4 <  N_2$                 & $2N_2-N_1-N_3$    & $2N_2-N_1-N_3$ \\
    $N_3 <  N_2 <  N_1 \leqslant  N_4$                 & $N_4-N_3$         & $2N_4-N_2-N_3$ \\
    $N_3 <  N_2 \leqslant  N_4 <  N_1$                 & $N_1-N_3$         & $2N_1-N_2-N_3$ \\
    $N_3 \leqslant  N_4 <  N_1 \leqslant  N_2$         & $2N_2-N_1-N_3$    & $2N_2-N_1-N_3$ \\
    $N_3 \leqslant  N_4 <  N_2 <  N_1$                 & $N_1-N_3$         & $2N_1-N_2-N_3$ \\
    $N_4 <  N_1 \leqslant  N_2 \leqslant  N_3$         & $2N_3-N_1-N_4$    & $2N_3-N_1-N_4$ \\
    $N_4 <  N_1 \leqslant  N_3 <  N_2$                 & $2N_2-N_1-N_4$    & $2N_2-N_1-N_4$ \\
    $N_4 <  N_2 <  N_1 \leqslant  N_3$                 & $2N_3-N_2-N_4$    & $2N_3-N_2-N_4$ \\
    $N_4 <  N_2 \leqslant  N_3 <  N_1$                 & $N_1+N_3-N_2-N_4$ & $2N_1-N_2-N_4$ \\
    $N_4 <  N_3 <  N_1 \leqslant  N_2$                 & $2N_2-N_1-N_4$    & $2N_2-N_1-N_4$ \\
    $N_4 <  N_3 <  N_2 <  N_1$                         & $N_1-N_4$         & $2N_1-N_2-N_4$ \\
    \hline
\end{tabular}
\caption{Explicit expressions for the optimal and doubling dephasing for all possible relations between arguments in the case of $n=4$.} \label{tbl:1}
\end{table}

\section{Optimality for three segments}\label{app:Optimality 3 segments}

Here we will compare all possible schemes for a 3-segment repeater, when swapping is applied as soon as possible. We will not consider any scheme, which swaps only at the end or delays the entanglement swapping, as this increases the dephasing even further. For each scheme we calculate the random variables for the waiting time and the dephasing. In case of the dephasing the probability generating function is most useful, whereas for the waiting time we will only state the expectation value.

Moreover, we will consider two different types of schemes. The first type, which we will indicate by ``imm'', describes schemes where Alice and Bob measure their qubits immediately. This scenario is especially useful in QKD applications. The second type of schemes we consider is indicated by a subscript ``non'' and describes schemes, where Alice and Bob do not measure immediately and these types of schemes are important in non-QKD applications. A possible case of usage for those schemes is transferring quantum information between quantum computers by exchanging entangled photons. Here Alice and Bob will not measure their qubits until they share entanglement between each other.

\subsection{Sequential schemes}

\begin{figure}[ht]
	\subfloat[Sequential a]{\includegraphics[width=0.98\linewidth]{"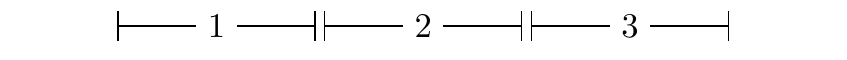"} \label{fig:Sequential a}} \\
	\subfloat[Sequential b]{\includegraphics[width=0.98\linewidth]{"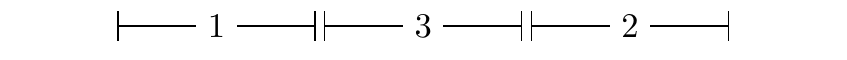"} \label{fig:Sequential b}} \\
	\subfloat[Sequential c]{\includegraphics[width=0.98\linewidth]{"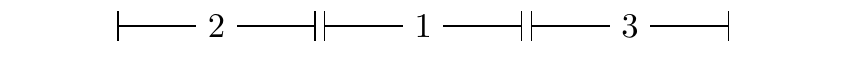"} \label{fig:Sequential c}}
	\caption{Sequential arrangements of entanglement generation in a three-segment repeater. The number in each segment corresponds to the moment when it starts.}
	\label{fig:Different schemes 3 segments seq}
\end{figure}

Let us start with sequential schemes, where entanglement generation only takes places in one segment after another. There are three possibilities. First, one starts generating entanglement in Alice's or Bob's segment and always connects adjacent segments after the previous one has finished successfully. Note that here entanglement swapping is performed as soon as possible. We will call this scheme ``sequential a'', see Fig.~\ref{fig:Sequential a}.

The second possibility is given by starting with the left or right segment, followed by the segment on the opposite side. Thus, no entanglement swapping is possible. Finally, the middle segment is connected. Let us call this scheme ``sequential b'', see Fig.~\ref{fig:Sequential b}.

The third possible arrangement is given by starting in the middle, continuing with the left or right segment and finishing of with the remaining segment on the opposing side, see Fig.~\ref{fig:Sequential c}. All other sequential arrangements for three segments are equivalent to those three schemes.

These three sequential schemes share the same waiting time, which is
\begin{equation}
	K^{\mathrm{seq}}_3=N_1+ N_2 +N_3,
\end{equation}
and has the expectation value
\begin{equation}
	\mathbf{E}[K^{\mathrm{seq}}_3]=\frac{3}{p}.
\end{equation}

Obviously, the dephasing of the schemes differs, and we also have to distinguish between schemes measuring immediately and non-immediately. At first, let us consider immediate schemes, as it will turn out the random variables of the non-immediate schemes are just scaled by a factor of two, although it might not be the random variable of the same scheme. We find
\begin{equation}
\begin{split}
	D^{\mathrm{seq,a}}_{3,\mathrm{imm}} &=  N_2 + N_3  ,\\
	D^{\mathrm{seq,b}}_{3,\mathrm{imm}} &=  2 N_2 + N_3, \\
	D^{\mathrm{seq,c}}_{3,\mathrm{imm}} &= 2 N_1 + N_3 .
\end{split}
\end{equation}

Since \(N_2\) and \(N_3\) are i.i.d., the probability generating function (PGF) of \(D^{\mathrm{seq,a}}_{3,\mathrm{imm}}\) is given by
\begin{equation}
	\tilde{G}^{\mathrm{seq,a}}_{3, \mathrm{imm}}(t)=g_{N_2}(t) \cdot g_{N_3}(t) = \left( \frac{pt}{1-qt} \right)^2.
\end{equation}
Due to the general relation 
\begin{equation}
    g_{2X}(t)=\mathbf{E}[t^{2X}]=\mathbf{E}[(t^2)^X]=g_X(t^2) 
\end{equation}	
valid for any discrete random variable \(X\), we have
\begin{align}
	\tilde{G}^{\mathrm{seq,b}}_{3,\mathrm{imm}}(t)= g_{N_2}(t^2) \cdot g_{N_3}(t) = \frac{p^2 t^3}{(1-qt)(1-qt^2)}.
\end{align}
The same holds true for the PGF of the immediate measurement scheme ``sequential c'', because 
\(N_1\) and \(N_2\) are i.i.d.. Thus, its PGF is also given by
\begin{equation}
	\tilde{G}^{\mathrm{seq,c}}_{3,\mathrm{imm}}(t)= \frac{p^2 t^3}{(1-qt)(1-qt^2)},
\end{equation}
which shows, that this scheme is actually equivalent to ``sequential b'' and will not be considered separately in the later comparison.

On the other hand, for non-immediate measurements we find the random variables to be 
\begin{equation}
\begin{split}
	D^{\mathrm{seq,a}}_{3,\mathrm{non}} &= 2 D^{\mathrm{seq,a}}_{3,\mathrm{imm}} = 2\left( N_2 + N_3 \right), \\
	D^{\mathrm{seq,b}}_{3,\mathrm{non}} &= 2 D^{\mathrm{seq,b}}_{3,\mathrm{imm}} = 2\left( 2 N_2 + N_3 \right), \\
	D^{\mathrm{seq,c}}_{3,\mathrm{non}} &= 2 D^{\mathrm{seq,a}}_{3,\mathrm{imm}} = 2\left( N_1 + N_3 \right).
\end{split}
\end{equation}

By using the same argument as before, we find the corresponding PGFs
\begin{equation}
\begin{split}
	G^{\mathrm{seq,a}}_{3,\mathrm{non}}(t) &= G^{\mathrm{seq,a}}_{3,\mathrm{imm}}(t^2) ,\\
	G^{\mathrm{seq,b}}_{3,\mathrm{non}}(t) &=  G^{\mathrm{seq,b}}_{3,\mathrm{imm}}(t^2) , \\
	G^{\mathrm{seq,c}}_{3,\mathrm{non}}(t) &= G^{\mathrm{seq,a}}_{3,\mathrm{imm}}(t^2).
\end{split}
\end{equation}
Again, the scheme ``sequential c'' is equivalent to another scheme, but now it is ``sequential a''. Therefore, the non-immediate version of ``sequential c'' will not be treated separately from ``sequential a''.

\subsection{Two segments simultaneously at the start}

\begin{figure}[ht]
	\subfloat[Two segments simultaneously at the start a]{\includegraphics[width=0.98\linewidth]{"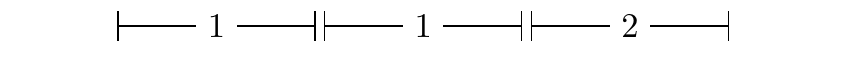"} \label{fig:two start a}} \\
	\subfloat[Two segments simultaneously at the start b]{\includegraphics[width=0.98\linewidth]{"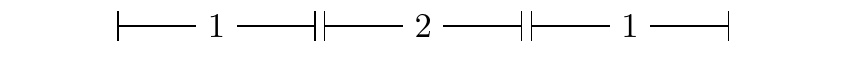"} \label{fig:two start b}}
	\caption{Possible arrangements of entanglement generation in a three-segment repeater, when two segments start simultaneously. The number in each segment corresponds to the moment when it starts.}
	\label{fig:Different schemes 3 segments sim start}
\end{figure}

When we generate entanglement in two segments simultaneously, we can do that by starting with these two segments or by finishing with these two. Here we will consider the case where one starts with them and we only have two different arrangements. However, we still have to distinguish between measuring immediately or not. 

For the first scheme in consideration, the middle and the left (or equivalently right) segment start generating entanglement at once. They swap as soon as both are done and then the last segment starts generating entanglement, see Fig.~\ref{fig:two start a}. Let us call this scheme ``start a''.
The dephasing random variables in this case are
\begin{equation}
\begin{split}
    D^{\mathrm{start,a}}_{3,\mathrm{imm}}&= 
    \begin{cases}
        N_2-N_1+N_3 & N_1 \leq N_2\\ 
        2\left(N_1-N_2\right)+N_3 & N_2 < N_1
    \end{cases}, \\
	D^{\mathrm{start,a}}_{3,\mathrm{non}}&=2|N_1-N_2|+2N_3.
\end{split}
\end{equation}
The PGF of \(D^{\mathrm{start,a}}_{3,\mathrm{non}}\) is obviously reads as
\begin{equation}
	\tilde{G}^{\mathrm{start,a}}_{3,\mathrm{non}}(t)=\tilde{G}_{2}(t^2) \cdot g_{N_3}(t^2) = \frac{p^3t^2(1+qt^2)}{(1-q^2)(1-qt^2)^2}.
\end{equation}
For immediate measurements use the methods presented in the previous section and derive the PGF of \(D^{\mathrm{start,a}}_{3,\mathrm{imm}}\)
\begin{equation}
	\tilde{G}^{\mathrm{start,a}}_{3,\mathrm{imm}}(t) = \frac{p^3 t (1-q^2 t^3)}{(1-q^2) (1-q t)^2 (1-q t^2)}.
\end{equation}

The second scheme is realised when we start with both the left and the right segment at once. As in the second sequential scheme there is no swapping possible, when both segments finished and one has to wait for the middle segment. We will call this scheme ``start b''. In pictures, it can be seen in Fig.~\ref{fig:two start b}. Here we have for the dephasing random variables
\begin{equation}
\begin{split}
	D^{\mathrm{start,b}}_{3,\mathrm{imm}} &= |N_1-N_3|+2N_2, \\
	D^{\mathrm{start,b}}_{3,\mathrm{non}} &= 2|N_1-N_3|+4N_2 = 2D^{\mathrm{start,b}}_{3,\mathrm{imm}}.
\end{split}
\end{equation}
We can simplify the calculation, by considering the immediate scheme first and using \(g_{2X}(t)=g_X(t^2)\). The PGF is given by
\begin{displaymath}
	\tilde{G}^{\mathrm{start,b}}_{3,\mathrm{imm}}(t) = \tilde{G}_{2}(t) \cdot g_{2N_3}(t) 
	= \frac{p^3 t^2 (1+qt)}{(1-q^2)(1-qt)(1-qt^2)}.
\end{displaymath}
Hence, the PGF of the non-immediate version is simply
\begin{align}
    \tilde{G}^{\mathrm{start,b}}_{3,\mathrm{non}}(t) = \tilde{G}^{\mathrm{start,b}}_{3,\mathrm{imm}}(t^2).
\end{align}

The waiting time is the same for both schemes in this subsection and amounts to
\begin{equation}
	K^{\mathrm{simult.}}_3=\max(N_1,N_2)+N_3,
\end{equation}
with an expectation value of
\begin{equation}
	\mathbf{E}[K^{\mathrm{simult.}}_3]=\frac{5-3p}{(2-p)p}.
\end{equation}

\subsection{Two segments simultaneously at the end}

\begin{figure}[ht]
	\subfloat[Two segments simultaneously at the end a]{\includegraphics[width=0.98\linewidth]{"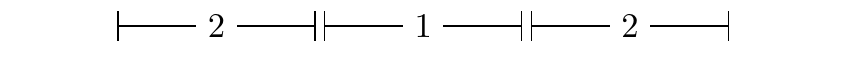"} \label{fig:two end a}} \\
	\subfloat[Two segments simultaneously at the end b]{\includegraphics[width=0.98\linewidth]{"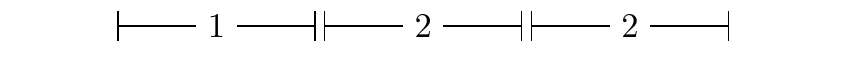"} \label{fig:two end b}}
	\caption{Possible arrangements of entanglement generation in a three-segment repeater, when only one segment starts and the rest finishes simultaneously. The number in each segment corresponds to the moment when it starts.}
	\label{fig:Different schemes 3 segments sim end}
\end{figure}

Finally, the last possible arrangement of two simultaneous segments is to start them in the last step. The waiting time stays the same as in the previous case, but again, there are two possibilities for the dephasing and two to perform measurements,i.e. immediate or non-immediate. The first scheme is realised, when we start with the segment in the middle and when it finishes, the left and right segment start generating entanglement simultaneously. We will call this scheme ``end a'' and it is shown schematically in Fig.~\ref{fig:two end a}. In this case the dephasing random variables are given by
\begin{equation}
\begin{split}
    D^{\mathrm{end,a}}_{3,\mathrm{imm}} &= N_1 + N_3 , \\
	D^{\mathrm{end,a}}_{3,\mathrm{non}} &=2\max(N_1,N_3), 
\end{split}
\end{equation}
with the PGFs
\begin{equation}
\begin{split}
    \tilde{G}^{\mathrm{end,a}}_{3,\mathrm{imm}}(t) &= \tilde{G}^{\mathrm{seq,a}}_{3, \mathrm{imm}}(t) = \left( \frac{pt}{1-qt} \right)^2, \\
	\tilde{G}^{\mathrm{end,a}}_{3,\mathrm{non}}(t) &= G^{\mathrm{par}}_n(t^2) = \frac{p^2 t^2 (1+qt^2)}{(1-qt^2)(1-q^2t^2)}.
\end{split}
\end{equation}

The second possibility is to start with the left or right segment and after it finished generate entanglement
simultaneously in the remaining segments. The schemes and random variables are equivalent independent whether one starts
with the left or right segment. We will call this scheme ``end b'' and its schematic representation, when starting with
the left segment, is shown in Fig.~\ref{fig:two end b}. Similarly to the scheme ``start a'', the dephasing random
variables depended on the order of successful entanglement generation.

Let us consider the scheme where we do not measure immediately as an example. First, assume that we started with the
left segment and it finished successfully after \(N_1\) attempts. Then both the middle and the right segment start
generating entanglement simultaneously. If the middle segments succeeds first after \(N_2\) attempts, we can swap
immediately and again have only one segment waiting. Eventually, the right segment will succeed after \(N_3\) attempts,
and we can also swap it. In total the dephasing will equal \(D^{\mathrm{end,b}}_{3,\mathrm{non}}=2N_3\), because
\(2N_2\) cancels out. This is the optimal case of this scheme.

Alternatively, it could also happen that the right segment finishes first, and we have two segments waiting for the
middle to succeed. In this case, we have \(D^{\mathrm{end,b}}_{3,\mathrm{non}}=4N_2-2N_3\). Hence, in total the
dephasing is
\begin{equation}
	D^{\mathrm{end,b}}_{3,\mathrm{non}}=
    \begin{cases}
		2N_3 & N_3 \geq N_2 \\
		4N_2-2N_3 & N_3 < N_2 
	\end{cases}.
\end{equation}
A similar consideration yields the dephasing random variable of the immediate measurement scheme to be \begin{equation}
	D^{\mathrm{end,b}}_{3,\mathrm{imm}}=
    \begin{cases}
		N_3 & N_3 \geq N_2 \\
		2N_2-N_3 & N_3 < N_2 
	\end{cases}.
\end{equation}
As mentioned a few times so far, we can exploit that \(g_{2X}(t)=g_X(t^2)\), and thus we calculate the PGF of the immediate scheme first, which reads as
\begin{equation}
	\tilde{G}^{\mathrm{end,b}}_{3,\mathrm{imm}}(t) = \frac{p^2 t \left(1-q^2 t^3\right)}{(1-qt) \left(1- q^2 t\right) \left(1-q t^2\right)}.
\end{equation}
Therefore, the PGF of of \(D^{\mathrm{end,b}}_{3,\mathrm{non}}\) is given by
\begin{equation}
    \tilde{G}^{\mathrm{end,b}}_{3,\mathrm{non}}(t) = \tilde{G}^{\mathrm{end,b}}_{3,\mathrm{imm}}(t^2),
\end{equation}
and we have covered all possibles schemes of this subsection.

\subsection{Overlapping schemes}
\begin{figure}[ht]
	\subfloat[Overlapping a]{\includegraphics[width=0.98\linewidth]{"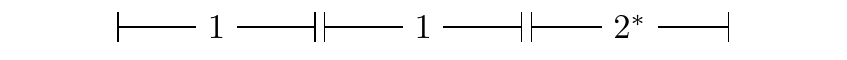"} \label{fig:Overlapping a}} \\
	\subfloat[Overlapping b]{\includegraphics[width=0.98\linewidth]{"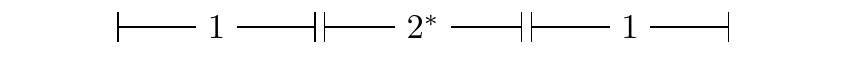"} \label{fig:Overlapping b}}
	\caption{Possible arrangements of entanglement generation in a three-segment repeater, when two segments start simultaneously and the remaining segment starts as soon as one is successful. The number in each segment corresponds to the moment when it starts and the star indicates that this segment starts as soon as one of the others finished.}
	\label{fig:Different overlapping schemes}
\end{figure}

Before, considering fully parallel schemes, we turn our attention to a mixture of the previous simultaneous schemes. We will call the schemes of this section overlapping schemes. The procedure is as follows, we start generating entanglement in two segments simultaneously and as soon as one of the two segments finishes, we start with the remaining one as well. Thus, the two processes of entanglement generation are overlapping, explaining the naming. In Fig.~\ref{fig:Different overlapping schemes} a schematic version of the overlapping schemes can be seen.

There are two different possible arrangements presented in Fig.~\ref{fig:Overlapping a} and Fig.~\ref{fig:Overlapping
b}. In the former one the left (or equivalently the right) and the middle segment start from the beginning. This scheme
will be called ``overlapping, a''. The latter scheme starts with both outer segments and will be called ``overlapping,
b".

For the scheme ``overlapping, a'' we find with immediate measurements the dephasing random variable to be
\begin{equation}
    D^{\mathrm{over,a}}_{3,\mathrm{imm}}= 
    \begin{cases} 
    N_3 &\Omega_1 \\ 
    2(N_2-N_1)-N_3 & \Omega_2 \\ 
    N_1-N_2+N_3 &\Omega_3 \\ 
    \end{cases}
\end{equation}
where we have chosen the partition $\Omega = \mathbb{N}^3 = \Omega_1 \sqcup \Omega_2 \sqcup \Omega_3$ given by the following inequalities:
\begin{equation}
\begin{split}
    \Omega_1 &= N_1 \leq N_2, N_2-N_1 \leq N_3, \\
    \Omega_2 &= N_1 < N_2, N_2-N_1 > N_3, \\
    \Omega_3 &= N_2< N_1.
\end{split}
\end{equation}
The dephasing varies depending on the order in which the segments finish, since one cannot swap or measure depending on
which segment is done first. Thus, we have three different cases. One can calculate the full PGF of the dephasing in a
similar way to the previous schemes and finds
\begin{align}
    \tilde{G}^{\mathrm{over,a}}_{3,\mathrm{imm}}(t) = \frac{p^3 t (1 + q - 2 q^2 t - q t^2 + q^4 t^4)}
    { (1 - q^2) (1 -q t)^2 (1 - q^2 t) (1\! -q t^2)}.
\end{align}

For the non-immediate version of the scheme ``overlapping, a'', we do not have to take the measurements into account,
but this still does not result in more symmetries simplifying the expression. Hence, one has to consider all possible
orders separately and we find the dephasing to be
\begin{equation}
    D^{\mathrm{over,a}}_{3,\mathrm{non}}= 
	\begin{cases} 
    2N_3 & \Omega_1 \\  
    2\left(2\left(N_2-N_1\right)-N_3\right) & \Omega_2 \\
    2N_3 & \Omega_3 \\ 
    2\left(N_1-N_2\right) & \Omega_4 \\
    \end{cases}
\end{equation}
where the partition in this case is given by
\begin{equation}
\begin{split}
    \Omega_1 &= N_1 \leq N_2, N_2-N_1 \leq N_3, \\
    \Omega_2 &= N_1 < N_2, N_2-N_1 > N_3, \\
    \Omega_3 &= N_2< N_1, N_1-N_2 \leq N_3, \\
    \Omega_4 &= N_2< N_1, N_1-N_2 > N_3.
\end{split}
\end{equation}
The resulting PGF reads as
\begin{displaymath}
    \tilde{G}^{\mathrm{over,a}}_{3,\mathrm{non}}(t) = \frac{p^3 t^2 (1 +2 q - q (1 + q) t^4 - q^3 t^6)}
    {(1 - q^2) (1 - q t^2) (1 -  q^2 t^2) (1 -  q t^4)}.
\end{displaymath}

The other overlapping scheme possesses more symmetry, thus we find more compact expressions for the random variables. It mainly depends on the relative difference of steps between the outer segments. We find for the immediate and non-immediate scheme
\begin{equation}
\begin{split}
    D^{\mathrm{over,b}}_{3,\mathrm{imm}} &= 
    \begin{cases}
		2N_2 - \abs{N_1 - N_3} & \abs{N_1-N_3} < N_2 \\
		\abs{N_1-N_3} & \abs{N_1-N_3} \geq N_2
	\end{cases}, \\
	D^{\mathrm{over,b}}_{3,\mathrm{non}} &= 
	\begin{cases}
		4N_2 - 2\abs{N_1- N_3} & \abs{N_1-N_3} < N_2 \\
		2\abs{N_1\!-\!N_3} \! & \abs{N_1-N_3} \geq N_2
	\end{cases}.
\end{split}
\end{equation}
By case analysis we derive the PGFs
\begin{equation}
\begin{split}
    \tilde{G}^{\mathrm{over,b}}_{3,\mathrm{imm}}(t) &= \frac{p^3 t (t + q (2 - t^2 (1 + q + q^2 t)))}{(1 - q^2) (1 - q t) (1 - q^2 t) (1 - q t^2)}, \\
	\tilde{G}^{\mathrm{over,b}}_{3,\mathrm{non}}(t) &= \tilde{G}^{\mathrm{over,b}}_{3,\mathrm{imm}}(t^2).
\end{split}
\end{equation}

Finally, the only missing piece is the waiting time of the overlapping schemes and its expectation value. The random variable of the waiting time is
\begin{equation}
    K^{\mathrm{over}}_3 = \min(N_1,N_2) + \max(|N_1-N_2|,N_3).
\end{equation}
Its expectation value is found to be
\begin{equation}
    E[K^{\mathrm{over}}_3] = \frac{8 - 3p \left(3 - p \right)}{p\left(2 - p\right)^2}.
\end{equation}

\subsection{Parallel schemes}
Here we only consider the potentially optimal scheme, since all parallel schemes posses the same raw rate, but differ in dephasing. In the optimal scheme the dephasing is minimized, such that it has the best secret key rate of all schemes of this class. 

\begin{table}[ht]
\begin{tabular}{|l|l|l|}
	\hline
	\hfil Domain & \hfil $D^\star_{3, \mathrm{non}}$ & \hfil $D^\star_{3, \mathrm{imm}}$ \\ 
	\hline
	$N_1 \leqslant N_2 \leqslant N_3$ & $2(N_3 - N_1)$        & $N_3 - N_1$ \\
	$N_1 \leqslant N_3 < N_2$         & $2(2N_2 - N_1 - N_3)$ & $2N_2 - N_3 - N_1$ \\
	$N_2 < N_1 \leqslant N_3$         & $2(N_3 - N_2)$        & $N_1 + N_3 - 2N_2$ \\
	$N_2 \leqslant N_3 < N_1$         & $2(N_1 - N_2)$        & $N_1 + N_3 - 2N_2$ \\
	$N_3 < N_1 \leqslant N_2$         & $2(2N_2 - N_1 - N_3)$ & $2N_2 - N_3 - N_1$ \\
	$N_3 < N_2 < N_1$                 & $2(N_1 - N_3)$        & $N_1 - N_3$ \\
	\hline
\end{tabular}
\caption{The values of $D^\star_{3, \mathrm{non}}$ and $D^\star_{3, \mathrm{imm}}$ on the domains of the partition.}\label{tbl:D3nm}
\end{table}

The waiting time is \(K^{\mathrm{par}}_3=\max(N_1,N_2,N_3)\) and following \eqref{eq:Knpar} or Appendix~\ref{app:GKn} its expectation value is
\begin{equation}
	E[K^{\mathrm{par}}_3]=\frac{1 + q \left(4 + 3 q \left(1 + q\right)\right)}{1 + q - q^3 - q^4}.
\end{equation}
The dephasing PGF can be computed with our partitioning approach. The six domains and the values of the dephasing variables in these domains are given in Table~\ref{tbl:D3nm}. The final result reads as
\begin{displaymath}
\begin{split}
	\tilde{G}^\star_{3,\mathrm{non}}(t) &= \frac{p^3}{1-q^3} \frac{1+q(1+2q)t^2-q^2(2+q)t^6-q^4t^8}{(1-qt^2)(1-q^2t^2)(1-qt^4)} \\
	\tilde{G}^\star_{3,\mathrm{imm}}(t) &= \frac{p^3}{1 - q^3} \frac{1+q^2t-2q^3t^2-2q^2t^3+q^3t^4+q^5t^5}{(1-qt)^2(1-q^2t)(1-qt^2)}
\end{split}
\end{displaymath}

\subsection{Comparisons}
Finally, as we have calculated all necessary statistical quantities we are able to compare the previously discussed schemes. Again as a remark, we only considered schemes here, which swap as soon as possible, as delaying the entanglement swapping increases the dephasing, which in turn decreases the SKR.

\begin{figure*}[ht]
	\centering
	\subfloat[a][$\tau_{\mathrm{coh}}={\unit[0.1]{s}}$, $p_{\mathrm{link}}=0.05$, $\mu = \mu_0=0.97$]{\includegraphics[width=0.33\linewidth]{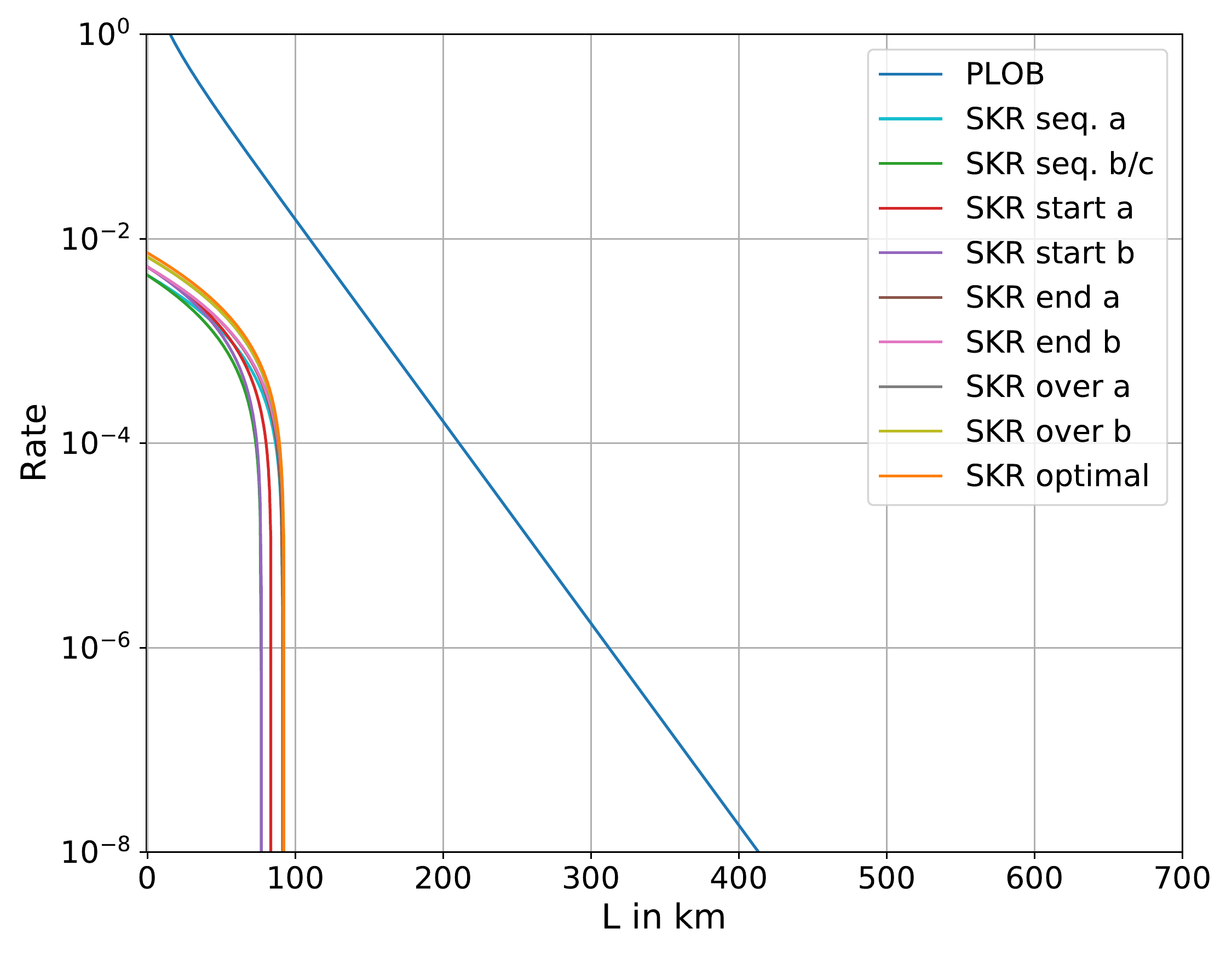}}
	\subfloat[][$\tau_{\mathrm{coh}}={\unit[0.1]{s}}$, $p_{\mathrm{link}}=0.05$, $\mu = \mu_0=1$]{\includegraphics[width=0.33\linewidth]{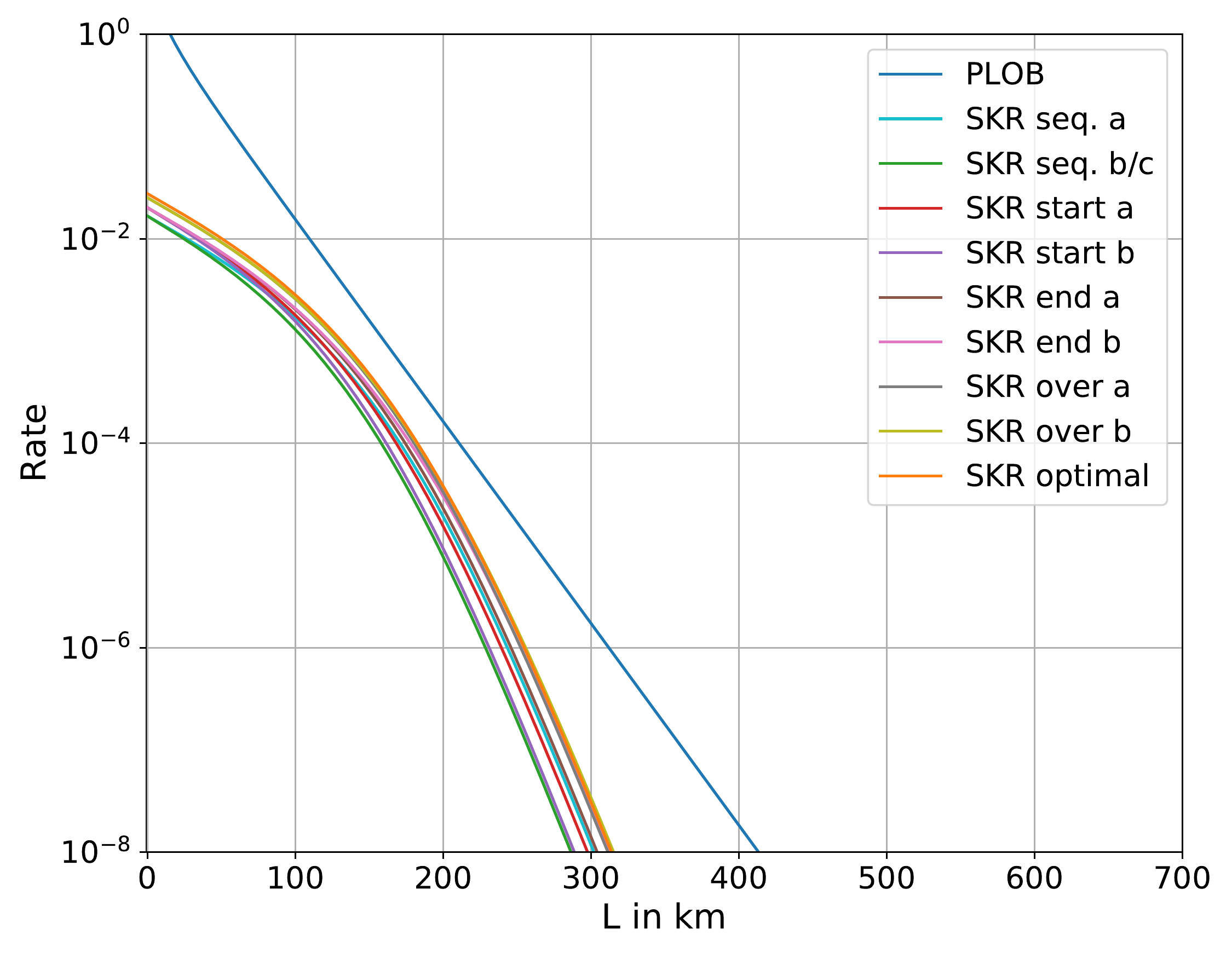}}
	\subfloat[][$\tau_{\mathrm{coh}}={\unit[0.1]{s}}$, $p_{\mathrm{link}}=0.7$, $\mu = \mu_0=0.97$]{\includegraphics[width=0.33\linewidth]{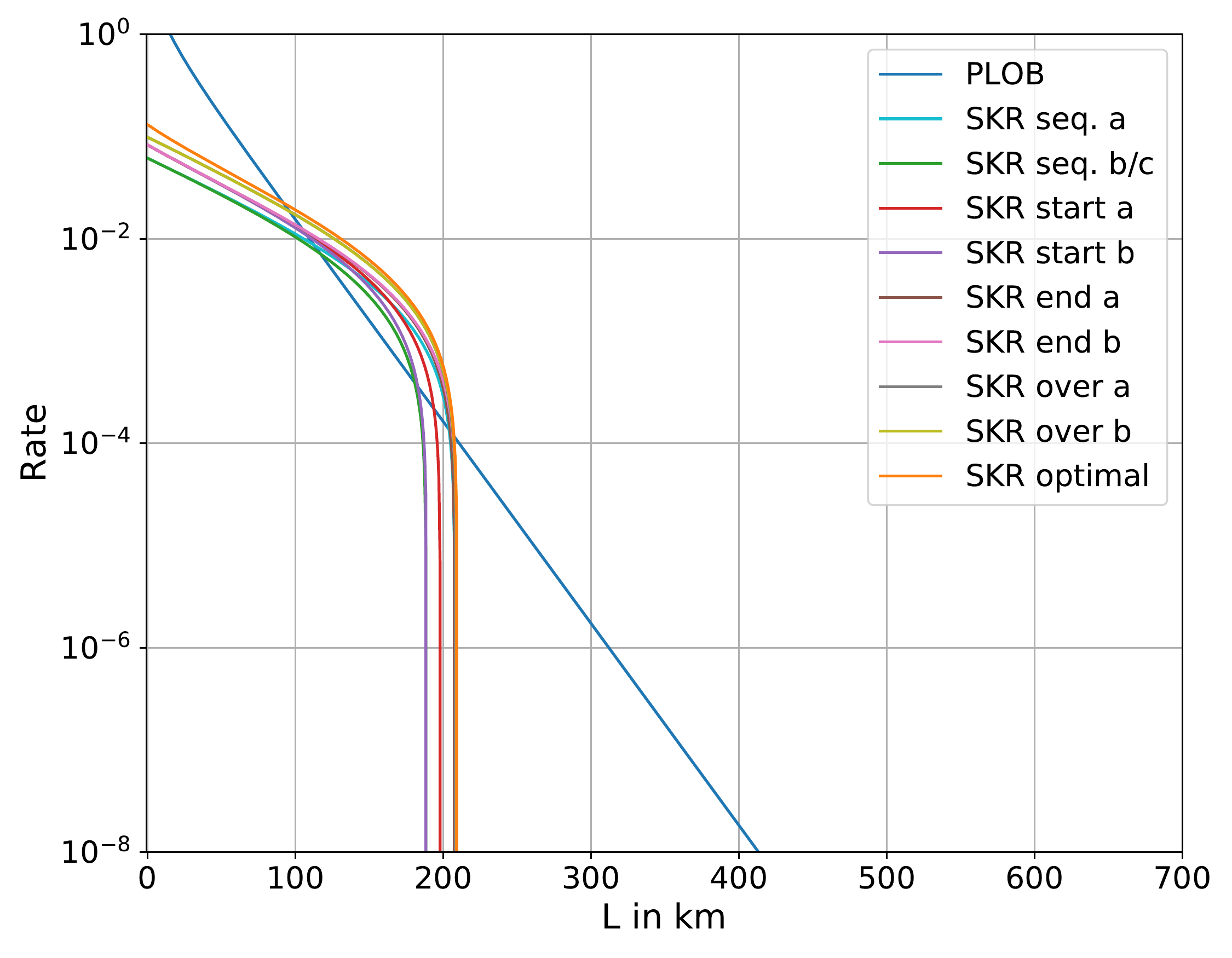}} \\
	\subfloat[][$\tau_{\mathrm{coh}}={\unit[0.1]{s}}$, $p_{\mathrm{link}}=0.7$, $\mu = \mu_0=1$]{\includegraphics[width=0.33\linewidth]{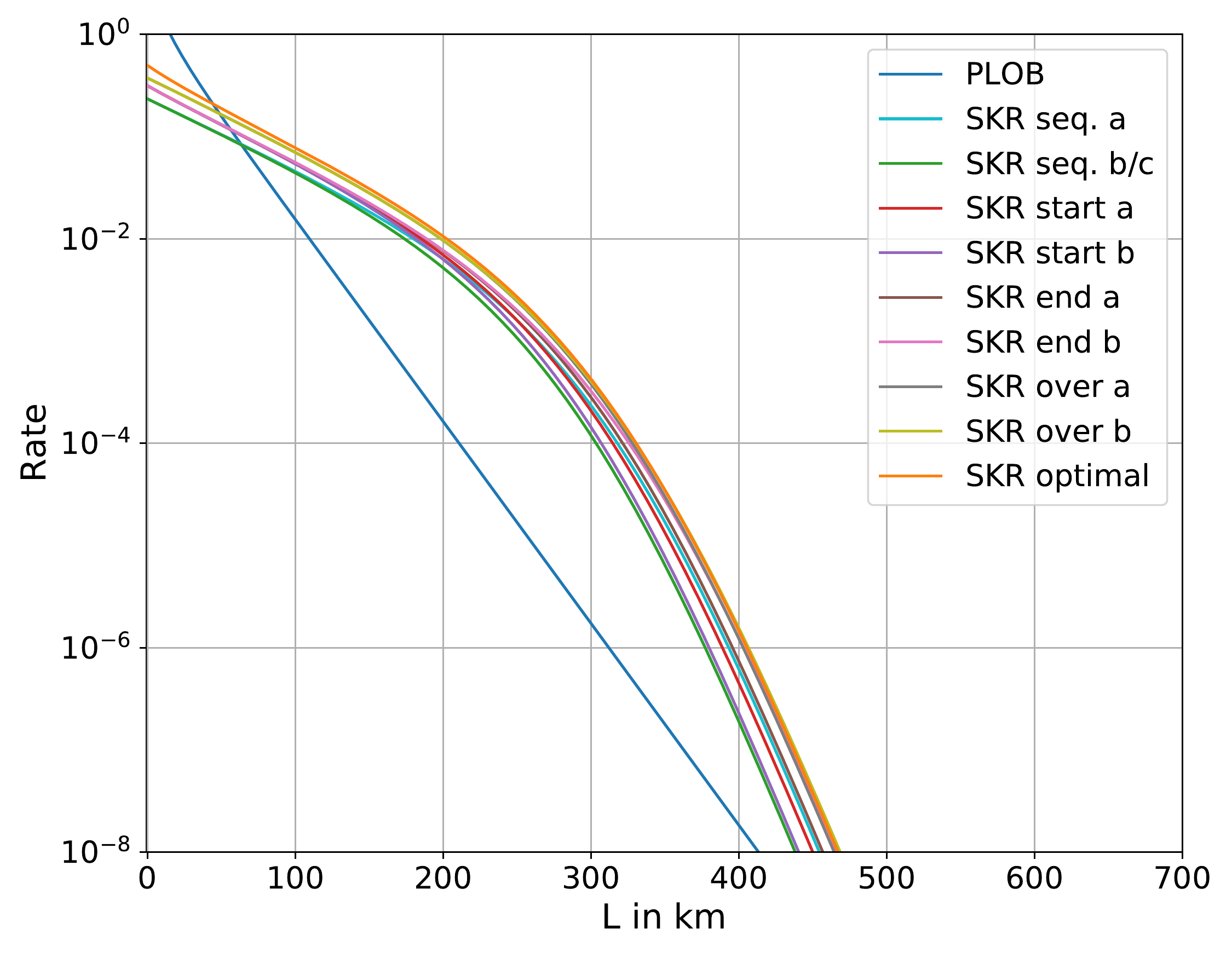}}
	\subfloat[][$\tau_{\mathrm{coh}}={\unit[0.1]{s}}$, $p_{\mathrm{link}}= \mu = \mu_0=1$]{\includegraphics[width=0.33\linewidth]{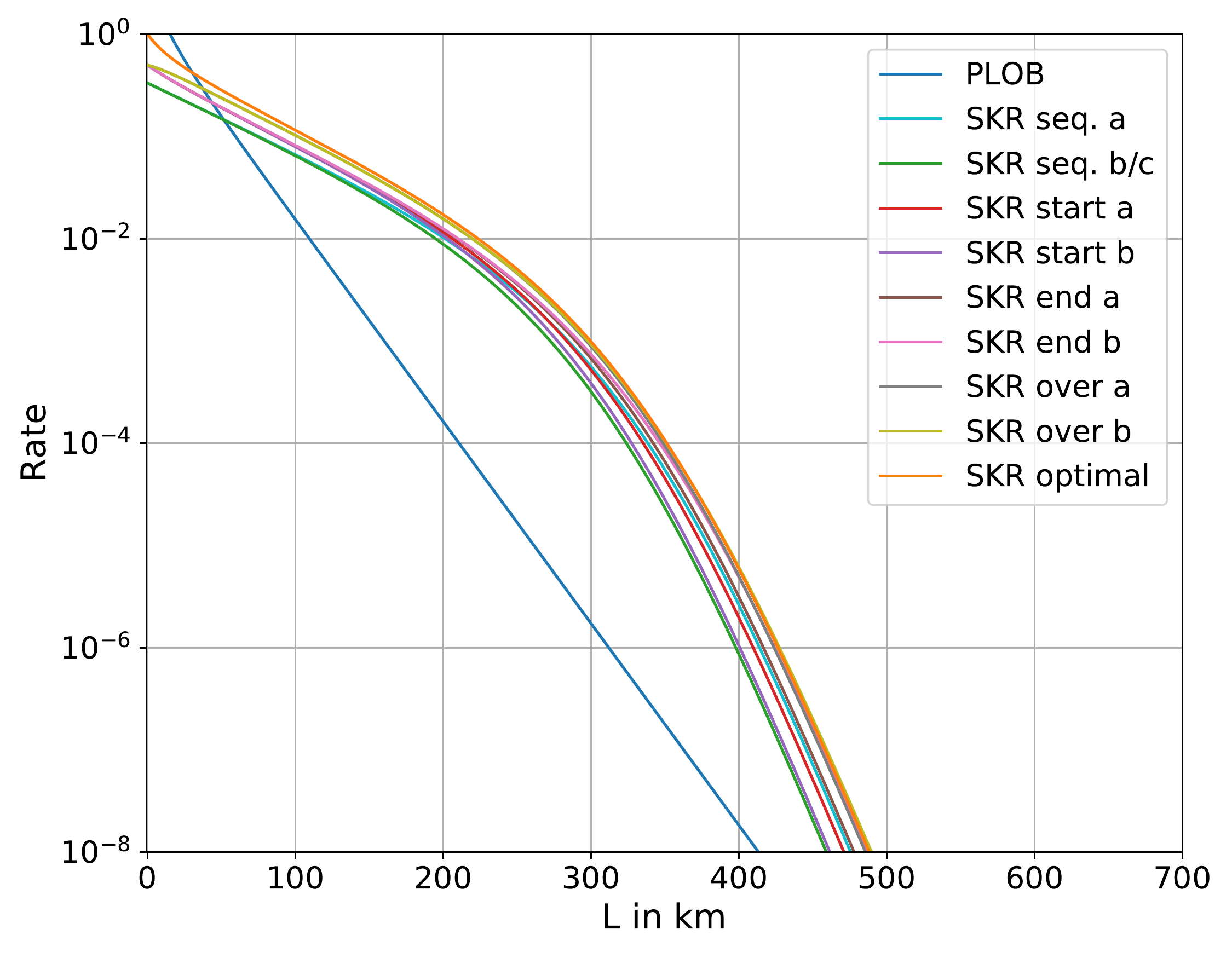}}
	\caption{Comparison of secret key rates of three-segment repeaters performing \emph{immediate} measurements for a total distance \(L\) and different experimental parameters. For all figures, a coherence time of $\tau_{\mathrm{coh}}={\unit[0.1]{s}}$ has been used.}
	\label{fig:Comparison_3_segments_immediate tau=0.1}
\end{figure*}

\begin{figure*}
    \centering
    \subfloat[][$\tau_{\mathrm{coh}}={\unit[10]{s}}$, $p_{\mathrm{link}}=0.05$, $\mu = \mu_0=0.97$]{\includegraphics[width=0.33\linewidth]{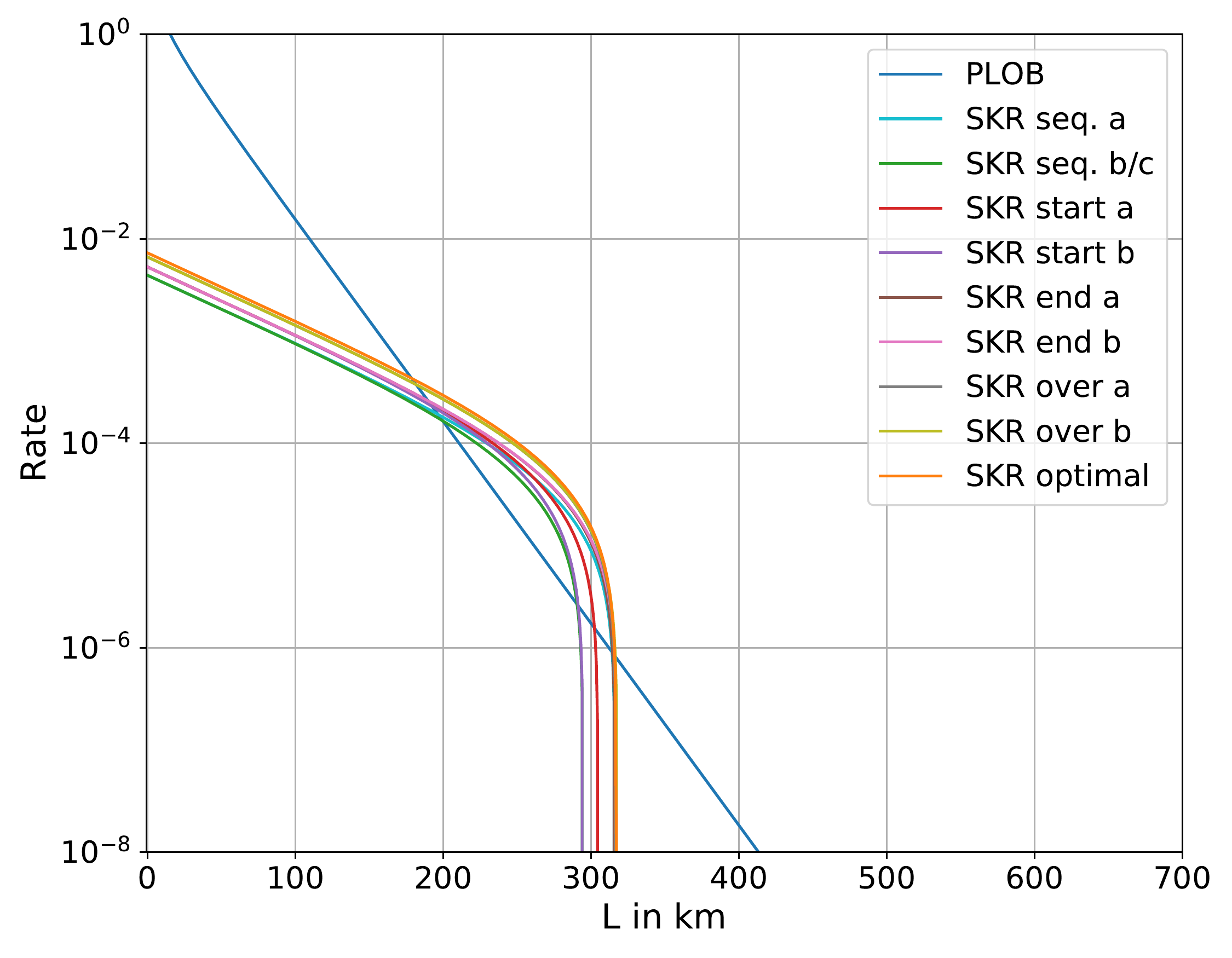}}
	\subfloat[][$\tau_{\mathrm{coh}}={\unit[10]{s}}$, $p_{\mathrm{link}}=0.05$, $\mu = \mu_0=1$]{\includegraphics[width=0.33\linewidth]{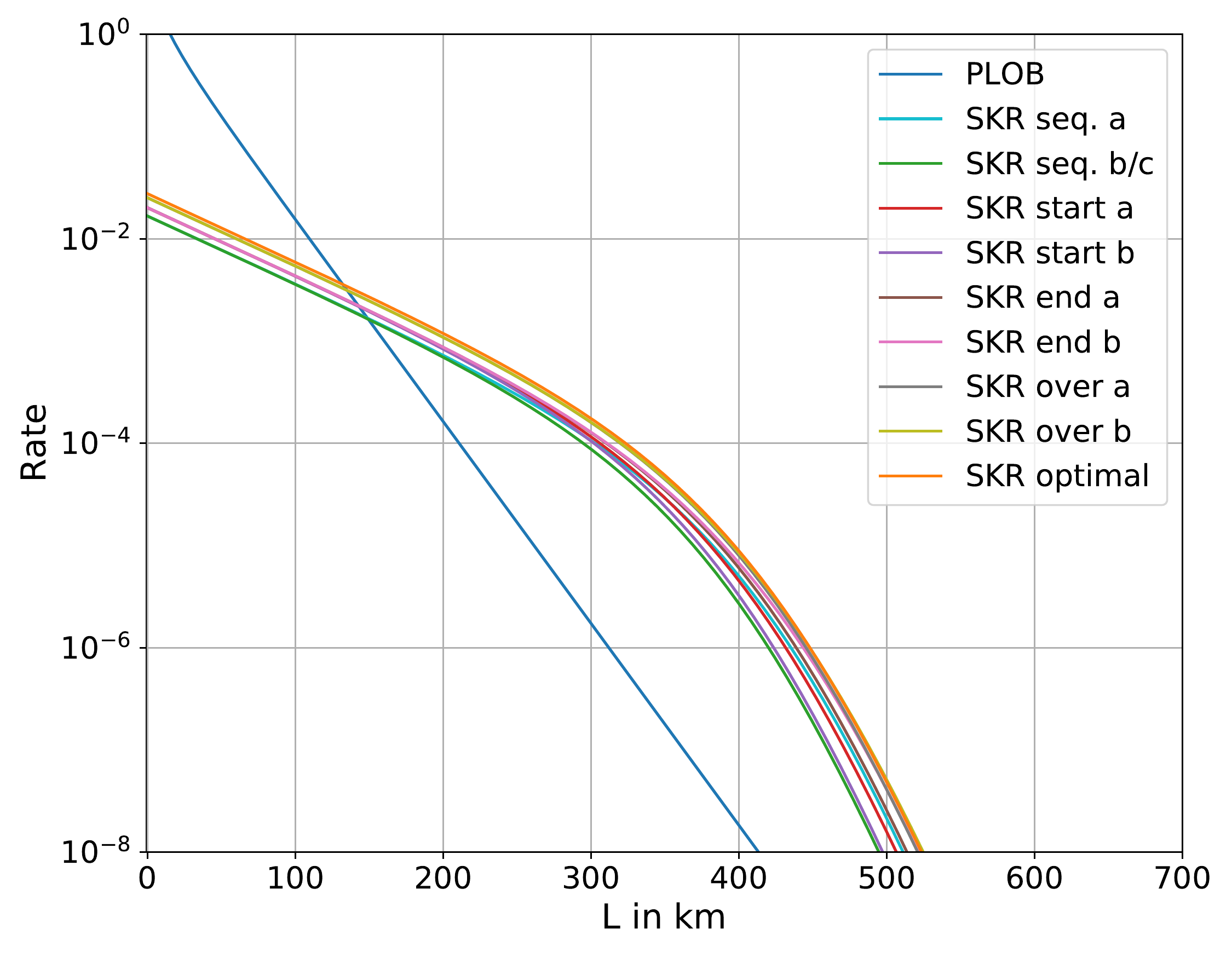}}
	\subfloat[][$\tau_{\mathrm{coh}}={\unit[10]{s}}$, $p_{\mathrm{link}}=0.7$, $\mu = \mu_0=0.97$]{\includegraphics[width=0.33\linewidth]{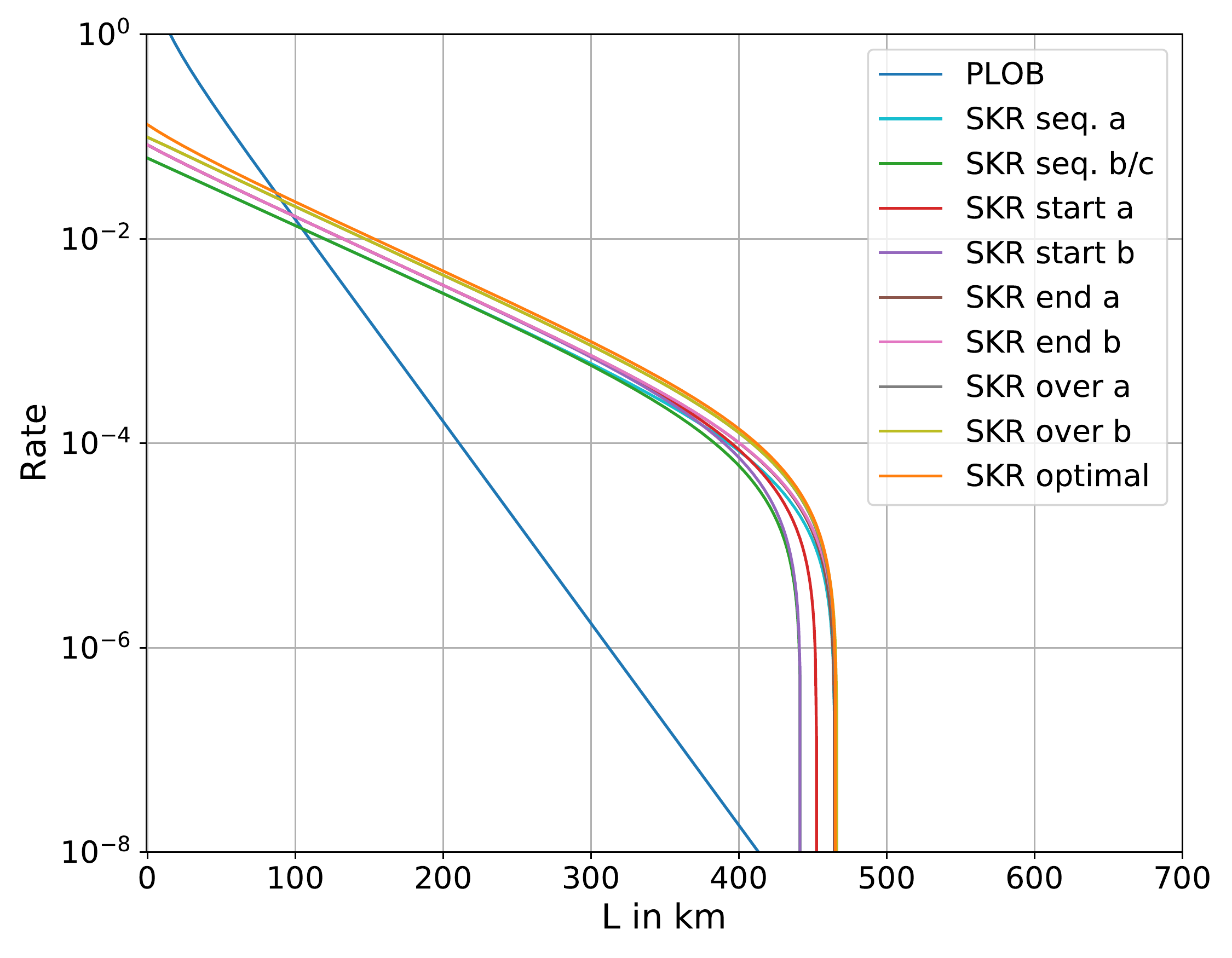}} \\
	\subfloat[][$\tau_{\mathrm{coh}}={\unit[10]{s}}$, $p_{\mathrm{link}}=0.7$, $\mu = \mu_0=1$]{\includegraphics[width=0.33\linewidth]{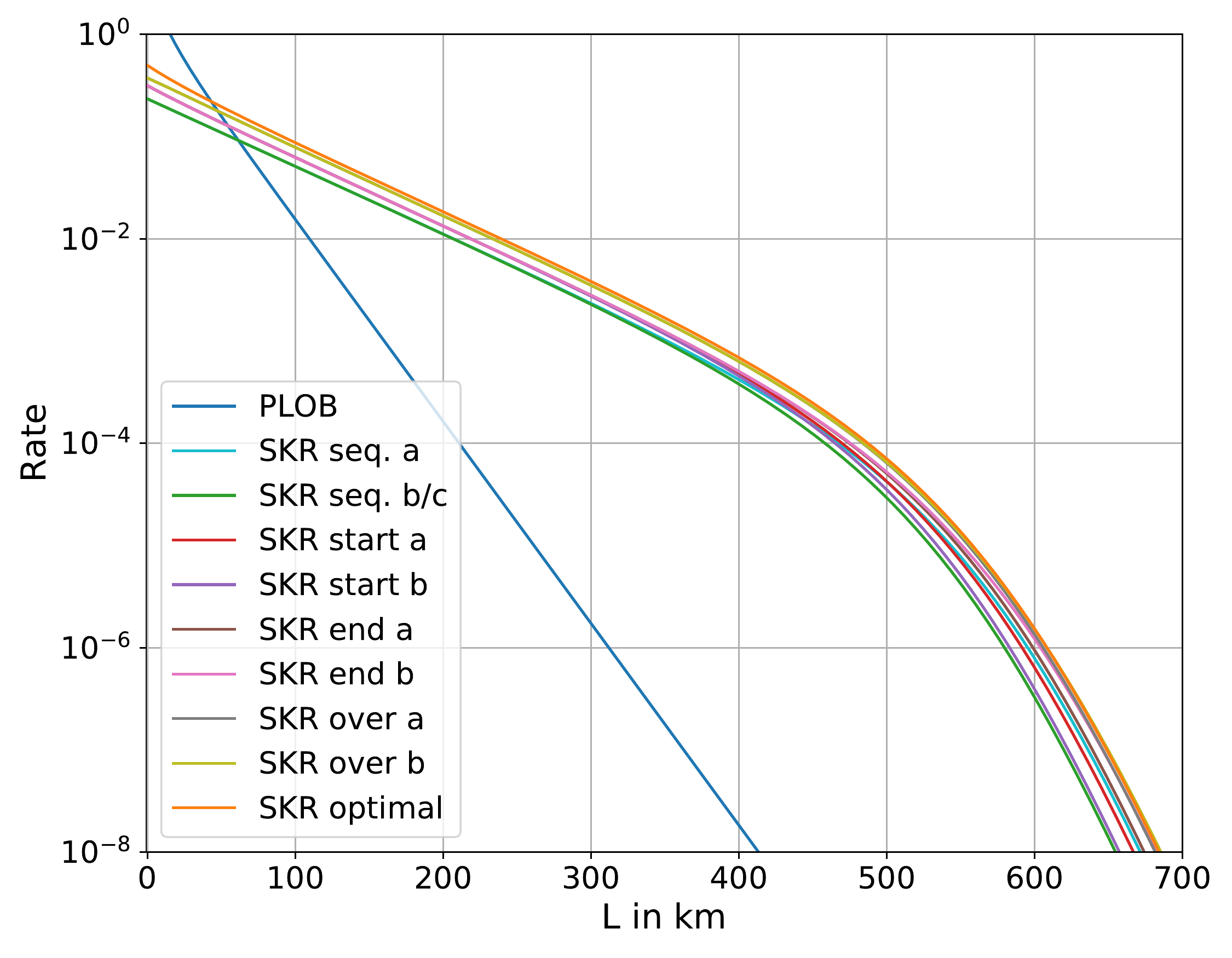}}
	\subfloat[][$\tau_{\mathrm{coh}}={\unit[10]{s}}$,  $p_{\mathrm{link}}= \mu = \mu_0=1$]{\includegraphics[width=0.33\linewidth]{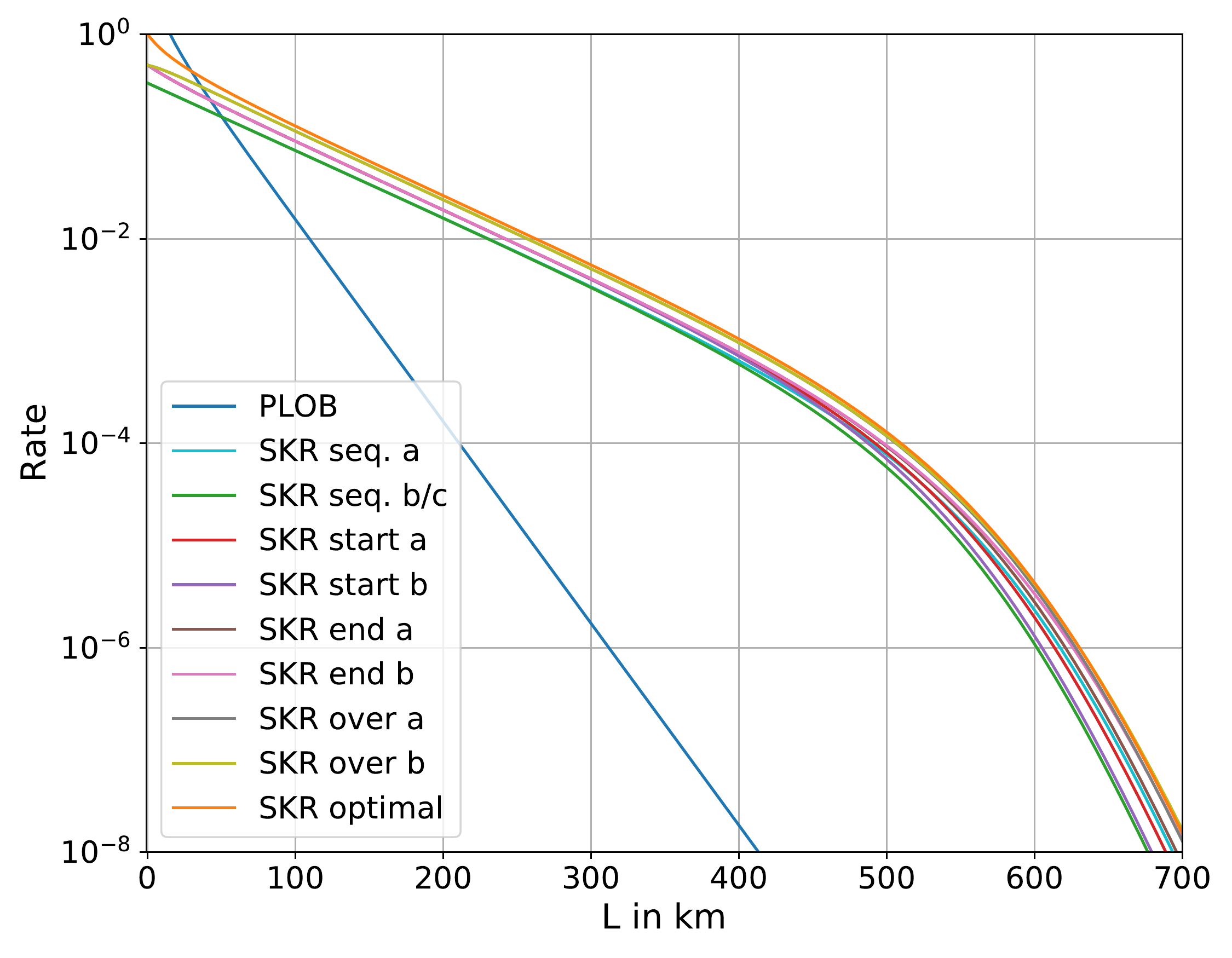}}
	\caption{Comparison of secret key rates of three-segment repeaters performing \emph{immediate} measurements for a total distance \(L\) and different experimental parameters. For all figures, a coherence time of $\tau_{\mathrm{coh}}={\unit[10]{s}}$ has been used.}
	\label{fig:Comparison_3_segments_immediate tau=10}
\end{figure*}

First, we consider the immediate measurement schemes. In Fig.~\ref{fig:Comparison_3_segments_immediate tau=0.1} ($\tau_{\mathrm{coh}}= \unit[0.1]{s}$) and Fig.~\ref{fig:Comparison_3_segments_immediate tau=10} ($\tau_{\mathrm{coh}}= \unit[10]{s}$), one can see a comparison of all immediate measurement schemes for a three-segment repeater using the previously discussed schemes. In both figures the SKR of the ``optimal'' scheme is represented in orange. As mentioned earlier, the scheme ``seq, c'' is equivalent to ``seq, b'' in this setting and thus not considered separately. For both coherence times the optimal schemes outperforms all other schemes. Especially for shorter distances, the optimal scheme performs clearly better than others. Only for longer distances, where the rate of any three-segment repeater drops, the schemes ``over, b'', ``over, a'' and ``end, b'' catch up, but do not surpass it. Typically, one would not use this regime of a repeater, as the rates are too low. Additionally, in the limit of increasing hardware resources, i.e. \( p_{\mathrm{link}} \rightarrow 1 ,\; \mu \rightarrow 1 , \; \mu_0 \rightarrow 1 \), the optimal scheme keeps performing the best. Thus, we conclude that the immediate measurement version of the optimal scheme is truly optimal for \(n \leq 3\).

\begin{figure*}[ht]
	\centering
	\subfloat[a][$\tau_{\mathrm{coh}}={\unit[0.1]{s}}$, $p_{\mathrm{link}}=0.05$, $\mu = \mu_0=0.97$]{\includegraphics[width=0.33\linewidth]{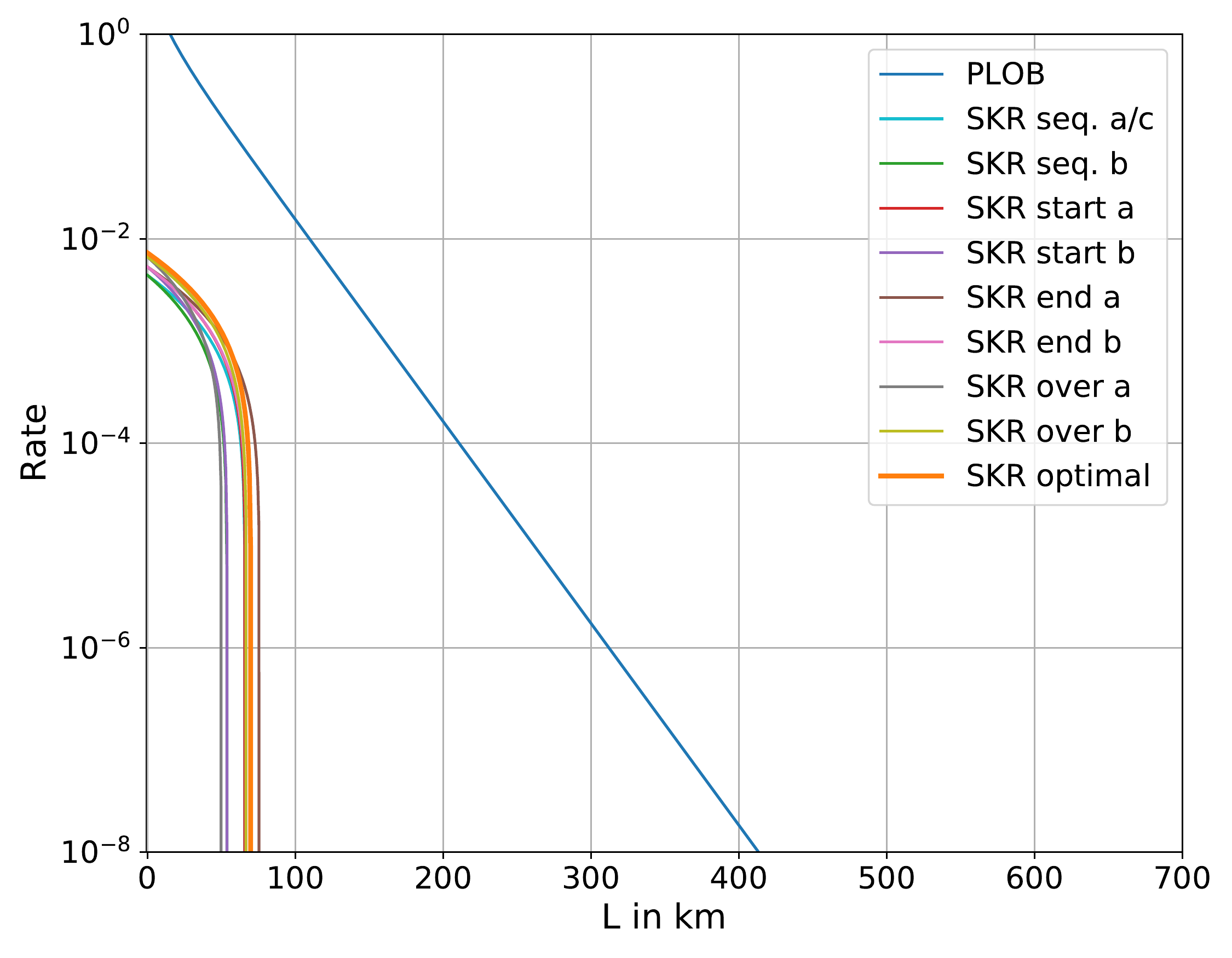}}
	\subfloat[][$\tau_{\mathrm{coh}}={\unit[0.1]{s}}$, $p_{\mathrm{link}}=0.05$, $\mu = \mu_0=1$]{\includegraphics[width=0.33\linewidth]{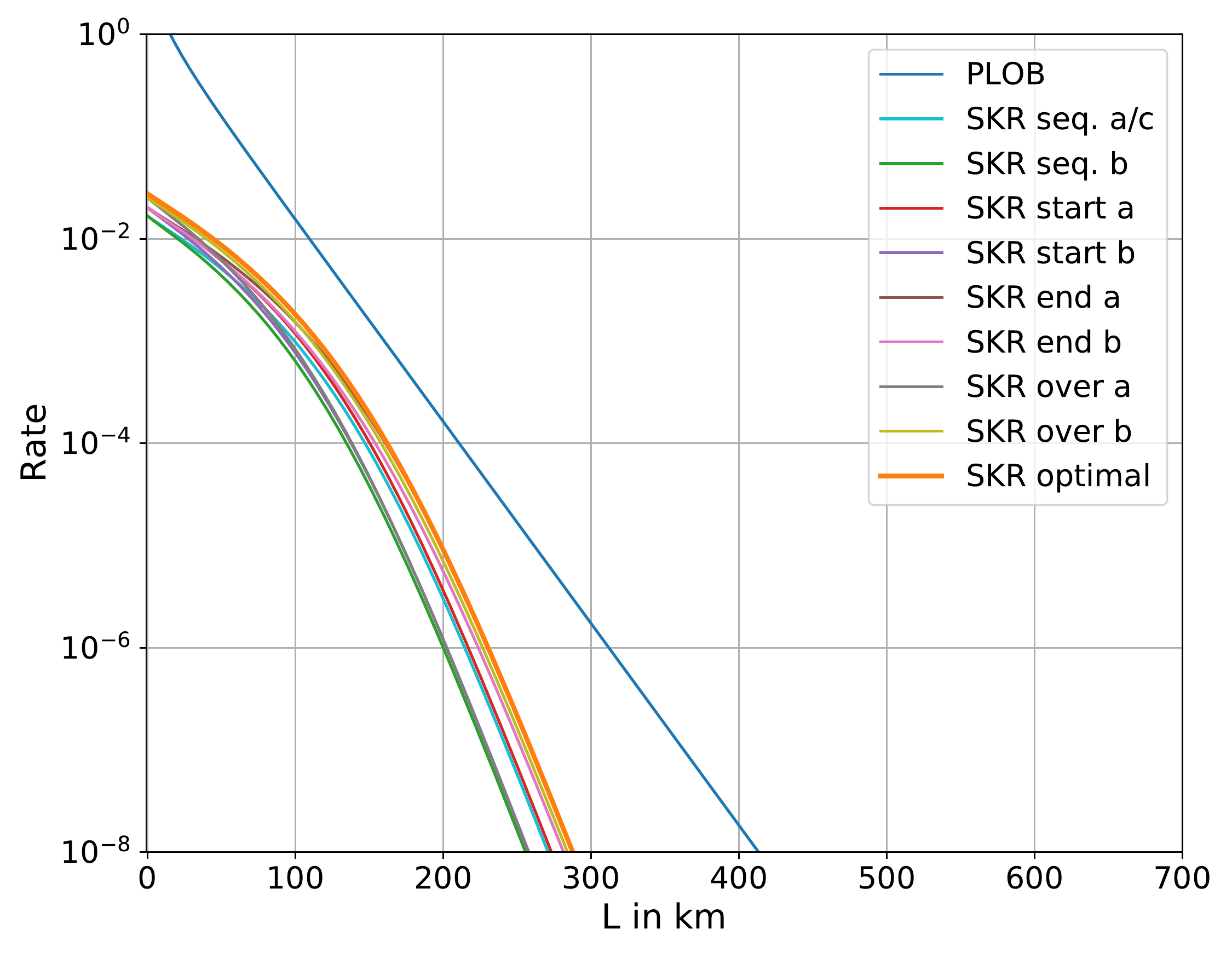}}
	\subfloat[][$\tau_{\mathrm{coh}}={\unit[0.1]{s}}$, $p_{\mathrm{link}}=0.7$, $\mu = \mu_0=0.97$]{\includegraphics[width=0.33\linewidth]{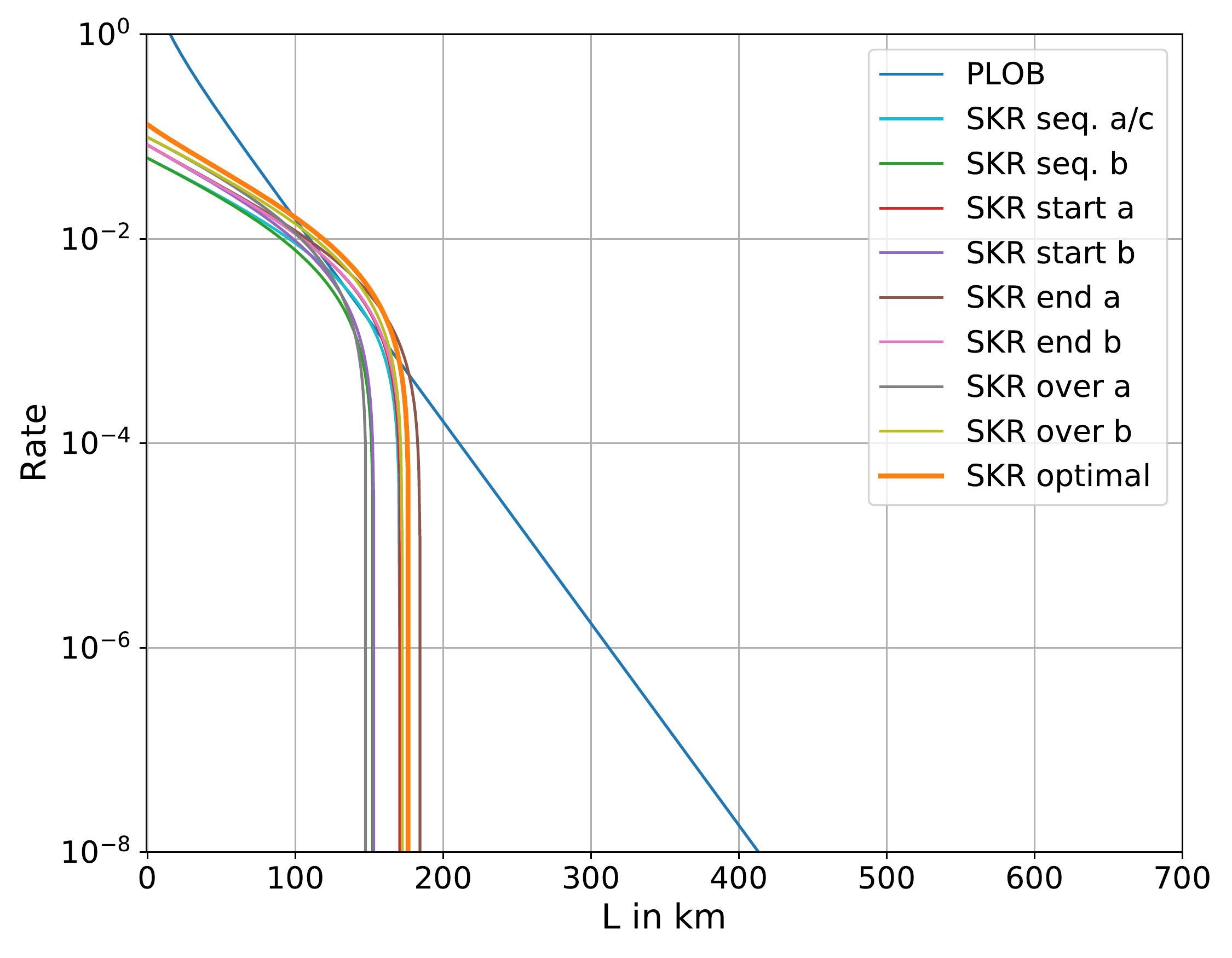}} \\
	\subfloat[][$\tau_{\mathrm{coh}}={\unit[0.1]{s}}$, $p_{\mathrm{link}}=0.7$, $\mu = \mu_0=1$]{\includegraphics[width=0.33\linewidth]{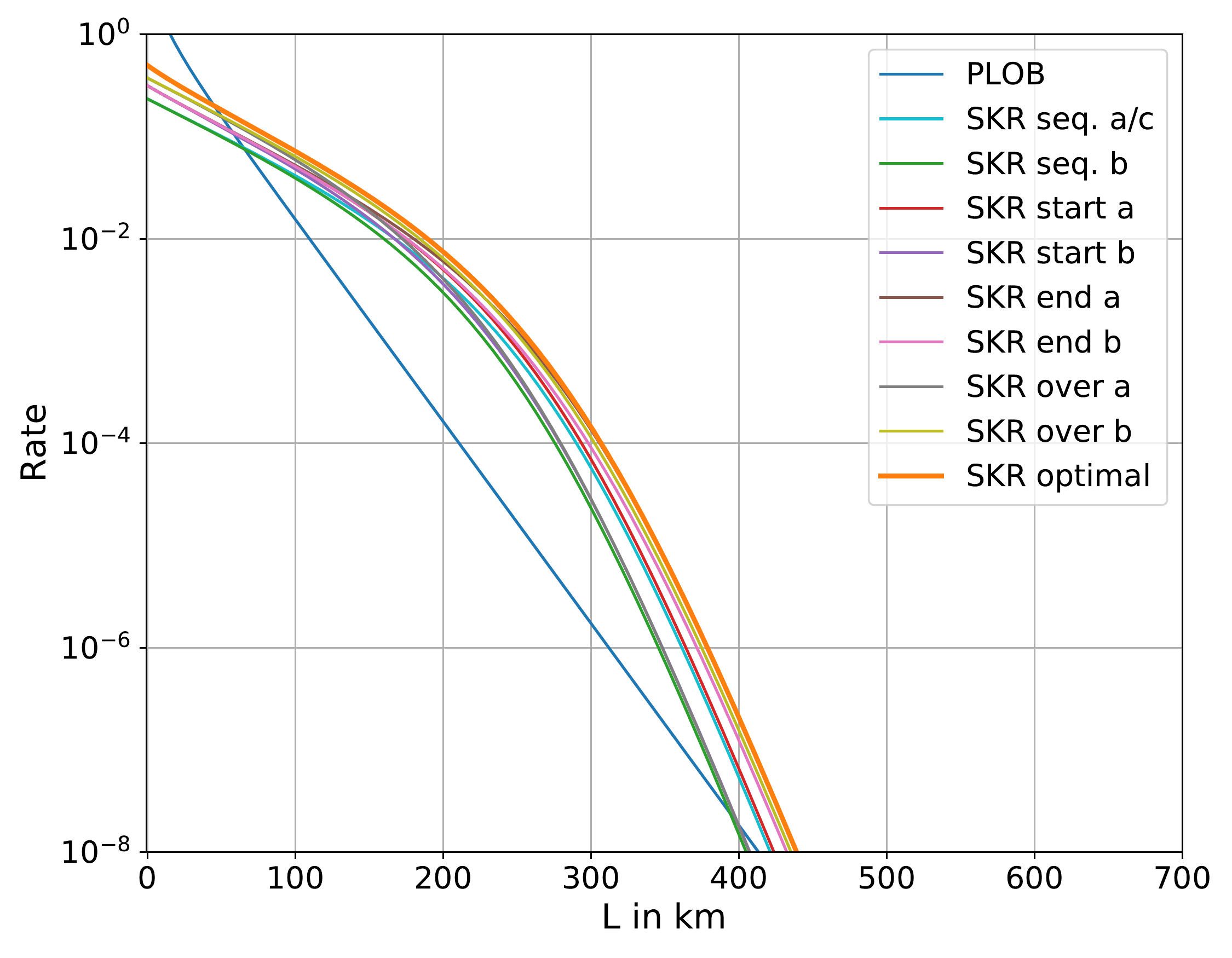}}
	\subfloat[][$\tau_{\mathrm{coh}}={\unit[0.1]{s}}$, $p_{\mathrm{link}}= \mu = \mu_0=1$]{\includegraphics[width=0.33\linewidth]{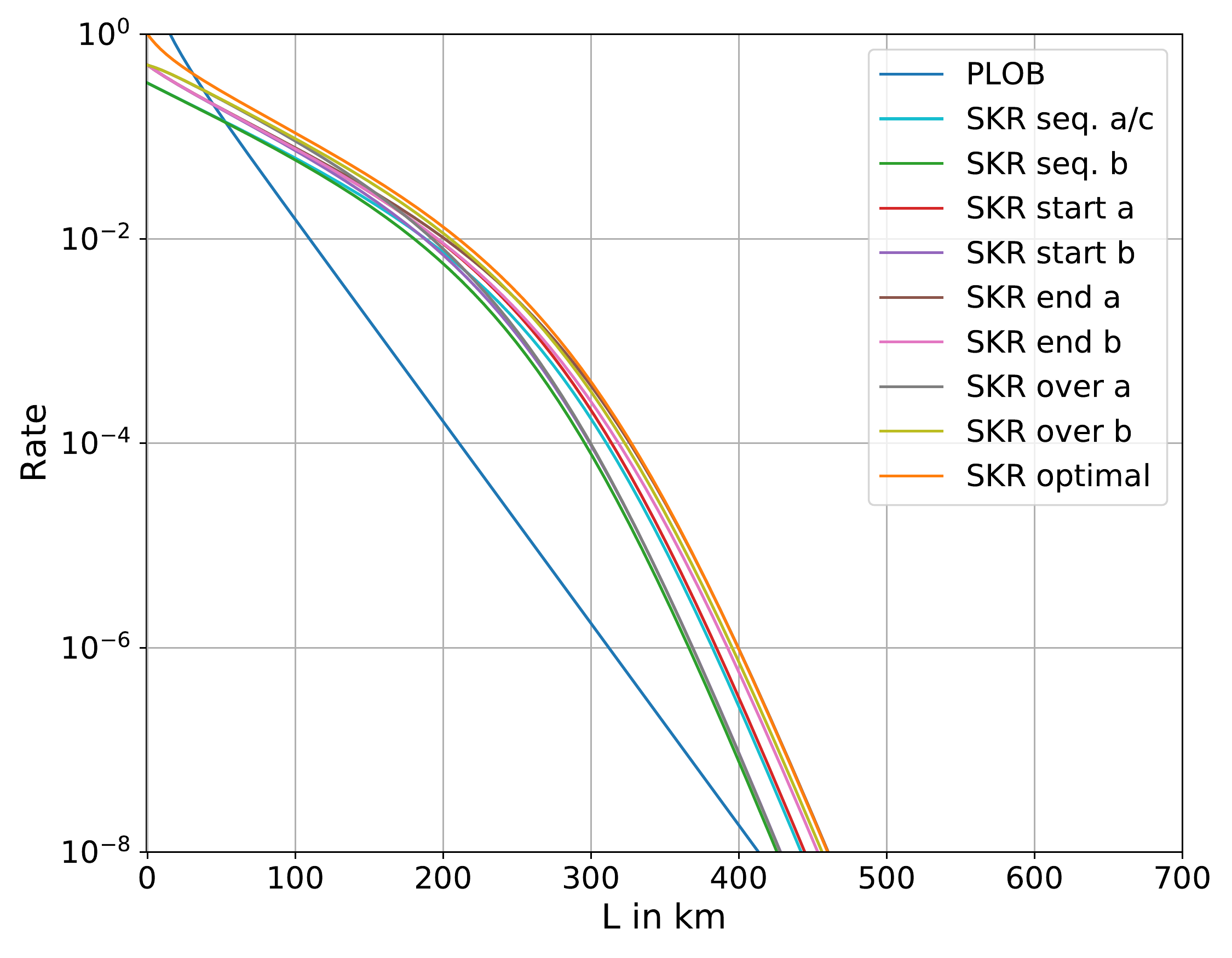}}
	\caption{Comparison of secret key rates of three-segment repeaters performing \emph{non-immediate} measurements for a total distance \(L\) and different experimental parameters. A coherence time of $\tau_{\mathrm{coh}}={\unit[0.1]{s}}$ has been used throughout.}
	\label{fig:Comparison_3_segments_non tau=0.1}
\end{figure*}

\begin{figure*}
    \centering
    \subfloat[a][$\tau_{\mathrm{coh}}={\unit[10]{s}}$, $p_{\mathrm{link}}=0.05$, $\mu = \mu_0=0.97$]{\includegraphics[width=0.33\linewidth]{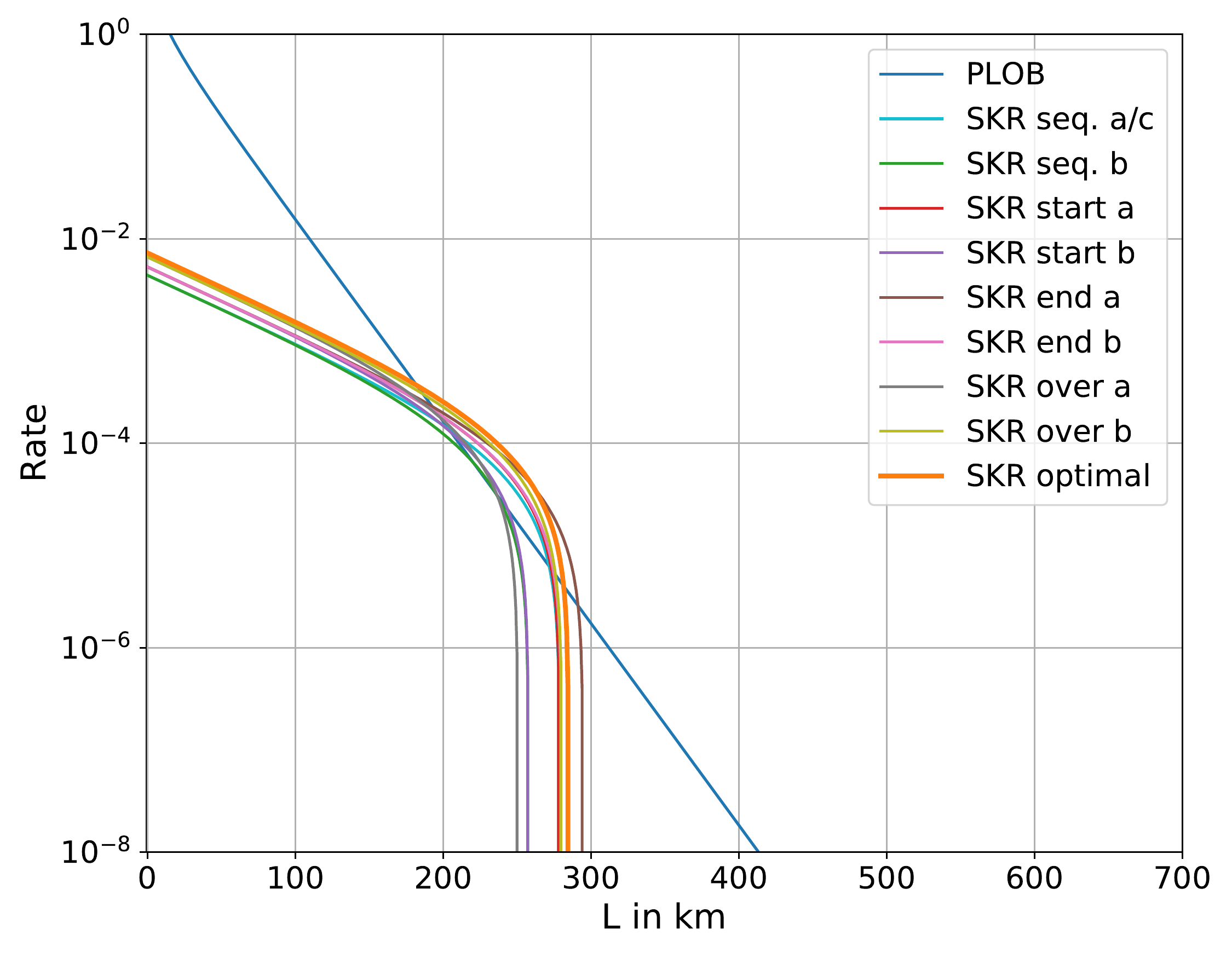}}
	\subfloat[][$\tau_{\mathrm{coh}}={\unit[10]{s}}$, $p_{\mathrm{link}}=0.05$, $\mu = \mu_0=1$]{\includegraphics[width=0.33\linewidth]{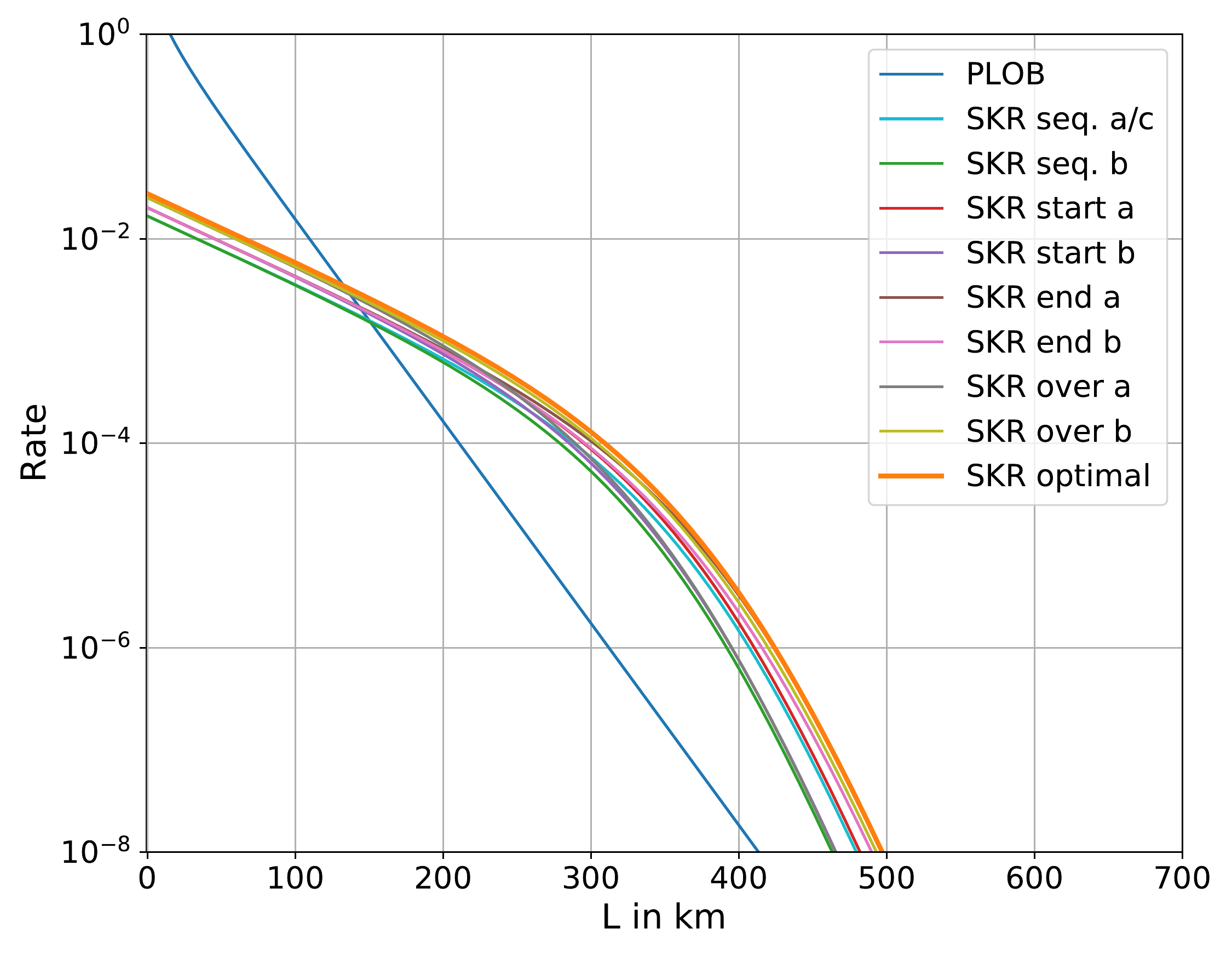}}
	\subfloat[][$\tau_{\mathrm{coh}}={\unit[10]{s}}$, $p_{\mathrm{link}}=0.7$, $\mu = \mu_0=0.97$]{\includegraphics[width=0.33\linewidth]{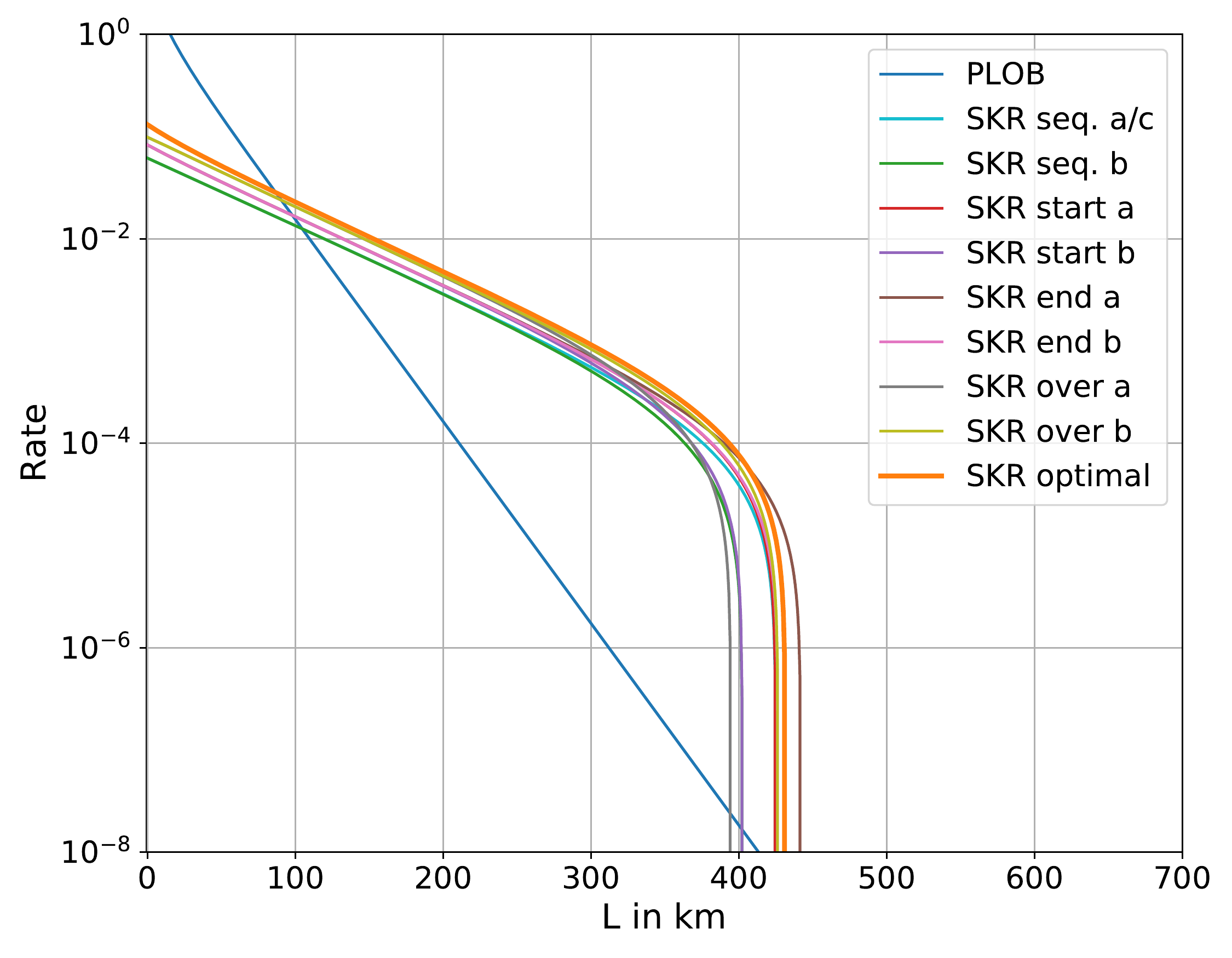}}\\
	\subfloat[][$\tau_{\mathrm{coh}}={\unit[10]{s}}$, $p_{\mathrm{link}}=0.7$, $\mu = \mu_0=1$]{\includegraphics[width=0.33\linewidth]{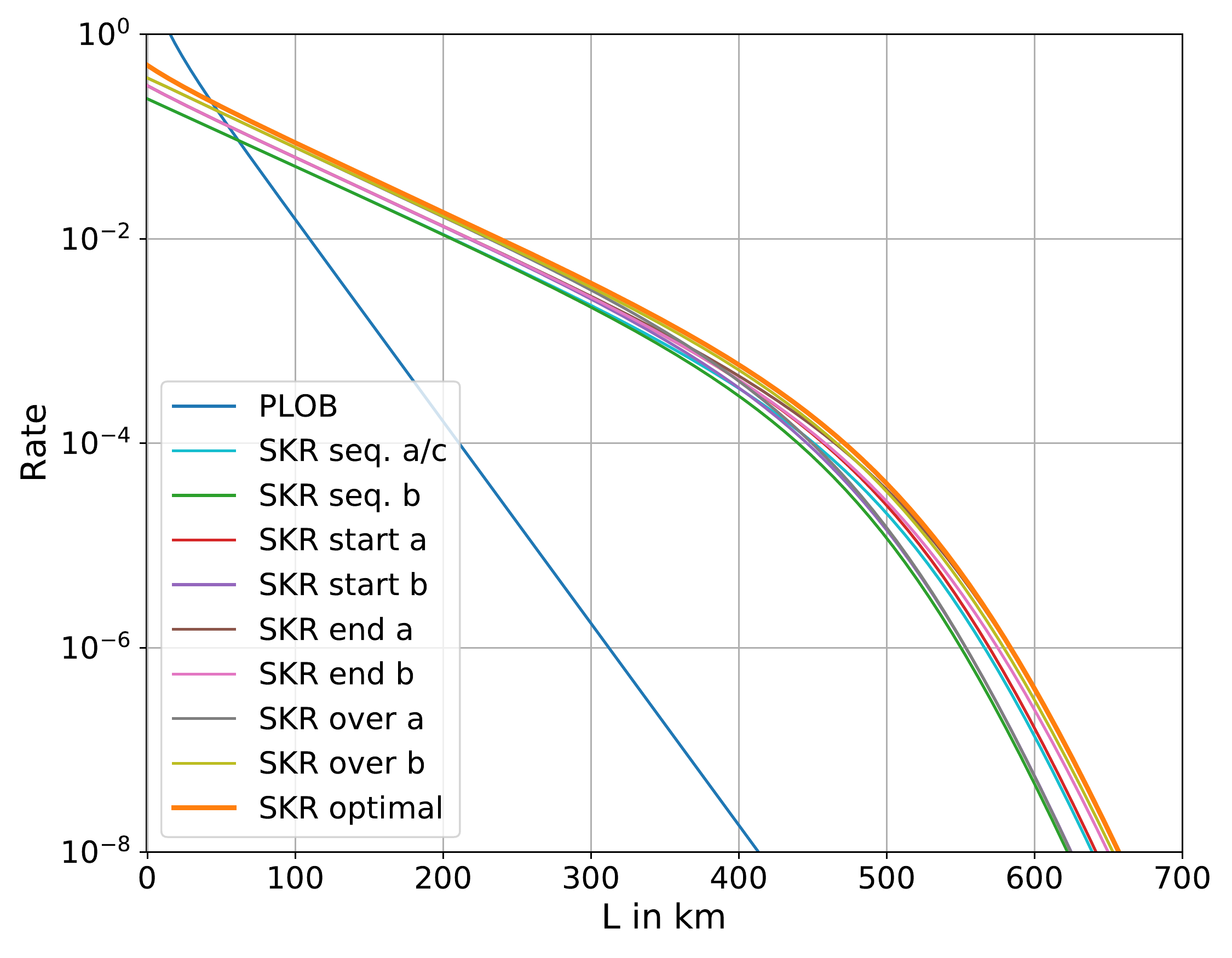}}
	\subfloat[][$\tau_{\mathrm{coh}}={\unit[10]{s}}$,  $p_{\mathrm{link}}= \mu = \mu_0=1$]{\includegraphics[width=0.33\linewidth]{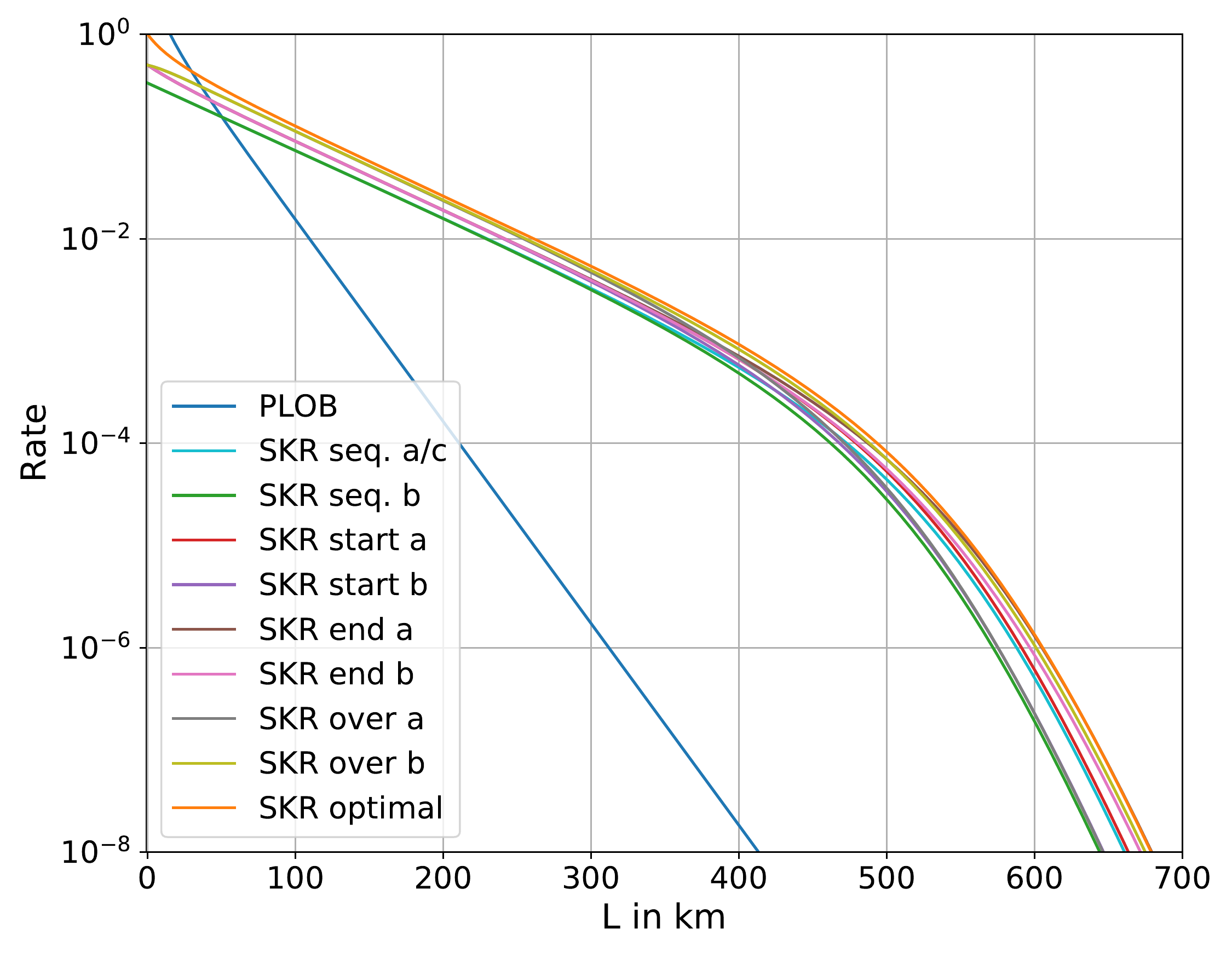}}
	\caption{Comparison of secret key rates of three-segment repeaters performing \emph{non-immediate} measurements for a total distance \(L\) and different experimental parameters. A coherence time of $\tau_{\mathrm{coh}}={\unit[10]{s}}$ has been used throughout.}
	\label{fig:Comparison_3_segments_non tau=10}
\end{figure*}

Next, in Fig.~\ref{fig:Comparison_3_segments_non tau=0.1} ($\tau_{\mathrm{coh}}= \unit[0.1]{s}$) and Fig.~\ref{fig:Comparison_3_segments_non tau=10} ($\tau_{\mathrm{coh}}= \unit[10]{s}$) one can see the same comparison of different swapping schemes using non-immediate measurements. Again, the ``optimal" scheme is presented in orange. This time the sequential schemes ``seq, a'' and ``seq, c'' are equivalent and thus are not considered separately.
As one can see, the optimal scheme outperforms all other schemes in the ideal case when \(\mu=\mu_0=1\) for all choices of \(\tau_{\mathrm{coh}}\) and \(p_{\mathrm{link}}\). Furthermore, it also provides the highest secret key rate in the non-ideal case until close to the drop-off. The scheme ``end a'' surpasses it only at those distances either close to or after both start declining dramatically, thus increasing the achievable distance. As discussed before, one typically would not use the regime of an repeater. However, if the main goal is to achieve the longest achievable distance possible, then the scheme ``end a'' performs the best.

In the end, the optimal scheme provides the best secret key rate under most realistic use scenarios. Moreover, it is truly optimal in the limit of increasing hardware parameters,  i.e. \( p_{\mathrm{link}} \rightarrow 1 ,\; \mu \rightarrow 1 , \; \mu_0 \rightarrow 1 \). Thus, it will be beneficial to use the ``optimal'' scheme as technology progresses and the hardware resources increase. Hence, our conclusion for non-immediate schemes is that the ``optimal" scheme is optimal under improving hardware parameters for \(n \leq 3\).

We conjecture that the same is true for both immediate and non-immediate measurement schemes for all \(n\geq 3\)-segment repeaters. This should be investigated in future research.

\section{Comparison of ``optimal" with fully sequential and Alice immediately measuring (n=8)}\label{app:8segmentsimmediate}

The fully sequential scheme, in which repeater segments are sequentially filled with entangled pairs from, for example, left to right is the overall slowest scheme leading to the smallest raw rates. However, a potential benefit is that parallel qubit storage can be almost entirely avoided. More specifically, when the first segment on the left is filled and waiting for the second segment to be filled too, the first segment waits for a random number of $N_2$ steps, whereas the second segment always only waits for one constant dephasing unit
(for each distribution attempt in the second segment). Thus, omitting the constant dephasing in each segment, the accumulated time-dependent random dephasing of the fully sequential scheme has only contributions from a single memory pair subject to memory dephasing at any elementary time step.
On average, this gives a total dephasing of $(n-1)/p$ which is the sum of the average waiting time in one segment for segments 2 through $n$, as discussed
in detail in the main text.

\begin{figure*}[ht]
	\centering
	\subfloat[a][$\tau_{\mathrm{coh}}={\unit[0.1]{s}}$, $p_{\mathrm{link}}=0.05$, $\mu = \mu_0=0.99$]{\includegraphics[width=0.33\linewidth]{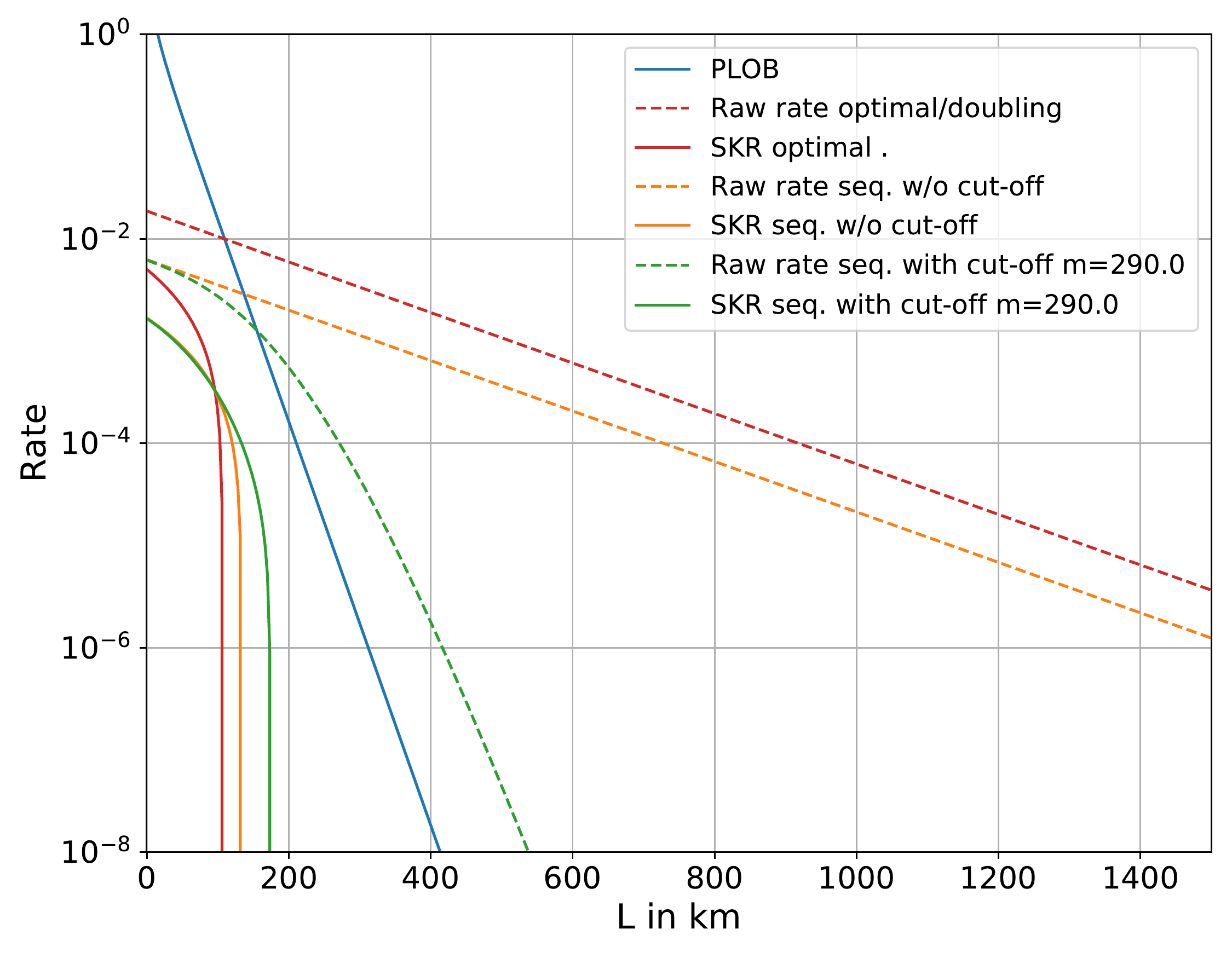}}
	\subfloat[][$\tau_{\mathrm{coh}}={\unit[0.1]{s}}$, $p_{\mathrm{link}}=0.05$, $\mu = \mu_0=1$]{\includegraphics[width=0.33\linewidth]{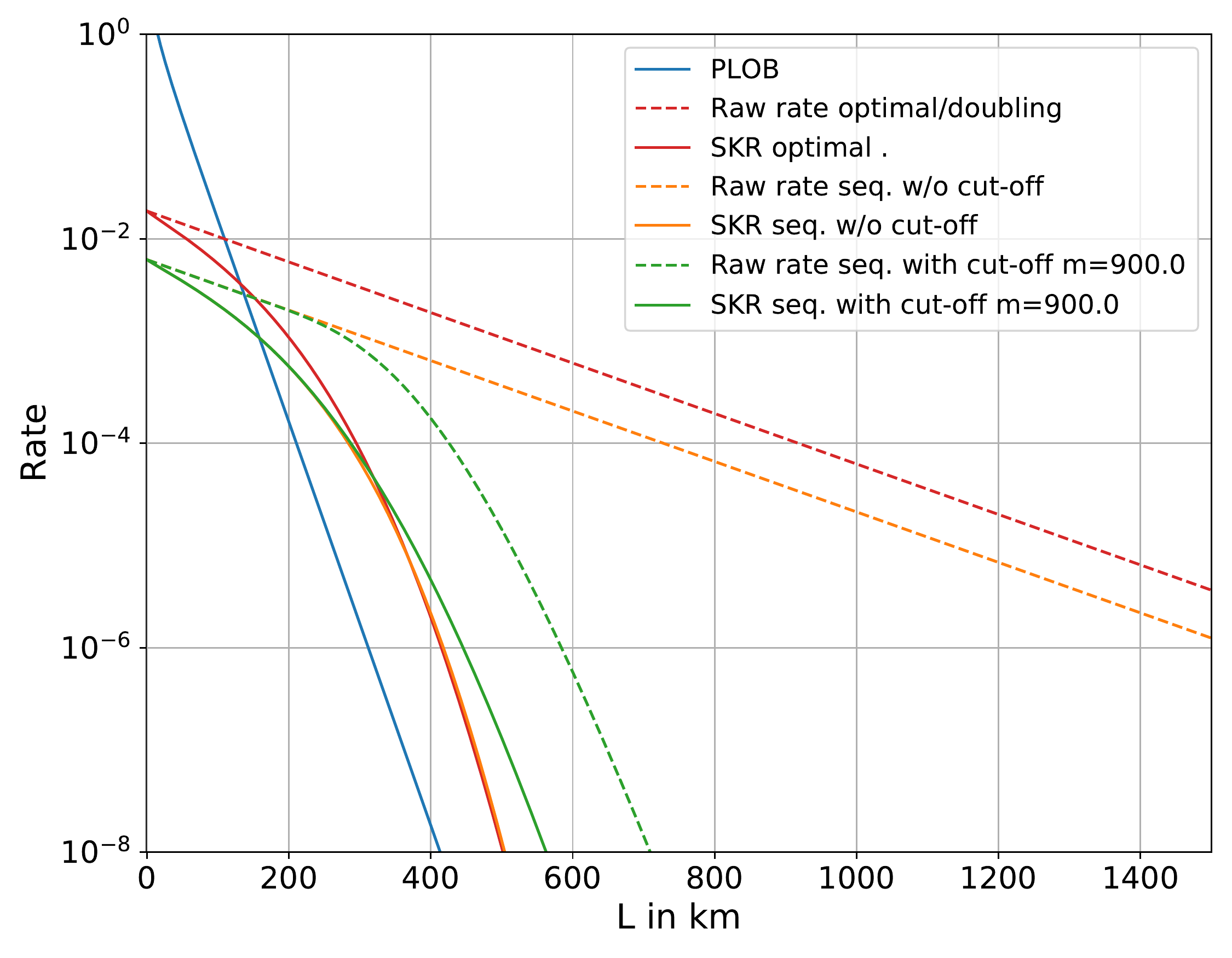}}
	\subfloat[][$\tau_{\mathrm{coh}}={\unit[0.1]{s}}$, $p_{\mathrm{link}}=0.7$, $\mu = \mu_0=0.99$]{\includegraphics[width=0.33\linewidth]{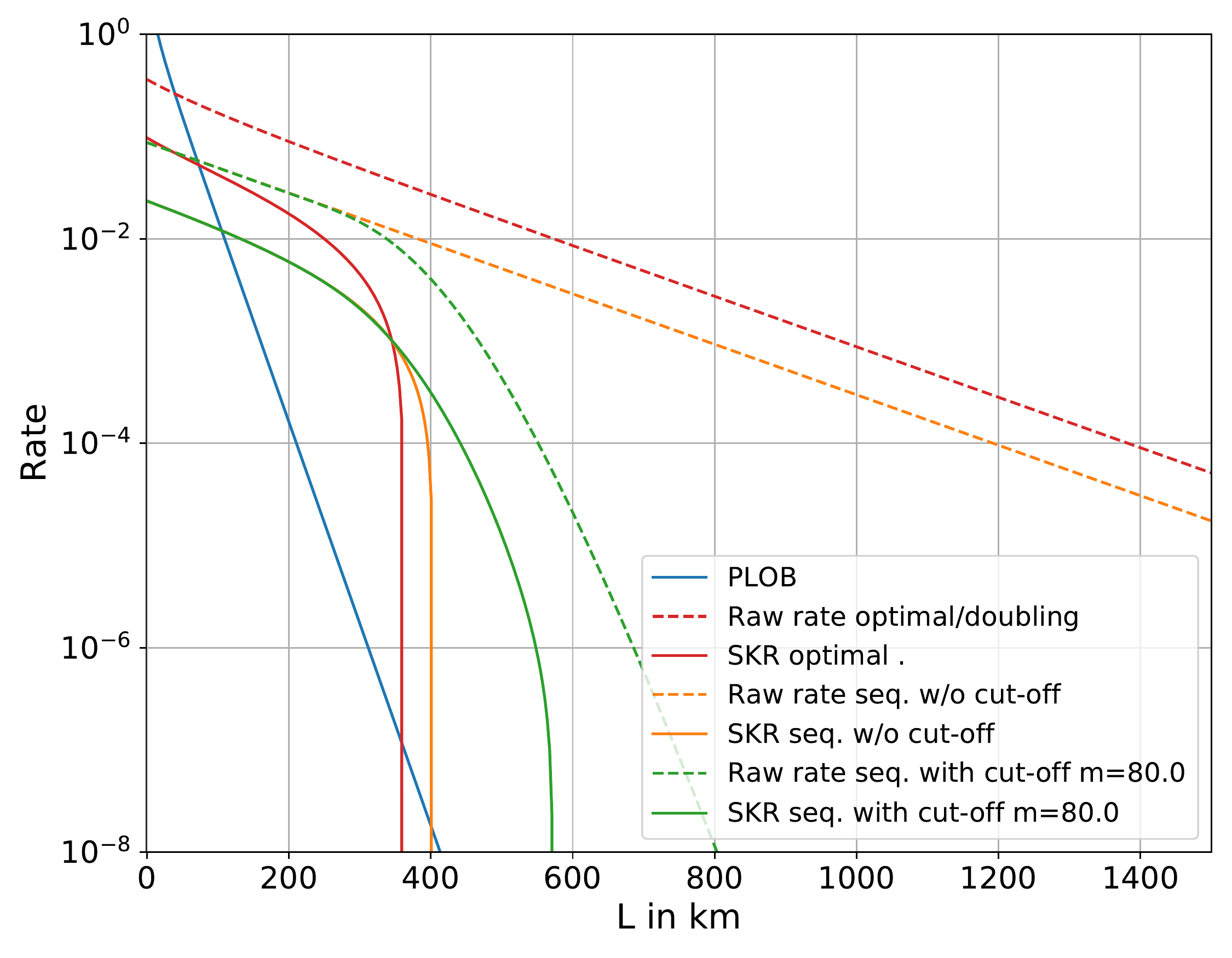}}\\
	\subfloat[][$\tau_{\mathrm{coh}}={\unit[0.1]{s}}$, $p_{\mathrm{link}}=0.7$, $\mu = \mu_0=1$]{\includegraphics[width=0.33\linewidth]{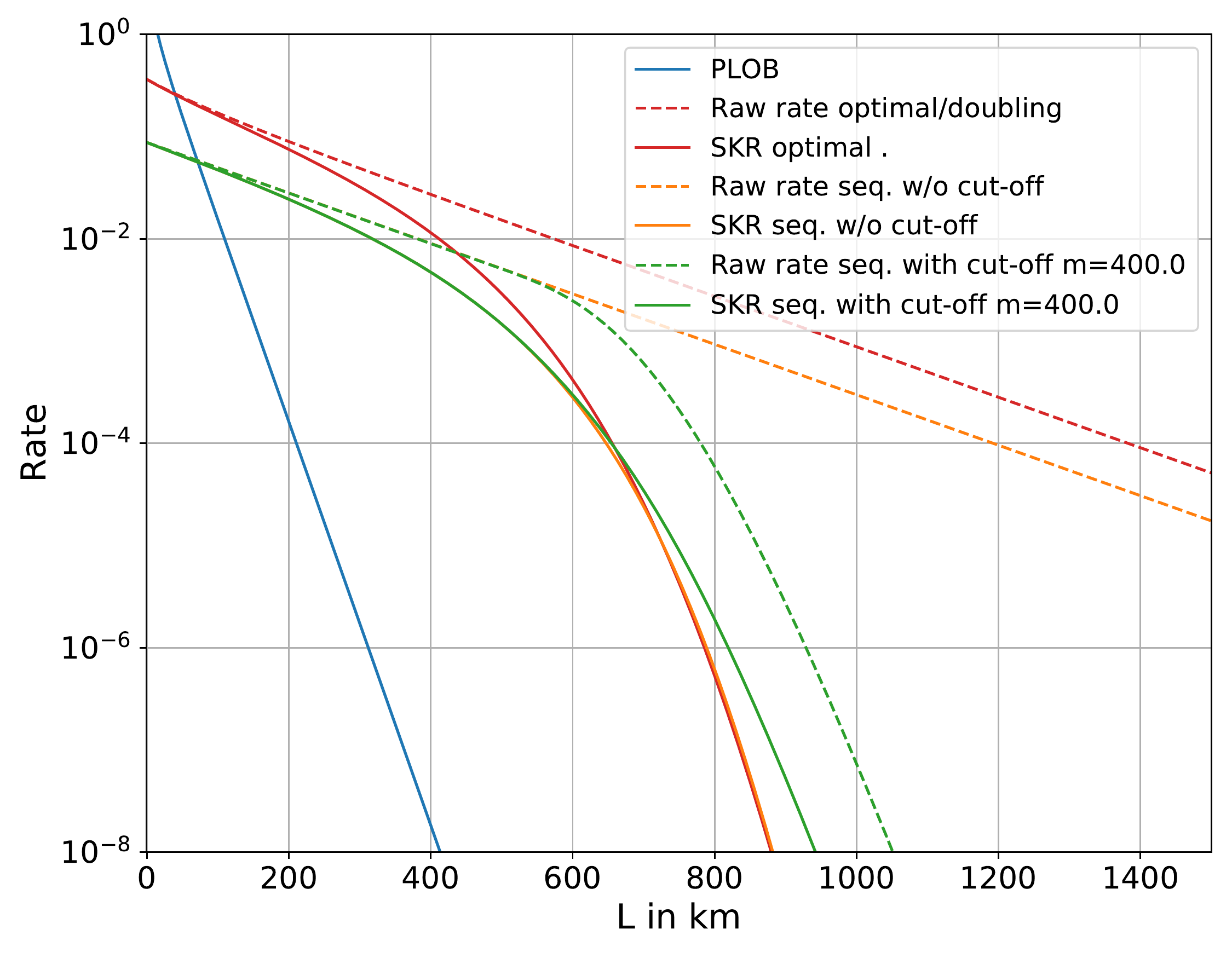}}
	\subfloat[][$\tau_{\mathrm{coh}}={\unit[10]{s}}$, $p_{\mathrm{link}}=0.05$, $\mu = \mu_0=0.99$]{\includegraphics[width=0.33\linewidth]{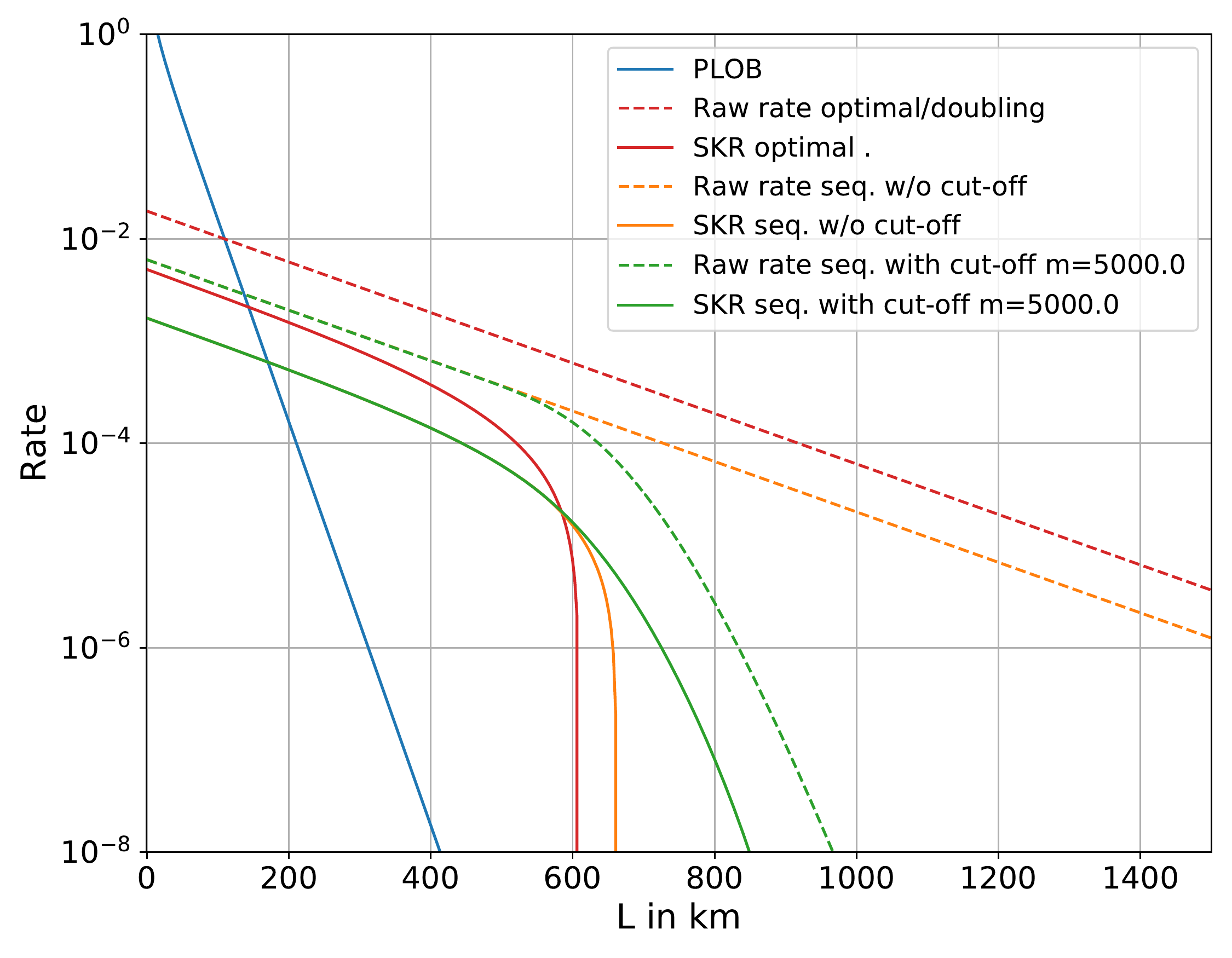}}
	\subfloat[][$\tau_{\mathrm{coh}}={\unit[10]{s}}$, $p_{\mathrm{link}}=0.05$, $\mu = \mu_0=1$]{\includegraphics[width=0.33\linewidth]{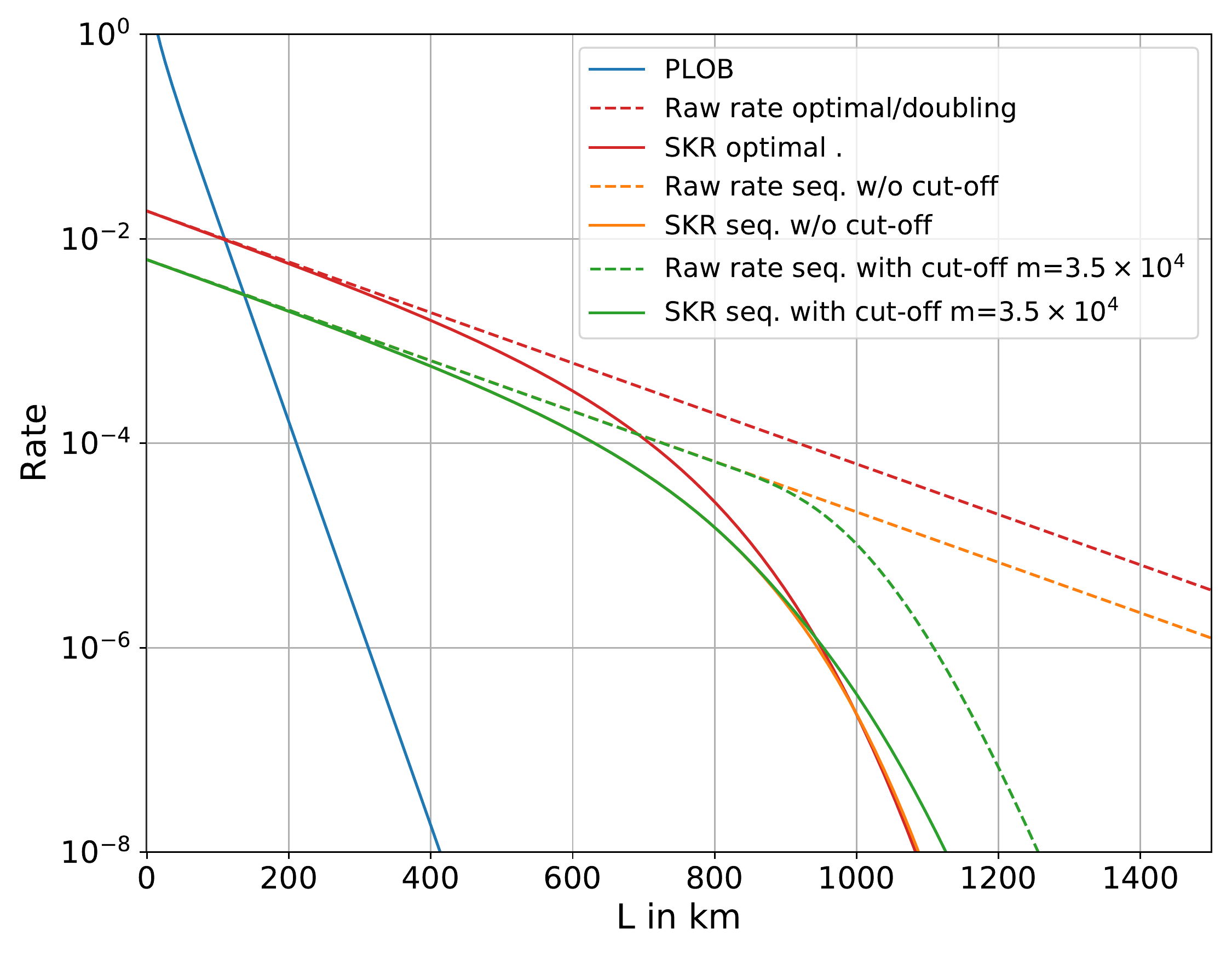}} \\
	\subfloat[][$\tau_{\mathrm{coh}}={\unit[10]{s}}$, $p_{\mathrm{link}}=0.7$, $\mu = \mu_0=0.99$]{\includegraphics[width=0.33\linewidth]{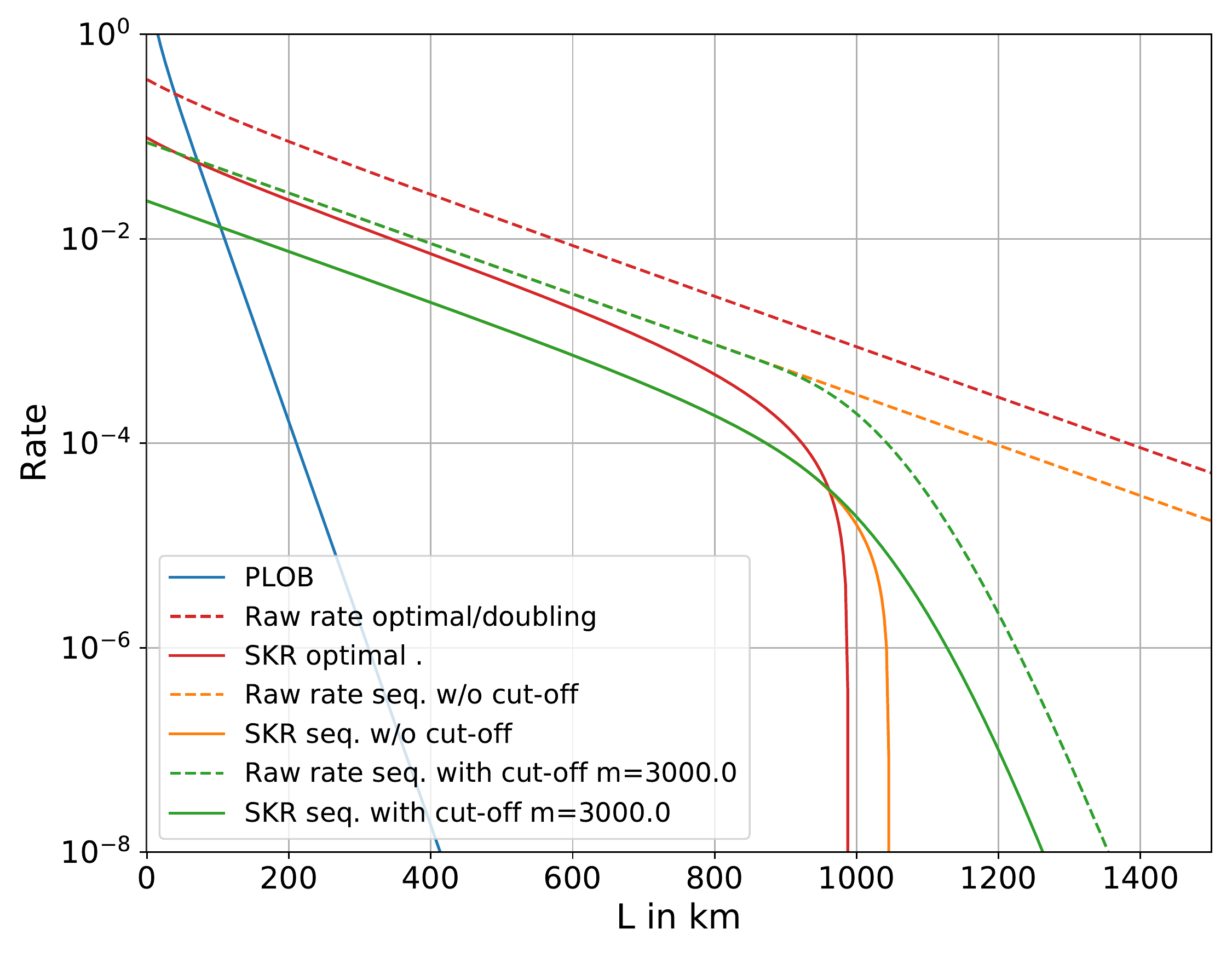}} 
	\subfloat[][$\tau_{\mathrm{coh}}={\unit[10]{s}}$, $p_{\mathrm{link}}=0.7$, $\mu = \mu_0=1$]{\includegraphics[width=0.33\linewidth]{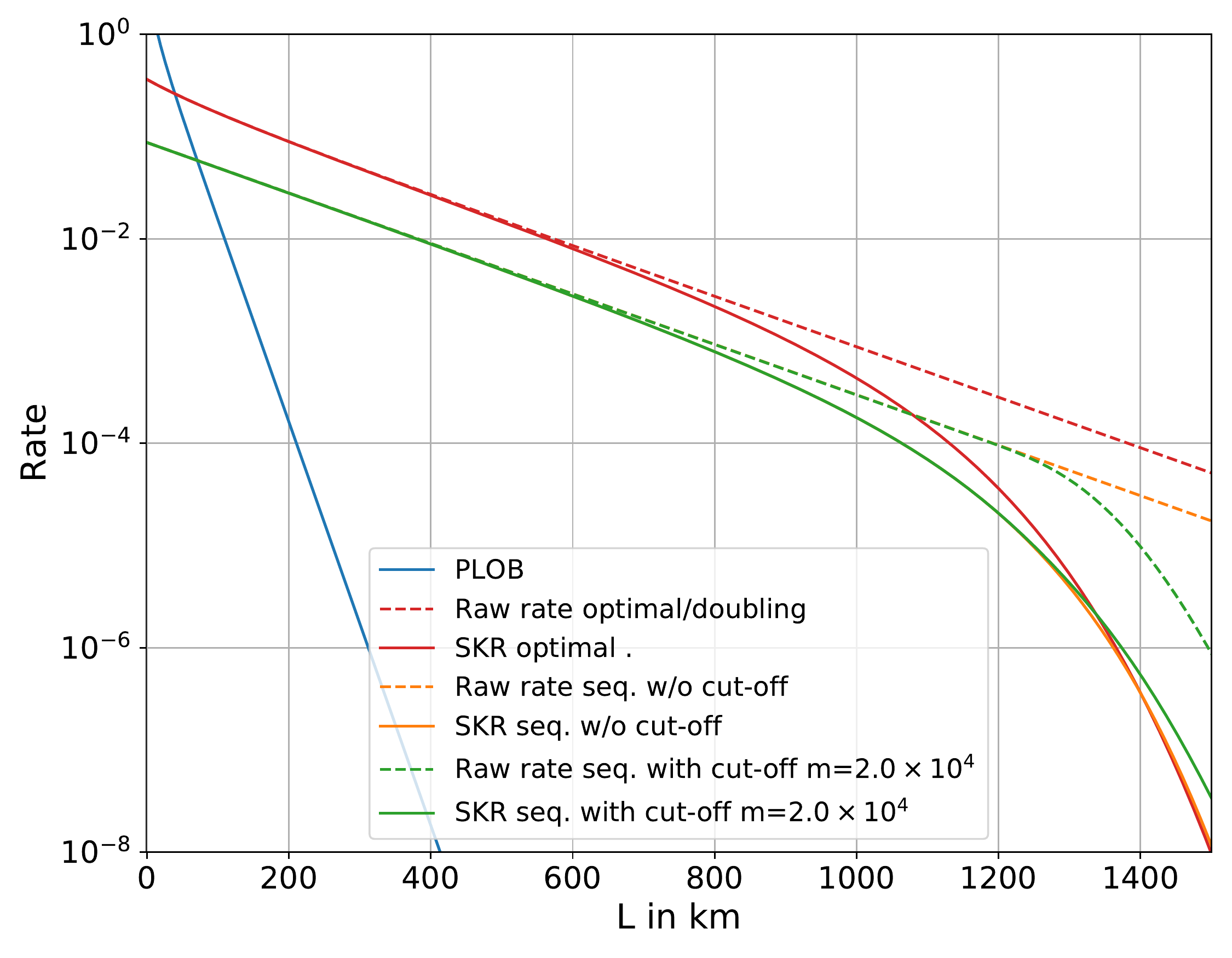}}
	\caption{Comparison of eight-segment repeaters for a total distance \(L\) and different experimental parameters. The ``optimal'' scheme (red) performing BB84 measurements at the end is compared with the fully sequential scheme (orange without memory cut-off, green with cut-off) performing immediate measurements on Alice's / Bob's sides.}
	\label{fig:Comparison_8_segments_imm_vs_non_imm}
\end{figure*}

In a QKD application, Alice's qubit can be measured immediately 
(and so can Bob's qubit at the very end when the entangled pair of the most right segment is being distributed). This way there is another factor of $1/2$ improvement possible for the effective dephasing, since at any elementary time step there is always only a single memory qubit dephasing instead of a qubit pair. In Fig.~\ref{fig:Comparison_8_segments_imm_vs_non_imm}, for eight repeater segments, we compare this fully sequential scheme and immediate measurements by Alice and Bob with the ``optimal'' scheme (parallel distribution and swap as soon as possible) where Alice and Bob store their qubits during the whole long-distance distribution procedure to do the BB84 measurements only at the very end. We see that a QKD protocol in which Alice and Bob measure their qubits immediately can be useful in order to go a bit farther. However, note that in the ``optimal'' scheme Alice and Bob may also measure their qubits immediately, resulting in higher rates but also requiring a more complicated rate analysis.

\section{Mixed strategies for distribution and swapping}\label{app:mixedstr}

In this appendix we shall illustrate that our formalism based on the calculation
of PGFs for the two basic random variables is so versatile that we can also obtain the rates for all kinds of mixed strategies. This applies to both the initial entanglement distributions and the entanglement swappings. In fact, for the case of three repeater segments ($n=3$), we have already explicitly calculated the secret key rates for all possible schemes with swapping as soon as possible, but with variations in the initial distribution strategies, see
App.~\ref{app:Optimality 3 segments}. This enabled us to consider schemes that are overall slower (exhibiting smaller raw rates), but can have a smaller accumulated dephasing. While swapping as soon as possible is optimal with regards to a minimal dephasing time, it may sometimes also be useful to consider a different swapping strategy. The most commonly considered swapping strategy is doubling which implies that it can sometimes happen that neighboring, ready segments will not be connected, as this would be inconsistent with a doubling of the covered repeater distances at each step. A conceptual argument for doubling could be that for a scalable (nested) repeater system one can incorporate entanglement distillation steps in a systematic way. A theoretical motivation to focus on doubling has been that rates are more easy to calculate -- a motivation that is rendered obsolete through the present work, at least for repeaters of size up to $n=8$. Nonetheless we shall give a few examples for mixed strategies for $n=4$ and $n=8$ segments.

\begin{figure}
    \includegraphics[width=\linewidth]{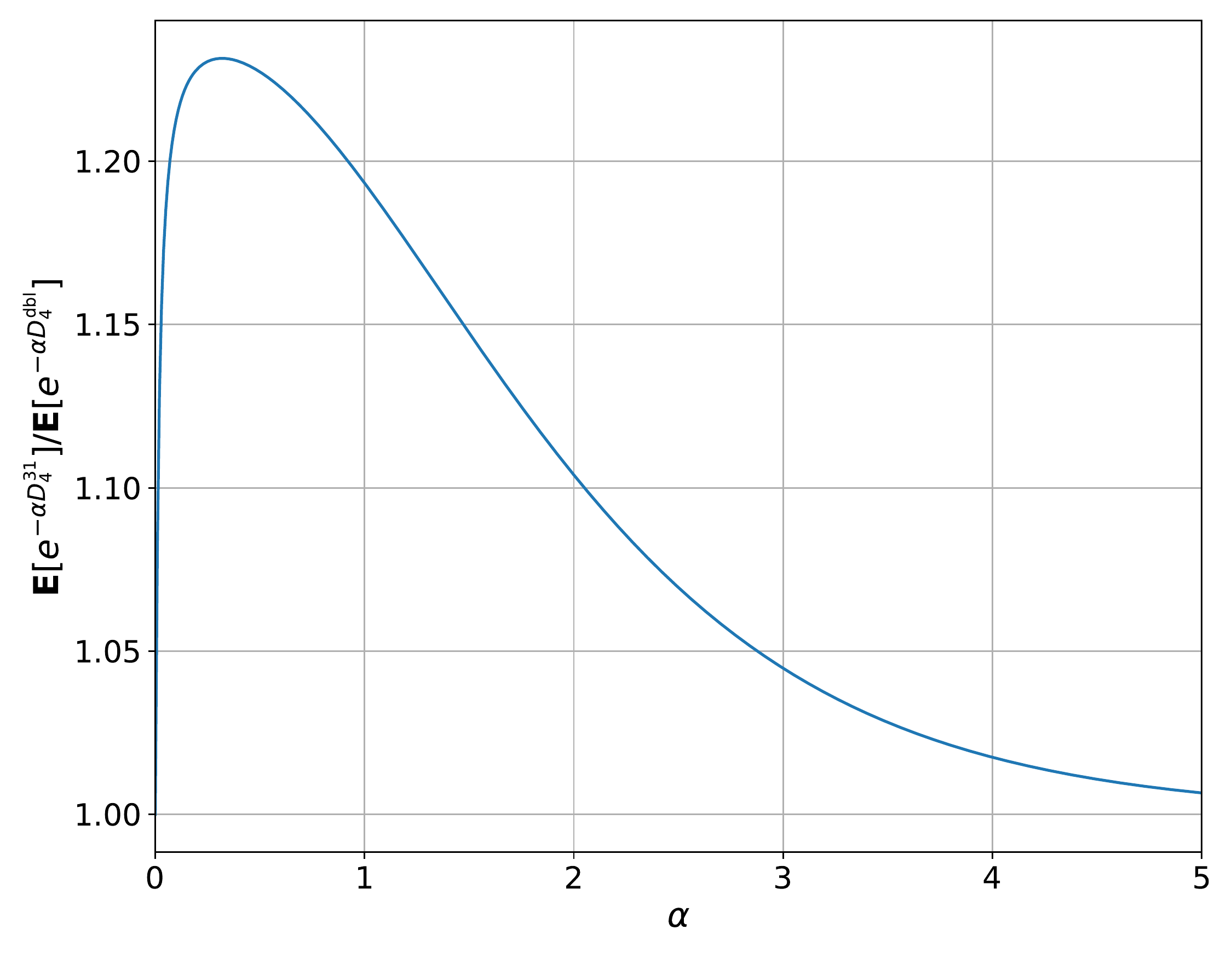}
	\caption{The ratio given by Eq.~\eqref{eq:r} as a function of $\alpha$ for $p = 0.01$ (corresponding to a segment length of 100km for ideal link coupling).}
	\label{fig:DD}
\end{figure}

For $n=4$ segments, in addition to those schemes discussed in the main text, let us consider another possibility where we distribute entanglement over the first three segments in the optimal way and then extend it
over the last segment. Note that this scheme is a variation of the swapping strategy, while the initial distributions still occur in parallel. As a consequence, it can happen that either segment 4 waits for the first three segments to accomplish their distributions and connections or the first three segments have to wait for segment 4. This part of the dephasing corresponds to the last term in the next equation below. The scheme serves as an illustration of the rich choice of possibilities for the swapping
strategies even when only $n=4$. We have
\begin{equation}\label{eq:D4-3-1}
	\begin{split}
		D^{31}_4(N_1, N_2, N_3, N_4) &= D^\star_3(N_1, N_2, N_3) \\
		&+ |\max(N_1, N_2, N_3) - N_4|.
\end{split}
\end{equation}
The PGF of this random variable reads as
\begin{equation}
	\tilde{G}^{31}_4(t) = \frac{p^4}{1-q^4} \frac{P^{31}_4(q, t)}{Q^{31}_4(q, t)},
\end{equation}
where the numerator and denominator are given by
\begin{displaymath}
\begin{split}
	P^{31}_4(q, &t) = 1 + (q^2 + 3 q^3) t + (q + 3 q^2 - q^4 - q^5 ) t^2 \\
	&+ (-2 q^2 - 4 q^3 - 4 q^4 + q^5 + q^6) t^3 \\
	&+ (-q^2 - 3 q^3 - q^4 -3 q^6 - 3 q^7 ) t^4 \\
	&+ (-2 q^2 - q^3 + 2 q^4 - 2 q^6 + q^7 + 2 q^8 ) t^5 \\
	&+ (3 q^3 + 3 q^4 + q^6 + 3 q^7 + q^8 ) t^6 \\
	&+ (-q^4 - q^5 + 4 q^6 + 4 q^7 + 2 q^8 ) t^7 \\
	&+ (q^5 + q^6 - 3 q^8 - q^9 ) t^8 - (3 q^7 + q^8 ) t^9 - q^{10} t^{10}, \\
	Q^{31}_4(q, &t) = (1-qt)(1-q^2t)(1-q^3t)(1-qt^2)\\
	&\times (1-q^2t^2)(1-qt^3).
\end{split}
\end{displaymath}
If we take the derivatives (see Eq.~\eqref{eq:PGF}), 
we can obtain the following relation,
\begin{equation}
	\mathbf{E}[D^{\mathrm{dbl}}_4] = \mathbf{E}[D^{31}_4].
\end{equation}
This means that the two random variables have the same expectation values, even though their distributions are different. For
the secret key fraction we need the averages of the exponential of these variables, which essentially leads to the values of the
corresponding PGFs (see Eq.~\eqref{eq:PGF_2}). These do differ, as Fig.~\ref{fig:DD} illustrates. It shows the ratio
\begin{equation}\label{eq:r}
	\frac{\mathbf{E}[e^{-\alpha D^{31}_4}]}{\mathbf{E}[e^{-\alpha D^{\mathrm{dbl}}_4}]} =
	\frac{\tilde{G}^{31}_4(e^{-\alpha})}{\tilde{G}^{\mathrm{dbl}}_4(e^{-\alpha})},
\end{equation}
as a function of $\alpha$. The two random variables have the same average, but the average $\mathbf{E}[e^{-\alpha
D^{31}_4}]$ is larger than the other, so in the scheme corresponding to the random variable given by Eq.~\eqref{eq:D4-3-1}, the
distributed state has a higher fidelity than the final state in the doubling scheme.

\begin{figure}[ht]
	\includegraphics[width=\linewidth]{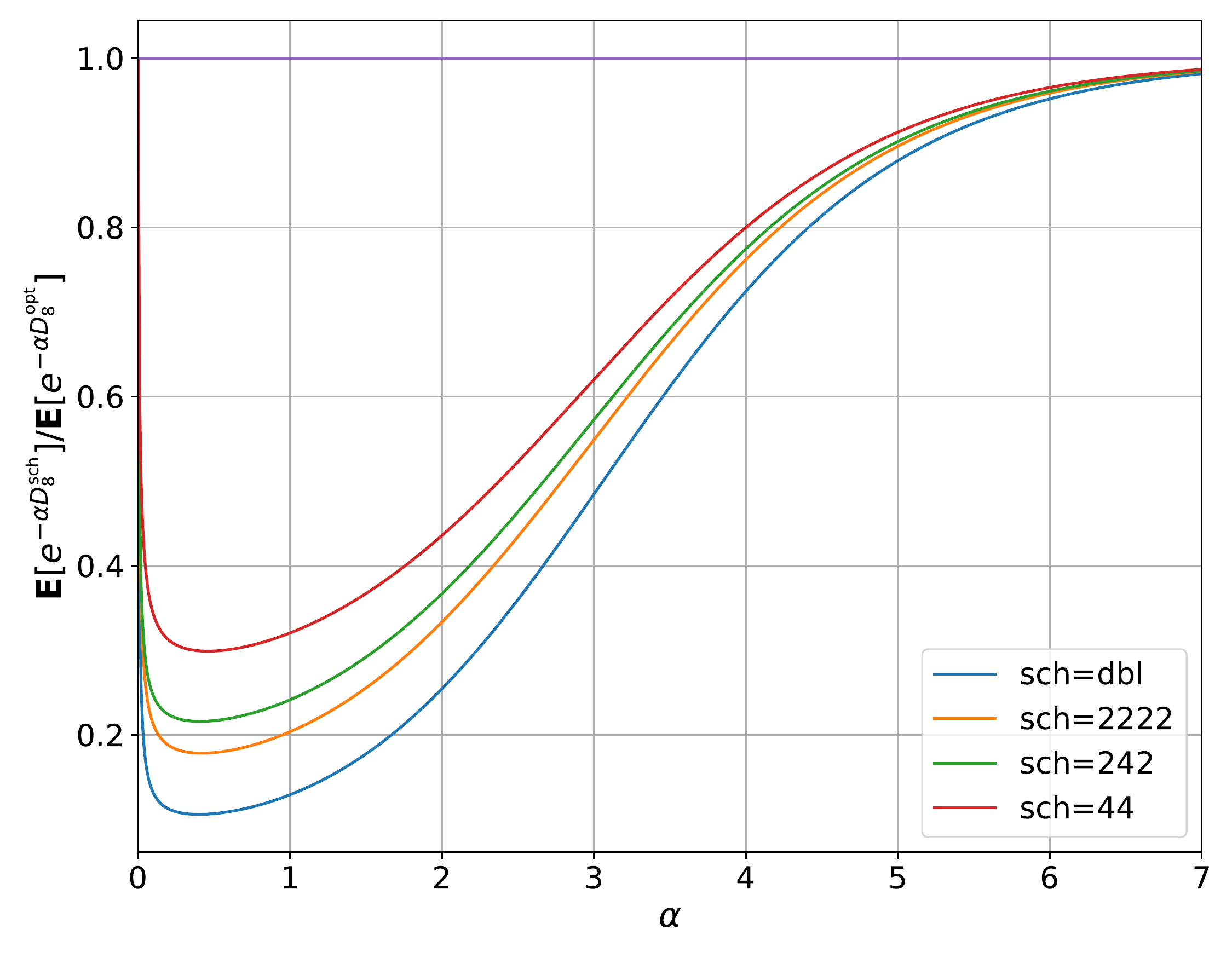}
    \caption{The ratio in Eq.~\eqref{eq:r2} for $\mathrm{sch} = \mathrm{dbl}$ (blue), 2222 (orange), 242 (green) and 44 (red), as a function of $\alpha$ and for $p = 0.01$ (100km segment length).}\label{fig:Ee}
\end{figure}

\begin{figure}[b]
	\includegraphics[width=\linewidth]{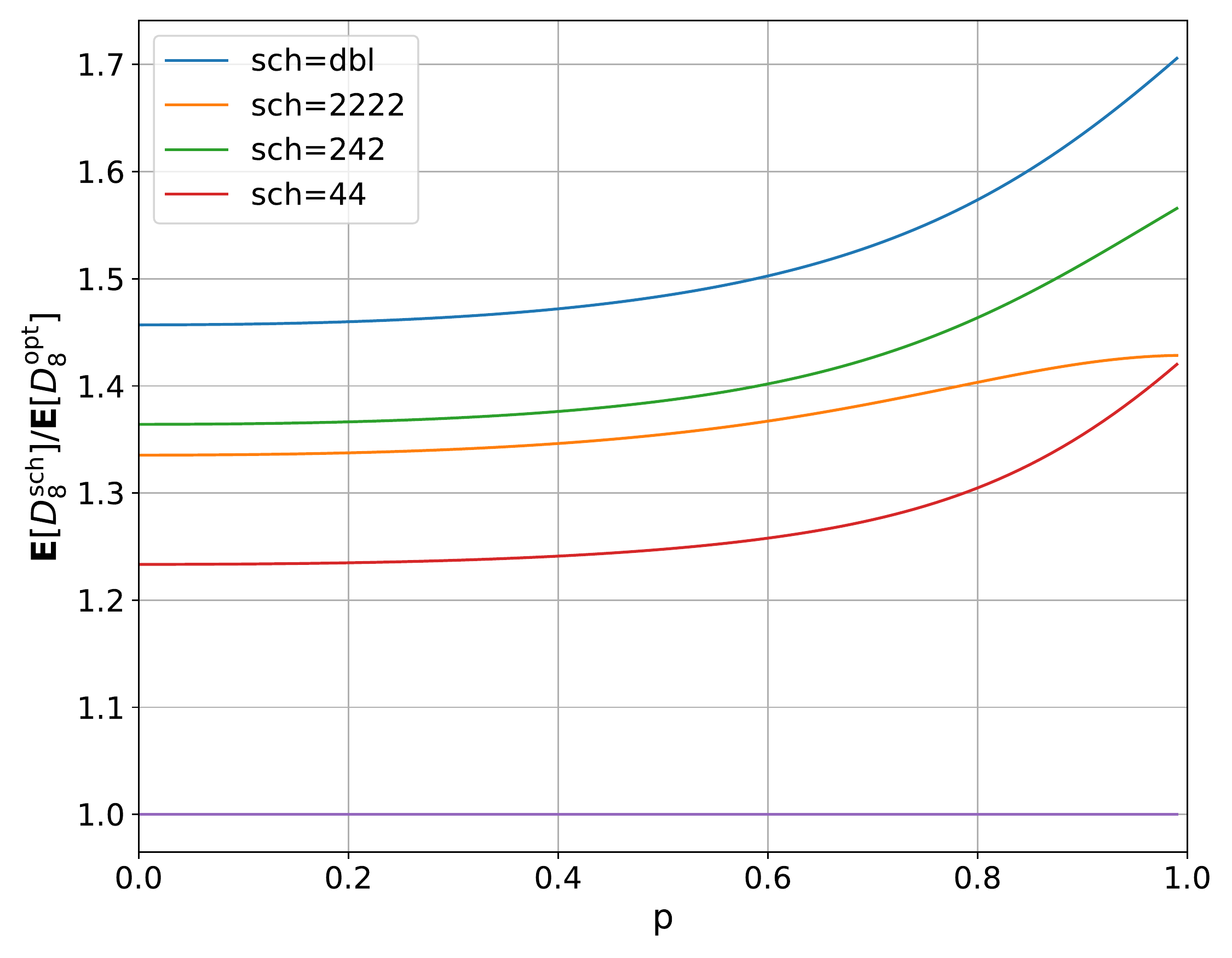}
	\caption{The ratio in Eq.~\eqref{eq:r3} for $\mathrm{sch} = \mathrm{dbl}$ (blue), 2222 (orange), 242 (green) and 44 (red), as a function of $p$.}\label{fig:Ea}
\end{figure}

For the case $n=8$, among a large number of other possibilities to swap the segments, we consider the following three (in addition, the doubling and optimal schemes are discussed in the main text). The
first scheme is to swap the two halves of the repeater in the optimal way (for four segments) and then swap the two larger
segments. We loosely denote the dephasing variable of these scheme as $D^{44}_8$, whose definition reads as
\begin{equation}
\begin{split}
	D^{44}_8&(N_1, \ldots, N_8) = D^\star_4(N_1, \ldots, N_4) \\
	&+ D^\star_4(N_5, \ldots, N_8) \\
	&+ |\max(N_1, \ldots, N_4) - \max(N_5, \ldots, N_8)|.
\end{split}
\end{equation}
Another possibility is to divide the repeater in four pairs, swap them and then swap the four larger segments optimally.
The expression for this dephasing variable $D^{2222}_8$ is a straightforward translation of this description:
\begin{equation}
\begin{split}
	D^{2222}_8&(N_1, \ldots, N_8) = |N_1 - N_2| + \ldots + |N_7 - N_8| \\
	&+ D^\star_4(\max(N_1, N_2), \ldots, \max(N_7, N_8)).
\end{split}
\end{equation}
Finally, we can divide the segments into three groups, consisting of two, four, and two segments. The middle group we
swap optimally (for four segments), and then we swap the three larger segments in the optimal way (for three segments). The
definition of the corresponding random variable $D^{242}_8$ reads as
\begin{equation}
\begin{split}
	D^{242}_8&(N_1, \ldots, N_8) = |N_1 - N_2| + |N_7 - N_8| \\
	&+ D^\star_4(N_3, \ldots, N_6) + D^\star_3(\max(N_1, N_2), \\
	&\max(N_3, \ldots, N_6), \max(N_7, N_8)).
\end{split}
\end{equation}
The PGFs of all these variables have all the same form,
\begin{equation}
	\frac{p^8}{1 - q^8} \frac{P(q, t)}{Q(q, t)},
\end{equation}
with appropriate polynomials $P(q, t)$ and $Q(q, t)$. The numerator polynomials $P(q, t)$ are quite large and contain
around one thousand terms, so we do not present them here.

We can compare the performances of different schemes by plotting the ratios
\begin{equation}\label{eq:r2}
	\frac{\mathbf{E}[e^{-\alpha D^{\mathrm{sch}}_8}]}{\mathbf{E}[e^{-\alpha D^{\mathrm{opt}}_8}]} =
	\frac{\tilde{G}^{\mathrm{sch}}_8(e^{-\alpha})}{\tilde{G}^{\mathrm{opt}}_8(e^{-\alpha})},
\end{equation}
similar to Eq.~\eqref{eq:r}, for $\mathrm{sch} = \mathrm{dbl}, 2222, 242, 44$. We see that among the five
schemes the doubling scheme is the worst with regards to dephasing, and the scheme 44 is the closest to the optimal scheme, see Fig.~\ref{fig:Ee}.
This means that the commonly used parallel-distribution doubling scheme, though fast in terms of $K_8$, is really inefficient in terms of dephasing $D_8$
by disallowing to swap when neighboring segments are ready
on all ``nesting'' levels \cite{Shchukin2021}.

\section{Two-Segment ``Node-Receives-Photon'' Repeaters}
\label{app:nrp}

\begin{figure*}[ht]
	\centering
	\subfloat[a][$\tau_{\mathrm{coh}}={\unit[0.1]{s}}$, $p_{\mathrm{link}}=0.05$, $m=1500$]{\includegraphics[width=0.5\linewidth]{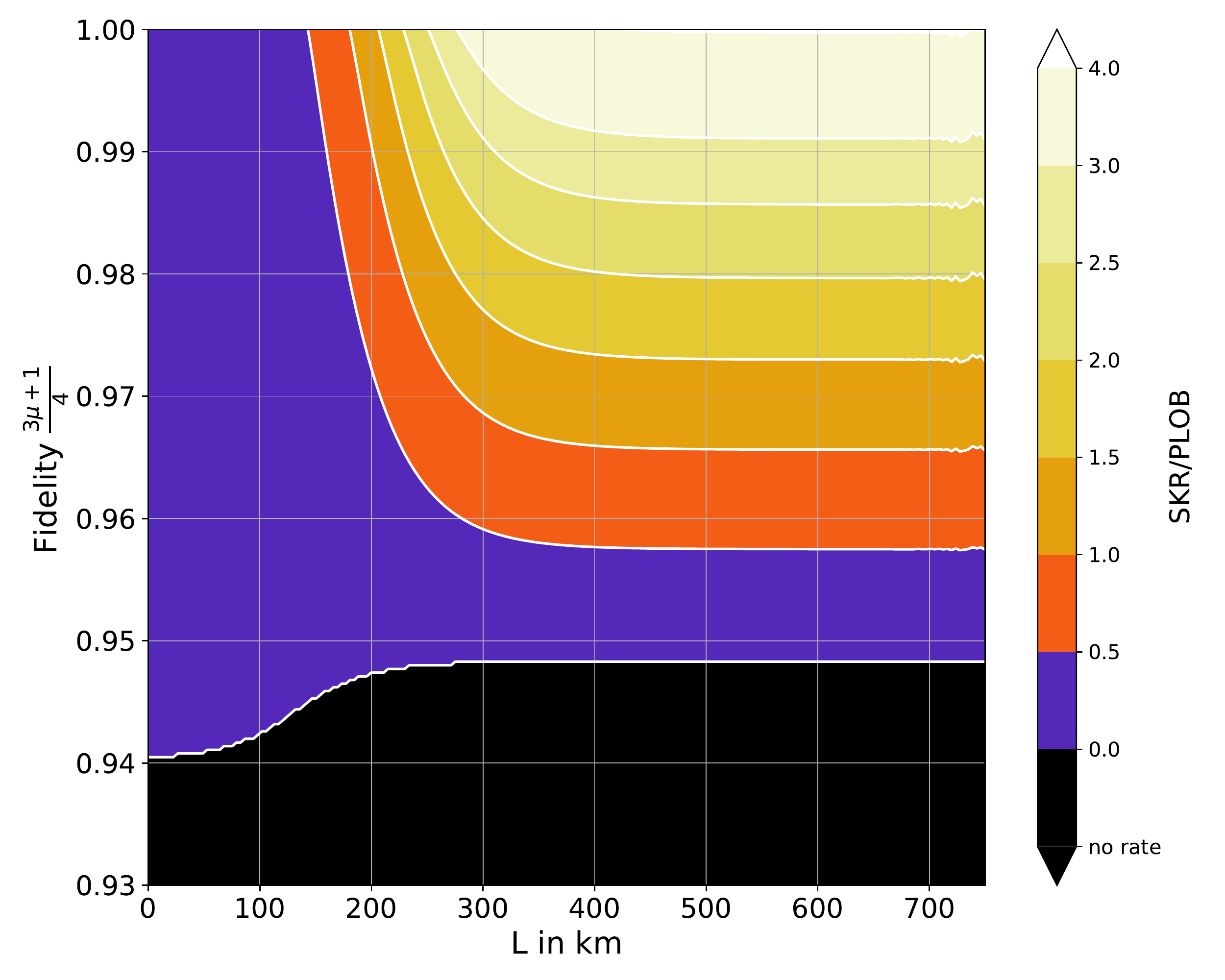}}
	\subfloat[][$\tau_{\mathrm{coh}}={\unit[0.1]{s}}$, $p_{\mathrm{link}}=0.7$, $m=1500$]{\includegraphics[width=0.5\linewidth]{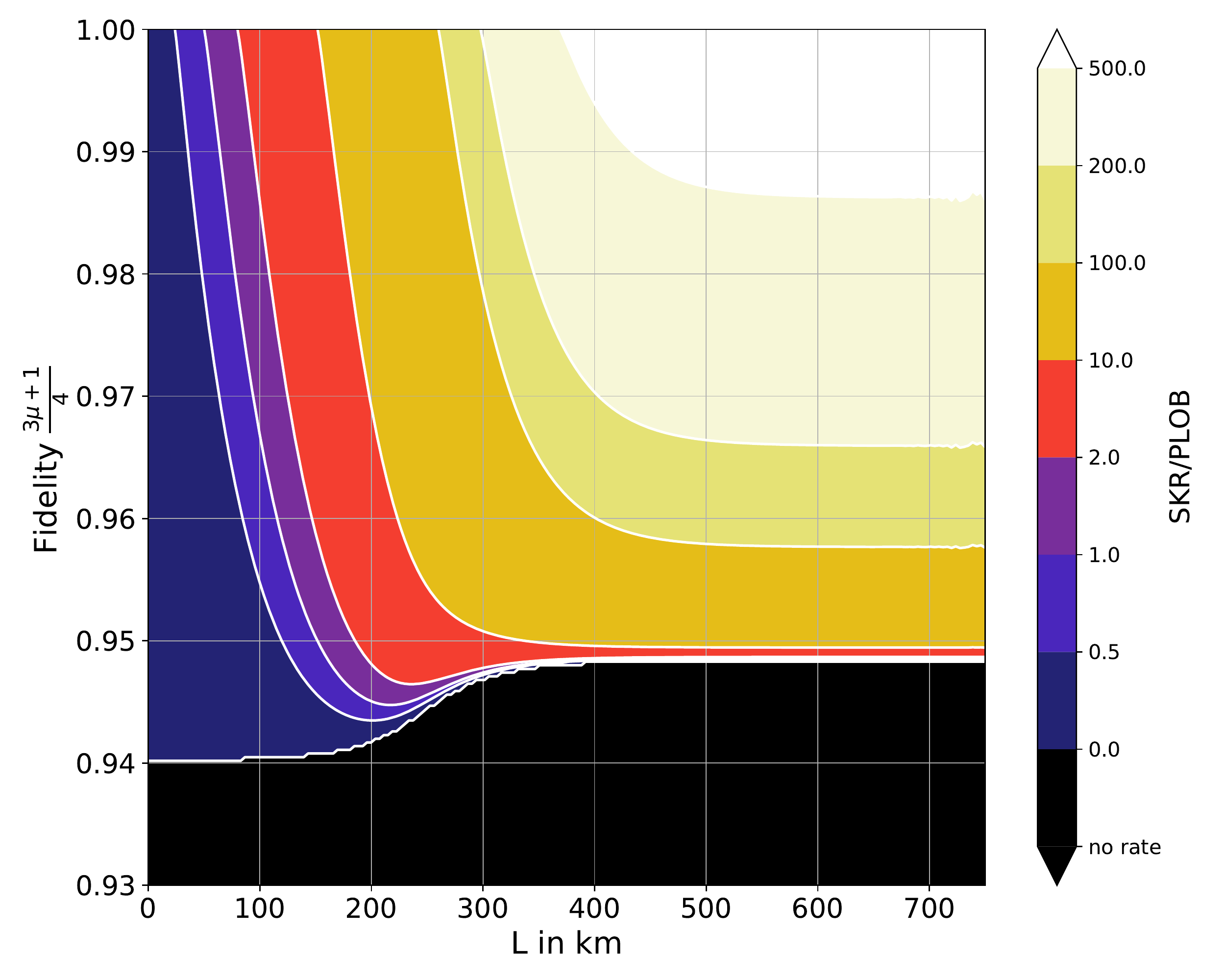}} \\
	\subfloat[][$\tau_{\mathrm{coh}}={\unit[10]{s}}$, $p_{\mathrm{link}}=0.05$, $m=1.5\times 10^5$]{\includegraphics[width=0.5\linewidth]{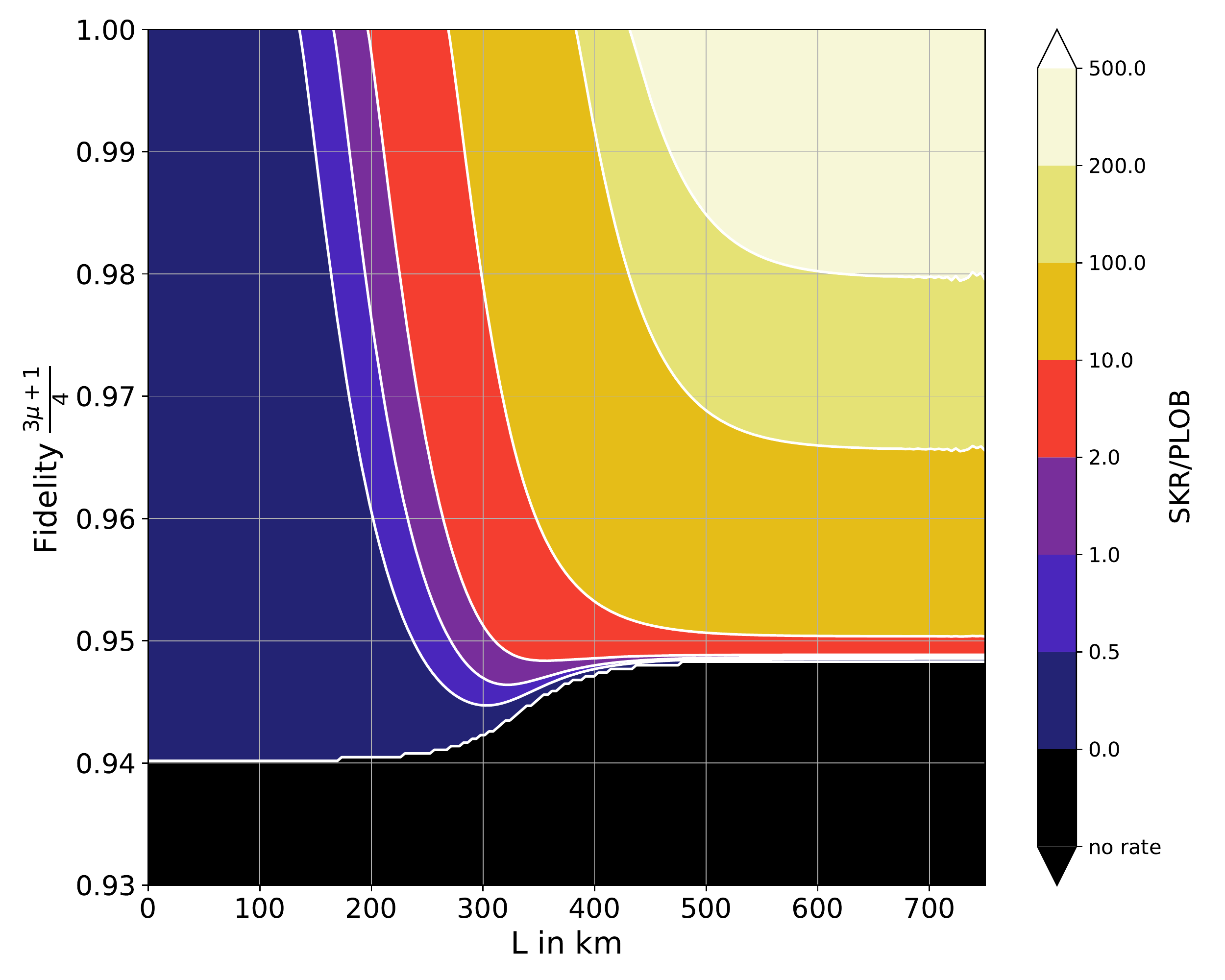}} 
	\subfloat[][$\tau_{\mathrm{coh}}={\unit[10]{s}}$, $p_{\mathrm{link}}=0.7$, $m=1.5\times 10^5$]{\includegraphics[width=0.5\linewidth]{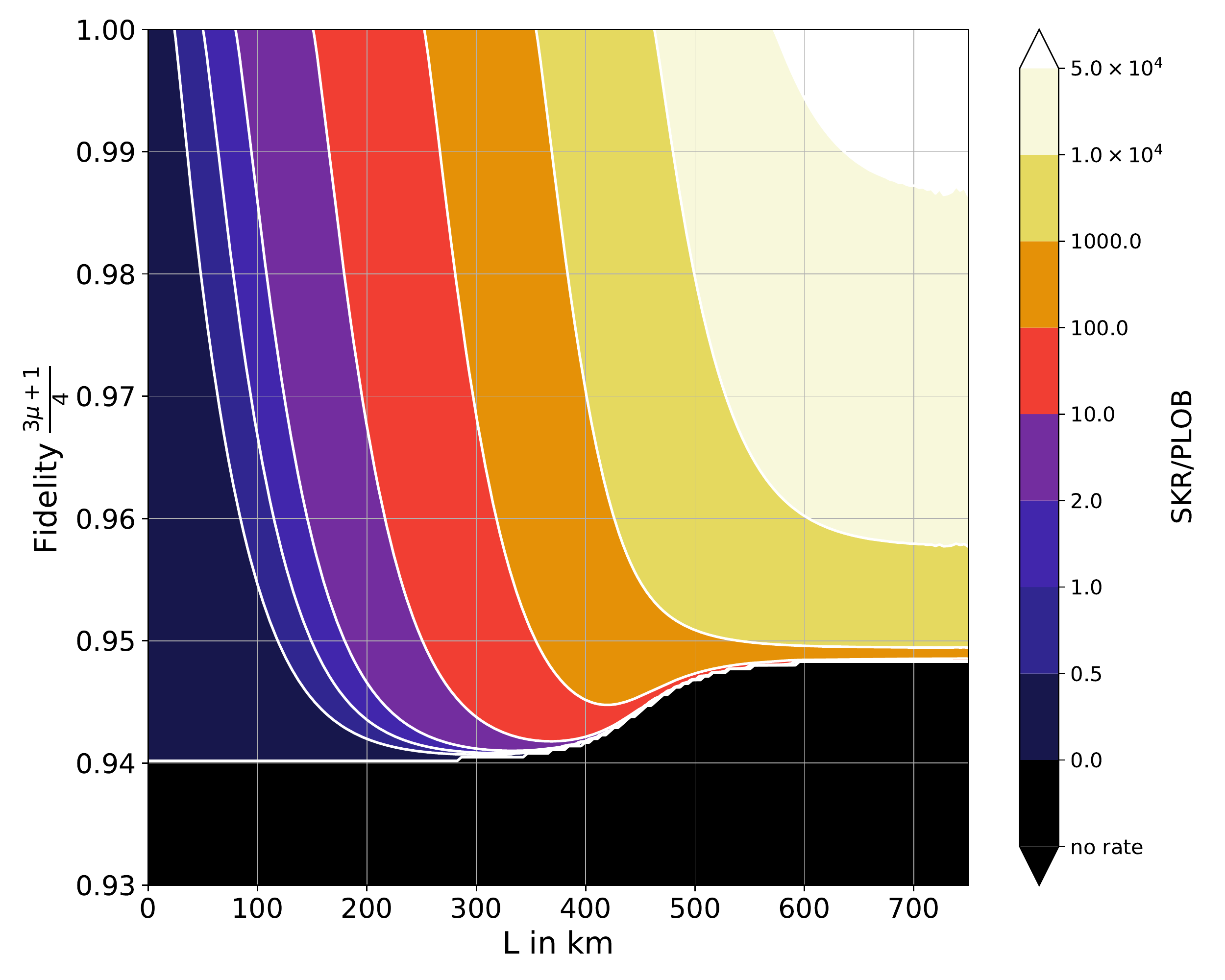}}
	\caption{Contour plots illustrating the minimal fidelity requirements to overcome the PLOB bound by a two-segment NRP repeater for different parameter sets. In all contour plots, \(\mu = \mu_0\), \(\tau_{\mathrm{clock}}=\unit[10]{MHz}\) and \(F_0=1\) has been used.}
	\label{fig:NRP_Contour_2_segments}
\end{figure*}

Figure~\ref{fig:NRP_Contour_2_segments}
shows the BB84 rates in a two-segment quantum repeater 
based on the NRP concept with one middle station 
receiving optical quantum signals sent from 
two outer stations at Alice and Bob.
By circumventing the need for extra classical communication
and thus significantly reducing the effective memory dephasing,
the minimal state and gate fidelity values can even be kept constant 
over large distance regimes.
For the experimental clock rate we have chosen $\tau_{\mathrm{clock}}=\unit[10]{MHz}$,
limited by the local interaction and processing times 
of the light-matter interface at the middle station.

\section{Calculation for Cabrillo's scheme}\label{app:cabrillo}
	First, we consider two entangled states of a single-rail qubit with a quantum memory ($\gamma\in\mathbb{R}$)
	\begin{align}
		\frac{1}{1+\gamma^2} &\left[ \ket{\uparrow,\uparrow,0,0}  +\gamma \ket{\uparrow,\downarrow,0,1} \right. \nonumber \\ 
		 &\hspace{20pt} + \left. \gamma\ket{\downarrow,\uparrow,1,0}+\gamma^2\ket{\downarrow,\downarrow,1,1} \right].
	\end{align}
	After applying a lossy channel with transmission parameter $\eta=p_{\mathrm{link}}\exp(-\frac{L_0}{2L_{\mathrm{att}}})$ to both optical modes, we obtain the following state after introducing two additional environmental modes
	\begin{widetext}

	\begin{align}
		\frac{1}{1+\gamma^2} &\Bigg[ \gamma^2 \ket{\downarrow,\downarrow} \otimes \left(\eta\ket{1,1,0,0} + \sqrt{\eta(1-\eta)} \left(\ket{1,0,0,1} + \ket{0,1,1,0}\right) + (1-\eta) \ket{0,0,1,1} \right) \nonumber\\
		&\hspace{20pt} + \left. \gamma \ket{\uparrow,\downarrow} \otimes \left(\sqrt{\eta} \ket{0,1,0,0} + \sqrt{1-\eta} \ket{0,0,0,1} \right) \right. \nonumber\\
		&\hspace{20pt} + \left. \gamma \ket{\downarrow,\uparrow} \otimes \left(\sqrt{\eta} \ket{1,0,0,0} + \sqrt{1-\eta} \ket{0.0,1,0} \right) \right.\\
		&\hspace{20pt} + \ket{\uparrow,\uparrow,0,0,0,0} \Bigg]\nonumber
	\end{align}
		\end{widetext}
	We apply a 50:50 beam splitter to the (non-environmental) optical mode and obtain the state
	\begin{widetext}
		\begin{align}
		\frac{1}{1+\gamma^2} \Bigg[&\gamma^2 \ket{\downarrow,\downarrow} \otimes \sqrt{\frac{\eta(1-\eta)}{2}} \left(\ket{1,0,0,1}+\ket{0,1,0,1}+\ket{1,0,1,0}-\ket{0,1,1,0}\right) \nonumber\\
		&\hspace{20pt} + \gamma^2 \ket{\downarrow,\downarrow} \otimes \frac{\eta}{2}\left(\ket{2,0,0,0}-\ket{0,2,0,0}\right) \nonumber \\
		&\hspace{20pt} +\gamma^2 \ket{\downarrow,\downarrow} \otimes (1-\eta)\ket{0,0,1,1} \nonumber \\
		&\hspace{20pt} \left. + \gamma \ket{\uparrow,\downarrow} \otimes \left(\sqrt{\frac{\eta}{2}} \left(\ket{1,0,0,0}-\ket{0,1,0,0}\right) + \sqrt{1-\eta} \ket{0,0,0,1} \right)\right.\\
		&\hspace{20pt} \left. + \gamma \ket{\downarrow,\uparrow} \otimes \left(\sqrt{\frac{\eta}{2}} \left(\ket{1,0,0,0}+\ket{0,1,0,0}\right)  + \sqrt{1-\eta}\ket{0,0,1,0}\right) \right.\nonumber\\
		&\hspace{20pt} +\ket{\uparrow,\uparrow,0,0,0,0} \Bigg]\nonumber\,.
		\end{align}
	\end{widetext}
		We can obtain entangled memory states by post-selecting single photon events at the detectors. If we detect a single photon at the first detector and no photon at the other, we obtain the following (unnormalized) 2-memory reduced density operator (see \cite[App. E]{tf_repeater}) 
		\begin{align}
			\frac{\gamma^2\eta}{(1+\gamma)^2}\left[\ket{\Psi^+}\bra{\Psi^+}+\gamma^2(1-\eta)\ket{\downarrow,\downarrow}\bra{\downarrow,\downarrow}\right].
		\end{align}
		When using simple on/off detectors instead of photon number resolving detectors (PNRD) two-photon events will also lead to a detection event. The two-memory state after a two-photon event is given by
		\begin{align}
			\frac{\gamma^4\eta^2}{4(1+\gamma^2)^2}\ket{\downarrow,\downarrow}\bra{\downarrow,\downarrow}\,.
		\end{align}
		Thus, the probability of a successful entanglement generation is given by $p_{\mathrm{PNRD}}=\frac{2\gamma^2\eta}{(1+\gamma^2)^2}(1+\gamma^2(1-\eta))$, when using PNRD, and $p_{\mathrm{on/off}}=\frac{2\gamma^2\eta}{(1+\gamma^2)^2}(1+\gamma^2(1-\frac{3}{4}\eta))$, when using on/off detectors. The factor 2 comes from the possibility to detect the photon at the other detector instead, although in this case the memory state differs by a single-qubit $Z$-operation. After a suitable twirling, we can find a one-qubit Pauli channel which maps the state $\ket{\Psi^+}\bra{\Psi^+}$ to the actual memory state, i.e. we can claim that the loss channel acting on the optical modes induces a Pauli channel on the memories. We can parametrize this Pauli channel by the tuple of error probabilities $(p_I,p_X,p_Y,p_Z)$ and for the case with PNRDs this tuple is given by 
		\begin{align}
			\frac{1}{1+\gamma^2(1-\eta)}\left(1,\frac{\gamma^2}{2}(1-\eta),\frac{\gamma^2}{2}(1-\eta),0\right)
		\end{align}
		and for on/off detectors it is given by 
		\begin{align}
			\frac{1}{1+\gamma^2(1-\frac{3}{4}\eta)}\left(1,\frac{\gamma^2}{2}(1-\frac{3}{4}\eta),\frac{\gamma^2}{2}(1-\frac{3}{4}\eta),0\right)\,.
		\end{align}
		When we consider an $n$-segment repeater, we have to consider a concatenation of $n$ such Pauli channels and we finally obtain the error rates
		\begin{align}
			e_x&=\frac{1}{2}\left(1-\mu^{n-1}\mu_0^{n}\frac{(2F_0-1)^n\mathbf{E}[e^{-\alpha D_n}]}{(1+\gamma^2(1-\eta))^n}\right),\\
			e_z&=\frac{1}{2}\left(1-\mu^{n-1}\mu_0^{n}\left(\frac{1-\gamma^2(1-\eta)}{1+\gamma^2(1-\eta)}\right)^n\right)
		\end{align}
		in the case of PNRDs. When we consider on/off detectors, we can simply replace $\eta$ by $\frac{3}{4}\eta$ in the error rates.


\begin{thebibliography}{66}%
	\makeatletter
	\providecommand \@ifxundefined [1]{%
	 \@ifx{#1\undefined}
	}%
	\providecommand \@ifnum [1]{%
	 \ifnum #1\expandafter \@firstoftwo
	 \else \expandafter \@secondoftwo
	 \fi
	}%
	\providecommand \@ifx [1]{%
	 \ifx #1\expandafter \@firstoftwo
	 \else \expandafter \@secondoftwo
	 \fi
	}%
	\providecommand \natexlab [1]{#1}%
	\providecommand \enquote  [1]{``#1''}%
	\providecommand \bibnamefont  [1]{#1}%
	\providecommand \bibfnamefont [1]{#1}%
	\providecommand \citenamefont [1]{#1}%
	\providecommand \href@noop [0]{\@secondoftwo}%
	\providecommand \href [0]{\begingroup \@sanitize@url \@href}%
	\providecommand \@href[1]{\@@startlink{#1}\@@href}%
	\providecommand \@@href[1]{\endgroup#1\@@endlink}%
	\providecommand \@sanitize@url [0]{\catcode `\\12\catcode `\$12\catcode
	  `\&12\catcode `\#12\catcode `\^12\catcode `\_12\catcode `\%12\relax}%
	\providecommand \@@startlink[1]{}%
	\providecommand \@@endlink[0]{}%
	\providecommand \url  [0]{\begingroup\@sanitize@url \@url }%
	\providecommand \@url [1]{\endgroup\@href {#1}{\urlprefix }}%
	\providecommand \urlprefix  [0]{URL }%
	\providecommand \Eprint [0]{\href }%
	\providecommand \doibase [0]{https://doi.org/}%
	\providecommand \selectlanguage [0]{\@gobble}%
	\providecommand \bibinfo  [0]{\@secondoftwo}%
	\providecommand \bibfield  [0]{\@secondoftwo}%
	\providecommand \translation [1]{[#1]}%
	\providecommand \BibitemOpen [0]{}%
	\providecommand \bibitemStop [0]{}%
	\providecommand \bibitemNoStop [0]{.\EOS\space}%
	\providecommand \EOS [0]{\spacefactor3000\relax}%
	\providecommand \BibitemShut  [1]{\csname bibitem#1\endcsname}%
	\let\auto@bib@innerbib\@empty
	\bibitem [{\citenamefont {Arute}\ \emph {et~al.}(2019)\citenamefont {Arute},
	  \citenamefont {Arya},\ and\ \citenamefont {et. al.}}]{Arute2019}%
	  \BibitemOpen
	  \bibfield  {author} {\bibinfo {author} {\bibfnamefont {F.}~\bibnamefont
	  {Arute}}, \bibinfo {author} {\bibfnamefont {K.}~\bibnamefont {Arya}},\ and\
	  \bibinfo {author} {\bibfnamefont {R.~B.}\ \bibnamefont {et. al.}},\
	  }\bibfield  {title} {\bibinfo {title} {Quantum supremacy using a programmable
	  superconducting processor},\ }\href
	  {https://doi.org/10.1038/s41586-019-1666-5} {\bibfield  {journal} {\bibinfo
	  {journal} {Nature}\ }\textbf {\bibinfo {volume} {574}},\ \bibinfo {pages}
	  {505} (\bibinfo {year} {2019})}\BibitemShut {NoStop}%
	\bibitem [{\citenamefont {ANIS}\ \emph {et~al.}(2021)\citenamefont {ANIS},
	  \citenamefont {Abraham},\ and\ \citenamefont {et. al.}}]{Qiskit}%
	  \BibitemOpen
	  \bibfield  {author} {\bibinfo {author} {\bibfnamefont {M.~S.}\ \bibnamefont
	  {ANIS}}, \bibinfo {author} {\bibfnamefont {H.}~\bibnamefont {Abraham}},\ and\
	  \bibinfo {author} {\bibfnamefont {A.}~\bibnamefont {et. al.}},\ }\href
	  {https://doi.org/10.5281/zenodo.2573505} {\bibinfo {title} {Qiskit: An
	  open-source framework for quantum computing}} (\bibinfo {year}
	  {2021})\BibitemShut {NoStop}%
	\bibitem [{\citenamefont {Zhong}\ \emph {et~al.}(2020)\citenamefont {Zhong},
	  \citenamefont {Wang}, \citenamefont {Deng}, \citenamefont {Chen},
	  \citenamefont {Peng}, \citenamefont {Luo}, \citenamefont {Qin}, \citenamefont
	  {Wu}, \citenamefont {Ding}, \citenamefont {Hu}, \citenamefont {Hu},
	  \citenamefont {Yang}, \citenamefont {Zhang}, \citenamefont {Li},
	  \citenamefont {Li}, \citenamefont {Jiang}, \citenamefont {Gan}, \citenamefont
	  {Yang}, \citenamefont {You}, \citenamefont {Wang}, \citenamefont {Li},
	  \citenamefont {Liu}, \citenamefont {Lu},\ and\ \citenamefont
	  {Pan}}]{Pan2020}%
	  \BibitemOpen
	  \bibfield  {author} {\bibinfo {author} {\bibfnamefont {H.-S.}\ \bibnamefont
	  {Zhong}}, \bibinfo {author} {\bibfnamefont {H.}~\bibnamefont {Wang}},
	  \bibinfo {author} {\bibfnamefont {Y.-H.}\ \bibnamefont {Deng}}, \bibinfo
	  {author} {\bibfnamefont {M.-C.}\ \bibnamefont {Chen}}, \bibinfo {author}
	  {\bibfnamefont {L.-C.}\ \bibnamefont {Peng}}, \bibinfo {author}
	  {\bibfnamefont {Y.-H.}\ \bibnamefont {Luo}}, \bibinfo {author} {\bibfnamefont
	  {J.}~\bibnamefont {Qin}}, \bibinfo {author} {\bibfnamefont {D.}~\bibnamefont
	  {Wu}}, \bibinfo {author} {\bibfnamefont {X.}~\bibnamefont {Ding}}, \bibinfo
	  {author} {\bibfnamefont {Y.}~\bibnamefont {Hu}}, \bibinfo {author}
	  {\bibfnamefont {P.}~\bibnamefont {Hu}}, \bibinfo {author} {\bibfnamefont
	  {X.-Y.}\ \bibnamefont {Yang}}, \bibinfo {author} {\bibfnamefont {W.-J.}\
	  \bibnamefont {Zhang}}, \bibinfo {author} {\bibfnamefont {H.}~\bibnamefont
	  {Li}}, \bibinfo {author} {\bibfnamefont {Y.}~\bibnamefont {Li}}, \bibinfo
	  {author} {\bibfnamefont {X.}~\bibnamefont {Jiang}}, \bibinfo {author}
	  {\bibfnamefont {L.}~\bibnamefont {Gan}}, \bibinfo {author} {\bibfnamefont
	  {G.}~\bibnamefont {Yang}}, \bibinfo {author} {\bibfnamefont {L.}~\bibnamefont
	  {You}}, \bibinfo {author} {\bibfnamefont {Z.}~\bibnamefont {Wang}}, \bibinfo
	  {author} {\bibfnamefont {L.}~\bibnamefont {Li}}, \bibinfo {author}
	  {\bibfnamefont {N.-L.}\ \bibnamefont {Liu}}, \bibinfo {author} {\bibfnamefont
	  {C.-Y.}\ \bibnamefont {Lu}},\ and\ \bibinfo {author} {\bibfnamefont {J.-W.}\
	  \bibnamefont {Pan}},\ }\bibfield  {title} {\bibinfo {title} {Quantum
	  computational advantage using photons},\ }\href
	  {https://doi.org/10.1126/science.abe8770} {\bibfield  {journal} {\bibinfo
	  {journal} {Science}\ }\textbf {\bibinfo {volume} {370}},\ \bibinfo {pages}
	  {1460} (\bibinfo {year} {2020})}\BibitemShut {NoStop}%
	\bibitem [{\citenamefont {Scarani}\ \emph
	  {et~al.}(2009{\natexlab{a}})\citenamefont {Scarani}, \citenamefont
	  {Bechmann-Pasquinucci}, \citenamefont {Cerf}, \citenamefont
	  {Du\ifmmode~\check{s}\else \v{s}\fi{}ek}, \citenamefont {L\"utkenhaus},\ and\
	  \citenamefont {Peev}}]{NLRMP}%
	  \BibitemOpen
	  \bibfield  {author} {\bibinfo {author} {\bibfnamefont {V.}~\bibnamefont
	  {Scarani}}, \bibinfo {author} {\bibfnamefont {H.}~\bibnamefont
	  {Bechmann-Pasquinucci}}, \bibinfo {author} {\bibfnamefont {N.~J.}\
	  \bibnamefont {Cerf}}, \bibinfo {author} {\bibfnamefont {M.}~\bibnamefont
	  {Du\ifmmode~\check{s}\else \v{s}\fi{}ek}}, \bibinfo {author} {\bibfnamefont
	  {N.}~\bibnamefont {L\"utkenhaus}},\ and\ \bibinfo {author} {\bibfnamefont
	  {M.}~\bibnamefont {Peev}},\ }\bibfield  {title} {\bibinfo {title} {The
	  security of practical quantum key distribution},\ }\href
	  {https://doi.org/10.1103/RevModPhys.81.1301} {\bibfield  {journal} {\bibinfo
	  {journal} {Rev. Mod. Phys.}\ }\textbf {\bibinfo {volume} {81}},\ \bibinfo
	  {pages} {1301} (\bibinfo {year} {2009}{\natexlab{a}})}\BibitemShut {NoStop}%
	\bibitem [{\citenamefont {Pirandola}\ \emph {et~al.}(2020)\citenamefont
	  {Pirandola}, \citenamefont {Andersen}, \citenamefont {Banchi}, \citenamefont
	  {Berta}, \citenamefont {Bunandar}, \citenamefont {Colbeck}, \citenamefont
	  {Englund}, \citenamefont {Gehring}, \citenamefont {Lupo}, \citenamefont
	  {Ottaviani}, \citenamefont {Pereira}, \citenamefont {Razavi}, \citenamefont
	  {Shaari}, \citenamefont {Tomamichel}, \citenamefont {Usenko}, \citenamefont
	  {Vallone}, \citenamefont {Villoresi},\ and\ \citenamefont
	  {Wallden}}]{PirRMP}%
	  \BibitemOpen
	  \bibfield  {author} {\bibinfo {author} {\bibfnamefont {S.}~\bibnamefont
	  {Pirandola}}, \bibinfo {author} {\bibfnamefont {U.~L.}\ \bibnamefont
	  {Andersen}}, \bibinfo {author} {\bibfnamefont {L.}~\bibnamefont {Banchi}},
	  \bibinfo {author} {\bibfnamefont {M.}~\bibnamefont {Berta}}, \bibinfo
	  {author} {\bibfnamefont {D.}~\bibnamefont {Bunandar}}, \bibinfo {author}
	  {\bibfnamefont {R.}~\bibnamefont {Colbeck}}, \bibinfo {author} {\bibfnamefont
	  {D.}~\bibnamefont {Englund}}, \bibinfo {author} {\bibfnamefont
	  {T.}~\bibnamefont {Gehring}}, \bibinfo {author} {\bibfnamefont
	  {C.}~\bibnamefont {Lupo}}, \bibinfo {author} {\bibfnamefont {C.}~\bibnamefont
	  {Ottaviani}}, \bibinfo {author} {\bibfnamefont {J.~L.}\ \bibnamefont
	  {Pereira}}, \bibinfo {author} {\bibfnamefont {M.}~\bibnamefont {Razavi}},
	  \bibinfo {author} {\bibfnamefont {J.~S.}\ \bibnamefont {Shaari}}, \bibinfo
	  {author} {\bibfnamefont {M.}~\bibnamefont {Tomamichel}}, \bibinfo {author}
	  {\bibfnamefont {V.~C.}\ \bibnamefont {Usenko}}, \bibinfo {author}
	  {\bibfnamefont {G.}~\bibnamefont {Vallone}}, \bibinfo {author} {\bibfnamefont
	  {P.}~\bibnamefont {Villoresi}},\ and\ \bibinfo {author} {\bibfnamefont
	  {P.}~\bibnamefont {Wallden}},\ }\bibfield  {title} {\bibinfo {title}
	  {Advances in quantum cryptography},\ }\href
	  {https://doi.org/10.1364/AOP.361502} {\bibfield  {journal} {\bibinfo
	  {journal} {Advances in Optics and Photonics}\ }\textbf {\bibinfo {volume}
	  {12}},\ \bibinfo {pages} {1012} (\bibinfo {year} {2020})}\BibitemShut
	  {NoStop}%
	\bibitem [{\citenamefont {Bennett}\ and\ \citenamefont
	  {Brassard}(2014)}]{BB84}%
	  \BibitemOpen
	  \bibfield  {author} {\bibinfo {author} {\bibfnamefont {C.~H.}\ \bibnamefont
	  {Bennett}}\ and\ \bibinfo {author} {\bibfnamefont {G.}~\bibnamefont
	  {Brassard}},\ }\bibfield  {title} {\bibinfo {title} {Quantum cryptography:
	  Public key distribution and coin tossing},\ }\href
	  {https://doi.org/https://doi.org/10.1016/j.tcs.2014.05.025} {\bibfield
	  {journal} {\bibinfo  {journal} {Theoretical Computer Science}\ }\textbf
	  {\bibinfo {volume} {560}},\ \bibinfo {pages} {7} (\bibinfo {year} {2014})},\
	  \bibinfo {note} {theoretical Aspects of Quantum Cryptography – celebrating
	  30 years of BB84}\BibitemShut {NoStop}%
	\bibitem [{\citenamefont {Hwang}(2002)}]{Hwang2002}%
	  \BibitemOpen
	  \bibfield  {author} {\bibinfo {author} {\bibfnamefont {W.~Y.}\ \bibnamefont
	  {Hwang}},\ }\bibfield  {title} {\bibinfo {title} {Quantum key distribution
	  with high loss: Toward global secure communication},\ }\href
	  {https://doi.org/10.1103/PhysRevLett.91.057901} {\bibfield  {journal}
	  {\bibinfo  {journal} {Physical Review Letters}\ }\textbf {\bibinfo {volume}
	  {91}},\ \bibinfo {pages} {057901} (\bibinfo {year} {2002})}\BibitemShut
	  {NoStop}%
	\bibitem [{\citenamefont {Lo}\ \emph {et~al.}(2004)\citenamefont {Lo},
	  \citenamefont {Ma},\ and\ \citenamefont {Chen}}]{Lo2004}%
	  \BibitemOpen
	  \bibfield  {author} {\bibinfo {author} {\bibfnamefont {H.-K.}\ \bibnamefont
	  {Lo}}, \bibinfo {author} {\bibfnamefont {X.}~\bibnamefont {Ma}},\ and\
	  \bibinfo {author} {\bibfnamefont {K.}~\bibnamefont {Chen}},\ }\bibfield
	  {title} {\bibinfo {title} {Decoy state quantum key distribution},\ }\href
	  {https://doi.org/10.1103/PhysRevLett.94.230504} {\bibfield  {journal}
	  {\bibinfo  {journal} {Physical Review Letters}\ }\textbf {\bibinfo {volume}
	  {94}},\ \bibinfo {pages} {230504} (\bibinfo {year} {2004})}\BibitemShut
	  {NoStop}%
	\bibitem [{\citenamefont {Lucamarini}\ \emph {et~al.}(2018)\citenamefont
	  {Lucamarini}, \citenamefont {Yuan}, \citenamefont {Dynes},\ and\
	  \citenamefont {Shields}}]{Lucamarini}%
	  \BibitemOpen
	  \bibfield  {author} {\bibinfo {author} {\bibfnamefont {M.}~\bibnamefont
	  {Lucamarini}}, \bibinfo {author} {\bibfnamefont {Z.~L.}\ \bibnamefont
	  {Yuan}}, \bibinfo {author} {\bibfnamefont {J.~F.}\ \bibnamefont {Dynes}},\
	  and\ \bibinfo {author} {\bibfnamefont {A.~J.}\ \bibnamefont {Shields}},\
	  }\bibfield  {title} {\bibinfo {title} {Overcoming the rate--distance limit of
	  quantum key distribution without quantum repeaters},\ }\href
	  {https://doi.org/10.1038/s41586-018-0066-6} {\bibfield  {journal} {\bibinfo
	  {journal} {Nature}\ }\textbf {\bibinfo {volume} {557}},\ \bibinfo {pages}
	  {400} (\bibinfo {year} {2018})}\BibitemShut {NoStop}%
	\bibitem [{\citenamefont {Lo}\ \emph {et~al.}(2012)\citenamefont {Lo},
	  \citenamefont {Curty},\ and\ \citenamefont {Qi}}]{LoCurty}%
	  \BibitemOpen
	  \bibfield  {author} {\bibinfo {author} {\bibfnamefont {H.-K.}\ \bibnamefont
	  {Lo}}, \bibinfo {author} {\bibfnamefont {M.}~\bibnamefont {Curty}},\ and\
	  \bibinfo {author} {\bibfnamefont {B.}~\bibnamefont {Qi}},\ }\bibfield
	  {title} {\bibinfo {title} {Measurement-device-independent quantum key
	  distribution},\ }\href {https://doi.org/10.1103/PhysRevLett.108.130503}
	  {\bibfield  {journal} {\bibinfo  {journal} {Phys. Rev. Lett.}\ }\textbf
	  {\bibinfo {volume} {108}},\ \bibinfo {pages} {130503} (\bibinfo {year}
	  {2012})}\BibitemShut {NoStop}%
	\bibitem [{\citenamefont {Braunstein}\ and\ \citenamefont
	  {Pirandola}(2011)}]{PirBraun}%
	  \BibitemOpen
	  \bibfield  {author} {\bibinfo {author} {\bibfnamefont {S.~L.}\ \bibnamefont
	  {Braunstein}}\ and\ \bibinfo {author} {\bibfnamefont {S.}~\bibnamefont
	  {Pirandola}},\ }\bibfield  {title} {\bibinfo {title} {Side-channel-free
	  quantum key distribution},\ }\href
	  {https://doi.org/10.1103/PhysRevLett.108.130502} {\bibfield  {journal}
	  {\bibinfo  {journal} {Physical Review Letters}\ }\textbf {\bibinfo {volume}
	  {108}},\ \bibinfo {pages} {130502} (\bibinfo {year} {2011})}\BibitemShut
	  {NoStop}%
	\bibitem [{\citenamefont {Wootters}\ and\ \citenamefont
	  {Zurek}(1982)}]{NoCloning}%
	  \BibitemOpen
	  \bibfield  {author} {\bibinfo {author} {\bibfnamefont {W.~K.}\ \bibnamefont
	  {Wootters}}\ and\ \bibinfo {author} {\bibfnamefont {W.~H.}\ \bibnamefont
	  {Zurek}},\ }\bibfield  {title} {\bibinfo {title} {A single quantum cannot be
	  cloned},\ }\href {https://doi.org/10.1038/299802a0} {\bibfield  {journal}
	  {\bibinfo  {journal} {Nature}\ }\textbf {\bibinfo {volume} {299}},\ \bibinfo
	  {pages} {802} (\bibinfo {year} {1982})}\BibitemShut {NoStop}%
	\bibitem [{\citenamefont {Dieks}(1982)}]{Dieks1982}%
	  \BibitemOpen
	  \bibfield  {author} {\bibinfo {author} {\bibfnamefont {D.}~\bibnamefont
	  {Dieks}},\ }\bibfield  {title} {\bibinfo {title} {Communication by epr
	  devices},\ }\href {https://doi.org/10.1016/0375-9601(82)90084-6} {\bibfield
	  {journal} {\bibinfo  {journal} {Physics Letters A}\ }\textbf {\bibinfo
	  {volume} {92}},\ \bibinfo {pages} {271} (\bibinfo {year} {1982})}\BibitemShut
	  {NoStop}%
	\bibitem [{\citenamefont {Briegel}\ \emph {et~al.}(1998)\citenamefont
	  {Briegel}, \citenamefont {D\"ur}, \citenamefont {Cirac},\ and\ \citenamefont
	  {Zoller}}]{BriegelDur}%
	  \BibitemOpen
	  \bibfield  {author} {\bibinfo {author} {\bibfnamefont {H.-J.}\ \bibnamefont
	  {Briegel}}, \bibinfo {author} {\bibfnamefont {W.}~\bibnamefont {D\"ur}},
	  \bibinfo {author} {\bibfnamefont {J.~I.}\ \bibnamefont {Cirac}},\ and\
	  \bibinfo {author} {\bibfnamefont {P.}~\bibnamefont {Zoller}},\ }\bibfield
	  {title} {\bibinfo {title} {Quantum repeaters: The role of imperfect local
	  operations in quantum communication},\ }\href
	  {https://doi.org/10.1103/PhysRevLett.81.5932} {\bibfield  {journal} {\bibinfo
	   {journal} {Phys. Rev. Lett.}\ }\textbf {\bibinfo {volume} {81}},\ \bibinfo
	  {pages} {5932} (\bibinfo {year} {1998})}\BibitemShut {NoStop}%
	\bibitem [{\citenamefont {Dür}\ \emph {et~al.}(1999)\citenamefont {Dür},
	  \citenamefont {Briegel}, \citenamefont {Cirac},\ and\ \citenamefont
	  {Zoller}}]{Dur1999}%
	  \BibitemOpen
	  \bibfield  {author} {\bibinfo {author} {\bibfnamefont {W.}~\bibnamefont
	  {Dür}}, \bibinfo {author} {\bibfnamefont {H.-J.}\ \bibnamefont {Briegel}},
	  \bibinfo {author} {\bibfnamefont {J.~I.}\ \bibnamefont {Cirac}},\ and\
	  \bibinfo {author} {\bibfnamefont {P.}~\bibnamefont {Zoller}},\ }\bibfield
	  {title} {\bibinfo {title} {Quantum repeaters based on entanglement
	  purification},\ }\href {https://doi.org/10.1103/PhysRevA.59.169} {\bibfield
	  {journal} {\bibinfo  {journal} {Physical Review A}\ }\textbf {\bibinfo
	  {volume} {59}},\ \bibinfo {pages} {169} (\bibinfo {year} {1999})}\BibitemShut
	  {NoStop}%
	\bibitem [{\citenamefont {Hartmann}\ \emph {et~al.}(2007)\citenamefont
	  {Hartmann}, \citenamefont {Kraus}, \citenamefont {Briegel},\ and\
	  \citenamefont {Dür}}]{Hartmann2007}%
	  \BibitemOpen
	  \bibfield  {author} {\bibinfo {author} {\bibfnamefont {L.}~\bibnamefont
	  {Hartmann}}, \bibinfo {author} {\bibfnamefont {B.}~\bibnamefont {Kraus}},
	  \bibinfo {author} {\bibfnamefont {H.-J.}\ \bibnamefont {Briegel}},\ and\
	  \bibinfo {author} {\bibfnamefont {W.}~\bibnamefont {Dür}},\ }\bibfield
	  {title} {\bibinfo {title} {Role of memory errors in quantum repeaters},\
	  }\href {https://doi.org/10.1103/PhysRevA.75.032310} {\bibfield  {journal}
	  {\bibinfo  {journal} {Physical Review A}\ }\textbf {\bibinfo {volume} {75}},\
	  \bibinfo {pages} {032310} (\bibinfo {year} {2007})}\BibitemShut {NoStop}%
	\bibitem [{\citenamefont {Duan}\ \emph {et~al.}(2001)\citenamefont {Duan},
	  \citenamefont {Lukin}, \citenamefont {Cirac},\ and\ \citenamefont
	  {Zoller}}]{DLCZ}%
	  \BibitemOpen
	  \bibfield  {author} {\bibinfo {author} {\bibfnamefont {L.~M.}\ \bibnamefont
	  {Duan}}, \bibinfo {author} {\bibfnamefont {M.~D.}\ \bibnamefont {Lukin}},
	  \bibinfo {author} {\bibfnamefont {J.~I.}\ \bibnamefont {Cirac}},\ and\
	  \bibinfo {author} {\bibfnamefont {P.}~\bibnamefont {Zoller}},\ }\bibfield
	  {title} {\bibinfo {title} {Long-distance quantum communication with atomic
	  ensembles and linear optics},\ }\href {https://doi.org/10.1038/35106500}
	  {\bibfield  {journal} {\bibinfo  {journal} {Nature 2001 414:6862}\ }\textbf
	  {\bibinfo {volume} {414}},\ \bibinfo {pages} {413} (\bibinfo {year}
	  {2001})}\BibitemShut {NoStop}%
	\bibitem [{\citenamefont {Sangouard}\ \emph {et~al.}(2011)\citenamefont
	  {Sangouard}, \citenamefont {Simon}, \citenamefont {de~Riedmatten},\ and\
	  \citenamefont {Gisin}}]{Sangouard}%
	  \BibitemOpen
	  \bibfield  {author} {\bibinfo {author} {\bibfnamefont {N.}~\bibnamefont
	  {Sangouard}}, \bibinfo {author} {\bibfnamefont {C.}~\bibnamefont {Simon}},
	  \bibinfo {author} {\bibfnamefont {H.}~\bibnamefont {de~Riedmatten}},\ and\
	  \bibinfo {author} {\bibfnamefont {N.}~\bibnamefont {Gisin}},\ }\bibfield
	  {title} {\bibinfo {title} {Quantum repeaters based on atomic ensembles and
	  linear optics},\ }\href {https://doi.org/10.1103/RevModPhys.83.33} {\bibfield
	   {journal} {\bibinfo  {journal} {Rev. Mod. Phys.}\ }\textbf {\bibinfo
	  {volume} {83}},\ \bibinfo {pages} {33} (\bibinfo {year} {2011})}\BibitemShut
	  {NoStop}%
	\bibitem [{\citenamefont {Childress}\ \emph {et~al.}(2006)\citenamefont
	  {Childress}, \citenamefont {Taylor}, \citenamefont {Sørensen},\ and\
	  \citenamefont {Lukin}}]{ChildressNV}%
	  \BibitemOpen
	  \bibfield  {author} {\bibinfo {author} {\bibfnamefont {L.}~\bibnamefont
	  {Childress}}, \bibinfo {author} {\bibfnamefont {J.~M.}\ \bibnamefont
	  {Taylor}}, \bibinfo {author} {\bibfnamefont {A.~S.}\ \bibnamefont
	  {Sørensen}},\ and\ \bibinfo {author} {\bibfnamefont {M.~D.}\ \bibnamefont
	  {Lukin}},\ }\bibfield  {title} {\bibinfo {title} {Fault-tolerant quantum
	  communication based on solid-state photon emitters},\ }\href
	  {https://doi.org/10.1103/PhysRevLett.96.070504} {\bibfield  {journal}
	  {\bibinfo  {journal} {Physical Review Letters}\ }\textbf {\bibinfo {volume}
	  {96}},\ \bibinfo {pages} {070504} (\bibinfo {year} {2006})}\BibitemShut
	  {NoStop}%
	\bibitem [{\citenamefont {Humphreys}\ \emph {et~al.}(2017)\citenamefont
	  {Humphreys}, \citenamefont {Kalb}, \citenamefont {Morits}, \citenamefont
	  {Schouten}, \citenamefont {Vermeulen}, \citenamefont {Twitchen},
	  \citenamefont {Markham},\ and\ \citenamefont {Hanson}}]{Humphreys2017}%
	  \BibitemOpen
	  \bibfield  {author} {\bibinfo {author} {\bibfnamefont {P.~C.}\ \bibnamefont
	  {Humphreys}}, \bibinfo {author} {\bibfnamefont {N.}~\bibnamefont {Kalb}},
	  \bibinfo {author} {\bibfnamefont {J.~P.~J.}\ \bibnamefont {Morits}}, \bibinfo
	  {author} {\bibfnamefont {R.~N.}\ \bibnamefont {Schouten}}, \bibinfo {author}
	  {\bibfnamefont {R.~F.~L.}\ \bibnamefont {Vermeulen}}, \bibinfo {author}
	  {\bibfnamefont {D.~J.}\ \bibnamefont {Twitchen}}, \bibinfo {author}
	  {\bibfnamefont {M.}~\bibnamefont {Markham}},\ and\ \bibinfo {author}
	  {\bibfnamefont {R.}~\bibnamefont {Hanson}},\ }\bibfield  {title} {\bibinfo
	  {title} {Deterministic delivery of remote entanglement on a quantum
	  network},\ }\href {https://doi.org/10.1038/s41586-018-0200-5} {\bibfield
	  {journal} {\bibinfo  {journal} {Nature 2018 558:7709}\ }\textbf {\bibinfo
	  {volume} {558}},\ \bibinfo {pages} {268} (\bibinfo {year}
	  {2017})}\BibitemShut {NoStop}%
	\bibitem [{\citenamefont {van Loock}\ \emph {et~al.}(2006)\citenamefont {van
	  Loock}, \citenamefont {Ladd}, \citenamefont {Sanaka}, \citenamefont
	  {Yamaguchi}, \citenamefont {Nemoto}, \citenamefont {Munro},\ and\
	  \citenamefont {Yamamoto}}]{HybridPRL}%
	  \BibitemOpen
	  \bibfield  {author} {\bibinfo {author} {\bibfnamefont {P.}~\bibnamefont {van
	  Loock}}, \bibinfo {author} {\bibfnamefont {T.~D.}\ \bibnamefont {Ladd}},
	  \bibinfo {author} {\bibfnamefont {K.}~\bibnamefont {Sanaka}}, \bibinfo
	  {author} {\bibfnamefont {F.}~\bibnamefont {Yamaguchi}}, \bibinfo {author}
	  {\bibfnamefont {K.}~\bibnamefont {Nemoto}}, \bibinfo {author} {\bibfnamefont
	  {W.~J.}\ \bibnamefont {Munro}},\ and\ \bibinfo {author} {\bibfnamefont
	  {Y.}~\bibnamefont {Yamamoto}},\ }\bibfield  {title} {\bibinfo {title} {Hybrid
	  quantum repeater using bright coherent light},\ }\href
	  {https://doi.org/10.1103/PhysRevLett.96.240501} {\bibfield  {journal}
	  {\bibinfo  {journal} {Physical Review Letters}\ }\textbf {\bibinfo {volume}
	  {96}},\ \bibinfo {pages} {240501} (\bibinfo {year} {2006})}\BibitemShut
	  {NoStop}%
	\bibitem [{\citenamefont {Schmidt}\ and\ \citenamefont {van
	  Loock}(2020)}]{tf_repeater}%
	  \BibitemOpen
	  \bibfield  {author} {\bibinfo {author} {\bibfnamefont {F.}~\bibnamefont
	  {Schmidt}}\ and\ \bibinfo {author} {\bibfnamefont {P.}~\bibnamefont {van
	  Loock}},\ }\bibfield  {title} {\bibinfo {title} {Memory-assisted
	  long-distance phase-matching quantum key distribution},\ }\href
	  {https://doi.org/10.1103/PhysRevA.102.042614} {\bibfield  {journal} {\bibinfo
	   {journal} {Phys. Rev. A}\ }\textbf {\bibinfo {volume} {102}},\ \bibinfo
	  {pages} {042614} (\bibinfo {year} {2020})}\BibitemShut {NoStop}%
	\bibitem [{\citenamefont {Muralidharan}\ \emph {et~al.}(2016)\citenamefont
	  {Muralidharan}, \citenamefont {Li}, \citenamefont {Kim}, \citenamefont
	  {Lütkenhaus}, \citenamefont {Lukin},\ and\ \citenamefont
	  {Jiang}}]{JiangRvw}%
	  \BibitemOpen
	  \bibfield  {author} {\bibinfo {author} {\bibfnamefont {S.}~\bibnamefont
	  {Muralidharan}}, \bibinfo {author} {\bibfnamefont {L.}~\bibnamefont {Li}},
	  \bibinfo {author} {\bibfnamefont {J.}~\bibnamefont {Kim}}, \bibinfo {author}
	  {\bibfnamefont {N.}~\bibnamefont {Lütkenhaus}}, \bibinfo {author}
	  {\bibfnamefont {M.~D.}\ \bibnamefont {Lukin}},\ and\ \bibinfo {author}
	  {\bibfnamefont {L.}~\bibnamefont {Jiang}},\ }\bibfield  {title} {\bibinfo
	  {title} {Optimal architectures for long distance quantum communication},\
	  }\href {https://doi.org/10.1038/srep20463} {\bibfield  {journal} {\bibinfo
	  {journal} {Scientific Reports 2016 6:1}\ }\textbf {\bibinfo {volume} {6}},\
	  \bibinfo {pages} {1} (\bibinfo {year} {2016})}\BibitemShut {NoStop}%
	\bibitem [{\citenamefont {Zhihao}(2017)}]{ChinaDaily}%
	  \BibitemOpen
	  \bibfield  {author} {\bibinfo {author} {\bibfnamefont {Z.}~\bibnamefont
	  {Zhihao}},\ }\href
	  {http://www.chinadaily.com.cn/china/2017-09/30/content_32669593.htm}
	  {\bibinfo {title} {Beijing-shanghai quantum link a 'new era'}} (\bibinfo
	  {year} {2017})\BibitemShut {NoStop}%
	\bibitem [{\citenamefont {Yin}\ \emph {et~al.}(2017)\citenamefont {Yin},
	  \citenamefont {Cao},\ and\ \citenamefont {et. al.}}]{Yin2017}%
	  \BibitemOpen
	  \bibfield  {author} {\bibinfo {author} {\bibfnamefont {J.}~\bibnamefont
	  {Yin}}, \bibinfo {author} {\bibfnamefont {Y.}~\bibnamefont {Cao}},\ and\
	  \bibinfo {author} {\bibfnamefont {Y.-H.~L.}\ \bibnamefont {et. al.}},\
	  }\bibfield  {title} {\bibinfo {title} {Satellite-based entanglement
	  distribution over 1200 kilometers},\ }\href
	  {https://doi.org/10.1126/science.aan3211} {\bibfield  {journal} {\bibinfo
	  {journal} {Science}\ }\textbf {\bibinfo {volume} {356}},\ \bibinfo {pages}
	  {1140} (\bibinfo {year} {2017})}\BibitemShut {NoStop}%
	\bibitem [{\citenamefont {Vallone}\ \emph {et~al.}(2015)\citenamefont
	  {Vallone}, \citenamefont {Bacco}, \citenamefont {Dequal}, \citenamefont
	  {Gaiarin}, \citenamefont {Luceri}, \citenamefont {Bianco},\ and\
	  \citenamefont {Villoresi}}]{Vallone2015}%
	  \BibitemOpen
	  \bibfield  {author} {\bibinfo {author} {\bibfnamefont {G.}~\bibnamefont
	  {Vallone}}, \bibinfo {author} {\bibfnamefont {D.}~\bibnamefont {Bacco}},
	  \bibinfo {author} {\bibfnamefont {D.}~\bibnamefont {Dequal}}, \bibinfo
	  {author} {\bibfnamefont {S.}~\bibnamefont {Gaiarin}}, \bibinfo {author}
	  {\bibfnamefont {V.}~\bibnamefont {Luceri}}, \bibinfo {author} {\bibfnamefont
	  {G.}~\bibnamefont {Bianco}},\ and\ \bibinfo {author} {\bibfnamefont
	  {P.}~\bibnamefont {Villoresi}},\ }\bibfield  {title} {\bibinfo {title}
	  {Experimental satellite quantum communications},\ }\href
	  {https://doi.org/10.1103/PhysRevLett.115.040502} {\bibfield  {journal}
	  {\bibinfo  {journal} {Physical Review Letters}\ }\textbf {\bibinfo {volume}
	  {115}},\ \bibinfo {pages} {040502} (\bibinfo {year} {2015})}\BibitemShut
	  {NoStop}%
	\bibitem [{\citenamefont {Wehner}\ \emph {et~al.}(2018)\citenamefont {Wehner},
	  \citenamefont {Elkouss},\ and\ \citenamefont {Hanson}}]{WehnerHanson}%
	  \BibitemOpen
	  \bibfield  {author} {\bibinfo {author} {\bibfnamefont {S.}~\bibnamefont
	  {Wehner}}, \bibinfo {author} {\bibfnamefont {D.}~\bibnamefont {Elkouss}},\
	  and\ \bibinfo {author} {\bibfnamefont {R.}~\bibnamefont {Hanson}},\
	  }\bibfield  {title} {\bibinfo {title} {Quantum internet: A vision for the
	  road ahead},\ }\bibfield  {journal} {\bibinfo  {journal} {Science}\ }\textbf
	  {\bibinfo {volume} {362}},\ \href {https://doi.org/10.1126/science.aam9288}
	  {10.1126/science.aam9288} (\bibinfo {year} {2018})\BibitemShut {NoStop}%
	\bibitem [{\citenamefont {Bhaskar}\ \emph {et~al.}(2020)\citenamefont
	  {Bhaskar}, \citenamefont {Riedinger}, \citenamefont {Machielse},
	  \citenamefont {Levonian}, \citenamefont {Nguyen}, \citenamefont {Knall},
	  \citenamefont {Park}, \citenamefont {Englund}, \citenamefont {Lončar},
	  \citenamefont {Sukachev},\ and\ \citenamefont {Lukin}}]{Lukin}%
	  \BibitemOpen
	  \bibfield  {author} {\bibinfo {author} {\bibfnamefont {M.~K.}\ \bibnamefont
	  {Bhaskar}}, \bibinfo {author} {\bibfnamefont {R.}~\bibnamefont {Riedinger}},
	  \bibinfo {author} {\bibfnamefont {B.}~\bibnamefont {Machielse}}, \bibinfo
	  {author} {\bibfnamefont {D.~S.}\ \bibnamefont {Levonian}}, \bibinfo {author}
	  {\bibfnamefont {C.~T.}\ \bibnamefont {Nguyen}}, \bibinfo {author}
	  {\bibfnamefont {E.~N.}\ \bibnamefont {Knall}}, \bibinfo {author}
	  {\bibfnamefont {H.}~\bibnamefont {Park}}, \bibinfo {author} {\bibfnamefont
	  {D.}~\bibnamefont {Englund}}, \bibinfo {author} {\bibfnamefont
	  {M.}~\bibnamefont {Lončar}}, \bibinfo {author} {\bibfnamefont {D.~D.}\
	  \bibnamefont {Sukachev}},\ and\ \bibinfo {author} {\bibfnamefont {M.~D.}\
	  \bibnamefont {Lukin}},\ }\bibfield  {title} {\bibinfo {title} {Experimental
	  demonstration of memory-enhanced quantum communication},\ }\href
	  {https://doi.org/10.1038/s41586-020-2103-5} {\bibfield  {journal} {\bibinfo
	  {journal} {Nature 2020 580:7801}\ }\textbf {\bibinfo {volume} {580}},\
	  \bibinfo {pages} {60} (\bibinfo {year} {2020})}\BibitemShut {NoStop}%
	\bibitem [{\citenamefont {Langenfeld}\ \emph {et~al.}(2021)\citenamefont
	  {Langenfeld}, \citenamefont {Thomas}, \citenamefont {Morin},\ and\
	  \citenamefont {Rempe}}]{Rempe}%
	  \BibitemOpen
	  \bibfield  {author} {\bibinfo {author} {\bibfnamefont {S.}~\bibnamefont
	  {Langenfeld}}, \bibinfo {author} {\bibfnamefont {P.}~\bibnamefont {Thomas}},
	  \bibinfo {author} {\bibfnamefont {O.}~\bibnamefont {Morin}},\ and\ \bibinfo
	  {author} {\bibfnamefont {G.}~\bibnamefont {Rempe}},\ }\bibfield  {title}
	  {\bibinfo {title} {Quantum repeater node demonstrating unconditionally secure
	  key distribution},\ }\href {https://doi.org/10.1103/PhysRevLett.126.230506}
	  {\bibfield  {journal} {\bibinfo  {journal} {Phys. Rev. Lett.}\ }\textbf
	  {\bibinfo {volume} {126}},\ \bibinfo {pages} {230506} (\bibinfo {year}
	  {2021})}\BibitemShut {NoStop}%
	\bibitem [{\citenamefont {van Loock}\ \emph {et~al.}(2020)\citenamefont {van
	  Loock}, \citenamefont {Alt}, \citenamefont {Becher}, \citenamefont {Benson},
	  \citenamefont {Boche}, \citenamefont {Deppe}, \citenamefont {Eschner},
	  \citenamefont {Höfling}, \citenamefont {Meschede}, \citenamefont {Michler},
	  \citenamefont {Schmidt},\ and\ \citenamefont {Weinfurter}}]{White}%
	  \BibitemOpen
	  \bibfield  {author} {\bibinfo {author} {\bibfnamefont {P.}~\bibnamefont {van
	  Loock}}, \bibinfo {author} {\bibfnamefont {W.}~\bibnamefont {Alt}}, \bibinfo
	  {author} {\bibfnamefont {C.}~\bibnamefont {Becher}}, \bibinfo {author}
	  {\bibfnamefont {O.}~\bibnamefont {Benson}}, \bibinfo {author} {\bibfnamefont
	  {H.}~\bibnamefont {Boche}}, \bibinfo {author} {\bibfnamefont
	  {C.}~\bibnamefont {Deppe}}, \bibinfo {author} {\bibfnamefont
	  {J.}~\bibnamefont {Eschner}}, \bibinfo {author} {\bibfnamefont
	  {S.}~\bibnamefont {Höfling}}, \bibinfo {author} {\bibfnamefont
	  {D.}~\bibnamefont {Meschede}}, \bibinfo {author} {\bibfnamefont
	  {P.}~\bibnamefont {Michler}}, \bibinfo {author} {\bibfnamefont
	  {F.}~\bibnamefont {Schmidt}},\ and\ \bibinfo {author} {\bibfnamefont
	  {H.}~\bibnamefont {Weinfurter}},\ }\bibfield  {title} {\bibinfo {title}
	  {Extending quantum links: Modules for fiber‐ and memory‐based quantum
	  repeaters},\ }\href {https://doi.org/10.1002/qute.201900141} {\bibfield
	  {journal} {\bibinfo  {journal} {Advanced Quantum Technologies}\ }\textbf
	  {\bibinfo {volume} {3}},\ \bibinfo {pages} {1900141} (\bibinfo {year}
	  {2020})}\BibitemShut {NoStop}%
	\bibitem [{\citenamefont {Luong}\ \emph {et~al.}(2016)\citenamefont {Luong},
	  \citenamefont {Jiang}, \citenamefont {Kim},\ and\ \citenamefont
	  {L{\"u}tkenhaus}}]{NL}%
	  \BibitemOpen
	  \bibfield  {author} {\bibinfo {author} {\bibfnamefont {D.}~\bibnamefont
	  {Luong}}, \bibinfo {author} {\bibfnamefont {L.}~\bibnamefont {Jiang}},
	  \bibinfo {author} {\bibfnamefont {J.}~\bibnamefont {Kim}},\ and\ \bibinfo
	  {author} {\bibfnamefont {N.}~\bibnamefont {L{\"u}tkenhaus}},\ }\bibfield
	  {title} {\bibinfo {title} {Overcoming lossy channel bounds using a single
	  quantum repeater node},\ }\href {https://doi.org/10.1007/s00340-016-6373-4}
	  {\bibfield  {journal} {\bibinfo  {journal} {Applied Physics B}\ }\textbf
	  {\bibinfo {volume} {122}},\ \bibinfo {pages} {96} (\bibinfo {year}
	  {2016})}\BibitemShut {NoStop}%
	\bibitem [{\citenamefont {Rozpedek}\ \emph {et~al.}(2017)\citenamefont
	  {Rozpedek}, \citenamefont {Goodenough}, \citenamefont {Ribeiro},
	  \citenamefont {Kalb}, \citenamefont {Caprara~Vivoli}, \citenamefont
	  {Reiserer}, \citenamefont {Hanson}, \citenamefont {Wehner},\ and\
	  \citenamefont {Elkouss}}]{WehPar}%
	  \BibitemOpen
	  \bibfield  {author} {\bibinfo {author} {\bibfnamefont {F.}~\bibnamefont
	  {Rozpedek}}, \bibinfo {author} {\bibfnamefont {K.}~\bibnamefont
	  {Goodenough}}, \bibinfo {author} {\bibfnamefont {J.}~\bibnamefont {Ribeiro}},
	  \bibinfo {author} {\bibfnamefont {N.}~\bibnamefont {Kalb}}, \bibinfo {author}
	  {\bibfnamefont {V.}~\bibnamefont {Caprara~Vivoli}}, \bibinfo {author}
	  {\bibfnamefont {A.}~\bibnamefont {Reiserer}}, \bibinfo {author}
	  {\bibfnamefont {R.}~\bibnamefont {Hanson}}, \bibinfo {author} {\bibfnamefont
	  {S.}~\bibnamefont {Wehner}},\ and\ \bibinfo {author} {\bibfnamefont
	  {D.}~\bibnamefont {Elkouss}},\ }\bibfield  {title} {\bibinfo {title}
	  {Realistic parameter regimes for a single sequential quantum repeater},\
	  }\href {https://doi.org/10.1088/2058-9565/aab31b} {\bibfield  {journal}
	  {\bibinfo  {journal} {Quantum Science and Technology}\ }\textbf {\bibinfo
	  {volume} {3}} (\bibinfo {year} {2017})}\BibitemShut {NoStop}%
	\bibitem [{\citenamefont {Coopmans}\ \emph
	  {et~al.}(2021{\natexlab{a}})\citenamefont {Coopmans}, \citenamefont
	  {Knegjens}, \citenamefont {Dahlberg}, \citenamefont {Maier}, \citenamefont
	  {Nijsten}, \citenamefont {de~Oliveira~Filho}, \citenamefont {Papendrecht},
	  \citenamefont {Rabbie}, \citenamefont {Rozpedek}, \citenamefont {Skrzypczyk},
	  \citenamefont {Wubben}, \citenamefont {de~Jong}, \citenamefont {Podareanu},
	  \citenamefont {Torres-Knoop}, \citenamefont {Elkouss},\ and\ \citenamefont
	  {Wehner}}]{Coopmans}%
	  \BibitemOpen
	  \bibfield  {author} {\bibinfo {author} {\bibfnamefont {T.}~\bibnamefont
	  {Coopmans}}, \bibinfo {author} {\bibfnamefont {R.}~\bibnamefont {Knegjens}},
	  \bibinfo {author} {\bibfnamefont {A.}~\bibnamefont {Dahlberg}}, \bibinfo
	  {author} {\bibfnamefont {D.}~\bibnamefont {Maier}}, \bibinfo {author}
	  {\bibfnamefont {L.}~\bibnamefont {Nijsten}}, \bibinfo {author} {\bibfnamefont
	  {J.}~\bibnamefont {de~Oliveira~Filho}}, \bibinfo {author} {\bibfnamefont
	  {M.}~\bibnamefont {Papendrecht}}, \bibinfo {author} {\bibfnamefont
	  {J.}~\bibnamefont {Rabbie}}, \bibinfo {author} {\bibfnamefont
	  {F.}~\bibnamefont {Rozpedek}}, \bibinfo {author} {\bibfnamefont
	  {M.}~\bibnamefont {Skrzypczyk}}, \bibinfo {author} {\bibfnamefont
	  {L.}~\bibnamefont {Wubben}}, \bibinfo {author} {\bibfnamefont
	  {W.}~\bibnamefont {de~Jong}}, \bibinfo {author} {\bibfnamefont
	  {D.}~\bibnamefont {Podareanu}}, \bibinfo {author} {\bibfnamefont
	  {A.}~\bibnamefont {Torres-Knoop}}, \bibinfo {author} {\bibfnamefont
	  {D.}~\bibnamefont {Elkouss}},\ and\ \bibinfo {author} {\bibfnamefont
	  {S.}~\bibnamefont {Wehner}},\ }\bibfield  {title} {\bibinfo {title}
	  {Netsquid, a network simulator for quantum information using discrete
	  events},\ }\bibfield  {journal} {\bibinfo  {journal} {Communications
	  Physics}\ }\textbf {\bibinfo {volume} {4}},\ \href
	  {https://doi.org/10.1038/s42005-021-00647-8} {10.1038/s42005-021-00647-8}
	  (\bibinfo {year} {2021}{\natexlab{a}})\BibitemShut {NoStop}%
	\bibitem [{\citenamefont {Kuzmin}\ and\ \citenamefont
	  {Vasilyev}(2021)}]{Kuzmin2021}%
	  \BibitemOpen
	  \bibfield  {author} {\bibinfo {author} {\bibfnamefont {V.~V.}\ \bibnamefont
	  {Kuzmin}}\ and\ \bibinfo {author} {\bibfnamefont {D.~V.}\ \bibnamefont
	  {Vasilyev}},\ }\bibfield  {title} {\bibinfo {title} {Diagrammatic technique
	  for simulation of large-scale quantum repeater networks with dissipating
	  quantum memories},\ }\href {https://doi.org/10.1103/PhysRevA.103.032618}
	  {\bibfield  {journal} {\bibinfo  {journal} {Physical Review A}\ }\textbf
	  {\bibinfo {volume} {103}},\ \bibinfo {pages} {032618} (\bibinfo {year}
	  {2021})}\BibitemShut {NoStop}%
	\bibitem [{\citenamefont {Kuzmin}\ \emph {et~al.}(2019)\citenamefont {Kuzmin},
	  \citenamefont {Vasilyev}, \citenamefont {Sangouard}, \citenamefont {Dür},\
	  and\ \citenamefont {Muschik}}]{Kuzmin2019}%
	  \BibitemOpen
	  \bibfield  {author} {\bibinfo {author} {\bibfnamefont {V.~V.}\ \bibnamefont
	  {Kuzmin}}, \bibinfo {author} {\bibfnamefont {D.~V.}\ \bibnamefont
	  {Vasilyev}}, \bibinfo {author} {\bibfnamefont {N.}~\bibnamefont {Sangouard}},
	  \bibinfo {author} {\bibfnamefont {W.}~\bibnamefont {Dür}},\ and\ \bibinfo
	  {author} {\bibfnamefont {C.~A.}\ \bibnamefont {Muschik}},\ }\bibfield
	  {title} {\bibinfo {title} {Scalable repeater architectures for multi-party
	  states},\ }\href {https://doi.org/10.1038/s41534-019-0230-3} {\bibfield
	  {journal} {\bibinfo  {journal} {npj Quantum Information}\ }\textbf {\bibinfo
	  {volume} {5}},\ \bibinfo {pages} {115} (\bibinfo {year} {2019})}\BibitemShut
	  {NoStop}%
	\bibitem [{\citenamefont {Shchukin}\ \emph {et~al.}(2019)\citenamefont
	  {Shchukin}, \citenamefont {Schmidt},\ and\ \citenamefont {van Loock}}]{PvL}%
	  \BibitemOpen
	  \bibfield  {author} {\bibinfo {author} {\bibfnamefont {E.}~\bibnamefont
	  {Shchukin}}, \bibinfo {author} {\bibfnamefont {F.}~\bibnamefont {Schmidt}},\
	  and\ \bibinfo {author} {\bibfnamefont {P.}~\bibnamefont {van Loock}},\
	  }\bibfield  {title} {\bibinfo {title} {Waiting time in quantum repeaters with
	  probabilistic entanglement swapping},\ }\href
	  {https://doi.org/10.1103/PhysRevA.100.032322} {\bibfield  {journal} {\bibinfo
	   {journal} {Phys. Rev. A}\ }\textbf {\bibinfo {volume} {100}},\ \bibinfo
	  {pages} {032322} (\bibinfo {year} {2019})}\BibitemShut {NoStop}%
	\bibitem [{\citenamefont {Shchukin}\ and\ \citenamefont {van
	  Loock}(2021)}]{Shchukin2021}%
	  \BibitemOpen
	  \bibfield  {author} {\bibinfo {author} {\bibfnamefont {E.}~\bibnamefont
	  {Shchukin}}\ and\ \bibinfo {author} {\bibfnamefont {P.}~\bibnamefont {van
	  Loock}},\ }\href {https://arxiv.org/abs/2109.00793} {\bibinfo {title}
	  {{Optimal entanglement swapping in quantum repeaters}}} (\bibinfo {year}
	  {2021}),\ \bibinfo {note} {arXiv:2109.00793}\BibitemShut {NoStop}%
	\bibitem [{\citenamefont {Vinay}\ and\ \citenamefont {Kok}(2019)}]{Vinay2019}%
	  \BibitemOpen
	  \bibfield  {author} {\bibinfo {author} {\bibfnamefont {S.~E.}\ \bibnamefont
	  {Vinay}}\ and\ \bibinfo {author} {\bibfnamefont {P.}~\bibnamefont {Kok}},\
	  }\bibfield  {title} {\bibinfo {title} {Statistical analysis of
	  quantum-entangled-network generation},\ }\href
	  {https://doi.org/10.1103/PhysRevA.99.042313} {\bibfield  {journal} {\bibinfo
	  {journal} {Physical Review A}\ }\textbf {\bibinfo {volume} {99}},\ \bibinfo
	  {pages} {042313} (\bibinfo {year} {2019})}\BibitemShut {NoStop}%
	\bibitem [{\citenamefont {Khatri}\ \emph {et~al.}(2019)\citenamefont {Khatri},
	  \citenamefont {Matyas}, \citenamefont {Siddiqui},\ and\ \citenamefont
	  {Dowling}}]{Khatri2019}%
	  \BibitemOpen
	  \bibfield  {author} {\bibinfo {author} {\bibfnamefont {S.}~\bibnamefont
	  {Khatri}}, \bibinfo {author} {\bibfnamefont {C.~T.}\ \bibnamefont {Matyas}},
	  \bibinfo {author} {\bibfnamefont {A.~U.}\ \bibnamefont {Siddiqui}},\ and\
	  \bibinfo {author} {\bibfnamefont {J.~P.}\ \bibnamefont {Dowling}},\
	  }\bibfield  {title} {\bibinfo {title} {Practical figures of merit and
	  thresholds for entanglement distribution in quantum networks},\ }\href
	  {https://doi.org/10.1103/PhysRevResearch.1.023032} {\bibfield  {journal}
	  {\bibinfo  {journal} {Physical Review Research}\ }\textbf {\bibinfo {volume}
	  {1}},\ \bibinfo {pages} {023032} (\bibinfo {year} {2019})}\BibitemShut
	  {NoStop}%
	\bibitem [{\citenamefont {Sipahigil}\ \emph {et~al.}(2016)\citenamefont
	  {Sipahigil}, \citenamefont {Evans}, \citenamefont {Sukachev}, \citenamefont
	  {Burek}, \citenamefont {Borregaard}, \citenamefont {Bhaskar}, \citenamefont
	  {Nguyen}, \citenamefont {Pacheco}, \citenamefont {Atikian}, \citenamefont
	  {Meuwly}, \citenamefont {Camacho}, \citenamefont {Jelezko}, \citenamefont
	  {Bielejec}, \citenamefont {Park}, \citenamefont {Lončar},\ and\
	  \citenamefont {Lukin}}]{LukinSiV}%
	  \BibitemOpen
	  \bibfield  {author} {\bibinfo {author} {\bibfnamefont {A.}~\bibnamefont
	  {Sipahigil}}, \bibinfo {author} {\bibfnamefont {R.~E.}\ \bibnamefont
	  {Evans}}, \bibinfo {author} {\bibfnamefont {D.~D.}\ \bibnamefont {Sukachev}},
	  \bibinfo {author} {\bibfnamefont {M.~J.}\ \bibnamefont {Burek}}, \bibinfo
	  {author} {\bibfnamefont {J.}~\bibnamefont {Borregaard}}, \bibinfo {author}
	  {\bibfnamefont {M.~K.}\ \bibnamefont {Bhaskar}}, \bibinfo {author}
	  {\bibfnamefont {C.~T.}\ \bibnamefont {Nguyen}}, \bibinfo {author}
	  {\bibfnamefont {J.~L.}\ \bibnamefont {Pacheco}}, \bibinfo {author}
	  {\bibfnamefont {H.~A.}\ \bibnamefont {Atikian}}, \bibinfo {author}
	  {\bibfnamefont {C.}~\bibnamefont {Meuwly}}, \bibinfo {author} {\bibfnamefont
	  {R.~M.}\ \bibnamefont {Camacho}}, \bibinfo {author} {\bibfnamefont
	  {F.}~\bibnamefont {Jelezko}}, \bibinfo {author} {\bibfnamefont
	  {E.}~\bibnamefont {Bielejec}}, \bibinfo {author} {\bibfnamefont
	  {H.}~\bibnamefont {Park}}, \bibinfo {author} {\bibfnamefont {M.}~\bibnamefont
	  {Lončar}},\ and\ \bibinfo {author} {\bibfnamefont {M.~D.}\ \bibnamefont
	  {Lukin}},\ }\bibfield  {title} {\bibinfo {title} {An integrated diamond
	  nanophotonics platform for quantum-optical networks},\ }\href
	  {https://doi.org/10.1126/science.aah6875} {\bibfield  {journal} {\bibinfo
	  {journal} {Science}\ }\textbf {\bibinfo {volume} {354}},\ \bibinfo {pages}
	  {847} (\bibinfo {year} {2016})}\BibitemShut {NoStop}%
	\bibitem [{\citenamefont {Childress}\ and\ \citenamefont
	  {Hanson}(2013)}]{HansonNV}%
	  \BibitemOpen
	  \bibfield  {author} {\bibinfo {author} {\bibfnamefont {L.}~\bibnamefont
	  {Childress}}\ and\ \bibinfo {author} {\bibfnamefont {R.}~\bibnamefont
	  {Hanson}},\ }\bibfield  {title} {\bibinfo {title} {Diamond nv centers for
	  quantum computing and quantum networks},\ }\href
	  {https://doi.org/10.1557/mrs.2013.20} {\bibfield  {journal} {\bibinfo
	  {journal} {MRS Bulletin}\ }\textbf {\bibinfo {volume} {38}},\ \bibinfo
	  {pages} {134} (\bibinfo {year} {2013})}\BibitemShut {NoStop}%
	\bibitem [{\citenamefont {Collins}\ \emph {et~al.}(2007)\citenamefont
	  {Collins}, \citenamefont {Jenkins}, \citenamefont {Kuzmich},\ and\
	  \citenamefont {Kennedy}}]{CollinsPrl}%
	  \BibitemOpen
	  \bibfield  {author} {\bibinfo {author} {\bibfnamefont {O.~A.}\ \bibnamefont
	  {Collins}}, \bibinfo {author} {\bibfnamefont {S.~D.}\ \bibnamefont
	  {Jenkins}}, \bibinfo {author} {\bibfnamefont {A.}~\bibnamefont {Kuzmich}},\
	  and\ \bibinfo {author} {\bibfnamefont {T.~A.~B.}\ \bibnamefont {Kennedy}},\
	  }\bibfield  {title} {\bibinfo {title} {Multiplexed memory-insensitive quantum
	  repeaters},\ }\href {https://doi.org/10.1103/PhysRevLett.98.060502}
	  {\bibfield  {journal} {\bibinfo  {journal} {Phys. Rev. Lett.}\ }\textbf
	  {\bibinfo {volume} {98}},\ \bibinfo {pages} {060502} (\bibinfo {year}
	  {2007})}\BibitemShut {NoStop}%
	\bibitem [{\citenamefont {Santra}\ \emph {et~al.}(2019)\citenamefont {Santra},
	  \citenamefont {Jiang},\ and\ \citenamefont {Malinovsky}}]{Jiang}%
	  \BibitemOpen
	  \bibfield  {author} {\bibinfo {author} {\bibfnamefont {S.}~\bibnamefont
	  {Santra}}, \bibinfo {author} {\bibfnamefont {L.}~\bibnamefont {Jiang}},\ and\
	  \bibinfo {author} {\bibfnamefont {V.~S.}\ \bibnamefont {Malinovsky}},\
	  }\bibfield  {title} {\bibinfo {title} {Quantum repeater architecture with
	  hierarchically optimized memory buffer times},\ }\href
	  {https://doi.org/10.1088/2058-9565/ab0bc2} {\bibfield  {journal} {\bibinfo
	  {journal} {Quantum Science and Technology}\ }\textbf {\bibinfo {volume}
	  {4}},\ \bibinfo {pages} {025010} (\bibinfo {year} {2019})}\BibitemShut
	  {NoStop}%
	\bibitem [{\citenamefont {Coopmans}\ \emph
	  {et~al.}(2021{\natexlab{b}})\citenamefont {Coopmans}, \citenamefont {Brand},\
	  and\ \citenamefont {Elkouss}}]{Elkouss2021}%
	  \BibitemOpen
	  \bibfield  {author} {\bibinfo {author} {\bibfnamefont {T.}~\bibnamefont
	  {Coopmans}}, \bibinfo {author} {\bibfnamefont {S.}~\bibnamefont {Brand}},\
	  and\ \bibinfo {author} {\bibfnamefont {D.}~\bibnamefont {Elkouss}},\
	  }\bibfield  {title} {\bibinfo {title} {Improved analytical bounds on delivery
	  times of long-distance entanglement},\ }\href
	  {https://doi.org/10.1103/PhysRevA.105.012608} {\bibfield  {journal} {\bibinfo
	   {journal} {Physical Review A}\ }\textbf {\bibinfo {volume} {105}},\ \bibinfo
	  {pages} {012608} (\bibinfo {year} {2021}{\natexlab{b}})}\BibitemShut
	  {NoStop}%
	\bibitem [{\citenamefont {Ladd}\ \emph {et~al.}(2006)\citenamefont {Ladd},
	  \citenamefont {van Loock}, \citenamefont {Nemoto}, \citenamefont {Munro},\
	  and\ \citenamefont {Yamamoto}}]{HybridNJP}%
	  \BibitemOpen
	  \bibfield  {author} {\bibinfo {author} {\bibfnamefont {T.~D.}\ \bibnamefont
	  {Ladd}}, \bibinfo {author} {\bibfnamefont {P.}~\bibnamefont {van Loock}},
	  \bibinfo {author} {\bibfnamefont {K.}~\bibnamefont {Nemoto}}, \bibinfo
	  {author} {\bibfnamefont {W.~J.}\ \bibnamefont {Munro}},\ and\ \bibinfo
	  {author} {\bibfnamefont {Y.}~\bibnamefont {Yamamoto}},\ }\bibfield  {title}
	  {\bibinfo {title} {Hybrid quantum repeater based on dispersive cqed
	  interactions between matter qubits and bright coherent light},\ }\href
	  {https://doi.org/10.1088/1367-2630/8/9/184} {\bibfield  {journal} {\bibinfo
	  {journal} {New Journal of Physics}\ }\textbf {\bibinfo {volume} {8}},\
	  \bibinfo {pages} {184} (\bibinfo {year} {2006})}\BibitemShut {NoStop}%
	\bibitem [{\citenamefont {Pirandola}\ \emph {et~al.}(2017)\citenamefont
	  {Pirandola}, \citenamefont {Laurenza}, \citenamefont {Ottaviani},\ and\
	  \citenamefont {Banchi}}]{PLOB}%
	  \BibitemOpen
	  \bibfield  {author} {\bibinfo {author} {\bibfnamefont {S.}~\bibnamefont
	  {Pirandola}}, \bibinfo {author} {\bibfnamefont {R.}~\bibnamefont {Laurenza}},
	  \bibinfo {author} {\bibfnamefont {C.}~\bibnamefont {Ottaviani}},\ and\
	  \bibinfo {author} {\bibfnamefont {L.}~\bibnamefont {Banchi}},\ }\bibfield
	  {title} {\bibinfo {title} {Fundamental limits of repeaterless quantum
	  communications},\ }\bibfield  {journal} {\bibinfo  {journal} {Nature
	  Communications}\ }\href {https://doi.org/{10.1038/ncomms15043}}
	  {{10.1038/ncomms15043}} (\bibinfo {year} {2017})\BibitemShut {NoStop}%
	\bibitem [{\citenamefont {Schuck}\ \emph {et~al.}(2013)\citenamefont {Schuck},
	  \citenamefont {Pernice},\ and\ \citenamefont {Tang}}]{Schuck2013}%
	  \BibitemOpen
	  \bibfield  {author} {\bibinfo {author} {\bibfnamefont {C.}~\bibnamefont
	  {Schuck}}, \bibinfo {author} {\bibfnamefont {W.~H.~P.}\ \bibnamefont
	  {Pernice}},\ and\ \bibinfo {author} {\bibfnamefont {H.~X.}\ \bibnamefont
	  {Tang}},\ }\bibfield  {title} {\bibinfo {title} {Waveguide integrated low
	  noise nbtin nanowire single-photon detectors with milli-hz dark count rate},\
	  }\href {https://doi.org/10.1038/srep01893} {\bibfield  {journal} {\bibinfo
	  {journal} {Scientific Reports}\ }\textbf {\bibinfo {volume} {3}},\ \bibinfo
	  {pages} {1893} (\bibinfo {year} {2013})}\BibitemShut {NoStop}%
	\bibitem [{\citenamefont {Bru\ss{}}(1998)}]{sixstate}%
	  \BibitemOpen
	  \bibfield  {author} {\bibinfo {author} {\bibfnamefont {D.}~\bibnamefont
	  {Bru\ss{}}},\ }\bibfield  {title} {\bibinfo {title} {Optimal eavesdropping in
	  quantum cryptography with six states},\ }\href
	  {https://doi.org/10.1103/PhysRevLett.81.3018} {\bibfield  {journal} {\bibinfo
	   {journal} {Phys. Rev. Lett.}\ }\textbf {\bibinfo {volume} {81}},\ \bibinfo
	  {pages} {3018} (\bibinfo {year} {1998})}\BibitemShut {NoStop}%
	\bibitem [{\citenamefont {Scarani}\ \emph
	  {et~al.}(2009{\natexlab{b}})\citenamefont {Scarani}, \citenamefont
	  {Bechmann-Pasquinucci}, \citenamefont {Cerf}, \citenamefont
	  {Du\ifmmode~\check{s}\else \v{s}\fi{}ek}, \citenamefont {L\"utkenhaus},\ and\
	  \citenamefont {Peev}}]{RevModPhys.81.1301}%
	  \BibitemOpen
	  \bibfield  {author} {\bibinfo {author} {\bibfnamefont {V.}~\bibnamefont
	  {Scarani}}, \bibinfo {author} {\bibfnamefont {H.}~\bibnamefont
	  {Bechmann-Pasquinucci}}, \bibinfo {author} {\bibfnamefont {N.~J.}\
	  \bibnamefont {Cerf}}, \bibinfo {author} {\bibfnamefont {M.}~\bibnamefont
	  {Du\ifmmode~\check{s}\else \v{s}\fi{}ek}}, \bibinfo {author} {\bibfnamefont
	  {N.}~\bibnamefont {L\"utkenhaus}},\ and\ \bibinfo {author} {\bibfnamefont
	  {M.}~\bibnamefont {Peev}},\ }\bibfield  {title} {\bibinfo {title} {The
	  security of practical quantum key distribution},\ }\href
	  {https://doi.org/10.1103/RevModPhys.81.1301} {\bibfield  {journal} {\bibinfo
	  {journal} {Rev. Mod. Phys.}\ }\textbf {\bibinfo {volume} {81}},\ \bibinfo
	  {pages} {1301} (\bibinfo {year} {2009}{\natexlab{b}})}\BibitemShut {NoStop}%
	\bibitem [{\citenamefont {Rozpedek}\ \emph {et~al.}(2019)\citenamefont
	  {Rozpedek}, \citenamefont {Yehia}, \citenamefont {Goodenough}, \citenamefont
	  {Ruf}, \citenamefont {Humphreys}, \citenamefont {Hanson}, \citenamefont
	  {Wehner},\ and\ \citenamefont {Elkouss}}]{WehnerNV}%
	  \BibitemOpen
	  \bibfield  {author} {\bibinfo {author} {\bibfnamefont {F.}~\bibnamefont
	  {Rozpedek}}, \bibinfo {author} {\bibfnamefont {R.}~\bibnamefont {Yehia}},
	  \bibinfo {author} {\bibfnamefont {K.}~\bibnamefont {Goodenough}}, \bibinfo
	  {author} {\bibfnamefont {M.}~\bibnamefont {Ruf}}, \bibinfo {author}
	  {\bibfnamefont {P.~C.}\ \bibnamefont {Humphreys}}, \bibinfo {author}
	  {\bibfnamefont {R.}~\bibnamefont {Hanson}}, \bibinfo {author} {\bibfnamefont
	  {S.}~\bibnamefont {Wehner}},\ and\ \bibinfo {author} {\bibfnamefont
	  {D.}~\bibnamefont {Elkouss}},\ }\bibfield  {title} {\bibinfo {title}
	  {Near-term quantum-repeater experiments with nitrogen-vacancy centers:
	  Overcoming the limitations of direct transmission},\ }\href
	  {https://doi.org/10.1103/PhysRevA.99.052330} {\bibfield  {journal} {\bibinfo
	  {journal} {Physical Review A}\ }\textbf {\bibinfo {volume} {99}},\ \bibinfo
	  {pages} {052330} (\bibinfo {year} {2019})}\BibitemShut {NoStop}%
	\bibitem [{\citenamefont {Bernardes}\ \emph {et~al.}(2011)\citenamefont
	  {Bernardes}, \citenamefont {Praxmeyer},\ and\ \citenamefont {van
	  Loock}}]{PhysRevA.83.012323}%
	  \BibitemOpen
	  \bibfield  {author} {\bibinfo {author} {\bibfnamefont {N.~K.}\ \bibnamefont
	  {Bernardes}}, \bibinfo {author} {\bibfnamefont {L.}~\bibnamefont
	  {Praxmeyer}},\ and\ \bibinfo {author} {\bibfnamefont {P.}~\bibnamefont {van
	  Loock}},\ }\bibfield  {title} {\bibinfo {title} {Rate analysis for a hybrid
	  quantum repeater},\ }\href {https://doi.org/10.1103/PhysRevA.83.012323}
	  {\bibfield  {journal} {\bibinfo  {journal} {Phys. Rev. A}\ }\textbf {\bibinfo
	  {volume} {83}},\ \bibinfo {pages} {012323} (\bibinfo {year}
	  {2011})}\BibitemShut {NoStop}%
	\bibitem [{\citenamefont {Pirandola}(2019)}]{PLOB_QR}%
	  \BibitemOpen
	  \bibfield  {author} {\bibinfo {author} {\bibfnamefont {S.}~\bibnamefont
	  {Pirandola}},\ }\bibfield  {title} {\bibinfo {title} {End-to-end capacities
	  of a quantum communication network},\ }\href
	  {https://doi.org/10.1038/s42005-019-0147-3} {\bibfield  {journal} {\bibinfo
	  {journal} {Communications Physics}\ }\textbf {\bibinfo {volume} {2}},\
	  \bibinfo {pages} {51} (\bibinfo {year} {2019})}\BibitemShut {NoStop}%
	\bibitem [{\citenamefont {Gottesman}\ and\ \citenamefont
	  {Lo}(2003)}]{twowayqkd}%
	  \BibitemOpen
	  \bibfield  {author} {\bibinfo {author} {\bibfnamefont {D.}~\bibnamefont
	  {Gottesman}}\ and\ \bibinfo {author} {\bibfnamefont {H.-K.}\ \bibnamefont
	  {Lo}},\ }\bibfield  {title} {\bibinfo {title} {Proof of security of quantum
	  key distribution with two-way classical communications},\ }\href
	  {https://doi.org/10.1109/TIT.2002.807289} {\bibfield  {journal} {\bibinfo
	  {journal} {IEEE Transactions on Information Theory}\ }\textbf {\bibinfo
	  {volume} {49}},\ \bibinfo {pages} {457} (\bibinfo {year} {2003})}\BibitemShut
	  {NoStop}%
	\bibitem [{Note1()}]{Note1}%
	  \BibitemOpen
	  \bibinfo {note} {For $a<1$, regimes exist where in terms of the raw rates
	  ``doubling'' performs strictly worse than ``swap as soon as possible'' \cite
	  {Shchukin2021}, similar to regimes here for the full secret key rates with
	  $a=1$ when the dephasing becomes dominant.}\BibitemShut {Stop}%
	\bibitem [{\citenamefont {Laurenza}\ \emph {et~al.}(2021)\citenamefont
	  {Laurenza}, \citenamefont {Walk}, \citenamefont {Eisert},\ and\ \citenamefont
	  {Pirandola}}]{PirEisert}%
	  \BibitemOpen
	  \bibfield  {author} {\bibinfo {author} {\bibfnamefont {R.}~\bibnamefont
	  {Laurenza}}, \bibinfo {author} {\bibfnamefont {N.}~\bibnamefont {Walk}},
	  \bibinfo {author} {\bibfnamefont {J.}~\bibnamefont {Eisert}},\ and\ \bibinfo
	  {author} {\bibfnamefont {S.}~\bibnamefont {Pirandola}},\ }\href
	  {https://arxiv.org/abs/2110.10168v2} {\bibinfo {title} {Rate limits in
	  quantum networks with lossy repeaters}} (\bibinfo {year} {2021}),\ \bibinfo
	  {note} {arXiv:2110.10168}\BibitemShut {NoStop}%
	\bibitem [{\citenamefont {Eisenberg}(2008)}]{Eisenberg}%
	  \BibitemOpen
	  \bibfield  {author} {\bibinfo {author} {\bibfnamefont {B.}~\bibnamefont
	  {Eisenberg}},\ }\bibfield  {title} {\bibinfo {title} {On the expectation of
	  the maximum of iid geometric random variables},\ }\href
	  {https://doi.org/https://doi.org/10.1016/j.spl.2007.05.011} {\bibfield
	  {journal} {\bibinfo  {journal} {Statistics \& Probability Letters}\ }\textbf
	  {\bibinfo {volume} {78}},\ \bibinfo {pages} {135} (\bibinfo {year}
	  {2008})}\BibitemShut {NoStop}%
	\bibitem [{\citenamefont {Azuma}\ \emph {et~al.}(2015)\citenamefont {Azuma},
	  \citenamefont {Tamaki},\ and\ \citenamefont {Munro}}]{Azuma2015}%
	  \BibitemOpen
	  \bibfield  {author} {\bibinfo {author} {\bibfnamefont {K.}~\bibnamefont
	  {Azuma}}, \bibinfo {author} {\bibfnamefont {K.}~\bibnamefont {Tamaki}},\ and\
	  \bibinfo {author} {\bibfnamefont {W.~J.}\ \bibnamefont {Munro}},\ }\bibfield
	  {title} {\bibinfo {title} {All-photonic intercity quantum key distribution},\
	  }\href {https://doi.org/10.1038/ncomms10171} {\bibfield  {journal} {\bibinfo
	  {journal} {Nature Communications}\ }\textbf {\bibinfo {volume} {6}},\
	  \bibinfo {pages} {10171} (\bibinfo {year} {2015})}\BibitemShut {NoStop}%
	\bibitem [{\citenamefont {Munro}\ \emph {et~al.}(2010)\citenamefont {Munro},
	  \citenamefont {Harrison}, \citenamefont {Stephens}, \citenamefont {Devitt},\
	  and\ \citenamefont {Nemoto}}]{MunroNatPhot}%
	  \BibitemOpen
	  \bibfield  {author} {\bibinfo {author} {\bibfnamefont {W.~J.}\ \bibnamefont
	  {Munro}}, \bibinfo {author} {\bibfnamefont {K.~A.}\ \bibnamefont {Harrison}},
	  \bibinfo {author} {\bibfnamefont {A.~M.}\ \bibnamefont {Stephens}}, \bibinfo
	  {author} {\bibfnamefont {S.~J.}\ \bibnamefont {Devitt}},\ and\ \bibinfo
	  {author} {\bibfnamefont {K.}~\bibnamefont {Nemoto}},\ }\bibfield  {title}
	  {\bibinfo {title} {From quantum multiplexing to high-performance quantum
	  networking},\ }\href {https://doi.org/10.1038/nphoton.2010.213} {\bibfield
	  {journal} {\bibinfo  {journal} {Nature Photonics}\ }\textbf {\bibinfo
	  {volume} {4}},\ \bibinfo {pages} {792} (\bibinfo {year} {2010})}\BibitemShut
	  {NoStop}%
	\bibitem [{\citenamefont {Tr\'enyi}\ and\ \citenamefont
	  {L\"utkenhaus}(2020)}]{LutPRA}%
	  \BibitemOpen
	  \bibfield  {author} {\bibinfo {author} {\bibfnamefont {R.}~\bibnamefont
	  {Tr\'enyi}}\ and\ \bibinfo {author} {\bibfnamefont {N.}~\bibnamefont
	  {L\"utkenhaus}},\ }\bibfield  {title} {\bibinfo {title} {Beating direct
	  transmission bounds for quantum key distribution with a multiple quantum
	  memory station},\ }\href {https://doi.org/10.1103/PhysRevA.101.012325}
	  {\bibfield  {journal} {\bibinfo  {journal} {Phys. Rev. A}\ }\textbf {\bibinfo
	  {volume} {101}},\ \bibinfo {pages} {012325} (\bibinfo {year}
	  {2020})}\BibitemShut {NoStop}%
	\bibitem [{\citenamefont {Razavi}\ \emph
	  {et~al.}(2009{\natexlab{a}})\citenamefont {Razavi}, \citenamefont {Piani},\
	  and\ \citenamefont {L\"utkenhaus}}]{RazLut}%
	  \BibitemOpen
	  \bibfield  {author} {\bibinfo {author} {\bibfnamefont {M.}~\bibnamefont
	  {Razavi}}, \bibinfo {author} {\bibfnamefont {M.}~\bibnamefont {Piani}},\ and\
	  \bibinfo {author} {\bibfnamefont {N.}~\bibnamefont {L\"utkenhaus}},\
	  }\bibfield  {title} {\bibinfo {title} {Quantum repeaters with imperfect
	  memories: Cost and scalability},\ }\href
	  {https://doi.org/10.1103/PhysRevA.80.032301} {\bibfield  {journal} {\bibinfo
	  {journal} {Phys. Rev. A}\ }\textbf {\bibinfo {volume} {80}},\ \bibinfo
	  {pages} {032301} (\bibinfo {year} {2009}{\natexlab{a}})}\BibitemShut
	  {NoStop}%
	\bibitem [{\citenamefont {Jones}\ \emph {et~al.}(2016)\citenamefont {Jones},
	  \citenamefont {Kim}, \citenamefont {Rakher}, \citenamefont {Kwiat},\ and\
	  \citenamefont {Ladd}}]{Cody_Jones}%
	  \BibitemOpen
	  \bibfield  {author} {\bibinfo {author} {\bibfnamefont {C.}~\bibnamefont
	  {Jones}}, \bibinfo {author} {\bibfnamefont {D.}~\bibnamefont {Kim}}, \bibinfo
	  {author} {\bibfnamefont {M.~T.}\ \bibnamefont {Rakher}}, \bibinfo {author}
	  {\bibfnamefont {P.~G.}\ \bibnamefont {Kwiat}},\ and\ \bibinfo {author}
	  {\bibfnamefont {T.~D.}\ \bibnamefont {Ladd}},\ }\bibfield  {title} {\bibinfo
	  {title} {Design and analysis of communication protocols for quantum repeater
	  networks},\ }\href {https://doi.org/10.1088/1367-2630/18/8/083015} {\bibfield
	   {journal} {\bibinfo  {journal} {New Journal of Physics}\ }\textbf {\bibinfo
	  {volume} {18}},\ \bibinfo {pages} {083015} (\bibinfo {year}
	  {2016})}\BibitemShut {NoStop}%
	\bibitem [{\citenamefont {Razavi}\ \emph
	  {et~al.}(2009{\natexlab{b}})\citenamefont {Razavi}, \citenamefont {Thompson},
	  \citenamefont {Farmanbar}, \citenamefont {Piani},\ and\ \citenamefont
	  {Lütkenhaus}}]{RazaviProc}%
	  \BibitemOpen
	  \bibfield  {author} {\bibinfo {author} {\bibfnamefont {M.}~\bibnamefont
	  {Razavi}}, \bibinfo {author} {\bibfnamefont {K.}~\bibnamefont {Thompson}},
	  \bibinfo {author} {\bibfnamefont {H.}~\bibnamefont {Farmanbar}}, \bibinfo
	  {author} {\bibfnamefont {M.}~\bibnamefont {Piani}},\ and\ \bibinfo {author}
	  {\bibfnamefont {N.}~\bibnamefont {Lütkenhaus}},\ }\bibfield  {title}
	  {\bibinfo {title} {Physical and architectural considerations in quantum
	  repeaters}\ }(\bibinfo  {publisher} {SPIE},\ \bibinfo {year} {2009})\ p.\
	  \bibinfo {pages} {723603}\BibitemShut {NoStop}%
	\bibitem [{Note2()}]{Note2}%
	  \BibitemOpen
	  \bibinfo {note} {When considering $p_{\protect \mathrm {link}}<1$ one can
	  incorporate this as an additional length of $-\ln (p_{\protect \mathrm
	  {link}})L_{\protect \mathrm {att}}$ regarding $L_0$.}\BibitemShut {Stop}%
	\bibitem [{\citenamefont {Takeoka}\ \emph {et~al.}(2015)\citenamefont
	  {Takeoka}, \citenamefont {Guha},\ and\ \citenamefont {Wilde}}]{TGW}%
	  \BibitemOpen
	  \bibfield  {author} {\bibinfo {author} {\bibfnamefont {M.}~\bibnamefont
	  {Takeoka}}, \bibinfo {author} {\bibfnamefont {S.}~\bibnamefont {Guha}},\ and\
	  \bibinfo {author} {\bibfnamefont {M.~M.}\ \bibnamefont {Wilde}},\ }\bibfield
	  {title} {\bibinfo {title} {Fundamental rate-loss tradeoff for optical quantum
	  key distribution},\ }\href {https://doi.org/10.1038/ncomms6235} {\bibfield
	  {journal} {\bibinfo  {journal} {Nature Communications 2014 5:1}\ }\textbf
	  {\bibinfo {volume} {5}},\ \bibinfo {pages} {1} (\bibinfo {year}
	  {2015})}\BibitemShut {NoStop}%
	\bibitem [{\citenamefont {Cabrillo}\ \emph {et~al.}(1999)\citenamefont
	  {Cabrillo}, \citenamefont {Cirac}, \citenamefont {Garc\'{\i}a-Fern\'andez},\
	  and\ \citenamefont {Zoller}}]{cabrillo}%
	  \BibitemOpen
	  \bibfield  {author} {\bibinfo {author} {\bibfnamefont {C.}~\bibnamefont
	  {Cabrillo}}, \bibinfo {author} {\bibfnamefont {J.~I.}\ \bibnamefont {Cirac}},
	  \bibinfo {author} {\bibfnamefont {P.}~\bibnamefont
	  {Garc\'{\i}a-Fern\'andez}},\ and\ \bibinfo {author} {\bibfnamefont
	  {P.}~\bibnamefont {Zoller}},\ }\bibfield  {title} {\bibinfo {title} {Creation
	  of entangled states of distant atoms by interference},\ }\href
	  {https://doi.org/10.1103/PhysRevA.59.1025} {\bibfield  {journal} {\bibinfo
	  {journal} {Phys. Rev. A}\ }\textbf {\bibinfo {volume} {59}},\ \bibinfo
	  {pages} {1025} (\bibinfo {year} {1999})}\BibitemShut {NoStop}%
	\bibitem [{\citenamefont {Klenke}(2020)}]{Klenke2020}%
	  \BibitemOpen
	  \bibfield  {author} {\bibinfo {author} {\bibfnamefont {A.}~\bibnamefont
	  {Klenke}},\ }\bibinfo {title} {Erzeugendenfunktion},\ in\ \href
	  {https://doi.org/10.1007/978-3-662-62089-2_3} {\emph {\bibinfo {booktitle}
	  {Wahrscheinlichkeitstheorie}}}\ (\bibinfo  {publisher} {Springer Berlin
	  Heidelberg},\ \bibinfo {address} {Berlin, Heidelberg},\ \bibinfo {year}
	  {2020})\ pp.\ \bibinfo {pages} {85--93}\BibitemShut {NoStop}%
	\end{thebibliography}
\end{document}